%% file: ChaosTheory.tex
\documentclass[11pt,english,mathscr,psamsfonts,cmex10]{article}
\usepackage{ae}
\usepackage{aecompl}
\usepackage[T1]{fontenc}
\usepackage[latin1]{inputenc}
\usepackage{geometry}
\usepackage{array}
\usepackage{pifont}
\usepackage{amsmath}
\usepackage{setspace}
\usepackage{amssymb}
\usepackage[authoryear]{natbib}

\makeatletter

\newcommand{\noun}[1]{\textsc{#1}}
\newenvironment{LyXParagraphLeftIndent}[1]%
{
  \begin{list}{}{%
    \setlength\topsep{0pt}%
    \addtolength{\leftmargin}{#1}
    \setlength\parsep{0pt plus 1pt}%
  }
  \item[]
}
{\end{list}}
\providecommand{\tabularnewline}{\\}

\usepackage{amssymb}
\usepackage{geometry}
\allowdisplaybreaks[1]
\usepackage{setspace}
\usepackage{amsfonts}
\usepackage{latexsym}
\usepackage{amsthm}
\usepackage{graphicx}
\usepackage{multirow}
\usepackage{multicol}
\usepackage{color}
\usepackage{subfigure}
\usepackage{index}
\usepackage{fancybox}
\usepackage{oldgerm}
\usepackage{euscript}
\usepackage{exscale}
\usepackage{pifont}
\usepackage{showidx}
\usepackage{array}
\usepackage{afterpage}
\usepackage{bm}
\makeindex

\AtBeginDocument{
  
}

\makeatother
\begin{document}
\noindent \emph{\small Int. J. Bifurcation and Chaos} \textbf{\small 13}{\small ,
3147-3233, (2003). Tutorial and Review paper.}{\small \par}
\vspace{1cm}

\noindent \begin{center}\textbf{\LARGE TOWARD A THEORY OF CHAOS}\end{center}{\LARGE \par}

\noindent \begin{center}A. Sengupta\\Department of Mechanical Engineering\\Indian
Institute of Technology. Kanpur\\E-Mail: osegu@iitk.ac.in\end{center}

\noindent \begin{center}\textbf{\large ABSTRACT}\end{center}{\large \par}

\noindent {\small This paper formulates a new approach to the study of chaos
in discrete dynamical systems based on the notions of inverse ill-posed problems,
set-valued mappings, generalized and multivalued inverses, graphical convergence
of a net of functions in an extended multifunction space \citep{Sengupta2000},
and the topological theory of convergence. Order, chaos, and complexity are
described as distinct components of this unified mathematical structure that
can be viewed as an application of the theory of convergence in topological
spaces to increasingly nonlinear mappings, with the boundary between order and
complexity in the topology of graphical convergence being the region in $\textrm{Multi}(X)$
that is susceptible to chaos. The paper uses results from the discretized spectral
approximation in neutron transport theory \citep{Sengupta1988,Sengupta1995}
and concludes that the numerically exact results obtained by this approximation
of the Case singular eigenfunction solution is due to the graphical convergence
of the Poisson and conjugate Poisson kernels to the Dirac delta and the principal
value multifunctions respectively. In $\textrm{Multi}(X)$, the continuous spectrum
is shown to reduce to a point spectrum, and we introduce a notion of} \emph{\small latent
chaotic states} {\small to interpret superposition over generalized eigenfunctions.
Along with these latent states, spectral theory of nonlinear operators is used
to conclude that nature supports complexity to attain efficiently a multiplicity
of states that otherwise would remain unavailable to it. }{\small \par}

\smallskip{}

\vspace{1.25cm}
\noindent \textbf{\large Prologue }{\large \par}
\medskip{}

\textbf{1.} \textsf{\textsl{\small Generally speaking, the analysis of chaos
is extremely difficult. While a general definition for chaos applicable to most
cases of interest is still lacking, mathematicians agree that for the special
case of iteration of transformations there are three common characteristics
of chaos:}}{\small \par}

\smallskip{}
\textsf{\textsl{\small 1. Sensitive dependence on initial conditions,}}{\small \par}

\textsf{\textsl{\small 2. Mixing,}}{\small \par}

\textsf{\textsl{\small 3. Dense periodic points.}}{\small \par}

\begin{flushright}\citet{Peitgen1992}\end{flushright}
\vspace{-0.25cm}

\textbf{2.} \textsf{\textsl{\small The study of chaos is a part of a larger
program of study of so-called {}``strongly'' nonlinear system. $\cdots$ Linearity
means that the rule that determines what a piece of a system is going to do
next is not influenced by what it is doing now. More precisely this is intended
in a differential or incremental sense: For a linear spring, the increase of
its tension is proportional to the increment whereby it is stretched, with the
ratio of these increments exactly independent of how much it has already been
stretched. Such a spring can be stretched arbitrarily far $\cdots$. Accordingly
no real spring is linear. The mathematics of linear objects is particularly
felicitous. As it happens, linear objects enjoy an identical, simple geometry.
The simplicity of this geometry always allows a relatively easy mental image
to capture the essence of a problem, with the technicality, growing with the
number of parts, basically a detail. The historical prejudice against nonlinear
problems is that no so simple nor universal geometry usually exists.}}{\small \par}

\begin{flushright}Mitchell Feigenbaum's \emph{Foreword} (pp 1-7) in \citet{Peitgen1992}\end{flushright}
\vspace{-0.25cm}

\textbf{3.} \textsf{\textsl{\small The objective of this symposium is to explore
the impact of the emerging science of chaos on various disciplines and the broader
implications for science and society. The characteristic of chaos is its universality
and ubiquity. At this meeting, for example, we have scholars representing mathematics,
physics, biology, geophysics and geophysiology, astronomy, medicine, psychology,
meteorology, engineering, computer science, economics and social sciences}}%
\footnote{\label{Foot: UNConf}A partial listing of papers is as follows: \emph{Chaos
and Politics: Application of nonlinear dynamics to socio-political issues; Chaos
in Society: Reflections on the impact of chaos theory on sociology; Chaos in
neural networks; The impact of chaos on mathematics; The impact of chaos on
physics; The impact of chaos on economic theory; The impact of chaos on engineering;
The impact of chaos on biology; Dynamical disease:} and \emph{The impact of
nonlinear dynamics and chaos on cardiology and medicine. }%
}\textsf{\textsl{\small . Having so many disciplines meeting together, of course,
involves the risk that we might not always speak the same language, even if
all of us have come to talk about {}``chaos''. }}{\small \par}

\begin{flushright}Opening address of Heitor Gurgulino de Souza, Rector United
Nations University, Tokyo \citet{Grebogi1997}\end{flushright}
\vspace{-0.25cm}

\textbf{4.} \textsf{\textsl{\small The predominant approach (of how the different
fields of science relate to one other) is reductionist: Questions in physical
chemistry can be understood in terms of atomic physics, cell biology in terms
of how biomolecules work $\cdots$. We have the best of reasons for taking this
reductionist approach: it works. But shortfalls in reductionism are increasingly
apparent (and) there is something to be gained from supplementing the predominantly
reductionist approach with an integrative agenda. This special section on complex
systems is an initial scan (where) we have taken a {}``complex system'' to
be one whose properties are not fully explained by an understanding of its component
parts. Each Viewpoint author}}%
\footnote{\label{Foot: ScienceMag}The eight Viewpoint articles are titled: \emph{Simple
Lessons from Complexity; Complexity in Chemistry; Complexity in Biological Signaling
Systems; Complexity and the Nervous System; Complexity, Pattern, and Evolutionary
Trade-Offs in Animal Aggregation; Complexity in Natural Landform Patterns; Complexity
and Climate} and \emph{Complexity and the Economy}.%
} \textsf{\textsl{\small was invited to define {}``complex'' as it applied
to his or her discipline. }}{\small \par}

\begin{flushright}\citet{Gallagher1999}\end{flushright}
\vspace{-0.25cm}

\textbf{5.} \textsf{\textsl{\small One of the most striking aspects of physics
is the simplicity of its laws. Maxwell's equations, Schroedinger's equations,
and Hamilton mechanics can each be expressed in a few lines. $\cdots$ Everything
is simple and neat except, of course, the world. Every place we look outside
the physics classroom we see an world of amazing complexity. $\cdots$ So why,
if the laws are so simple, is the world so complicated? To us complexity means
that we have structure with variations. Thus a living organism is complicated
because it has many different working parts, each formed by variations in the
working out of the same genetic coding. Chaos is also found very frequently.
In a chaotic world it is hard to predict which variation will arise in a given
place and time. A complex world is interesting because it is highly structured.
A chaotic world is interesting because we do not know what is coming next. Our
world is both complex and chaotic. Nature can produce complex structures even
in simple situations and obey simple laws even in complex situations.}}{\small \par}

\begin{flushright}\citet{Goldenfeld1999}\end{flushright}
\vspace{-0.25cm}

\textbf{6.} \textsf{\textsl{\small Where chaos begins, classical science stops.
For as long as the world has had physicists inquiring into the laws of nature,
it has suffered a special ignorance about disorder in the atmosphere, in the
turbulent sea, in the fluctuations in the wildlife populations, in the oscillations
of the heart and the brain. But in the 1970s a few scientists began to find
a way through disorder. They were mathematicians, physicists, biologists, chemists
$\cdots$ (and) the insights that emerged led directly into the natural world:
the shapes of clouds, the paths of lightning, the microscopic intertwining of
blood vessels, the galactic clustering of stars. $\cdots$ Chaos breaks across
the lines that separate scientific disciplines, (and) has become a shorthand
name for a fast growing movement that is reshaping the fabric of the scientific
establishment.}}{\small \par}

\begin{flushright}\citet{Gleick1987}\end{flushright}
\vspace{-0.25cm}

\textbf{7.} \textsf{\textsl{\small order $\longrightarrow$ complexity $\longrightarrow$
chaos.}}{\small \par}

\begin{flushright}\citet{Waldrop1992}\end{flushright}
\vspace{-0.25cm}

\textbf{8.} \textsf{\textsl{\small Our conclusions based on these examples seems
simple: At present}} \textsf{\textsl{\emph{\small chaos}}} \textsf{\textsl{\small is
a philosophical term, not a rigorous mathematical term. It may be a subjective
notion illustrating the present day limitations of the human intellect or it
may describe an intrinsic property of nature such as the {}``randomness''
of the sequence of prime numbers. Moreover, chaos may be undecidable in the
sense of Godel in that no matter what definition is given for chaos, there is
some example of chaos which cannot be proven to be chaotic from the definition. }}{\small \par}

\begin{flushright}\citet{Brown1996}\end{flushright}
\vspace{-0.25cm}

\textbf{9.} \textsf{\textsl{\small My personal feeling is that the definition
of a {}``fractal'' should be regarded in the same way as the biologist regards
the definition of {}``life''. There is no hard and fast definition, but just
a list of properties characteristic of a living thing $\cdots$. Most living
things have most of the characteristics on the list, though there are living
objects that are exceptions to each of them. In the same way, it seems best
to regard a fractal as a set that has properties such as those listed below,
rather than to look for a precise definition which will certainly exclude some
interesting cases. }}{\small \par}

\begin{flushright}\citet{Falconer1990}\end{flushright}
\vspace{-0.25cm}

\medskip{}
\noindent \textbf{10.} \textsf{\textsl{\small Dynamical systems are often said
to exhibit chaos without a precise definition of what this means. }}{\small \par}

\noindent \begin{flushright}\citet{Robinson1999}\end{flushright}
\bigskip{}

\noindent \begin{flushleft}\textbf{\large 1. Introduction}\end{flushleft}{\large \par}

\smallskip{}
\noindent The purpose of this paper is to present an unified, self-contained
mathematical structure and physical understanding of the nature of chaos in
a discrete dynamical system and to suggest a plausible explanation of \emph{why}
natural systems tend to be chaotic. The somewhat extensive quotations with which
we begin above, bear testimony to both the increasingly significant --- and
perhaps all-pervasive --- role of nonlinearity in the world today as also our
imperfect state of understanding of its manifestations. The list of papers at
both the UN Conference \citep{Grebogi1997} and in \emph{Science} \citep{Gallagher1999}
is noteworthy if only to justify the observation of \citet{Gleick1987} that
{}``chaos seems to be everywhere''. Even as everybody appears to be finding
chaos and complexity in all likely and unlikely places, and possibly because
of it, it is necessary that we have a clear mathematically-physical understanding
of these notions that are supposedly reshaping our view of nature. This paper
is an attempt to contribute to this goal. To make this account essentially self-contained
we include here, as far as this is practicable, the basics of the background
material needed to understand the paper in the form of \emph{Tutorials} and
an extended \emph{Appendix.} 

The paradigm of chaos of the kneading of the dough is considered to provide
an intuitive basis of the mathematics of chaos \citep{Peitgen1992}, and one
of our fundamental objectives here is to recount the mathematical framework
of this process in terms of the theory of ill-posed problems arising from non-injectivity
\citep{Sengupta1997}, \emph{maximal ill-posedness,} and \emph{graphical convergence}
of functions \citep{Sengupta2000}. A natural mathematical formulation of the
kneading of the dough in the form of \emph{stretch-cut-and-paste} and \emph{stretch-cut-and-fold}
operations is in the ill-posed problem arising from the increasing non-injectivity
of the function $f$ modeling the kneading operation. 

\begin{flushright}\textbf{\textit{Begin Tutorial1: Functions and Multifunctions }}\end{flushright}

\begin{figure}[htbp]
\noindent \begin{center}{\small \input{functions.pstex_t}}\end{center}{\small \par}

\begin{singlespace}

\caption{{\footnotesize \label{Fig: functions}Functional and non-functional relations
between two sets $X$ and $Y$: while $f$ and $g$ are functional relations,
$\mathscr{M}$ is not. In figure (a) $f$ and $g$ are both injective and surjective
(that is they are bijective), in (b) $g$ is bijective but $f$ is only injective
and $f^{-1}(\{ y_{2}\}):=\emptyset$, in (c) $f$ is not $1:1$, $g$ is not
onto, while in (d) $\mathscr{M}$ is not a function but is a} \emph{\footnotesize multifunction.}}\end{singlespace}

\end{figure}
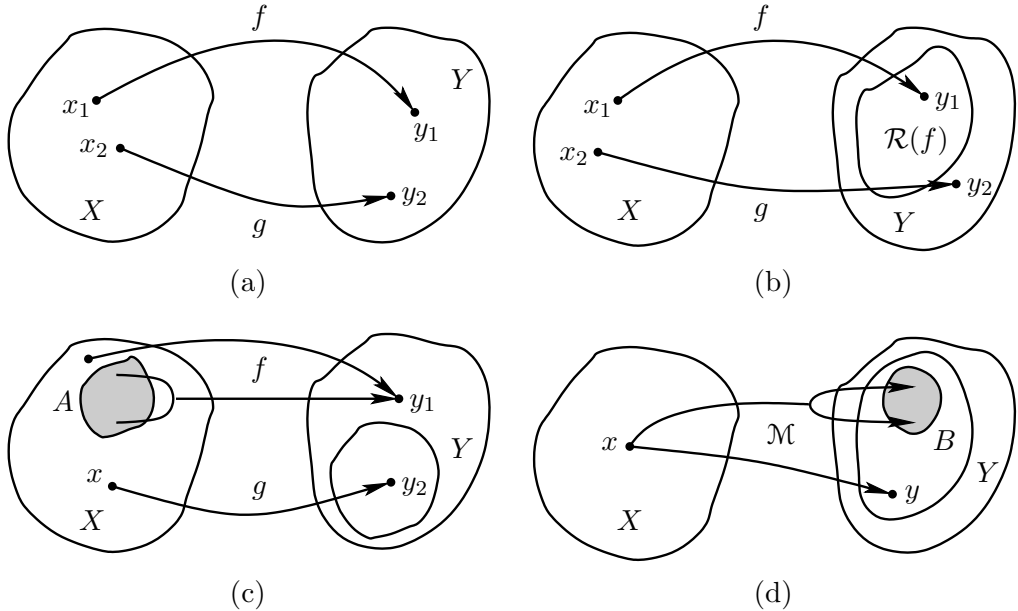

\noindent A \emph{relation,} or \emph{correspondence,} between two sets $X$
and $Y$, written $\mathscr{M}\!:X\qquad Y$, is basically a rule that associates
subsets of $X$ to subsets of $Y$; this is often expressed as $(A,B)\in\mathscr{M}$
where $A\subset X$ and $B\subset Y$ and $(A,B)$ is an ordered pair of sets.
The domain \[
\mathcal{D}(\mathscr{M})\overset{\textrm{def}}=\{ A\subset X\!:(\!\exists Z\in\mathscr{M})(\pi_{X}(Z)=A)\}\]
and range \[
\mathcal{R}(\mathscr{M})\overset{\textrm{def}}=\{ B\subset Y\!:(\!\exists Z\in\mathscr{M})(\pi_{Y}(Z)=B)\}\]

\noindent of $\mathscr{M}$ are respectively the sets of $X$ which under $\mathscr{M}$
corresponds to sets in $Y$; here $\pi_{X}$ and $\pi_{Y}$ are the projections
of $Z$ on $X$ and $Y$ respectively. Equivalently, $\mathcal{D}(\mathcal{M})=\{ x\in X\!:\mathscr{M}(x)\neq\emptyset\}$
and $\mathcal{R}(\mathscr{M})=\bigcup_{x\in\mathcal{D}(\mathcal{M})}\mathscr{M}(x)$.
The \emph{inverse} $\mathscr M^{-}$ of $\mathscr{M}$ is the relation \[
\mathscr M^{-}=\{(B,A)\!:(A,B)\!\in\mathscr{M}\}\]
 so that $\mathscr M^{-}$ assigns $A$ to $B$ iff $\mathscr{M}$ assigns $B$
to $A$. In general, a relation may assign many elements in its range to a single
element from its domain; of especial significance are \emph{functional relations}
$f$%
\footnote{{\small \label{Foot: reln&graph}We do not distinguish between a relation and
its graph although technically they are different objects. Thus although a functional
relation, strictly speaking, is the triple $(X,f,Y)$ written traditionally
as $f\!:X\rightarrow Y$, we use it synonymously with the graph $f$ itself.
Parenthetically, the word} \emph{\small functional} {\small in this work is
not necessarily employed for a scalar-valued function, but is used in a wider
sense to distinguish between a function and an arbitrary relation (that is a
multifunction). Formally, whereas an arbitrary relation from $X$ to $Y$ is
a subset of $X\times Y$, a functional relation must satisfy an additional restriction
that requires $y_{1}=y_{2}$ whenever $(x,y_{1})\in f$ and $(x,y_{2})\in f$.
In this subset notation, $(x,y)\in f\Leftrightarrow y=f(x)$. }%
} that can assigns only a unique element in $\mathcal{R}(f)$ to any element
in $\mathcal{D}(f)$. Fig. \ref{Fig: functions} illustrates the distinction
between arbitrary and functional relations $\mathscr{M}$ and $f$. This difference
between functions (or maps) and multifunctions is basic to our developments
and should be fully understood. Functions can again be classified as injections
(or $1:1$) and surjections (or onto). $f\!:X\rightarrow Y$ is said to be \emph{injective}
(or \emph{one-to-one}) if $x_{1}\neq x_{2}\Rightarrow f(x_{1})\neq f(x_{2})$
for all $x_{1},x_{2}\in X$, while it is \emph{surjective} (or \emph{onto})
if $Y=f(X)$. $f$ is \emph{bijective} if it is both $1:1$ and onto.

{\small }%
{\small \par}

Associated with a function $f\!:X\rightarrow Y$ is its inverse $f^{-1}\!:Y\supseteq\mathcal{R}(f)\rightarrow X$
that exists on $\mathcal{R}(f)$ iff $f$ is injective. Thus when $f$ is bijective,
$f^{-1}(y):=\{ x\in X\!:y=f(x)\}$ exists for every $y\in Y$; infact $f$ is
bijective iff $f^{-1}(\{ y\})$ is a singleton for each $y\in Y$. Non-injective
functions are not at all rare; if anything, they are very common even for linear
maps and it would be perhaps safe to conjecture that they are overwhelmingly
predominant in the nonlinear world of nature. Thus for example, the simple linear
homogeneous differential equation with constant coefficients of order $n>1$
has $n$ linearly independent solutions so that the operator $D^{n}$ of $D^{n}(y)=0$
has a $n$-dimensional null space. Inverses of non-injective, and in general
non-bijective, functions will be denoted by $f^{-}$. If $f$ is not injective
then 

\[
A\subset f^{-}f(A)\overset{\textrm{def}}=\textrm{sat}(A)\]
 where $\textrm{sat}(A)$ is the \emph{saturation of} $A\subseteq X$ \emph{induced
by} $f$; if $f$ is not surjective then \[
{\textstyle ff^{-}(B):=B\bigcap f(X)\subseteq B.}\]
 If $A=\textrm{sat}(A)$, then $A$ is said to be \emph{saturated,} and $B\subseteq\mathcal{R}(f)$
whenever $ff^{-}(B)=B$. Thus for non-injective $f$, $f^{-}f$ is not an identity
on $X$ just as $ff^{-}$ is not \textbf{$\mathbf{1}_{Y}$} if $f$ is not surjective.
However the set of relations \begin{equation}
ff^{-}f=f,\qquad f^{-}ff^{-}=f^{-}\label{Eqn: f_inv_f}\end{equation}
 that is always true will be of basic significance in this work. Following are
some equivalent statements on the injectivity and surjectivity of functions
$f\!:X\rightarrow Y$. 

(Injec) $f$ is $1:1$ $\Leftrightarrow$ there is a function $f_{\textrm{L }}\!:Y\rightarrow X$
called the left inverse of $f$, such that $f_{\textrm{L}}f=\mathbf{1}_{X}$
$\Leftrightarrow$ $A=f^{-}f(A)$ for all subsets $A$ of $X$$\Leftrightarrow$$f(\bigcap A_{i})=\bigcap f(A_{i})$. 

(Surjec) $f$ is onto $\Leftrightarrow$ there is a function $f_{\textrm{R }}\!:Y\rightarrow X$
called the right inverse of $f$, such that $ff_{\textrm{R}}=\mathbf{1}_{Y}$
$\Leftrightarrow$ $B=ff^{-}(B)$ for all subsets $B$ of $Y$. 

As we are primarily concerned with non-injectivity of functions, saturated sets
generated by equivalence classes of $f$ will play a significant role in our
discussions. A relation $\mathscr{E}\mathcal{E}$ on a set $X$ is said to be
an \emph{equivalence relation} if it is%
\footnote{{\small \label{Foot: EquivRel}An useful alternate way of expressing these properties
for a relation $\mathscr{M}$ on $X$ are}{\small \par}

\smallskip{}
{\small $\quad$(ER1) $\mathscr{M}$ is reflexive iff $\mathbf{1}_{X}\subseteq\mathscr{M}$}{\small \par}

{\small $\quad$(ER2) $\mathscr{M}$ is symmetric iff $\mathscr{M}=\mathscr M^{-}$ }{\small \par}

{\small $\quad$(ER3) $\mathscr{M}$ is transitive iff $\mathscr{M}\circ\mathscr{M}\subseteq\mathscr{M}$, }{\small \par}
\smallskip{}

\noindent {\small with $\mathscr{M}$ an equivalence relation only if $\mathscr{M}\circ\mathscr{M}=\mathscr{M}$,
where for $\mathscr{M}\subseteq X\times Y$ and $\mathscr{N}\subseteq Y\times Z$,
the composition $\mathscr{N}\circ\mathscr{M}:=\{(x,z)\in X\times Z\!:(\exists y\in Y)\textrm{ }((x,y)\in\mathscr{M})\wedge((y,z)\in\mathscr{N})\}$}%
} 

\smallskip{}
(ER1) Reflexive: $(\forall x\in X)(x\mathcal{E}x)$. 

(ER2) Symmetric: $(\forall x,y\in X)(x\mathcal{E}y\Longrightarrow y\mathcal{E}x)$.

(ER3) Transitive: $(\forall x,y,z\in X)(x\mathcal{E}y\wedge y\mathcal{E}z\Longrightarrow x\mathcal{E}z)$. 
\smallskip{}

\noindent Equivalence relations group together unequal elements $x_{1}\neq x_{2}$
of a set as equivalent according to the requirements of the relation. This is
expressed as $x_{1}\sim x_{2}\textrm{ }(\textrm{mod }\mathcal{E})$ and will
be represented here by the shorthand notation $x_{1}\sim_{\mathcal{E}}x_{2}$,
or even simply as $x_{1}\sim x_{2}$ if the specification of $\mathcal{E}$
is not essential. Thus for a noninjective map if $f(x_{1})=f(x_{2})$ for $x_{1}\neq x_{2}$,
then $x_{1}$ and $x_{2}$ can be considered to be equivalent to each other
since they map onto the same point under $f$; thus $x_{1}\sim_{f}x_{2}\Leftrightarrow f(x_{1})=f(x_{2})$
defines the equivalence relation $\sim_{f}$ induced by the map $f$. Given
an equivalence relation $\sim$ on a set $X$ and an element $x\in X$ the subset
\[
[x]\overset{\textrm{def}}=\{ y\in X\!:y\sim x\}\]
 is called the \emph{equivalence class of $x$;} thus $x\sim y\Leftrightarrow[x]=[y]$\emph{.}
In particular, equivalence classes generated by $f\!:X\rightarrow Y$, $[x]_{f}=\{ x_{\alpha}\in X\!:f(x_{\alpha})=f(x)\}$,
will be a cornerstone of our analysis of chaos generated by the iterates of
non-injective maps, and the equivalence relation $\sim_{f}:=\{(x,y)\!:f(x)=f(y)\}$
generated by $f$ is uniquely defined by the partition that $f$ induces on
$X$. Of course as $x\sim x$, $x\in[x]$. It is a simple matter to see that
any two equivalence classes are either disjoint or equal so that the equivalence
classes generated by an equivalence relation on $X$ form a disjoint cover of
$X.$ The \emph{quotient set of $X$ under $\sim$,} denoted by $X/\sim\;:=\{[x]\!:x\in X\}$,
has the equivalence classes $[x]$ as its elements; thus $[x]$ plays a dual
role either as subsets of $X$ or as elements of $X/\sim$. The rule $x\mapsto[x]$
defines a surjective function $Q\!:X\rightarrow X/\sim$ known as the \emph{quotient
map. }

\medskip{}
\noindent \textbf{Example 1.1.} Let \[
S^{1}=\{(x,y)\in\mathbb{R}^{2})\!:x^{2}+y^{2}=1\}\]

\noindent be the unit circle in $\mathbb{R}^{2}$. Consider $X=[0,1]$ as a
subspace of $\mathbb{R}$, define a map \[
q\!:X\rightarrow S^{1},\qquad s\longmapsto(\cos2\pi s,\sin2\pi s),\,\, s\in X,\]

\noindent from $\mathbb{R}$ to $\mathbb{R}^{2}$, and let $\sim$ be the equivalence
relation on $X$ \[
s\sim t\Longleftrightarrow(s=t)\vee(s=0,t=1)\vee(s=1,t=0).\]
 If we bend $X$ around till its ends touch, the resulting circle represents
the quotient set $Y=X/\sim$ whose points are equivalent under $\sim$ as follows
\[
[0]=\{0,1\}=[1],\qquad[s]=\{ s\}\,\textrm{for all }s\in(0,1).\]

{\small }%
\begin{figure}[htbp]
\noindent \begin{center}{\small \input{quotient.pstex_t}}\end{center}{\small \par}

\begin{singlespace}

\caption{{\footnotesize \label{Fig: quotient}The quotient map $Q$}}\end{singlespace}

\end{figure}
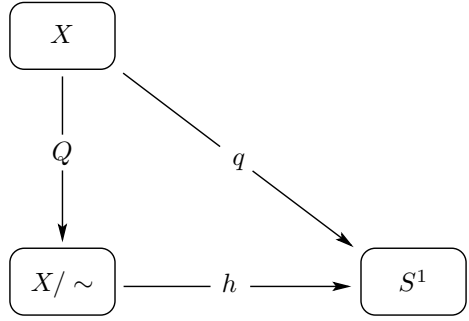
 Thus $q$ is bijective for $s\in(0,1)$ but two-to-one for the special values
$s=0\textrm{ and }1$, so that for $s,t\in X$,\[
s\sim t\Longleftrightarrow q(s)=q(t).\]
This yields a bijection $h\!:X/\sim\:\rightarrow S^{1}$ such that \[
q=h\circ Q\]

\noindent defines the quotient map $Q\!:X\rightarrow X/\sim$ by $h([s])=q(s)$
for all $s\in[0,1]$. The situation is illustrated by the commutative diagram
of Fig. \ref{Fig: quotient} that appears as an integral component in a different
and more general context in Sec. 2. It is to be noted that commutativity of
the diagram implies that if a given equivalence relation $\sim$ on $X$ is
completely determined by $q$ that associates the partitioning equivalence classes
in $X$ to unique points in $S^{1}$, then $\sim$ is identical to the equivalence
relation that is induced by $Q$ on $X$. Note that a larger size of the equivalence
classes can be obtained by considering $X=\mathbb{R}_{+}$ for which $s\sim t\Leftrightarrow|s-t|\in\mathbb{Z}_{+}$.$\qquad\blacksquare$

\noindent \begin{flushright}\textbf{\textit{End Tutorial1}}\end{flushright}
\medskip{}

\noindent One of the central concepts that we consider and employ in this work
is the inverse $f^{-}$ of a nonlinear, non-injective, function $f$; here the
equivalence classes $[x]_{f}=f^{-}f(x)$ of $x\in X$ are the saturated subsets
of $X$ that partition $X$. While a detailed treatment of this question in
the form of the non-linear ill-posed problem and its solution is given in Sec.
2 \cite{Sengupta1997}, it is sufficient to point out here from Figs. \ref{Fig: functions}(c)
and \ref{Fig: functions}(d), that the inverse of a noninjective function is
not a function but a multifunction while the inverse of a multifunction is a
noninjective function. Hence one has the general result that\begin{eqnarray}
f\textrm{ is a non injective function} & \Longleftrightarrow & f^{-}\textrm{ is a multifunction}.\label{Eqn: func-multi}\\
f\textrm{ is a multifunction} & \Longleftrightarrow & f^{-}\textrm{ is a non injective function}\nonumber \end{eqnarray}

\noindent The inverse of a multifunction $\mathscr{M}\!:X\qquad Y$ is a generalization
of the corresponding notion for a function $f\!:X\rightarrow Y$ such that \[
\mathscr M^{-}(y)\overset{\textrm{def}}=\{ x\in X\!:y\in\mathscr{M}(x)\}\]

\noindent leads to \[
{\textstyle \mathscr M^{-}(B)=\{ x\in X\!:\mathscr{M}(x)\bigcap B\neq\emptyset\}}\]

\noindent for any $B\subseteq Y$, while a more restricted inverse that we shall
not be concerned with is given as $\mathscr M^{+}(B)=\{ x\in X\!:\mathscr{M}(x)\subseteq B\}$.
Obviously, $\mathscr M^{+}(B)\subseteq\mathscr M^{-}(B)$. A multifunction is
injective if $x_{1}\neq x_{2}\Rightarrow\mathscr{M}(x_{1})\bigcap\mathscr{M}(x_{2})=\emptyset$,
and in common with functions it is true that \begin{align*}
\mathscr{M}\left(\bigcup_{\alpha\in\mathbb{{D}}}A_{\alpha}\right)= & \bigcup_{\alpha\in\mathbb{{D}}}\mathscr{M}(A_{\alpha})\\
\mathscr{M}\left(\bigcap_{\alpha\in\mathbb{{D}}}A_{\alpha}\right)\subseteq & \bigcap_{\alpha\in\mathbb{{D}}}\mathscr{M}(A_{\alpha})\end{align*}

\noindent and where $\mathbb{D}$ is an index set. The following illustrates
the difference between the two inverses of $\mathscr{M}$. Let $X$ be a set
that is partitioned into two disjoint $\mathscr{M}$-invariant subsets $X_{1}$
and $X_{2}$. If $x\in X_{1}$ (or $x\in X_{2}$) then $\mathscr{M}(x)$ represents
that part of $X_{1}$ (or of $X_{2}$ ) that is realized immediately after one
application of $\mathscr{M}$, while $\mathscr M^{-}(x)$ denotes the possible
precursors of $x$ in $X_{1}$ (or of $X_{2}$) and $\mathscr M^{+}(B)$ is
that subset of $X$ whose image lies in $B$ for any subset $B\subset X$. 

In this work the multifunctions we are explicitly concerned with arise as the
inverses of non-injective functions.

The second major component of our theory is the \emph{graphical convergence
of a net of functions to a multifunction.} In Tutorial2 below, we replace for
the sake of simplicity and without loss of generality, the net (which is basically
a sequence where the index set is not necessarily the positive integers; thus
every sequence is a net but the family%
\footnote{{\small \label{Foot: family}A function $\chi\!:\mathbb{D}\rightarrow X$ will
be called a} \emph{\small family} {\small in $X$ indexed by $\mathbb{D}$ when
reference to the domain $\mathbb{D}$ is of interest, and a} \emph{\small net}
{\small when it is required to focus attention on its values in $X$.}%
} indexed, for example, by $\mathbb{Z}$, the set of \emph{all} integers, is
a net and not a sequence) with a sequence and provide the necessary background
and motivation for the concept of graphical convergence. 

\medskip{}
\noindent \begin{flushright}\textbf{\textit{Begin Tutorial2: Convergence of
Functions}}\end{flushright}

\noindent This Tutorial reviews the inadequacy of the usual notions of convergence
of functions either to limit functions or to distributions and suggests the
motivation and need for introduction of the notion of graphical convergence
of functions to multifunctions. Here, we follow closely the exposition of \citet*{Korevaar1968},
and use the notation $(f_{k})_{k=1}^{\infty}$ to denote real or complex valued
functions on a bounded or unbounded interval $J$. 
\smallskip{}

A sequence of piecewise continuous functions $(f_{k})_{k=1}^{\infty}$ is said
to converge to the function $f$, notation $f_{k}\rightarrow f$, on a bounded
or unbounded interval $J$%
\footnote{{\small \label{Foot: extension}Observe that it is} \emph{\small not} {\small being
claimed that $f$ belongs to the same class as $(f_{k})$. This is the single
most important cornerstone on which this paper is based: the need to {}``complete''
spaces that are topologically {}``incomplete''. The classical high-school
example of the related problem of having to enlarge, or extend, spaces that
are not big enough is the solution space of algebraic equations with real coefficients
like $x^{2}+1=0$. }%
} 

(1) \emph{Pointwise} if\[
f_{k}(x)\longrightarrow f(x)\qquad\textrm{for all }x\in J,\]

\noindent that is: Given any arbitrary real number $\varepsilon>0$ there exists
a $K\in\mathbb{N}$ that may depend on $x$, such that $|f_{k}(x)-f(x)|<\varepsilon$
for all $k\geq K$. 

(2) \emph{Uniformly} if \[
\sup_{x\in J}|f(x)-f_{k}(x)|\longrightarrow0\qquad\textrm{as }k\longrightarrow\infty,\]

\noindent that is: Given any arbitrary real number $\varepsilon>0$ there exists
a $K\in\mathbb{N}$, such that $\sup_{x\in J}|f_{k}(x)-f(x)|<\varepsilon$ for
all $k\geq K$. 

(3) \emph{In the mean of order $p\geq1$} if $|f(x)-f_{k}(x)|^{p}$ is integrable
over $J$ for each $k$ \[
\int_{J}|f(x)-f_{k}(x)|^{p}\longrightarrow0\qquad\textrm{as }k\rightarrow\infty.\]

\noindent For $p=1$, this is the simple case of \emph{convergence in the mean. }

(4) \emph{In the mean $m$-integrally} if it is possible to select indefinite
integrals \[
f_{k}^{(-m)}(x)=\pi_{k}(x)+\int_{c}^{x}dx_{1}\int_{c}^{x_{1}}dx_{2}\cdots\int_{c}^{x_{m-1}}dx_{m}f_{k}(x_{m})\]

\noindent and

\[
f^{(-m)}(x)=\pi(x)+\int_{c}^{x}dx_{1}\int_{c}^{x_{1}}dx_{2}\cdots\int_{c}^{x_{m-1}}dx_{m}f(x_{m})\]

\noindent such that for some arbitrary real $p\geq1$, \[
\int_{J}|f^{(-m)}-f_{k}^{(-m)}|^{p}\longrightarrow0\qquad\textrm{as }k\rightarrow\infty.\]

\noindent where the polynomials $\pi_{k}(x)$ and $\pi(x)$ are of degree $<m$,
and $c$ is a constant to be chosen appropriately. 

(5) \emph{Relative to test functions $\varphi$} if $f\varphi$ and $f_{k}\varphi$
are integrable over $J$ and \[
\int_{J}(f_{k}-f)\varphi\longrightarrow0\qquad\textrm{for every }\varphi\in\mathcal{C}_{0}^{\infty}(J)\textrm{ as }k\longrightarrow\infty,\]

\noindent where $\mathcal{C}_{0}^{\infty}(J)$ is the class of infinitely differentiable
continuous functions that vanish throughout some neighbourhood of each of the
end points of $J$. For an unbounded $J$, a function is said to vanish in some
neighbourhood of $+\infty$ if it vanishes on some ray $(r,\infty)$. 

While pointwise convergence does not imply any other type of convergence, uniform
convergence on a bounded interval implies all the other convergences. 

It is to be observed that apart from pointwise and uniform convergences, all
the other modes listed above represent some sort of an averaged contribution
of the entire interval $J$ and are therefore not of much use when pointwise
behaviour of the limit $f$ is necessary. Thus while limits in the mean are
not unique, oscillating functions are tamed by $m$-integral convergence for
adequately large values of $m$, and convergence relative to test functions,
as we see below, can be essentially reduced to $m$-integral convergence. On
the contrary, our graphical convergence --- which may be considered as a pointwise
biconvergence with respect to both the direct and inverse images of $f$ just
as usual pointwise convergence is with respect to its direct image only ---
allows a sequence (in fact, a net) of functions to converge to an arbitrary
relation, unhindered by external influences such as the effects of integrations
and test functions. To see how this can indeed matter, consider the following

\medskip{}
\noindent \textbf{Example 1.2.} Let $f_{k}(x)=\sin kx,\, k=1,2,\cdots$ and
let $J$ be any bounded interval of the real line. Then $1$-integrally we have\[
f_{k}^{(-1)}(x)=-\frac{1}{k}\cos kx=-\frac{1}{k}+\int_{0}^{x}\sin kx_{1}dx_{1},\]

\noindent which obviously converges to $0$ uniformly (and therefore in the
mean) as $k\rightarrow\infty$. And herein lies the point: even though we cannot
conclude about the exact nature of $\sin kx$ as $k$ increases indefinitely
(except that its oscillations become more and more pronounced), we may very
definitely state that $\lim_{k\rightarrow\infty}(\cos kx)/k=0$ uniformly. Hence
from\[
f_{k}^{(-1)}(x)\longrightarrow0=0+\int_{0}^{x}\lim_{k\rightarrow\infty}\sin kx_{1}dx_{1}\]
 it follows that \begin{equation}
\lim_{k\rightarrow\infty}\sin kx=0\label{Eqn: intsin}\end{equation}
 $1$-integrally. 

Continuing with the same sequence of functions, we now examine its test-functional
convergence with respect to $\varphi\in\mathcal{C}_{0}^{1}(-\infty,\infty)$
that vanishes for all $x\notin(\alpha,\beta)$. Integrating by parts, \begin{align*}
{\displaystyle {\displaystyle \int_{-\infty}^{\infty}f_{k}\varphi}}= & {\displaystyle \int_{\alpha}^{\beta}\varphi(x_{1})\sin kx_{1}dx_{1}}\\
= & -\frac{1}{k}\left[\varphi(x_{1})\cos kx_{1}\right]_{\alpha}^{\beta}-\frac{1}{k}\int_{\alpha}^{\beta}\varphi^{\prime}(x_{1})\cos kx_{1}dx_{1}\end{align*}

\noindent The first integrated term is $0$ due to the conditions on $\varphi$
while the second also vanishes because $\varphi\in\mathcal{C}_{0}^{1}(-\infty,\infty)$.
Hence \[
\int_{-\infty}^{\infty}f_{k}\varphi\longrightarrow0=\int_{\alpha}^{\beta}\lim_{k\rightarrow\infty}\varphi(x_{1})\sin ksdx_{1}\]
 for all $\varphi$, and leading to the conclusion that \begin{equation}
\lim_{k\rightarrow\infty}\sin kx=0\label{Eqn: testsin}\end{equation}
 test-functionally.$\qquad\blacksquare$
\medskip{}

This example illustrates the fact that if $\textrm{Supp}(\varphi)=[\alpha,\beta]\subseteq J$%
\footnote{{\small \label{Foot: support}By definition, the support (or supporting interval)
of $\varphi(x)\in\mathcal{C}_{0}^{\infty}[\alpha,\beta]$ is $[\alpha,\beta]$
if $\varphi$ and all its derivatives vanish for $x\leq\alpha$ and $x\geq\beta$. }%
}, integrating by parts sufficiently large number of times so as to wipe out
the pathological behaviour of $(f_{k})$ gives \begin{align*}
\int_{J}f_{k}\varphi= & \int_{\alpha}^{\beta}f_{k}\varphi\\
= & \int_{\alpha}^{\beta}f_{k}^{(-1)}\varphi^{\prime}=\cdots=(-1)^{m}\int_{\alpha}^{\beta}f_{k}^{(-m)}\varphi^{m}\end{align*}

\noindent where $f_{k}^{(-m)}(x)=\pi_{k}(x)+\int_{c}^{x}dx_{1}\int_{c}^{x_{1}}dx_{2}\cdots\int_{c}^{x_{m-1}}dx_{m}f_{k}(x_{m})$
is an $m$-times arbitrary indefinite integral of $f_{k}$. If now it is true
that $\int_{\alpha}^{\beta}f_{k}^{(-m)}\rightarrow\int_{\alpha}^{\beta}f^{(-m)}$,
then it must also be true that $f_{k}^{(-m)}\varphi^{(m)}$ converges in the
mean to $f^{(-m)}\varphi^{(m)}$ so that \[
\int_{\alpha}^{\beta}f_{k}\varphi=(-1)^{m}\int_{\alpha}^{\beta}f_{k}^{(-m)}\varphi^{(m)}\longrightarrow(-1)^{m}\int_{\alpha}^{\beta}f^{(-m)}\varphi^{(m)}=\int_{\alpha}^{\beta}f\varphi.\]
 In fact the converse also holds leading to the following Equivalences between
$m$-convergence in the mean and convergence with respect to test-functions,
\cite{Korevaar1968}.

\smallskip{}
\noindent \textbf{Type 1 Equivalence.} If $f$ and $(f_{k})$ are functions
on $J$ that are integrable on every interior subinterval, then the following
are equivalent statements. 

(a) For every interior subinterval $I$ of $J$ there is an integer $m_{I}\geq0$,
and hence a smallest integer $m\geq0$, such that certain indefinite integrals
$f_{k}^{(-m)}$ of the functions $f_{k}$ converge in the mean on $I$ to an
indefinite integral $f^{(-m)}$; thus $\int_{I}|f_{k}^{(-m)}-f^{(-m)}|\rightarrow0.$

(b) $\int_{J}(f_{k}-f)\varphi\rightarrow0$ for every $\varphi\in\mathcal{C}_{0}^{\infty}(J)$. 

\noindent {\small }%
\begin{figure}[htbp]
\noindent \begin{center}{\small \input{FuncSpace.pstex_t}}\end{center}{\small \par}

\begin{singlespace}

\caption{{\footnotesize \label{Fig: FuncSpace}Incompleteness of function spaces. Figure
(a) demonstrates the classic example of non-completeness of the space of real-valued
continuous functions leading to the complete spaces $L_{n}[a,b]$ whose elements
are equivalence classes of functions with $f\sim g$ iff the Lebesgue integral
$\int_{a}^{b}|f-g|^{n}=0$. Figures (b) and (c) illustrate distributional convergence
of the functions $f_{k}(x)$ of Eq. (\ref{Eqn: Lp[a,b]}) to the Dirac delta
$\delta(x)$ leading to the complete space of generalized functions. In comparison,
note that the space of continuous functions in the uniform metric $C[a,b]$
is complete which suggests the importance of topologies in determining convergence
properties of spaces. }}\end{singlespace}

\end{figure}
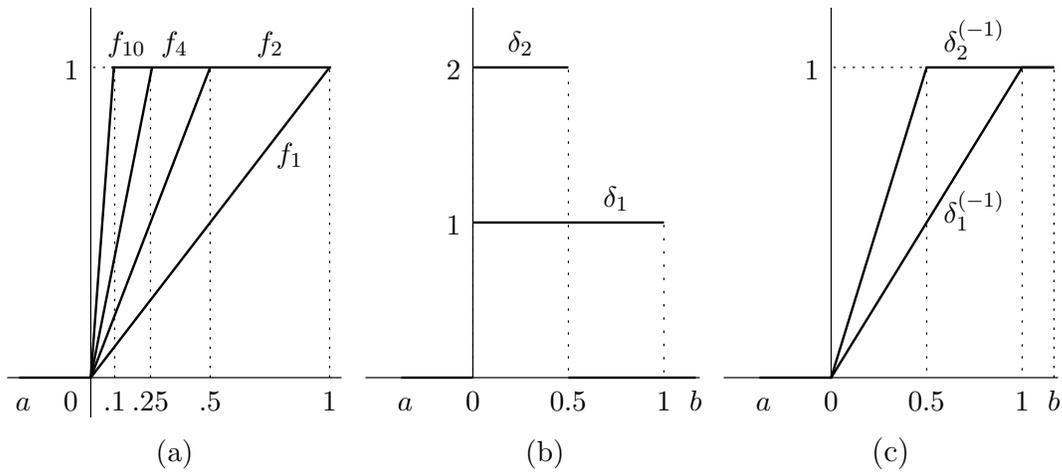
{\small \par}

A significant generalization of this Equivalence is obtained by dropping the
restriction that the limit object $f$ be a function. The need for this generalization
arises because metric function spaces are known not to be complete: Consider
the sequence of functions (Fig. \ref{Fig: FuncSpace}(a)) \begin{align}
f_{k}(x)= & \left\{ \begin{array}{lcl}
0 & \textrm{} & \textrm{if }a\leq x\leq0\\
kx & \textrm{} & \textrm{if }0\leq x\leq1/k\\
1 & \textrm{} & \textrm{if }1/k\leq x\leq b\end{array}\right.\label{Eqn: Lp[a,b]}\end{align}

\noindent which is not Cauchy in the uniform metric $\rho(f_{j},f_{k})=\sup_{a\leq x\leq b}|f_{j}(x)-f_{k}(x)|$
but is Cauchy in the mean $\rho(f_{j},f_{k})=\int_{a}^{b}|f_{j}(x)-f_{k}(x)|dx$,
or even pointwise. However in either case, $(f_{k})$ cannot converge in the
respective metrics to a \emph{continuous function} and the limit is a discontinuous
unit step function \[
\Theta(x)=\left\{ \begin{array}{lcl}
0 &  & \textrm{if }a\leq x\leq0\\
1 &  & \textrm{if }0<x\leq b\end{array}\right.\]
 with graph $([a,0],0)\bigcup((0,b],1)$, which is also integrable on $[a,b]$.
Thus even if the limit of the sequence of continuous functions is not continuous,
both the limit and the members of the sequence are integrable functions. This
Riemann integration is not sufficiently general, however, and this type of integrability
needs to be replaced by a much weaker condition resulting in the larger class
of the Lebesgue integrable complete space of functions $L[a,b]$.%
\footnote{{\small \label{Foot: integral}Both Riemann and Lebesgue integrals can be formulated
in terms of the so-called} \emph{\small step functions} {\small $s(x)$, which
are piecewise constant functions with values $(\sigma_{i})_{i=1}^{I}$on a finite
number of bounded subintervals $(J_{i})_{i=1}^{I}$ (which may reduce to a point
or may not contain one or both of the end-points) of a bounded or unbounded
interval $J$, with integral $\int_{J}s(x)dx\overset{\textrm{def}}=\sum_{i=1}^{I}\sigma_{i}|J_{i}|$.
While the Riemann integral of a bounded function $f(x)$ on a bounded interval
$J$ is defined with respect to sequences of step functions $(s_{j})_{j=1}^{\infty}$
and $(t_{j})_{j=1}^{\infty}$ satisfying $s_{j}(x)\leq f(x)\leq t_{j}(x)$ on
$J$ with $\int_{J}(s_{j}-t_{j})\rightarrow0$ as $j\rightarrow\infty$ as $R\int_{J}f(x)dx=\lim\int_{J}s_{j}(x)dx=\lim\int_{J}t_{j}(x)dx$,
the less restrictive Lebesgue integral is defined for arbitrary functions $f$
over bounded or unbounded intervals $J$ in terms of Cauchy sequences of step
functions $\int_{J}|s_{i}-s_{k}|\rightarrow0$, $i,k\rightarrow\infty$, converging
to $f(x)$ as \[
s_{j}(x)\rightarrow f(x)\textrm{ pointwise almost everywhere on }J,\]
}{\small \par}

\noindent {\small to be \[
\int_{J}f(x)dx\overset{\textrm{def}}=\lim_{j\rightarrow\infty}\int_{J}s_{j}(x)dx.\]
}{\small \par}

\noindent {\small That the Lebesgue integral is more general (and therefore
is the proper candidate for completion of function spaces) is illustrated by
the example of the function defined over $[0,1]$ to be $0$ on the rationals
and $1$ on the irrationals for which an application of the definitions verify
that whereas the Riemann integral is undefined, the Lebesgue integral exists
and has value $1$. The Riemann integral of a bounded function over a bounded
interval exists and is equal to its Lebesgue integral. Because it involves a
larger family of functions, all integrals in integral convergences are to be
understood in the Lebesgue sense. }%
} 

The functions in Fig \ref{Fig: FuncSpace}(b1), \[
\delta_{k}(x)=\left\{ \begin{array}{ccl}
k &  & \textrm{if }0<x<1/k\\
0 &  & x\in[a,b]-(0,1/k),\end{array}\right.\]

\noindent can be associated with the arbitrary indefinite integrals \[
\Theta_{k}(x)\overset{\textrm{def}}=\delta_{k}^{(-1)}(x)=\left\{ \begin{array}{lcl}
0 &  & a\leq x\leq0\\
kx &  & 0<x<1/k\\
1 &  & 1/k\leq x\leq b\end{array}\right.\]

\noindent of Fig. \ref{Fig: FuncSpace}(b2), which, as noted above, converge
in the mean to the unit step function $\Theta(x)$; hence $\int_{-\infty}^{\infty}\delta_{k}\varphi\equiv\int_{\alpha}^{\beta}\delta_{k}\varphi=-\int_{\alpha}^{\beta}\delta_{k}^{(-1)}\varphi^{\prime}\rightarrow-\int_{0}^{\beta}\varphi^{\prime}(x)dx=\varphi(0)$.
But there can be no \emph{functional relation $\delta(x)$} for which $\int_{\alpha}^{\beta}\delta(x)\varphi(x)dx=\varphi(0)$
for \emph{all} $\varphi\in C_{0}^{1}[\alpha,\beta]$, so that unlike in the
case in Type 1 Equivalence, the limit in the mean $\Theta(x)$ of the indefinite
integrals $\delta_{k}^{(-1)}(x)$ \emph{cannot be expressed as the indefinite
integral $\delta^{(-1)}(x)$ of some function $\delta(x)$ on any interval containing
the origin.} This leads to the second more general type of equivalence

\smallskip{}
\noindent \textbf{Type 2 Equivalence.} If $(f_{k})$ are functions on $J$ that
are integrable on every interior subinterval, then the following are equivalent
statements. 

(a) For every interior subinterval $I$ of $J$ there is an integer $m_{I}\geq0$,
and hence a smallest integer $m\geq0$, such that certain indefinite integrals
$f_{k}^{(-m)}$ of the functions $f_{k}$ converge in the mean on $I$ to an
integrable function $\Theta$ which, unlike in Type 1 Equivalence, need not
itself be an indefinite integral of some function $f$. 

(b) $c_{k}(\varphi)=\int_{J}f_{k}\varphi\rightarrow c(\varphi)$ for every $\varphi\in\mathcal{C}_{0}^{\infty}(J)$.
\smallskip{}

\noindent Since we are now given that $\int_{I}f_{k}^{(-m)}(x)dx\rightarrow\int_{I}\Psi(x)dx$,
it must also be true that $f_{k}^{(-m)}\varphi^{(m)}$ converges in the mean
to $\Psi\varphi^{(m)}$ whence \[
\int_{J}f_{k}\varphi=(-1)^{m}\int_{I}f_{k}^{(-m)}\varphi^{(m)}\longrightarrow(-1)^{m}\int_{I}\Psi\varphi^{(m)}\left(\neq(-1)^{m}\int_{I}f^{(-m)}\varphi^{(m)}\right).\]

\noindent The natural question that arises at this stage is then: What is the
nature of the relation (not function any more) $\Psi(x)$? For this it is now
stipulated, despite the non-equality in the equation above, that as in the mean
$m$-integral convergence of $(f_{k})$ to a \emph{function} $f$, \begin{equation}
\Theta(x):=\lim_{k\rightarrow\infty}\delta_{k}^{(-1)}(x)\overset{\textrm{def}}=\int_{-\infty}^{x}\delta(x^{\prime})dx^{\prime}\label{Eqn: delta1}\end{equation}

\noindent \emph{defines} the non-functional relation ({}``generalized function'')
$\delta(x)$ integrally as a solution of the integral equation (\ref{Eqn: delta1})
of the first kind; hence formally%
\footnote{{\small \label{Foot: delta}The observant reader cannot have failed to notice
how mathematical ingenuity successfully transferred the {}``troubles'' of
$(\delta_{k})_{k=1}^{\infty}$ to the sufficiently differentiable benevolent
receptor $\varphi$ so as to be able to work backward, via the resultant trouble
free $(\delta_{k}^{(-m)})_{k=1}^{\infty}$, to the final object $\delta$. This
necessarily hides the true character of $\delta$ to allow only a view of its
integral manifestation on functions. This unfortunately is not general enough
in the strongly nonlinear physical situations responsible for chaos, and is
the main reason for constructing the multifunctional extension of function spaces
that we use. }%
} \begin{equation}
\delta(x)=\frac{d\Theta}{dx}\label{Eqn: delta2}\end{equation}

\vspace{-0.15cm}
\noindent \begin{flushright}\textbf{\textit{End Tutorial2}}\end{flushright}
\medskip{}

The above tells us that the {}``delta function'' is not a function but its
indefinite integral is the piecewise continuous \emph{function} $\Theta$ obtained
as the mean (or pointwise) limit of a sequence of non-differentiable functions
with the integral of $d\Theta_{k}(x)/dx$ being preserved for all $k\in\mathbb{Z}_{+}$.
What then is the delta (and not its integral)? The answer to this question is
contained in our multifunctional extension $\textrm{Multi}(X,Y)$ of the function
space $\textrm{Map}(X,Y)$ considered in Sec. 3. Our treatment of ill-posed
problems is used to obtain an understanding and interpretation of the numerical
results of the discretized spectral approximation in neutron transport theory
\cite{Sengupta1988,Sengupta1995}. The main conclusions are the following: In
a one-dimensional discrete system that is governed by the iterates of a nonlinear
map, the dynamics is chaotic if and only if the system evolves to a state of
\emph{maximal ill-posedness.} The analysis is based on the non-injectivity,
and hence ill-posedness, of the map; this may be viewed as a mathematical formulation
of the \emph{stretch-and-fold} and \emph{stretch-cut-and-paste} kneading operations
of the dough that are well-established artifacts in the theory of chaos and
the concept of maximal ill-posedness helps in obtaining a \emph{physical understanding}
of the nature of chaos. We do this through the fundamental concept of the \emph{graphical
convergence} of a sequence (generally a net) of functions \cite{Sengupta2000}
that is allowed to converge graphically, when the conditions are right, to a
set-valued map or multifunction. Since ill-posed problems naturally lead to
multifunctional inverses through functional generalized inverses \cite{Sengupta1997},
it is natural to seek solutions of ill-posed problems in multifunctional space
$\textrm{Multi}(X,Y)$ rather than in spaces of functions $\textrm{Map}(X,Y)$;
here $\textrm{Multi}(X,Y)$ is an extension of $\textrm{Map}(X,Y)$ that is
generally larger than the smallest dense extension $\textrm{Multi}_{\mid}(X,Y)$.

Feedback and iteration are natural processes by which nature evolves itself.
Thus almost every process of evolution is a self-correction process by which
the system proceeds from the present to the future through a controlled mechanism
of input and evaluation of the past. Evolution laws are inherently nonlinear
and complex; here \emph{complexity} is to be understood as the natural manifestation
of the nonlinear laws that govern the evolution of the system. 

This work presents a mathematical description of complexity based on \cite{Sengupta1997}
and \cite{Sengupta2000} and is organized as follows. In Sec. 1, we follow \cite{Sengupta1997}
to give an overview of ill-posed problems and their solution that forms the
foundation of our approach. Secs. 2 to 4 apply these ideas by defining a chaotic
dynamical system as a \emph{maximally ill-posed problem;} by doing this we are
able to overcome the limitations of the three Devaney characterizations of chaos
\cite{Devaney1989} that apply to the specific case of iteration of transformations
in a metric space, and the resulting graphical convergence of functions to multifunctions
is the basic tool of our approach. Sec. 5 analyzes graphical convergence in
$\textrm{Multi}(X)$ for the discretized spectral approximation of neutron transport
theory, which suggests a natural link between ill-posed problems and spectral
theory of non-linear operators. This seems to offer an answer to the question
of \emph{why} a natural system should increase its complexity, and eventually
tend toward chaoticity, by becoming increasingly nonlinear. 

\vspace{1cm}
\noindent \begin{flushleft}\textbf{\large 2. Ill-Posed Problem and its solution}\end{flushleft}{\large \par}

\noindent This section based on \citet*{Sengupta1997} presents a formulation
and solution of ill-posed problems arising out of the non-injectivity of a function
$f\!:X\rightarrow Y$ between topological spaces $X$ and $Y$. A workable knowledge
of this approach is necessary as our theory of chaos leading to the characterization
of chaotic systems as being a \emph{maximally ill-posed} state of a dynamical
system is a direct application of these ideas and can be taken to constitute
a mathematical representation of the familiar \emph{stretch-cut-and paste} and
\emph{stretch-and-fold} paradigms of chaos. The problem of finding an $x\in X$
for a given $y\in Y$ from the functional relation $f(x)=y$ is an inverse problem
that is \emph{ill-posed} (or, the equation $f(x)=y$ is ill-posed) if any one
or more of the following conditions are satisfied. 

\smallskip{}
(IP1) $f$ \emph{is not injective.} This \emph{non-uniqueness} problem of the
solution for a given $y$ is the single most significant criterion of ill-posedness
used in this work. 

(IP2) \emph{$f$ is not surjective.} For a $y\in Y$, this is the \emph{existence}
problem of the given equation. 

(IP3) When $f$ \emph{is bijective,} the inverse \emph{$f^{-1}$} is not continuous,
which means that small changes in $y$ may lead to large changes in $x$. 

\smallskip{}
A problem $f(x)=y$ for which a solution exists, is unique, and small changes
in data $y$ lead to only small changes in the solution $x$ is said to be \emph{well-posed}
or \emph{properly posed.} This means that $f(x)=y$ is well-posed if $f$ is
bijective and the inverse $f^{-1}\!:Y\rightarrow X$ is continuous; otherwise
the equation is \emph{ill-posed} or \emph{improperly posed.} It is to be noted
that the three criteria are not, in general, independent of each other. Thus
if $f$ represents a bijective, bounded linear operator between Banach spaces
$X$ and $Y$, then the inverse mapping theorem guarantees that the inverse
$f^{-1}$ is continuous. Hence ill-posedness depends not only on the algebraic
structures of $X$, $Y$, $f$ but also on the topologies of $X$ and $Y$. 

\medskip{}
\noindent \textbf{Example 2.1.} As a non-trivial example of an inverse problem,
consider the heat equation\[
\frac{\partial\theta(x,t)}{\partial t}=c^{2}\frac{\partial^{2}\theta(x,t)}{\partial x^{2}}\]

\noindent for the temperature distribution $\theta(x,t)$ of a one-dimensional
homogeneous rod of length $L$ satisfying the initial condition $\theta(x,0)=\theta_{0}(x),\textrm{ }0\leq x\leq L$,
and boundary conditions $\theta(0,t)=0=\theta(L,t),\,0\leq t\leq T$, having
the Fourier sine-series solution \begin{equation}
\theta(x,t)=\sum_{n=1}^{\infty}A_{n}\sin\left(\frac{n\pi}{L}x\right)e^{-\lambda_{n}^{2}t}\label{Eqn: heat1}\end{equation}

\noindent where $\lambda_{n}=(c\pi/a)n$ and \[
A_{n}=\frac{2}{L}\int_{0}^{a}\theta_{0}(x^{\prime})\sin\left(\frac{n\pi}{L}x^{\prime}\right)dx^{\prime}\]
 are the Fourier expansion coefficients. While the direct problem evaluates
$\theta(x,t)$ from the differential equation and initial temperature distribution
$\theta_{0}(x)$, the inverse problem calculates $\theta_{0}(x)$ from the integral
equation \[
\theta_{T}(x)=\frac{2}{L}\int_{0}^{a}k(x,x^{\prime})\theta_{0}(x^{\prime})dx^{\prime},\qquad0\leq x\leq L,\]

\noindent when this final temperature $\theta_{T}$ is known, and \[
k(x,x^{\prime})=\sum_{n=1}^{\infty}\sin\left(\frac{n\pi}{L}x\right)\sin\left(\frac{n\pi}{L}x^{\prime}\right)e^{-\lambda_{n}^{2}T}\]
 is the kernel of the integral equation. In terms of the final temperature the
distribution becomes \begin{equation}
\theta_{T}(x)=\sum_{n=1}^{\infty}B_{n}\sin\left(\frac{n\pi}{L}x\right)e^{-\lambda_{n}^{2}(t-T)}\label{Eqn: heat2}\end{equation}

\noindent with Fourier coefficients \[
B_{n}=\frac{2}{L}\int_{0}^{a}\theta_{T}(x^{\prime})\sin\left(\frac{n\pi}{L}x^{\prime}\right)dx^{\prime}.\]

\noindent In $L^{2}[0,a]$, Eqs. (\ref{Eqn: heat1}) and (\ref{Eqn: heat2})
at $t=T$ and $t=0$ yield respectively \begin{equation}
\Vert\theta_{T}(x)\Vert^{2}=\frac{L}{2}\sum_{n=1}^{\infty}A_{n}^{2}e^{-2\lambda_{n}^{2}T}\leq e^{-2\lambda_{1}^{2}T}\Vert\theta_{0}\Vert^{2}\label{Eqn: heat3}\end{equation}
\begin{equation}
\Vert\theta_{0}\Vert^{2}=\frac{L}{2}\sum_{n=1}^{\infty}B_{n}^{2}e^{2\lambda_{n}^{2}T}.\label{Eqn: heat4}\end{equation}

\noindent The last two equations differ from each other in the significant respect
that whereas Eq. (\ref{Eqn: heat3}) shows that the direct problem is well-posed
according to (IP3), Eq. (\ref{Eqn: heat4}) means that in the absence of similar
bounds the inverse problem is ill-posed.%
\footnote{{\small \label{Foot: cont=3Dbound}Recall that for a linear operator continuity
and boundedness are equivalent concepts. }%
}$\qquad\blacksquare$

\medskip{}
\noindent \textbf{Example 2.2.} Consider the \textbf{}Volterra integral equation
of the first kind \[
y(x)=\int_{a}^{x}r(x^{\prime})dx^{\prime}=Kr\]

\noindent where $y,r\in C[a,b]$ and $K\!:C[0,1]\rightarrow C[0,1]$ is the
corresponding integral operator. Since the differential operator $D=d/dx$ under
the sup-norm $\Vert r\Vert=\sup_{0\leq x\leq1}|r(x)|$ is unbounded, the inverse
problem $r=Dy$ for a differentiable function $y$ on $[a,b]$ is ill-posed,
see Example 6.1. However, $y=Kr$ becomes well-posed if $y$ is considered to
be in $C^{1}[0,1]$ with norm $\Vert y\Vert=\sup_{0\leq x\leq1}|Dy|$. This
illustrates the importance of the topologies of $X$ and $Y$ in determining
the ill-posed nature of the problem when this is due to (IP3).$\qquad\blacksquare$
\medskip{}

Ill-posed problems in nonlinear mathematics of type (IP1) arising from the non-injectivity
of $f$ can be considered to be a generalization of non-uniqueness of solutions
of linear equations as, for example, in eigenvalue problems or in the solution
of a system of linear algebraic equations with a larger number of unknowns than
the number of equations. In both cases, for a given $y\in Y$, the solution
set of the equation $f(x)=y$ is given by \[
f^{-}(y)=[x]_{f}=\{ x^{\prime}\in X:f(x^{\prime})=f(x)=y\}.\]

\noindent A significant point of difference between linear and nonlinear problems
is that unlike the special importance of 0 in linear mathematics, there are
no preferred elements in nonlinear problems; this leads to a shift of emphasis
from the null space of linear problems to equivalence classes for nonlinear
equations. To motivate the role of equivalence classes, let us consider the
null spaces in the following linear problems. 

\smallskip{}
(a) Let $f:\mathbb{R}^{2}\rightarrow\mathbb{R}$ be defined by $f(x,y)=x+y$,
$(x,y)\in\mathbb{R}^{2}$. The null space of $f$ is generated by the equation
$y=-x$ on the $x$-$y$ plane, and the graph of $f$ is the plane passing through
the lines $\rho=x$ and $\rho=y.$ For each $\rho\in\textrm{R}$ the equivalence
classes $f^{-}(\rho)=\{(x,y)\in\textrm{R}^{2}\!:x+y=\rho\}$ are lines on the
graph parallel to the null set. 

(b) For a linear operator $A\!:\mathbb{R}^{n}\rightarrow\mathbb{R}^{m}$, $m<n$,
satisfying (1) and (2), the problem $Ax=y$ reduces $A$ to echelon form with
rank $r$ less than $\min\{ m,n\}$, when the given equations are consistent.
The solution however, produces a generalized inverse leading to a set-valued
inverse $A^{-}$ of $A$ for which the inverse images of $y\in\mathcal{R}(A)$
are multivalued because of the non-trivial null space of $A$ introduced by
assumption (1). Specifically, a null-space of dimension $n-r$ is generated
by the free variables $\{ x_{j}\}_{j=r+1}^{n}$ which are arbitrary: this is
illposedness of type (1). In addition, $m-r$ rows of the row reduced echelon
form of $A$ have all 0 entries that introduces restrictions on $m-r$ coordinates
$\{ y_{i}\}_{i=r+1}^{m}$ of $y$ which are now related to $\{ y_{i}\}_{i=1}^{r}$:
this illustrates illposedness of type (2). Inverse ill-posed problems therefore
generate multivalued solutions through a generalized inverse of the mapping. 

(c) The eigenvalue problem \[
\left(\frac{d^{2}}{dx^{2}}+\lambda^{2}\right)y=0\qquad y(0)=0=y(1)\]

\noindent has the following equivalence class of 0 \[
[0]_{D^{2}}=\{\sin(\pi mx)\}_{m=0}^{\infty},\qquad D^{2}=\left(d^{2}/dx^{2}+\lambda^{2}\right),\]

\noindent as its eigenfunctions corresponding to the eigenvalues $\lambda_{m}=\pi m$. 

\smallskip{}
Ill-posed problems are primarily of interest to us explicitly as noninjective
maps $f$, that is under the condition of (IP1). The two other conditions (IP2)
and (IP3) are not as significant and play only an implicit role in the theory.
In its application to iterative systems, the degree of non-injectivity of $f$
defined as the number of its injective branches, increases with iteration of
the map. A necessary (but not sufficient) condition for chaos to occur is the
increasing non-injectivity of $f$ that is expressed descriptively in the chaos
literature as \emph{stretch-and-fold} or \emph{stretch-cut-and-paste} operations.
This increasing noninjectivity that we discuss in the following sections, is
what causes a dynamical system to tend toward chaoticity. Ill-posedness arising
from non-surjectivity of (injective) $f$ in the form of \emph{regularization}
\cite{Tikhonov1977} \emph{}has received wide attention in the literature of
ill-posed problems; this however is not of much significance in our work. 

\medskip{}
\noindent \begin{flushright}\textbf{\textit{Begin Tutorial3: Generalized Inverse}}\end{flushright}
\vspace{-0.15cm}

\begin{figure}[htbp]
\noindent \begin{center}{\small \input{MP_Inverse.pstex_t}}\end{center}{\small \par}

\begin{singlespace}

\caption{{\footnotesize \label{Fig: MP_Inverse}(a) Moore-Penrose generalized inverse.
The decomposition of $X$ and $Y$ into the four fundamental subspaces of $A$
comprising the null space $\mathcal{N}(A)$, the column (or range) space $\mathcal{R}(A)$,
the row space $\mathcal{R}(A^{\textrm{T}})$ and $\mathcal{N}(A^{\textrm{T}})$,
the complement of $\mathcal{R}(A)$ in $Y$, is a basic result in the theory
of linear equations. The Moore-Penrose inverse takes advantage of the geometric
orthogonality of the row space $\mathcal{R}(A^{\textrm{T}})$ and $\mathcal{N}(A)$
in $\mathbb{R}^{n}$ and that of the column space and $\mathcal{N}(A^{\textrm{T}})$
in $\mathbb{R}^{m}$. (b) When $X$ and $Y$ are not inner-product spaces, a}
\emph{\footnotesize non-injective inverse} {\footnotesize can be defined by
extending $f$ to $Y-\mathcal{R}(f)$ suitably as shown by the dashed curve,
where $g(x):=r_{1}+((r_{2}-r_{1})/r_{1})f(x)$ for all $x\in\mathcal{D}(f)$
was taken to be a good definition of an extension that replicates $f$ in $Y-\mathcal{R}(f)$;
here $x_{1}\sim x_{2}$ under both $f$ and $g$, and $y_{1}\sim y_{2}$ under
$\{ f,g\}$ just as $b$ is equivalent to $b_{\Vert}$ in the Moore-Penrose
case. Note that both $\{ f,g\}$ and $\{ f^{-},g^{-}\}$ are both multifunctions
on $X$ and $Y$ respectively. Our inverse $G$, introduced later in this section,
is however injective with $G(Y-\mathcal{R}(f)):=0$. }}\end{singlespace}

\end{figure}
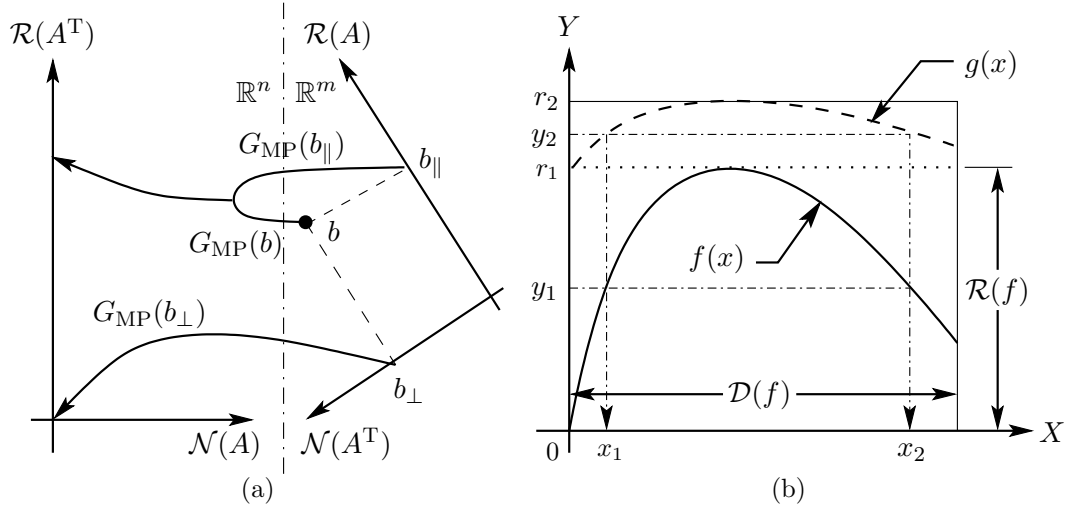

\noindent In this Tutorial, we take a quick look at the equation $a(x)=y$,
where $a\!:X\rightarrow Y$ is a linear map that need not be either one-one
or onto. Specifically, we will take $X$ and $Y$ to be the Euclidean spaces
$\mathbb{R}^{n}$ and $\mathbb{R}^{m}$ so that $a$ has a matrix representation
$A\in\mathbb{R}^{m\times n}$ where $\mathbb{R}^{m\times n}$ is the collection
of $m\times n$ matrices with real entries. The inverse $A^{-1}$ exists and
is unique iff $m=n$ and $\textrm{rank}(A)=n$; this is the situation depicted
in Fig. \ref{Fig: functions}(a). If $A$ is neither one-one or onto, then we
need to consider the multifunction $A^{-}$, a functional choice of which is
known as the \emph{generalized inverse} $G$ of $A$. A good introductory text
for generalized inverses is \citet*{Campbell1979}Figure \ref{Fig: MP_Inverse}(a)
introduces the following definition of the \emph{Moore-Penrose} generalized
inverse $G_{\textrm{MP}}$. 

\medskip{}
\noindent \textbf{Definition 2.1.} \textbf{\textit{Moore-Penrose Inverse.}}
\textsl{If $a\!:\mathbb{R}^{n}\rightarrow\mathbb{R}^{m}$ is a linear transformation
with matrix representation $A\in\mathbb{R}^{m\times n}$ then the} Moore-Penrose
inverse $G_{\textrm{MP}}\in\mathbb{R}^{n\times m}$ of $A$ \textsl{(we will
use the same notation} $G_{\textrm{MP}}\!:\mathbb{R}^{m}\rightarrow\mathbb{R}^{n}$
\textsl{for the inverse of the map $a$) is the noninjective map defined in
terms of the row and column spaces of $A$,} $\textrm{row}(A)=\mathcal{R}(A^{\textrm{T}})$,
$\textrm{col}(A)=\mathcal{R}(A)$\textsl{, as} \begin{equation}
G_{\textrm{MP}}(y)\overset{\textrm{def}}=\left\{ \begin{array}{lcl}
(a|_{\textrm{row}(A)})^{-1}(y), &  & \textrm{if }y\in\textrm{col}(A)\\
0 &  & \textrm{if }y\in\mathcal{N}(A^{\textrm{T}}).\end{array}\right.\qquad\square\label{Eqn: Def: Moore-Penrose}\end{equation}

{\small }%
Note that the restriction $a|_{\textrm{row}(A)}$ of $a$ to $\mathcal{R}(A^{\textrm{T}})$
is bijective so that the inverse $(a|_{\textrm{row}(A)})^{-1}$ is well-defined.
The role of the transpose matrix appears naturally, and the $G_{\textrm{MP}}$
of Eq. (\ref{Eqn: Def: Moore-Penrose}) is the unique matrix that satisfies
the conditions

\noindent \renewcommand{\arraystretch}{1.2}\begin{equation}
\begin{array}{c}
AG_{\textrm{MP}}A=A,\quad G_{\textrm{MP}}AG_{\textrm{MP}}=G_{\textrm{MP}},\\
(G_{\textrm{MP}}A)^{\textrm{T}}=G_{\textrm{MP}}A,\quad(AG_{\textrm{MP}})^{\textrm{T}}=AG_{\textrm{MP}}\end{array}\label{Eqn: MPInverse}\end{equation}

\noindent that follow immediately from the definition (\ref{Eqn: Def: Moore-Penrose});
hence $G_{\textrm{MP}}A$ and $AG_{\textrm{MP}}$ are orthogonal projections%
\footnote{{\small \label{Foot: OrthoMatrix}A real matrix $A$ is an orthogonal projector
iff $A^{2}=A$ and $A=A^{\textrm{T}}$. }%
} onto the subspaces $\mathcal{R}(A^{\textrm{T}})=\mathcal{R}(G_{\textrm{MP}})$
and $\mathcal{R}(A)$ respectively. Recall that the range space $\mathcal{R}(A^{\textrm{T}})$
of $A^{\textrm{T}}$ is the same as the \emph{row space} $\textrm{row}(A)$
of $A$, and $\mathcal{R}(A)$ is also known as the \emph{column space} of $A$,
$\textrm{col}(A)$. 

\medskip{}
\noindent \textbf{Example 2.3.} For $a\!:\mathbb{R}^{5}\rightarrow\mathbb{R}^{4}$,
let \[
A=\left(\begin{array}{rrrrr}
1 & -3 & 2 & 1 & 2\\
3 & -9 & 10 & 2 & 9\\
2 & -6 & 4 & 2 & 4\\
2 & -6 & 8 & 1 & 7\end{array}\right)\]

\noindent By reducing the augmented matrix $\left(A|y\right)$ to the row-reduced
echelon form, it can be verified that the null and range spaces of $A$ are
$3$- and $2$-dimensional respectively. A basis for the null space of $A^{\textrm{T}}$
and of the row and column space of $A$ obtained from the echelon form are respectively
\[
\left(\begin{array}{r}
-2\\
0\\
1\\
0\end{array}\right),\textrm{ }\left(\begin{array}{r}
1\\
-1\\
0\\
1\end{array}\right);\quad\textrm{and }\left(\begin{array}{r}
1\\
-3\\
0\\
3/2\\
1/2\end{array}\right),\textrm{ }\left(\begin{array}{r}
0\\
0\\
1\\
-1/4\\
3/4\end{array}\right);\textrm{ }\left(\begin{array}{r}
1\\
0\\
2\\
-1\end{array}\right),\textrm{ }\left(\begin{array}{r}
0\\
1\\
0\\
1\end{array}\right).\]

\noindent According to its definition Eq. (\ref{Eqn: Def: Moore-Penrose}),
the Moore-Penrose inverse maps the middle two of the above set to $(0,0,0,0,0)^{\textrm{T}}$,
and the $A$-image of the first two (which are respectively $(19,70,38,51)^{\textrm{T}}$
and $(70,275,140,205)^{\textrm{T}}$ lying, as they must, in the span of the
last two), to the span of $(1,-3,2,1,2)^{\textrm{T}}$ and $(3,-9,10,2,9)^{\textrm{T}}$
because $a$ restricted to this subspace of $\mathbb{R}^{5}$ is bijective.
Hence \[
G_{\textrm{MP}}\left(A\left(\begin{array}{r}
1\\
-3\\
0\\
3/2\\
1/2\end{array}\right)\textrm{ }A\left(\begin{array}{r}
0\\
0\\
1\\
-1/4\\
3/4\end{array}\right)\begin{array}{rr}
-2 & 1\\
0 & -1\\
1 & 0\\
0 & 1\end{array}\right)=\left(\begin{array}{rrrr}
1 & 0 & 0 & 0\\
-3 & 0 & 0 & 0\\
0 & 1 & 0 & 0\\
3/2 & -1/4 & 0 & 0\\
1/2 & 3/4 & 0 & 0\end{array}\right).\]

\noindent The second matrix on the left is invertible as its rank is $4$. This
gives \begin{equation}
{\displaystyle G_{\textrm{MP}}=\left(\begin{array}{rrrr}
9/275 & -1/275 & 18/275 & -2/55\\
-27/275 & 3/275 & -54/275 & 6/55\\
-10/143 & 6/143 & -20/143 & 16/143\\
238/3575 & -57/3575 & 476/3575 & -59/715\\
-129/3575 & 106/3575 & -258/3575 & 47/715\end{array}\right)}\label{Eqn: MPEx5}\end{equation}

\noindent as the Moore-Penrose inverse of $A$ that readily verifies all the
four conditions of Eqs. (\ref{Eqn: MPInverse}). The basic point here is that,
as in the case of a bijective map, $G_{\textrm{MP}}A$ and $AG_{\textrm{MP}}$
are identities on the row and column spaces of $A$ that define its rank. For
later use --- when we return to this example for a simpler inverse $G$ ---
given below are the orthonormal bases of the four fundamental subspaces with
respect to which $G_{\textrm{MP}}$ is a representation of the generalized inverse
of $A$; these calculations were done by MATLAB. The basis for 

\smallskip{}
(a) the column space of $A$ consists of the first $2$ columns of the eigenvectors
of $AA^{\textrm{T}}$: \[
\begin{array}{c}
(-1633/2585,-363/892,\textrm{ }3317/6387,\textrm{ }363/892)^{\textrm{T}}\\
(-929/1435,\textrm{ }709/1319,\textrm{ }346/6299,-709/1319)^{\textrm{T}}\end{array}\]

(b) the null space of $A^{\textrm{T}}$ consists of the last $2$ columns of
the eigenvectors of $AA^{\textrm{T}}$:\[
\begin{array}{c}
(-3185/8306,\textrm{ }293/2493,-3185/4153,\textrm{ }1777/3547)^{\textrm{T}}\\
(323/1732,\textrm{ }533/731,\textrm{ }323/866,\textrm{ }1037/1911)^{\textrm{T}}\end{array}\]

(c) the row space of $A$ consists of the first $2$ columns of the eigenvectors
of $A^{\textrm{T}}A$: \[
\begin{array}{c}
(421/13823,\textrm{ }44/14895,-569/918,-659/2526,\textrm{ }1036/1401)\\
(661/690,\textrm{ }412/1775,\textrm{ }59/2960,-1523/10221,-303/3974)\end{array}\]

(d) the null space of $A$ consists of the last $3$ columns of the of $A^{\textrm{T}}A$:\[
\begin{array}{c}
(-571/15469,-369/776,\textrm{ }149/25344,-291/350,-389/1365)\\
(-281/1313,\textrm{ }956/1489,\textrm{ }875/1706,-1279/2847,\textrm{ }409/1473)\\
(292/1579,-876/1579,\textrm{ }203/342,\textrm{ }621/4814,\textrm{ }1157/2152)\end{array}\]
\renewcommand{\arraystretch}{1}

\noindent The matrices $Q_{1}$ and $Q_{2}$ with these eigenvectors $(x_{i})$
satisfying $\Vert x_{i}\Vert=1$ and $(x_{i},x_{j})=0$ for $i\neq j$ as their
columns are \emph{orthogonal matrices} with the simple inverse criterion $Q^{-1}=Q^{\textrm{T}}$.$\qquad\blacksquare$

\vspace{-0.15cm}
\noindent \begin{flushright}\textbf{\textit{End Tutorial3}}\end{flushright}
\medskip{}

\noindent The basic issue in the solution of the inverse ill-posed problem is
its reduction to an well-posed one when restricted to suitable subspaces of
the domain and range of $A$. Considerations of geometry leading to their decomposition
into orthogonal subspaces is only an additional feature that is not central
to the problem: recall from Eq. (\ref{Eqn: f_inv_f}) that any function $f$
must necessarily satisfy the more general set-theoretic relations $ff^{-}f=f$
and $f^{-}ff^{-}=f^{-}$ of Eq. (\ref{Eqn: MPInverse}) for the multiinverse
$f^{-}$ of $f\!:X\rightarrow Y$. The second distinguishing feature of the
MP-inverse is that it is defined, by a suitable extension, on all of $Y$ and
not just on $f(X)$ which is perhaps more natural. The availability of orthogonality
in inner-product spaces allows this extension to be made in an almost normal
fashion. As we shall see below the additional geometric restriction of Eq. (\ref{Eqn: MPInverse})
is not essential to the solution process, and infact, only results in a less
canonical form of the inverse. 

\medskip{}
\noindent \begin{flushright}\textbf{\textit{Begin Tutorial4: Topological Spaces}}\end{flushright}
\vspace{-0.15cm}

\noindent This Tutorial is meant to familiarize the reader with the basic principles
of a topological space. A topological space $(X,\mathcal{U})$ is a set $X$
with a class%
\footnote{{\small \label{Foot: class}In this sense, a} \emph{\small class} {\small is
a set of sets. }%
} $\mathcal{U}$ of distinguished subsets, called \emph{open sets of $X$,} that
satisfy

(T1) The empty set $\emptyset$ and the whole $X$ belong to $\mathcal{U}$

(T2) Finite intersections of members of $\mathcal{U}$ belong to $\mathcal{U}$

(T3) Arbitrary unions of members of $\mathcal{U}$ belong to $\mathcal{U}$.

\medskip{}
\noindent \textbf{Example 2.4.} (1) The smallest topology possible on a set
$X$ is its \emph{indiscrete topology} when the only open sets are $\emptyset$
and $X$; the largest is the \emph{discrete topology} where every subset of
$X$ is open (and hence also closed). 

(2) In a metric space $(X,d)$, let $B_{\varepsilon}(x,d)=\{ y\in X\!:d(x,y)<\varepsilon\}$
be an open ball at $x$. Any subset $U$ of $X$ such that for each $x\in U$
there is a $d$-ball $B_{\varepsilon}(x,d)\subseteq U$ in $U$, is said to
be an open set of $(X,d)$. The collection of all these sets is the topology
induced by $d$. The topological space $(X,\mathcal{U})$ is then said to be
\emph{associated with (induced by)} $(X,d)$. 

(3) If $\sim$ is an equivalence relation on a set $X$, the set of all saturated
sets $[x]_{\sim}=\{ y\in X\!:y\sim x\}$ is a topology on $X;$ this topology
is called the \emph{topology of saturated sets. }

We argue in Sec. 4.2 that this constitutes the defining topology of a chaotic
system. 

(4) For any subset $A$ of the set $X$, the $A$-\emph{inclusion topology on
$X$} consists of $\emptyset$ and every superset of $A$, while the $A$-\emph{exclusion
topology on} $X$ consists of all subsets of $X-A$. Thus $A$ is open in the
inclusion topology and closed in the exclusion, and in general every open set
of one is closed in the other. 

The special cases of the \emph{$a$-inclusion} and \emph{$a$-exclusion} topologies
for $A=\{ a\}$ are defined in a similar fashion. 

(5) The \emph{cofinite} and \emph{cocountable topologies} in which the open
sets of an infinite (resp. uncountable) set $X$ are respectively the complements
of finite and countable subsets, are examples of topologies with some unusual
properties that are covered in Appendix A1. If $X$ is itself finite (respectively,
countable), then its cofinite (respectively, cocountable) topology is the  discrete
topology consisting of all its subsets. It is therefore useful to adopt the
convention, unless stated to the contrary, that cofinite and cocountable spaces
are respectively infinite and uncountable.$\qquad\blacksquare$
\medskip{}

In the space $(X,\mathcal{U})$, a \emph{neighbourhood of a point} $x\in X$
is a nonempty subset $N$ of $X$ that contains an open set $U$ containing
$x$; thus $N\subseteq X$ is a neighbourhood of $x$ iff \begin{equation}
x\in U\subseteq N\label{Eqn: Def: nbd1}\end{equation}
 for some $U\in\mathcal{U}$. The largest open set that can be used here is
$\textrm{Int}(N)$ (where, by definition, $\textrm{Int}(A)$ is the largest
open set that is contained in $A$) so that the above neighbourhood criterion
for a subset $N$ of $X$ can be expressed in the equivalent form \begin{equation}
N\subseteq X\textrm{ is a }\mathcal{U}-\textrm{neighbourhood of }x\textrm{ iff }x\in\textrm{Int}_{\mathcal{U}}(N)\label{Eqn: Def: nbd2}\end{equation}
implying that a subset of $(X,\mathcal{U})$ is a neighbourhood of all its interior
points, so that $N\in\mathcal{N}_{x}\Rightarrow N\in\mathcal{N}_{y}$ for all
$y\in\textrm{Int}(N)$. The collection of all neighbourhoods of $x$ \begin{equation}
\mathcal{N}_{x}\overset{\textrm{def}}=\{ N\subseteq X\!:x\in U\subseteq N\textrm{ for some }U\in\mathcal{U}\}\label{Eqn: Def: nbd system}\end{equation}
 \emph{}is the \emph{neighbourhood system} at $x$, and the subcollection $U$
of the topology used in this equation constitutes a \emph{neighbourhood} (\emph{local})
\emph{base} or \emph{basic neighbourhood system, at} $x$, see Def. A1.1 of
Appendix A1. The properties 

\smallskip{}
(N1) $x$ belongs to every member $N$ of \emph{$\mathcal{N}_{x}$, }

(N2) The intersection of any two neighbourhoods of \emph{$x$} is another neighbourhood
of $x$: $N,M\in\mathcal{N}_{x}\Rightarrow N\bigcap M\in\mathcal{N}_{x}$, 

(N3) Every superset of \emph{}any neighbourhood of $x$ is a neighbourhood of
$x$: $(M\in\mathcal{N}_{x})\wedge(M\subseteq N)\Rightarrow N\in\mathcal{N}_{x}$. 

\smallskip{}
\noindent that characterize \emph{$\mathcal{N}_{x}$} completely are a direct
consequence of the definition (\ref{Eqn: Def: nbd1}), (\ref{Eqn: Def: nbd2})
that may also be stated as 

\smallskip{}
(N0) Any neighbourhood $N\in\mathcal{N}_{x}$ contains another neighbourhood
$U$ of $x$ that is a \emph{neighbourhood of each of its point}s: $((\forall N\in\mathcal{N}_{x})(\exists U\in\mathcal{N}_{x})(U\subseteq N))\!:(\forall y\in U\Rightarrow U\in\mathcal{N}_{y})$. 

Property (N0) infact serves as the defining characteristic of an open set, and
\textsl{$U$} can be identified with the largest open set $\textrm{Int}(N)$
contained in $N$; hence \textsl{a set $G$ in a topological space is open iff
it is a neighbourhood of each of its points.} Accordingly if \emph{$\mathcal{N}_{x}$}
is a given class of subsets of $X$ associated with each $x\in X$ satisfying
$(\textrm{N}1)-(\textrm{N}3)$, then (N0) defines the special class of neighbourhoods
$G$ \begin{equation}
\mathcal{U}=\{ G\in\mathcal{N}_{x}\!:x\in B\subseteq G\textrm{ for all }x\in G\textrm{ and a basic nbd }B\in\mathcal{N}_{x}\}\label{Eqn: nbd-topology}\end{equation}
 as the unique topology on $X$ that contains a basic neighbourhood of each
of its points, for which the neighbourhood system at $x$ \emph{}\textit{\emph{coincides
exactly with the assigned collection}} \emph{$\mathcal{N}_{x}$}\textit{\emph{;
compare Def A1.1.}} Neighbourhoods in topological spaces are a generalization
of the familiar notion of distances of metric spaces that quantifies {}``closeness''
of points of $X$. 
\smallskip{}

A \emph{neighbourhood of a nonempty subset} $A$ of $X$ that will be needed
later on is defined in a similar manner: $N$ is a neighbourhood of $A$ iff
$A\subseteq\textrm{Int}(N)$, that is $A\subseteq U\subseteq N$; thus the neighbourhood
system at $A$ is given by $\mathcal{N}_{A}=\bigcap_{a\in A}\mathcal{N}_{a}:=\{ G\subseteq X\!:G\in\mathcal{N}_{a}\textrm{ for every }a\in A\}$
is the class of common neighbourhoods of each point of $A$. 

Some examples of neighbourhood systems at a point $x$ in $X$ are the following: 

\smallskip{}
(1) In an indiscrete space $(X,\mathcal{U})$, $X$ is the only neighbourhood
of every point of the space; in a discrete space any set containing $x$ is
a neighbourhood of the point. 

(2) In an infinite cofinite (or uncountable cocountable) space, every neighbourhood
of a point is an open neighbourhood of that point. 

(3) In the topology of saturated sets under the equivalence relation $\sim$,
the neighbourhood system at $x$ consists of all supersets of the equivalence
class $[x]_{\sim}$.

(4) Let $x\in X$. In the $x$-inclusion topology, $\mathcal{N}_{x}$ consists
of all the non-empty open sets of $X$ which are the supersets of $\{ x\}$.
For a point $y\neq x$ of $X$, $\mathcal{N}_{y}$ are the supersets of $\{ x,y\}$. 
\smallskip{}

For any given class $_{\textrm{T}}\mathcal{S}$ of subsets of $X$, a unique
topology $\mathcal{U}(_{\textrm{T}}\mathcal{S})$ can always be constructed
on $X$ by taking all \emph{finite} \emph{intersections} $_{\textrm{T}}\mathcal{S}_{\wedge}$
of members of $\mathcal{S}$ followed by \emph{arbitrary} \emph{unions} $_{\textrm{T}}\mathcal{S}_{\wedge\vee}$
of these finite intersections. $\mathcal{U}(_{\textrm{T}}\mathcal{S}):=\,_{\textrm{T}}\mathcal{S}_{\wedge\vee}$
is the smallest topology on $X$ that contains $_{\textrm{T}}\mathcal{S}$ and
is said to be \emph{generated by} $_{\textrm{T}}\mathcal{S}$. For a given topology
$\mathcal{U}$ on $X$ satisfying $\mathcal{U}=\mathcal{U}(_{\textrm{T}}\mathcal{S})$,
$_{\textrm{T}}\mathcal{S}$ is a \emph{subbasis,} and $_{\textrm{T}}\mathcal{S}_{\wedge}:=\,_{\textrm{T}}\mathcal{B}$
a \emph{basis, for the topology} $\mathcal{U}$; for more on topological basis,
see Appendix A1. The topology generated by a subbase essentially builds not
from the collection $_{\textrm{T}}\mathcal{S}$ itself but from the finite intersections
$_{\textrm{T}}\mathcal{S}_{\wedge}$ of its subsets; in comparison the base
generates a topology directly from a collection $_{\textrm{T}}\mathcal{S}$
of subsets by forming their unions. Thus whereas \emph{any} class of subsets
can be used as a subbasis, a given collection must meet certain qualifications
to pass the test of a base for a topology: these and related topics are covered
in Appendix A1. Different subbases, therefore, can be used to generate different
topologies on the same set $X$ as the following examples for the case of $X=\mathbb{R}$
demonstrates; here $(a,b)$, $[a,b)$, $(a,b]$ and $[a,b]$, for $a\leq b\in\mathbb{R}$,
are the usual open-closed intervals in $\mathbb{R}$%
\footnote{{\small \label{Foot: interval}By definition, an interval $I$ in a totally
ordered set $X$ is a subset of $X$ with the property \[
(x_{1},x_{2}\in I)\wedge(x_{3}\in X\!:x_{1}\prec x_{3}\prec x_{2})\Longrightarrow x_{3}\in I\]
}{\small \par}

\noindent {\small so that any element of $X$ lying between two elements of
$I$ also belongs to $I$.}%
}. The subbases $_{\textrm{T}}\mathcal{S}_{1}=\{(a,\infty),(-\infty,b)\}$, $_{\textrm{T}}\mathcal{S}_{2}=\{[a,\infty),(-\infty,b)\}$,
$_{\textrm{T}}\mathcal{S}_{3}=\{(a,\infty),(-\infty,b]\}$ and $_{\textrm{T}}\mathcal{S}_{4}=\{[a,\infty),(-\infty,b]\}$
give the respective bases $_{\textrm{T}}\mathcal{B}_{1}=\{(a,b)\}$, $_{\textrm{T}}\mathcal{B}_{2}=\{[a,b)\}$,
$_{\textrm{T}}\mathcal{B}_{3}=\{(a,b]\}$ and $_{\textrm{T}}\mathcal{B}_{4}=\{[a,b]\}$,
$a\leq b\in\mathbb{R}$, leading to the \emph{standard (usual})\emph{, lower
limit (Sorgenfrey})\emph{, upper limit,} and \emph{discrete} (take $a=b$) topologies
on $\mathbb{R}$. Bases of the type $(a,\infty)$ and $(-\infty,b)$ provide
the \emph{right} and \emph{left ray} topologies on $\mathbb{R}$. 

\smallskip{}
\emph{This feasibility of generating different topologies on a set can be of
great practical significance because open sets determine convergence characteristics
of nets and continuity characteristics of functions, thereby making it possible
for nature to play around with the structure of its working space in its kitchen
to its best possible advantage.}%
\footnote{{\small \label{Foot: entropy}Although we do not pursue this point of view here,
it is nonetheless tempting to speculate that the answer to the question} \emph{\small {}``Why}
{\small does the entropy of an isolated system increase?'' may be found by
exploiting this line of reasoning that seeks to explain the increase in terms
of a visible component associated with the usual topology as against a different
latent workplace topology that governs the dynamics of nature.}%
} \emph{}
\smallskip{}

\noindent Here are a few essential concepts and terminology for topological
spaces. 

\medskip{}
\noindent \textbf{Definition 2.2.} \textbf{\textit{Boundary, Closure, Interior}}\textbf{.}
\textsl{The} \emph{boundary of $A$ in $X$} \textsl{is the set of points $x\in X$
such that every neighbourhood $N$ of $x$ intersects both $A$ and $X-A$:}
\begin{equation}
{\textstyle \textrm{Bdy}(A)\overset{\textrm{def}}=\{ x\in X\!:(\forall N\in\mathcal{N}_{x})((N\bigcap A\neq\emptyset)\wedge(N\bigcap(X-A)\neq\emptyset))\}}\label{Eqn: Def: Boundary}\end{equation}
 \textsl{where $\mathcal{N}_{x}$ is the neighbourhood system of Eq. (\ref{Eqn: Def: nbd system})
at $x$. }

\textsl{The} \emph{closure of $A$} \textsl{is the set of all points $x\in X$
such that each neighbourhood of $x$ contains at least one point of $A$} \textbf{\textsl{that
may be $\boldmath{x}$ itself}}\textsl{. Thus the set} \begin{equation}
{\textstyle \textrm{Cl}(A)\overset{\textrm{def}}=\{ x\in X\!:(\forall N\in\mathcal{N}_{x})\textrm{ }(N\bigcap A\neq\emptyset)\}}\label{Eqn: Def: Closure}\end{equation}
 \emph{of all points in $X$ adherent} \textsl{to} \emph{$A$ is given by} \textsl{is
the union} \textsl{\emph{}}\textsl{of $A$ with its boundary. }

\textsl{The} \emph{interior of $A$} \begin{equation}
\textrm{Int}(A)\overset{\textrm{def}}=\{ x\in X\!:(\exists N\in\mathcal{N}_{x})\textrm{ }(N\subseteq A)\}\label{Eqn: Def: Interior}\end{equation}
 \textsl{consisting of those points of $X$ that are in $A$ but not in its
boundary,} $\textrm{Int}(A)=A-\textrm{Bdy}(A)$\textsl{, is the largest open
subset of $X$ that is contained in $A$. Hence it follows that} $\textrm{Int}(\textrm{Bdy}(A))=\emptyset$,
\textsl{the boundary of $A$ is the intersection of the closures of $A$ and
$X-A$,} \textsl{and a subset $N$ of $X$ is a neighbourhood of $x$ iff} $x\in\textrm{Int}(N)$\textsl{.$\qquad\square$}

\medskip{}
The three subsets $\textrm{Int}(A)$, $\textrm{Bdy}(A)$ and \emph{exterior}
of $A$ defined as $\textrm{Ext}(A):=\textrm{Int}(X-A)=X-\textrm{Cl}(A)$, are
pairwise disjoint and have the full space $X$ as their union. 

\medskip{}
\noindent \textbf{Definition 2.3.} \textbf{\textit{Derived and Isolated sets.}}
\textsl{Let $A$ be a subset of $X$. A point $x\in X$ (which may or may not
be a point of $A$) is a} \emph{cluster point of} $A$ \textsl{if every neighbourhood
$N\in\mathcal{N}_{x}$ contains atleast one point of $A$} \textbf{\textsl{different
from}} \textsl{$\mathbf{x}$. The} \emph{derived set of $A$} \begin{equation}
{\textstyle \textrm{Der}(A)\overset{\textrm{def}}=\{ x\in X\!:(\forall N\in\mathcal{N}_{x})\textrm{ }(N\bigcap(A-\{ x\})\neq\emptyset)\}}\label{Eqn: Def: Derived}\end{equation}
 \textsl{is the set of all cluster points of $A$. The complement of} $\textrm{Der}(A)$
in $A$ \begin{equation}
\textrm{Iso}(A)\overset{\textrm{def}}=A-\textrm{Der}(A)=\textrm{Cl}(A)-\textrm{Der}(A)\label{Eqn: Def: Isolated}\end{equation}
 \textsl{are the} \emph{isolated} \emph{points} \emph{of} $A$ \textsl{to which
no proper sequence in $A$ converges, that is there exists a neighbourhood of
any such point that contains no other point of $A$ so that} \emph{the only
sequence that converges to} $a\in\textrm{Iso}(A)$ \emph{is the constant sequence
$(a,a,a,\cdots)$.}\textsl{$\qquad\square$}
\medskip{}

Clearly, \renewcommand{\arraystretch}{1.25}\[
\begin{array}{ccl}
{\textstyle \textrm{Cl}(A)} & = & A\bigcup\textrm{Der}(A)=A\bigcup\textrm{Bdy}(A)\\
 & = & \textrm{Iso}(A)\bigcup\textrm{Der}(A)=\textrm{Int}(A)\bigcup\textrm{Bdy}(A)\end{array}\]
\renewcommand{\arraystretch}{1} with the last two being disjoint unions, and
$A$ is closed iff $A$ contains all its cluster points, $\textrm{Der}(A)\subseteq A$,
iff $A$ contains its closure. Hence \begin{multline*}
A=\textrm{Cl}(A)\Longleftrightarrow\textrm{Cl}(A)=\{ x\in A\!:((\exists N\in\mathcal{N}_{x})(N\subseteq A))\vee((\forall N\in\mathcal{N}_{x})(N\bigcap(X-A)\neq\emptyset))\}\end{multline*}
 Comparison of Eqs. (\ref{Eqn: Def: Boundary}) and (\ref{Eqn: Def: Derived})
also makes it clear that $\textrm{Bdy}(A)\subseteq\textrm{Der}(A)$. The special
case of $A=\textrm{Iso}(A)$ with $\textrm{Der}(A)\subseteq X-A$ is important
enough to deserve a special mention: 

\medskip{}
\noindent \textbf{Definition 2.4.} \textbf{\textit{Donor set.}} \textsl{A proper,
nonempty subset $A$ of $X$ such that} $\textrm{Iso}(A)=A$ \textsl{with} $\textrm{Der}(A)\subseteq X-A$
\textsl{will be called} \emph{self-isolated} \textsl{or} \emph{donor.} \textsl{Thus
sequences eventually in a donor set converges only in its complement; this is
the opposite of the characteristic of a closed set where all converging sequences
eventually in the set must necessarily converge in it. A closed-donor set with
a closed neighbour has no derived or boundary sets, and will be said to be}
\emph{isolated in $X$.}$\qquad\square$
\medskip{}

\noindent \textbf{Example 2.5.} In an isolated set sequences converge, if they
have to, simultaneously in the complement (because it is donor) and in it (because
it is closed). Convergent sequences in such a set can only be constant sequences.
Physically, if we consider adherents to be contributions made by the dynamics
of the corresponding sequences, then an isolated set is secluded from its neighbour
in the sense that it neither receives any contributions from its surroundings,
nor does it give away any. In this light and terminology, a closed set is a
\emph{selfish} set (recall that a set $A$ is closed in $X$ iff every convergent
net of $X$ that is eventually in $A$ converges in $A$; conversely a set is
open in $X$ iff the only nets that converge in $A$ are eventually in it),
\emph{}whereas a set with a derived set that intersects itself and its complement
may be considered to be \emph{neutral.} Appendix A3 shows the various possibilities
for the derived set and boundary of a subset $A$ of $X$.$\qquad\blacksquare$
\medskip{}

Some useful properties of these concepts for a subset $A$ of a topological
space $X$ are the following. 

\smallskip{}
(a) $\textrm{Bdy}_{X}(X)=\emptyset$, 

(b) $\textrm{Bdy}(A)=\textrm{Cl}(A)\bigcap\textrm{Cl}(X-A)$, 

(c) $\textrm{Int}(A)=X-\textrm{Cl}(X-A)=A-\textrm{Bdy}(A)=\textrm{Cl}(A)-\textrm{Bdy}(A)$, 

(d) $\textrm{Int}(A)\bigcap\textrm{Bdy}(A)=\emptyset$, 

(e) $X=\textrm{Int}(A)\bigcup\textrm{Bdy}(A)\bigcup\textrm{Int}(X-A)$, 

(f) \begin{equation}
{\textstyle \textrm{Int}(A)=\bigcup\{ G\subseteq X\!:G\textrm{ is an open set of }X\textrm{ contained in }A\}}\label{Eqn: interior}\end{equation}

(g) \begin{equation}
{\textstyle \textrm{Cl}(A)=\bigcap\{ F\subseteq X\!:F\textrm{ is a closed set of }X\textrm{ containing }A\}}\label{Eqn: closure}\end{equation}

A straightforward consequence of property (b) is that the boundary of any subset
$A$ of a topological space $X$ is closed in $X$; this significant result
may also be demonstrated as follows. If $x\in X$ is not in the boundary of
$A$ there is some neighbourhood $N$ of $x$ that does not intersect both $A$
and $X-A$. For each point $y\in N$, $N$ is a neighbourhood of that point
that does not meet $A$ and $X-A$ simultaneously so that $N$ is contained
wholly in $X-\textrm{Bdy}(A)$. We may now take $N$ to be open without any
loss of generality implying thereby that $X-\textrm{Bdy}(A)$ is an open set
of $X$ from which it follows that $\textrm{Bdy}(A)$ is closed in $X$. 

Further material on topological spaces relevant to our work can be found in
Appendix A3. 

\vspace{-0.15cm}
\noindent \begin{flushright}\textbf{\textit{End Tutorial4}}\end{flushright}
\medskip{}

Working in a general topological space, we now recall the solution of an ill-posed
problem $f(x)=y$ \cite{Sengupta1997} that leads to a multifunctional inverse
$f^{-}$ through the generalized inverse $G$. Let $f\!:(X,\mathcal{U})\rightarrow(Y,\mathcal{V})$
be a (nonlinear) function between two topological space $(X,\mathcal{U})$ and
$(Y,\mathcal{V})$ that is neither one-one or onto. Since $f$ is not one-one,
$X$ can be partitioned into disjoint equivalence classes with respect to the
equivalence relation $x_{1}\sim x_{2}\Leftrightarrow f(x_{1})=f(x_{2})$. Picking
a representative member from each of the classes (this is possible by the Axiom
of Choice; see the following Tutorial) produces a \emph{basic set} $X_{\textrm{B}}$
of $X$; it is basic as it corresponds to the row space in the linear matrix
example which is all that is needed for taking an inverse. $X_{\textrm{B}}$
is the counterpart of the quotient set $X/\sim$ of Sec. 1, with the important
difference that whereas the points of the quotient set are the equivalence classes
of $X$, $X_{\textrm{B}}$ \emph{is a subset of} $X$ with each of the classes
contributing a point to $X_{\textrm{B}}$. It then follows that $f_{\textrm{B}}\!:X_{\textrm{B}}\rightarrow f(X)$
is the bijective restriction $a|_{\textrm{row}(A)}$ that reduces the original
ill-posed problem to a well-posed one with $X_{\textrm{B}}$ and $f(X)$ corresponding
respectively to the row and column spaces of $A$, and $f_{\textrm{B}}^{-1}\!:f(X)\rightarrow X_{\textrm{B}}$
is the basic inverse from which the multiinverse $f^{-}$ is obtained through
$G$, which in turn corresponds to the Moore-Penrose inverse $G_{\textrm{MP}}$.
The topological considerations (obviously not for inner product spaces that
applies to the Moore-Penrose inverse) needed to complete the solution are discussed
below and in Appendix A1. 

\medskip{}
\noindent \begin{flushright}\textbf{\textit{Begin Tutorial5: Axiom of Choice
and Zorn's Lemma}}\end{flushright}
\vspace{-0.15cm}

\noindent Since some of our basic arguments depend on it, this Tutorial contains
a short description of the Axiom of Choice that has been described as {}``one
of the most important, and at the same time one of the most controversial, principles
of mathematics''. What this axiom states is this: For any set $X$ there exists
a function $f_{\textrm{C}}\!:\mathcal{P}_{0}(X)\rightarrow X$ such that $f_{\textrm{C}}(A_{\alpha})\in A_{\alpha}$
for every non-empty subset $A_{\alpha}$ of $X$; here $\mathcal{P}_{0}(X)$
is the class of all subsets of $X$ except $\emptyset$. Thus, if $X=\{ x_{1},x_{2},x_{3}\}$
is a three element set, a possible choice function is given by \renewcommand{\arraystretch}{1.2}\[
\begin{array}{c}
f_{\textrm{C}}(\{ x_{1},x_{2},x_{3}\})=x_{3},\quad f_{\textrm{C}}(\{ x_{1},x_{2}\})=x_{1},\quad f_{\textrm{C}}(\{ x_{2},x_{3}\})=x_{3},\quad f_{\textrm{C}}(\{ x_{3},x_{1}\})=x_{3},\\
f_{\textrm{C}}(\{ x_{1}\})=x_{1},\quad f_{\textrm{C}}(\{ x_{2}\})=x_{2},\quad f_{\textrm{C}}(\{ x_{3}\})=x_{3}.\end{array}\]
\renewcommand{\arraystretch}{1} It must be appreciated that the axiom is only
an existence result that asserts \emph{every set} to have a choice function,
even when nobody knows how to construct one in a specific case. Thus, for example,
how does one pick out the isolated irrationals $\sqrt{2}$ or $\pi$ from the
uncountable reals? There is no doubt that they do exist, for we can construct
a right angled triangle with sides of length $1$ or a circle of radius $1$.
The axiom tells us that these choices are possible even though we do not know
how exactly to do it; all that can be stated with confidence is that we can
actually pick up rationals arbitrarily close to these irrationals. 

The axiom of choice is essentially meaningful when $X$ is infinite as illustrated
in the last two examples. This is so because even when $X$ is denumerable,
it would be physically impossible to make an infinite number of selections either
all at a time or sequentially: the Axiom of Choice nevertheless tells us that
this is possible. The real strength and utility of the Axiom however is when
$X$ and some or all of its subsets are uncountable as in the case of the choice
of the \emph{single element} $\pi$ from the reals. To see this more closely
in the context of maps that we are concerned with, let $f\!:X\rightarrow Y$
be a non-injective, onto map. To construct a functional right inverse $f_{r}\!:Y\rightarrow X$
of $f$, we must choose, for each $y\in Y$ one \emph{representative} element
$x_{\textrm{rep}}$ from the set $f^{-}(y)$ and define $f_{r}(y)$ to be that
element according to $f\circ f_{r}(y)=f(x_{\textrm{rep}})=y$. If there is no
preferred or natural way to make this choice, the axiom of choice allows us
to make an arbitrary selection from the infinitely many that may be possible
from $f^{-}(y)$. When a natural choice is indeed available, as for example
in the case of the initial value problem $y^{\prime}(x)=x;\, y(0)=\alpha_{0}$
on $[0,a]$, the definite solution $\alpha_{0}+x^{2}/2$ may be selected from
the infinitely many $\int_{0}^{x}x^{\prime}dx^{\prime}=\alpha+x^{2}/2,\textrm{ }0\leq x\leq a$
that are permissible, and the axiom of choice sanctions this selection. In addition,
each $y\in Y$ gives rise to the family of solution sets $A_{y}=\{ f^{-}(y)\!:y\in Y\}$
and the real power of the axiom is its assertion that it is possible to make
a choice $f_{\textrm{C}}(A_{y})\in A_{y}$ on every $A_{y}$ simultaneously;
this permits the choice \emph{}on every $A_{y}$ of the collection to be made
at the same time.  

\vspace{-0.15cm}
\noindent \begin{flushright}\textbf{\textit{Pause Tutorial5}}\end{flushright}
\medskip{}

\begin{figure}[htbp]
\noindent \begin{center}{\small \input{GenInv.pstex_t}}\end{center}{\small \par}

\begin{singlespace}

\caption{{\footnotesize \label{Fig: GenInv}Solution of ill-posed problem $f(x)=y$,
$f\!:X\rightarrow Y$. $G\!:Y\rightarrow X_{\textrm{B}}$, a generalized inverse
of $f$ because of $fGf=f$ and $GfG=G$ which follows from the commutativity
of the diagrams, is a functional selection of the multiinverse $f^{-}\!:(Y,\mathcal{{V}})\;\quad(X,\mathcal{{U}})$.
$_{<}f$ and $f_{<}$ are the injective and surjective restrictions of $f$;
these will be topologically denoted by their generic notations $e$ and $q$
respectively.}}\end{singlespace}

\end{figure}
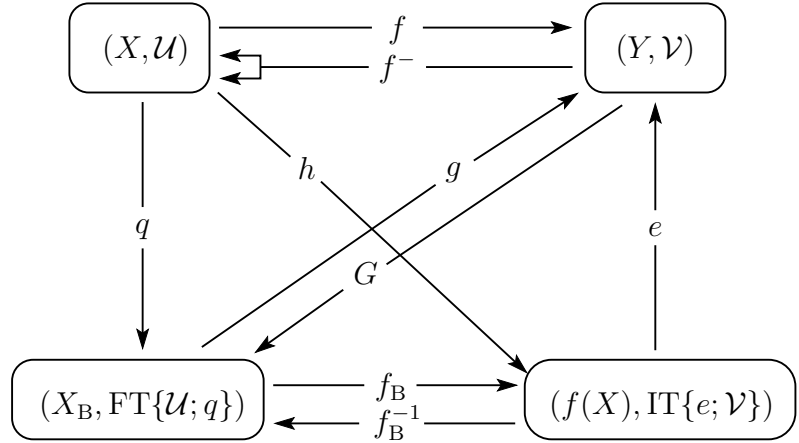

\noindent Figure \ref{Fig: GenInv} shows our \cite{Sengupta1997} formulation
and solution of the inverse ill-posed problem $f(x)=y$. In sub-diagram $X-X_{\textrm{B}}-f(X)$,
the surjection $p\!:X\rightarrow X_{\textrm{B}}$ is the counterpart of the
quotient map $Q$ of Fig. \ref{Fig: quotient} that is known in the present
context as the \emph{identification} of $X$ with $X_{\mathrm{B}}$ (as it \emph{identifies}
each saturated subset of $X$ with its representative point in $X_{\textrm{B}}$),
with the space $(X_{\textrm{B}},\textrm{FT}\{\mathcal{U};p\})$ carrying the
\emph{identification topology} $\textrm{FT}\{\mathcal{U};p\}$ being known as
an \emph{identification space.} By sub-diagram $Y-X_{\textrm{B}}-f(X)$, the
image $f(X)$ of $f$ gets the \emph{subspace topology}%
\footnote{{\small \label{Foot: subspace}In a subspace  $A$ of $X$, a subset $U_{A}$
of $A$ is open iff $U_{A}=A\bigcap U$ for some open set $U$ of $X$. The
notion of subspace topology can be formalized with the help of the inclusion
map $i\!:A\rightarrow(X,\mathcal{U})$ that puts every point of $A$ back to
where it came from, thus \[
\begin{array}{ccl}
\mathcal{U}_{A} & = & \{ U_{A}=A\bigcap U\!:U\in\mathcal{U}\}\\
 & = & \{ i^{-}(U)\!:U\in\mathcal{U}\}.\end{array}\]
}%
} $\textrm{IT}\{ j;\mathcal{V}\}$ from $(Y,\mathcal{V})$ by the inclusion $j\!:f(X)\rightarrow Y$
when its open sets are generated as, and only as, $j^{-1}(V)=V\bigcap f(X)$
for $V\in\mathcal{V}$. Furthermore if the bijection $f_{\textrm{B}}$ connecting
$X_{\textrm{B}}$ and $f(X)$ (which therefore acts as a $1:1$ correspondence
between their points, implying that these sets are set-theoretically identical
except for their names) is image continuous, then by Theorem A2.1 of Appendix
2, so is the \emph{association} $q=f_{\textrm{B}}\circ p\!:X\rightarrow f(X)$
that associates saturated sets of $X$ with elements of $f(X)$; this makes
$f(X)$ look like an identification space of $X$ by assigning to it the topology
$\textrm{FT}\{\mathcal{U};q\}$. On the other hand if $f_{\textrm{B}}$ happens
to be preimage continuous, then $X_{\textrm{B}}$ acquires, by Theorem A2.2,
the initial topology $\textrm{IT}\{ e;\mathcal{V}\}$ by the \emph{embedding}
$e\!:X_{\textrm{B}}\rightarrow Y$ that embeds $X_{\textrm{B}}$ into $Y$ through
$j\circ f_{\textrm{B}}$, making it look like a subspace of $Y$%
\footnote{{\small \label{Foot: assoc&embed}A surjective function is an} \emph{\small association}
{\small iff it is image continuous and an injective function is an} \emph{\small embedding}
{\small iff it is preimage continuous. }%
}. In this dual situation, $f_{\textrm{B}}$ has the highly interesting topological
property of being simultaneously image and preimage continuous when the open
sets of $X_{\textrm{B}}$ and $f(X)$ --- which are simply the $f_{\textrm{B}}^{-1}$-images
of the open sets of $f(X)$ which, in turn, are the $f_{\textrm{B}}$-images
of these saturated open sets --- can be considered to have been generated by
$f_{\textrm{B}}$, and are respectively the smallest and largest collection
of subsets of $X$ and $Y$ that makes $f_{\textrm{B}}$ \emph{ini}(tial-fi)\emph{nal
continuous} \cite{Sengupta1997}\emph{.} A bijective ininal function such as
$f_{\textrm{B}}$ is known as a \emph{homeomorphism} and ininality for functions
that are neither $1:1$ \emph{}nor onto is a generalization of homeomorphism
for bijections; refer Eqs. (\ref{Eqn: INI}) and (\ref{Eqn: HOM}) for a set-theoretic
formulation of this distinction. A homeomorphism $f\!:(X,\mathcal{U})\rightarrow(Y,\mathcal{V})$
renders the homeomorphic spaces $(X,\mathcal{U})$ and $(Y,\mathcal{V})$ topologically
indistinguishable which may be considered to be identical in as far as their
topological properties are concerned. 

{\small }%
{\small \par}

\smallskip{}
\noindent \textbf{Remark.} It may be of some interest here to speculate on the
significance of \emph{ininality} in our work. Physically, a map $f\!:(X,\mathcal{U})\rightarrow(Y,\mathcal{V})$
between two spaces can be taken to represent an interaction between them and
the algebraic and topological characters of $f$ determine the nature of this
interaction. A simple bijection merely sets up a correspondence, that is an
interaction, between every member of $X$ with some member $Y$, whereas a continuous
map establishes the correspondence among the special category of {}``open''
sets. Open sets, as we see in Appendix A1, are the basic ingredients in the
theory of convergence of sequences, nets and filters, and the characterization
of open sets in terms of convergence, namely that \textsl{a set $G$ in $X$
is open in it if every net or sequence that converges in $X$ to a point in
$G$ is eventually i}\emph{n $G$}, see Appendix A1, may be interpreted to mean
that such sets represent groupings of elements that require membership of the
group before permitting an element to belong it; an open set unlike its complement
the closed or \emph{selfish} set, however, does not forbid a net that has been
eventually in it to settle down in its selfish neighbour, who nonetheless will
never allow such a situation to develop in its own territory. An ininal map
forces these well-defined and definite groups in $(X,\mathcal{U})$ and $(Y,\mathcal{V})$
to interact with each other through $f$; this is not possible with simple continuity
as there may be open sets in $X$ that are not derived from those of $Y$ and
non-open sets in $Y$ whose inverse images are open in $X$. \emph{It is our
hypothesis that the driving force behind the evolution of a system represented
by the input-output relation $f(x)=y$ is the attainment of the ininal triple
state $(X,f,Y)$ for the system.} A preliminary analysis of this hypothesis
is to be found in Sec. 4.2.

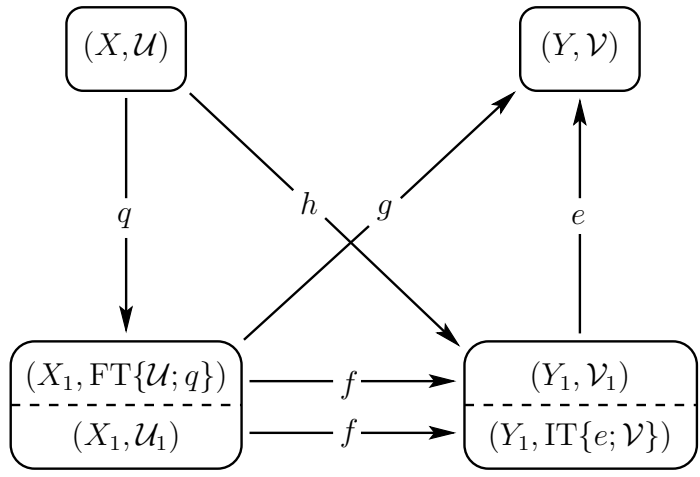
\begin{figure}[htbp]
\noindent \begin{center}\input{Initial-Final.pstex_t}\end{center}

\begin{singlespace}

\caption{\label{Fig: Initial-Final}{\small Continuity in final and initial topologies.}}\end{singlespace}

\end{figure}

\smallskip{}
For ininality of the interaction, it is therefore necessary to have\begin{eqnarray}
\textrm{FT}\{\mathcal{U};f_{<}\} & = & \textrm{IT}\{ j;\mathcal{V}\}\label{Eqn: ininal}\\
\textrm{IT}\{\,_{<}f;\mathcal{V}\} & = & \textrm{FT}\{\mathcal{U};p\}\};\nonumber \end{eqnarray}

\noindent in what follows we will refer to the injective and surjective restrictions
of $f$ by their generic topological symbols of embedding $e$ and association
$q$ respectively. What are the topological characteristics of $f$ in order
that the requirements of Eq. (\ref{Eqn: ininal}) be met? From Appendix A1,
it should be clear by superposing the two parts of Fig. \ref{Fig: Initial-Final}
over each other that given $q\!:(X,\mathcal{U})\rightarrow(f(X),\textrm{FT}\{\mathcal{U};q\})$
in the first of these equations, $\textrm{IT}\{ j;\mathcal{V}\}$ will equal
$\textrm{FT}\{\mathcal{U};q\}$ iff $j$ is an ininal open inclusion and $Y$
receives $\textrm{FT}\{\mathcal{U};f\}$. In a similar manner, preimage continuity
of $e$ requires $p$ to be open ininal and $f$ to be preimage continuous if
the second of Eq. (\ref{Eqn: ininal}) is to be satisfied. Thus under the restrictions
imposed by Eq. (\ref{Eqn: ininal}), the interaction $f$ between $X$ and $Y$
must be such as to give $X$ the smallest possible topology of $f$-saturated
sets and $Y$ the largest possible topology of images of all these sets: $f$,
under these conditions, is an ininal transformation. Observe that a direct application
of parts (b) of Theorems A2.1 and A2.2 to Fig. \ref{Fig: GenInv} implies that
Eq. (\ref{Eqn: ininal}) is satisfied iff $f_{\textrm{B}}$ is ininal, that
is iff it is a homeomorphism. Ininality of $f$ is simply a reflection of this
as it is neither $1:1$ nor onto. 

The $f$- and $p$-images of each saturated set of $X$ are singletons in $Y$
(these saturated sets in $X$ arose, in the first place, as $f^{-}(\{ y\})$
for $y\in Y$) and in $X_{\textrm{B}}$ respectively. This permits the embedding
$e=j\circ f_{\textrm{B}}$ to give $X_{\textrm{B}}$ the character of a virtual
subspace of $Y$ just as $i$ makes $f(X)$ a real subspace. Hence the inverse
images $p^{-}(x_{r})=f^{-}(e(x_{r}))$ with $x_{r}\in X_{\textrm{B}}$, and
$q^{-}(y)=f^{-}(i(y))$ with $y=f_{\textrm{B}}(x_{r})\in f(X)$ are the same,
and are just the corresponding $f^{-}$ images via the injections $e$ and $i$
respectively. $G$, a left inverse of $e$, is a generalized inverse of $f$.
$G$ is a generalized inverse because the two set-theoretic defining requirements
of $fGf=f$ and $GfG=G$ for the generalized inverse are satisfied, as Fig.
\ref{Fig: GenInv} shows, in the following forms \[
jf_{\textrm{B}}Gf=f\qquad Gjf_{\textrm{B}}G=G.\]
 In fact the commutativity embodied in these equalities is self evident from
the fact that $e=if_{\textrm{B}}$ is a left inverse of $G$, that is $eG=\bold1_{Y}$.
On putting back $X_{\textrm{B}}$ into $X$ by identifying each point of $X_{\textrm{B }}$
with the set it came from yields the required set-valued inverse $f^{-}$, and
$G$ may be viewed as a functional selection of the multiinverse $f^{-}$. 

An \emph{injective branch} of a function $f$ in this work refers to the restrictions
$f_{\mathrm{B}}$ and its associated inverse $f_{\mathrm{B}}^{-1}$. 

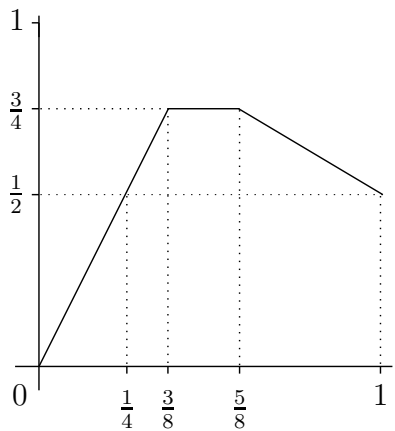
\begin{figure}[htbp]
\noindent \begin{center}{\small \input{gen-inv.pstex_t}}\end{center}{\small \par}

\begin{singlespace}

\caption{{\footnotesize \label{Fig: gen-inv}The function $f(x)=\left\{ \protect\begin{array}{lcl}
2x &  & 0\leq x<3/8\protect\\
3/4 &  & 3/8\leq x\leq5/8\protect\\
7/6-2x/3 &  & 5/8<x\leq1.\protect\end{array}\right.$}}\end{singlespace}

\end{figure}

\smallskip{}
The following example of an inverse ill-posed problem will be useful in fixing
the notations introduced above. Let $f$ on $[0,1]$ be the function of \ref{Fig: gen-inv}.

{\small }%
{\small \par}

\noindent Then $f(x)=y$ is well-posed for $[0,1/4)$, and ill-posed in {[}1/4,1{]}.
There are two injective branches of $f$ in $\{[1/4,3/8)$$\bigcup$ $(5/8,1]\}$,
and $f$ is constant ill-posed in $[3/8,5/8]$. Hence the basic component $f_{\textrm{B}}$
of $f$ can be taken to be $f_{\textrm{B}}(x)=2x$ for $x\in[0,3/8)$ having
the inverse $f_{\textrm{B}}^{-1}(y)=x/2$ with $y\in[0,3/4]$. The generalized
inverse is obtained by taking $[0,3/4]$ as a subspace of $[0,1]$, while the
multiinverse $f^{-}$ follows by associating with every point of the basic domain
$[0,1]_{\textrm{B}}=[0,3/8]$, the respective equivalent points $[3/8]_{f}=[3/8,5/8]$
and $[x]_{f}=\{ x,7/4-3x\}\textrm{ for }x\in[1/4,3/8)$. Thus the inverses $G$
and $f^{-}$ of $f$ are%
\footnote{{\small \label{Foot: 0=3Dphi}If $y\notin\mathcal{R}(f)$ then $f^{-}(\{ y\}):=\emptyset$
which is true for any subset of $Y-\mathcal{R}(f)$. However from the set-theoretic
definition of natural numbers that requires $0:=\emptyset$, $1=\{0\}$, $2=\{0,1\}$
to be defined recursively, it follows that $f^{-}(y)$ can be identified with
$0$ whenever $y$ is not in the domain of $f^{-}$. Formally, the successor
set $A^{+}=A\bigcup\{ A\}$ of $A$ can be used to write $0:=\emptyset$, $1=0^{+}=0\bigcup\{0\}$,
$2=1^{+}=1\bigcup\{1\}=\{0\}\bigcup\{1\}$ $3=2^{+}=2\bigcup\{2\}=\{0\}\bigcup\{1\}\bigcup\{2\}$
etc. Then the set of natural numbers $\mathbb{N}$ is defined to be the intersection
of all the successor sets, where a successor set $\mathcal{S}$ is any set that
contains $\emptyset$ and $A^{+}$ whenever $A$ belongs to $\mathcal{S}$.
Observe how in the successor notation, countable union of singleton integers
recursively define the corresponding sum of integers. }%
}

\noindent \[
G(y)=\left\{ \begin{array}{ccl}
y/2, &  & y\in[0,3/4]\\
0, &  & y\in(3/4,1]\end{array}\right.,\quad f^{-}(y)=\left\{ \begin{array}{ccl}
y/2, &  & y\in[0,1/2)\\
\{ y/2,7/4-3y/2\}, &  & y\in[1/2,3/4)\\
{}[3/8,5/8], &  & y=3/4\\
0, &  & y\in(3/4,1],\end{array}\right.\]

\noindent which shows that $f^{-}$ is multivalued. In order to avoid cumbersome
notations, an injective branch of $f$ will always refer to a representative
basic branch $f_{\textrm{B}}$, and its {}``inverse'' will mean either $f_{\textrm{B}}^{-1}$
or $G$. 

\medskip{}
\noindent \textbf{Example 2.3, Revisited.} The row reduced echelon form of the
augmented matrix $(A|b)$ of Example 2.3 is

\noindent \begin{equation}
{\displaystyle (A|b)\longrightarrow\left(\begin{array}{rrrrrcl}
1 & -3 & 0 & 3/2 & 1/2 &  & 5b_{1}/2-b_{2}/2\\
0 & 0 & 1 & -1/4 & 3/4 &  & -3b_{1}/4+b_{2}/4\\
0 & 0 & 0 & 0 & 0 &  & -2b_{1}+b_{3}\\
0 & 0 & 0 & 0 & 0 &  & b_{1}-b_{2}+b_{4}\end{array}\right)}\label{Eqn: RowReduce}\end{equation}

\noindent The multifunctional solution $x=A^{-}b$, with $b$ any element of
$Y=\mathbb{R}^{4}$ not necessarily in the of image of $a$, is\[
x=A^{-}b=Gb+x_{2}\left(\begin{array}{c}
3\\
1\\
0\\
0\\
0\end{array}\right)+x_{4}\left(\begin{array}{r}
-3/2\\
0\\
1/4\\
1\\
0\end{array}\right)+x_{5}\left(\begin{array}{r}
-1/2\\
0\\
-3/4\\
0\\
1\end{array}\right),\]

\noindent with its multifunctional character arising from the arbitrariness
of the coefficients $x_{2}$, $x_{4},$ and $x_{5}$. The generalized inverse
\begin{equation}
G=\left(\begin{array}{rrrr}
5/2 & -1/2 & 0 & 0\\
0 & 0 & 0 & 0\\
-3/4 & 1/4 & 0 & 0\\
0 & 0 & 0 & 0\\
0 & 0 & 0 & 0\end{array}\right)\!:Y\rightarrow X_{\textrm{B}}\label{Eqn: GenInvEx5}\end{equation}
 is the unique matrix representation of the functional inverse $a_{\textrm{B}}^{-1}\!:a(\mathbb{R}^{5})\rightarrow X_{\textrm{B}}$
extended to $Y$ defined according to%
\footnote{{\small See footnote \ref{Foot: 0=3Dphi} for a justification of the definition
when $b$ is not in $\mathcal{R}(a)$.}%
} \begin{equation}
g(b)=\left\{ \begin{array}{ccl}
a_{\textrm{B}}^{-1}(b), &  & \textrm{ if }b\in\mathcal{R}(a)\\
0, &  & \textrm{ if }b\in Y-\mathcal{R}(a),\end{array}\right.\label{Eqn: Def: GenInv}\end{equation}
that bears comparison with the basic inverse \[
A_{\textrm{B}}^{-1}(b^{*})=\left(\begin{array}{rrrr}
5/2 & -1/2 & 0 & 0\\
0 & 0 & 0 & 0\\
-3/4 & 1/4 & 0 & 0\\
0 & 0 & 0 & 0\\
0 & 0 & 0 & 0\end{array}\right)\left(\begin{array}{c}
b_{1}\\
b_{2}\\
2b_{1}\\
b_{2}-b_{1}\end{array}\right)\!:a(\mathbb{R}^{5})\rightarrow X_{\textrm{B}}\]
 between the $2$-dimensional column and row spaces of $A$ which is responsible
for the particular solution of $Ax=b$. Thus $G$ is simply $A_{\textrm{B}}^{-1}$
acting on its domain $a(X)$ considered a subspace of $Y$, suitably extended
to the whole of $Y$. That it is indeed a generalized inverse is readily seen
through the matrix multiplications $GAG$ and $AGA$ that can be verified to
reproduce $G$ and $A$ respectively. Comparison of Eqs. (\ref{Eqn: Def: Moore-Penrose})
and (\ref{Eqn: Def: GenInv}) shows that the Moore-Penrose inverse differs from
ours through the geometrical constraints imposed in its definition, Eqs. (\ref{Eqn: MPInverse}).
Of course, this results in a more complex inverse (\ref{Eqn: MPEx5}) as compared
to our very simple (\ref{Eqn: GenInvEx5}); nevertheless it is true that both
the inverses satisfy \begin{eqnarray*}
E((E(G_{\textrm{MP}}))^{\textrm{T}}) & = & \left(\begin{array}{ccccc}
1 & 0 & 0 & 0 & 0\\
0 & 1 & 0 & 0 & 0\\
0 & 0 & 0 & 0 & 0\\
0 & 0 & 0 & 0 & 0\end{array}\right)\\
\\ & = & E((E(G))^{\textrm{T}})\end{eqnarray*}

\noindent where $E(A)$ is the row-reduced echelon form of $A$. The canonical
simplicity of Eq. (\ref{Eqn: GenInvEx5}) as compared to Eq. (\ref{Eqn: MPEx5})
is a general feature that suggests a more natural choice of bases by the map
$a$ than the orthogonal set imposed by Moore and Penrose. This is to be expected
since the MP inverse, governed by Eq. (\ref{Eqn: MPInverse}), is a subset of
our less restricted inverse described by only the first two of (\ref{Eqn: MPInverse});
more specifically the difference is made clear in Fig. \ref{Fig: MP_Inverse}(a)
which shows that for any $b\notin\mathcal{R}(A)$, only $G_{\textrm{MP}}(b_{\bot})=0$
as compared to $G(b)=0$. This seems to imply that introducing extraneous topological
considerations into the purely set theoretic inversion process may not be a
recommended way of inverting, and the simple bases comprising the row and null
spaces of $A$ and $A^{\textrm{T}}$ --- that are mutually orthogonal just as
those of the Moore-Penrose --- are a better choice for the particular problem
$Ax=b$ than the general orthonormal bases that the MP inverse introduces. These
{}``good'' bases, with respect to which the generalized inverse $G$ has a
considerably simpler representation, are obtained in a straight forward manner
from the row-reduced forms of $A$ and $A^{\textrm{T}}$. These bases are 

\smallskip{}
(a) The column space of $A$ is spanned by the columns $(1,\textrm{ }3,\textrm{ }2,\textrm{ }2)^{\textrm{T}}$
and $(1,\textrm{ }5,\textrm{ }2,\textrm{ }4)^{\textrm{T}}$ of $A$ that correspond
to the basic columns containing the leading $1$'s in the row-reduced form of
$A$, 

(b) The null space of $A^{\textrm{T}}$ is spanned by the solutions $(-2,\textrm{ }0,\textrm{ }1,\textrm{ }0)^{\textrm{T}}$
and $(1,-1,\textrm{ }0,\textrm{ }1)^{\textrm{T}}$ of the equation $A^{\textrm{T}}b=0$, 

(c) The row space of $A$ is spanned by the rows $(1,-3,\textrm{ }2,\textrm{ }1,\textrm{ }2)$
and $(3,-9,\textrm{ }10,\textrm{ }2,\textrm{ }9)$ of $A$ corresponding to
the non-zero rows in the row-reduced form of $A$, 

(d) The null space of $A$ is spanned by the solutions $(3,\textrm{ }1,\textrm{ }0,\textrm{ }0,\textrm{ }0)$,
$(-6,\textrm{ }0,\textrm{ }1,\textrm{ }4,\textrm{ }0)$, and $(-2,\textrm{ }0,-3,\textrm{ }0,\textrm{ }4)$
of the equation $Ax=0$.$\qquad\blacksquare$
\medskip{}

The main differences between the natural {}``good'' bases and the MP-bases
that are responsible for the difference in form of the inverses, is that the
later have the additional restrictions of being orthogonal to each other (recall
the orthogonality property of the $Q$-matrices), and the more severe one of
basis vectors mapping onto basis vectors according to $Ax_{i}=\sigma_{i}b_{i}$,
$i=1,\cdots,r$, where the $\{ x_{i}\}_{i=1}^{n}$ and $\{ b_{j}\}_{j=1}^{m}$
are the eigenvectors of $A^{\textrm{T}}A$ and $AA^{\textrm{T}}$ respectively
and $(\sigma_{i})_{i=1}^{r}$ are the positive square roots of the non-zero
eigenvalues of $A^{\textrm{T}}A$ (or of $AA^{\textrm{T}}$), with $r$ denoting
the dimension of the row or column space. This is considered as a serious restriction
as the linear combination of the basis $\{ b_{j}\}$ that $Ax_{i}$ should otherwise
have been equal to, allows a greater flexibility in the matrix representation
of the inverse that shows up in the structure of $G$. These are, in fact, quite
general considerations in the matrix representation of linear operators; thus
the basis that diagonalizes an $n\times n$ matrix (when this is possible) is
not the standard {}``diagonal'' orthonormal basis of $\mathbb{R}^{n}$, but
a problem-dependent, less canonical, basis consisting of the $n$ eigenvectors
of the matrix. The $0$-rows of the inverse of Eq. (\ref{Eqn: GenInvEx5}) result
from the $3$-dimensional null-space variables $x_{2}$, $x_{4}$, and $x_{5}$,
while the $0$-columns come from the $2$-dimensional image-space dependency
of $b_{3}$, $b_{4}$ on $b_{1}$ and $b_{2}$, that is from the last two zero
rows of the reduced echelon form (\ref{Eqn: RowReduce}) of the augmented matrix. 

We will return to this theme of the generation of a most appropriate problem-dependent
topology for a given space in the more general context of chaos in Sec. 4.2. 

In concluding this introduction to generalized inverses we note that the inverse
$G$ of $f$ comes very close to being a right inverse: thus even though $AG\not\neq\bold1_{2}$
its row-reduced form \[
\left(\begin{array}{cccc}
1 & 0 & 0 & 0\\
0 & 1 & 0 & 0\\
0 & 0 & 0 & 0\\
0 & 0 & 0 & 0\end{array}\right)\]

\noindent is to be compared with the corresponding less satisfactory \[
\left(\begin{array}{cccr}
1 & 0 & 2 & -1\\
0 & 1 & 0 & 1\\
0 & 0 & 0 & 0\\
0 & 0 & 0 & 0\end{array}\right)\]

\noindent representation of $AG_{\textrm{MP}}$. 

\vspace{1cm}
\noindent \begin{flushleft}\textbf{\large 3. Multifunctional extension of function
spaces}\end{flushleft}{\large \par}

\noindent The previous section has considered the solution of ill-posed problems
as multifunctions and has shown how this solution may be constructed. Here we
introduce the multifunction space $\textrm{Multi}_{\mid}(X)$ as the first step
toward obtaining a smallest dense extension $\textrm{Multi}(X)$ of the function
space $\textrm{Map}(X)$. $\textrm{Multi}_{\mid}(X)$ is basic to our theory
of chaos \cite{Sengupta2000} in the sense that a chaotic state of a system
can be fully described by such an indeterminate multifunctional state. In fact,
multifunctions also enter in a natural way in describing the spectrum of nonlinear
functions that we consider in Section 6; this is required to complete the construction
of the smallest extension $\textrm{Multi}(X)$ of the function space $\textrm{Map}(X)$.
The main tool in obtaining the space $\textrm{Multi}_{\mid}(X)$ from $\textrm{Map}(X)$
is a generalization of the technique of pointwise convergence of continuous
functions to (discontinuous) functions. In the analysis below, we consider nets
instead of sequences as the spaces concerned, like the topology of pointwise
convergence, may not be first countable, Appendix A1. 

\bigskip{}
\noindent \begin{flushleft}\textbf{\emph{\large 3.1. Graphical convergence of
a net of functions}}\end{flushleft}{\large \par}

\noindent Let $(X,\mathcal{U})$ and $(Y,\mathcal{V})$ be Hausdorff spaces
and $(f_{\alpha})_{\alpha\in\mathbb{D}}:X\rightarrow Y$ be a net of piecewise
continuous functions, not necessarily with the same domain or range, and suppose
that for each $\alpha\in\mathbb{D}$ there is a finite set $I_{\alpha}=\{1,2,\cdots P_{\alpha}\}$
such that $f_{\alpha}^{-}$ has $P_{\alpha}$ functional branches possibly with
different domains; obviously $I_{\alpha}$ is a singleton iff $f$ is a injective.
For each $\alpha\in\mathbb{D}$, define functions $(g_{\alpha i})_{i\in I_{\alpha}}\!:Y\rightarrow X$
such that \[
f_{\alpha}g_{\alpha i}f_{\alpha}=f_{\alpha i}^{I}\qquad i=1,2,\cdots P_{\alpha,}\]

\noindent where $f_{\alpha i}^{I}$ is a basic injective branch of $f_{\alpha}$
on some subset of its domain: $g_{\alpha i}f_{\alpha i}^{I}=1_{X}$ on $\mathcal{D}(f_{\alpha i}^{I})$,
$f_{\alpha i}^{I}g_{\alpha i}=1_{Y}$ on $\mathcal{D}(g_{\alpha i})$ for each
$i\in I_{\alpha}$. The use of nets and filters is dictated by the fact that
we do not assume $X$ and $Y$ to be first countable. In the application to
the theory of dynamical systems that follows, $X$ and $Y$ are compact subsets
of $\mathbb{R}$ when the use of sequences suffice. 

In terms of the residual and cofinal subsets $\textrm{Res}(\mathbb{D})$ and
$\textrm{Cof}(\mathbb{D})$ of a directed set $\mathbb{D}$ (Def. A1.7), with
$x$ and $y$ in the equations below being taken to belong to the required domains,
define subsets $\mathcal{D}_{-}$ of $X$ and $\mathcal{R}_{-}$ of $Y$ as
\begin{equation}
\mathcal{D}_{-}=\{ x\in X\!:((f_{\nu}(x))_{\nu\in\mathbb{D}}\textrm{ converges in }(Y,\mathcal{V}))\}\label{Eqn: D-}\end{equation}
\begin{equation}
\mathcal{R}_{-}=\{ y\in Y\!:\textrm{ }(\exists i\in I_{\nu})((g_{\nu i}(y))_{\nu\in\mathbb{D}}\textrm{ converges in }(X,\mathcal{U}))\}\label{Eqn: R-}\end{equation}

\noindent Thus:

$\mathcal{D}_{-}$ is the set of points of $X$ on which the values of a given
net of functions $(f_{\alpha})_{\alpha\in\mathbb{D}}$ converge pointwise in
$Y$. Explicitly, this is the subset of $X$ on which subnets%
\footnote{{\small \label{Foot: subnet}A subnet is the generalized uncountable equivalent
of a subsequence; for the technical definition, see Appendix A1. }%
} in $\textrm{Map}(X,Y)$ combine to form a net of functions that converge pointwise
to a limit function $F:\mathcal{D}_{-}\rightarrow Y$.

$\mathcal{R}_{-}$ is the set of points of $Y$ on which the values of the nets
in $X$ generated by the injective branches of $(f_{\alpha})_{\alpha\in\mathbb{D}}$
converge pointwise in $Y$. Explicitly, this is the subset of $Y$ on which
subnets of injective branches of $(f_{\alpha})_{\alpha\in\mathbb{D}}$ in $\textrm{Map}(Y,X)$
combine to form a net of functions that converge pointwise to a family of limit
functions $G:\mathcal{R}_{-}\rightarrow X$. Depending on the nature of $(f_{\alpha})_{\alpha\in\mathbb{D}}$,
there may be more than one $\mathcal{R}_{-}$ with a corresponding family of
limit functions on each of them. To simplify the notation, we will usually let
$G:\mathcal{R}_{-}\rightarrow X$ denote all the limit functions on all the
sets $\mathcal{R}_{-}$. 

If we consider cofinal rather than residual subsets of $\mathbb{D}$ then corresponding
$\mathbb{D}_{+}$ and $\mathbb{R}_{+}$ can be expressed as \begin{equation}
\mathcal{D}_{+}=\{ x\in X\!:((f_{\nu}(x))_{\nu\in\textrm{Cof}(\mathbb{D})}\textrm{ converges in }(Y,\mathcal{V}))\}\label{Eqn: D+}\end{equation}
\begin{equation}
\mathcal{R}_{+}=\{ y\in Y\!:(\exists i\in I_{\nu})((g_{\nu i}(y))_{\nu\in\textrm{Cof}(\mathbb{D})}\textrm{ converges in }(X,\mathcal{U}))\}.\label{Eqn: R+}\end{equation}

\noindent It is to be noted that the conditions $\mathcal{D}_{+}=\mathcal{D}_{-}$
and $\mathcal{R}_{+}=\mathcal{R}_{-}$ are necessary and sufficient for the
Kuratowski convergence to exist. Since $\mathcal{D}_{+}$ and $\mathcal{R}_{+}$
differ from $\mathcal{D}_{-}$ and $\mathcal{R}_{-}$ only in having cofinal
subsets of $D$ replaced by residual ones, and since residual sets are also
cofinal, it follows that $\mathcal{D}_{-}\subseteq\mathcal{D}_{+}$ and $\mathcal{R}_{-}\subseteq\mathcal{R}_{+}$.
The sets $\mathcal{D}_{-}$ and $\mathcal{R}_{-}$ serve for the convergence
of a net of functions just as $\mathcal{D}_{+}$ and $\mathcal{R}_{+}$ are
for the convergence of subnets of the nets (\emph{adherence}). The later sets
are needed when subsequences are to be considered as sequences in their own
right as, for example, in dynamical systems theory in the case of $\omega$-limit
sets. 

As an illustration of these definitions, consider the sequence of injective
functions on the interval $[0,1]$ $f_{n}(x)=2^{n}x$, for $x\in\left[0,1/2^{n}\right],\textrm{ }n=0,1,2\cdots$.
Then $\mathbb{D}_{0.2}$ is the set $\{0,1,2\}$ and only $\mathbb{D}_{0}$
is eventual in $\mathbb{D}$. Hence $\mathcal{D}_{-}$ is the single point set
\{0\}. On the other hand $\mathbb{D}_{y}$ is eventual in $\mathbb{D}$ for
all $y$ and $\mathcal{R}_{-}$ is $[0,1]$. 

\medskip{}
\noindent \textbf{Definition} \textbf{\noun{3.1}}\textsl{\noun{.}} \textbf{\textit{Graphical
Convergence of a net of functions.}} \textsl{A net of functions $(f_{\alpha})_{\alpha\in D}\!:(X,\mathcal{U})\rightarrow(Y,\mathcal{V})$
is said to} \emph{converge graphically} \textsl{if either $\mathcal{D}_{-}\neq\emptyset$
or $\mathcal{R}_{-}\neq\emptyset$; in this case let $F\!:\mathcal{D}_{-}\rightarrow Y$
and $G:\mathcal{R}_{-}\rightarrow X$ be the entire collection of limit functions.
Because of the assumed Hausdorffness of $X$ and $Y$, these limits are well
defined. }

\textsl{The graph of the} \emph{graphical limit} $\mathscr{M}$ \emph{of the
net} $(f_{\alpha})\!:(X,\mathcal{U})\rightarrow(Y,\mathcal{V})$ \textsl{\emph{}}\textsl{denoted
by} \textsl{\emph{}}\textsl{$f_{\alpha}\overset{\mathbf{G}}\longrightarrow\mathscr{M}$,
is the subset of $\mathcal{D}_{-}\times\mathcal{R}_{-}$that is the union of
the graphs of the function $F$ and the multifunction $G^{-}$ \[
\mathbf{G}_{\mathscr{M}}=\mathbf{G}_{F}\bigcup\mathbf{G}_{G^{-}}\]
}

\noindent \textsl{where \[
\mathbf{G}_{G^{-}}=\{(x,y)\in X\times Y\!:(y,x)\in\mathbf{G}_{G}\subseteq Y\times X\}.\qquad\square\]
}

\noindent \begin{flushright}\textbf{\textit{Begin Tutorial6: Graphical Convergence}}\end{flushright}
\vspace{-0.15cm}

\noindent The following two examples are basic to the understanding of the graphical
convergence of functions to multifunctions and were the examples that motivated
our search of an acceptable technique that did not require vertical portions
of limit relations to disappear simply because they were non-functions: the
disturbing question that needed an answer was how not to mathematically sacrifice
these extremely significant physical components of the limiting correspondences.
Furthermore, it appears to be quite plausible to expect a physical interaction
between two spaces $X$ and $Y$ to be a consequence of both the direct interaction
represented by $f\!:X\rightarrow Y$ and also the inverse interaction $f^{-}\!:Y\rightarrow X$,
and our formulation of pointwise biconvergence is a formalization of this idea.
Thus the basic examples (1) and (2) below produce multifunctions instead of
discontinuous functions that would be obtained by the usual pointwise limit. 

\medskip{}
\noindent \textbf{Example 3.1.} (1)

\noindent \renewcommand{\arraystretch}{1.2}

\noindent \[
f_{n}(x)=\left\{ \begin{array}{lc}
0 & -1\leq x\leq0\\
nx & 0\leq x\leq1/n\\
1 & 1/n\leq x\leq1\end{array}\right.:\quad[-1,1]\rightarrow[0,1]\]
\[
g_{n}(y)=y/n:\quad[0,1]\rightarrow[0,1/n]\]

\noindent Then\[
F(x)=\left\{ \begin{array}{cc}
0 & -1\leq x\leq0\\
1 & 0<x\leq1\end{array}\right.\qquad\mathrm{on}\qquad\mathcal{D}_{-}=\mathcal{D}_{+}=[-1,0]\bigcup(0,1]\]

\renewcommand{\arraystretch}{1}

\noindent \[
G(y)=0\quad\mathrm{on}\quad\mathcal{R}_{-}=[0,1]=\mathcal{R}_{+}.\]

\noindent The graphical limit is $([-1,0],0)\bigcup(0,[0,1])\bigcup((0,1],1)$.

(2) $f_{n}(x)=nx$ for $x\in[0,1/n]$ gives $g_{n}(y)=y/n:[0,1]\rightarrow[0,1/n].$
Then \[
F(x)=0\quad\mathrm{on}\quad\mathcal{D}_{-}=\{0\}=\mathcal{D}_{+},\qquad G(y)=0\quad\mathrm{on}\quad\mathcal{R}_{-}=[0,1]=\mathcal{R}_{+}.\]

\noindent The graphical limit is $(0,[0,1])$.$\qquad\blacksquare$
\medskip{}

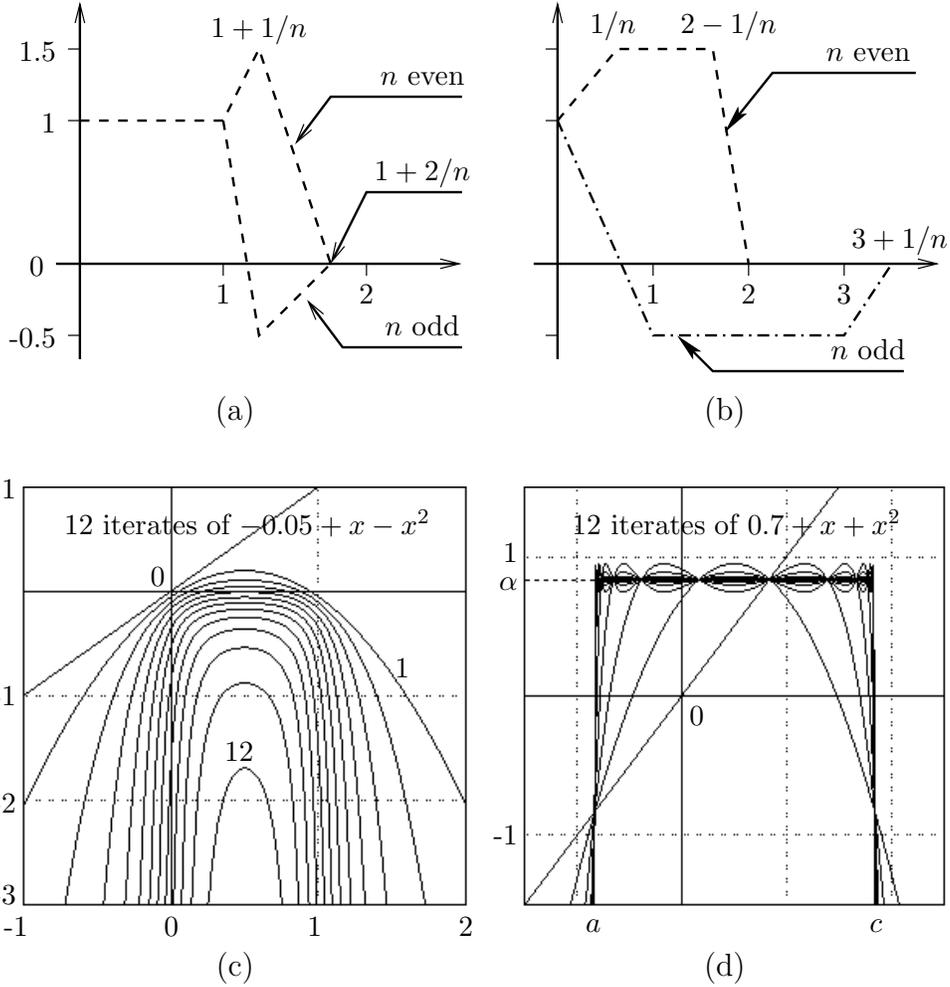
\begin{figure}[htbp]
\noindent \begin{center}{\small \input{Example2_1.pstex_t}}\end{center}{\small \par}

\begin{singlespace}

\caption{{\footnotesize \label{Fig: Example2_1}The graphical limits are: (a) $F(x)=\left\{ \protect\begin{array}{ccc}
1 & \textrm{for }0\leq x\leq1\textrm{ }\protect\\
0 & \textrm{for }1<x\leq2\protect\end{array}\right.$ on $\mathcal{D}_{-}=[0,1]\bigcup(1,2]$, and $G(y)=1$ on $\mathcal{R}_{-}=[0,1]$.
Also $G=\left\{ \protect\begin{array}{cl}
1 & \textrm{on }\mathcal{R}_{+}=[0,3/2]\protect\\
1 & \textrm{on }\mathcal{R}_{+}=[-1/2,1]\protect\end{array}\right.$.}}
\end{singlespace}

\begin{singlespace}
{\footnotesize $\quad$(b) $F(x)=1$ on $\mathcal{D}_{-}=\{0\}$ and $G(y)=0$
on $\mathcal{R}_{-}=\{1\}$. Also $F(x)=-1/2,\textrm{ }0,\textrm{ }1,\textrm{ }3/2$
respectively on $\mathcal{D}_{+}=(0,3],\textrm{ }\{2\},\textrm{ }\{0\},\textrm{ }(0,2)$
and $G(y)=0,\textrm{ }0,\textrm{ }2,\textrm{ }3$ respectively on $\mathcal{R}_{+}=(-1/2,1],\textrm{ }[1,3/2),\textrm{ }[0,3/2),\textrm{ }[-1/2,0)$. }{\footnotesize \par}

{\footnotesize $\quad$(c) For $f(x)=-0.05+x-x^{2}$, no graphical limit as
$\mathcal{D}_{-}=\emptyset=\mathcal{R}_{-}$.}{\footnotesize \par}
\end{singlespace}

\begin{singlespace}
{\footnotesize $\quad$(d) For $f(x)=0.7+x-x^{2}$, $F(x)=\alpha$ on $\mathcal{D}_{-}=[a,c]$,
$G_{1}(y)=a$ and $G_{2}(y)=c$ on $\mathcal{R}_{-}=(-\infty,\alpha]$. Notice
how the two fixed points and their equivalent images define the converged limit
rectangular multi. As in example (1) one has $\mathcal{D}_{-}=\mathcal{D}_{+}$;
also $\mathcal{R}_{-}=\mathcal{R}_{+}$.}\vspace{-0.2cm}
\end{singlespace}

\end{figure}

\begin{spacing}{1.4}
In these examples that we consider to be the prototypes of graphical convergence
of functions to multifunctions, $G(y)=0$ on $\mathcal{R}_{-}$ because $g_{n}(y)\rightarrow0$
for all $y\in\mathcal{R}_{-}$. Compare the graphical multifunctional limits
with the corresponding usual pointwise functional limits characterized by discontinuity
at $x=0$. Two more examples from \citet*{Sengupta2000} that illustrate this
new convergence principle tailored specifically to capture one-to-many relations
are shown in Fig. \ref{Fig: Example2_1} which also provides an example in Fig.
\ref{Fig: Example2_1}(c) of a function whose iterates do not converge graphically
because in this case both the sets $\mathcal{D}_{-}$ and $\mathcal{R}_{-}$are
empty. The power of graphical convergence in capturing multifunctional limits
is further demonstrated by the example of the sequence $(\sin n\pi x)_{n=1}^{\infty}$
that converges to $0$ both $1$-integrally and test-functionally, Eqs. (\ref{Eqn: intsin})
and (\ref{Eqn: testsin}). {\small }{\small \par}
\end{spacing}

{\small }%
{\small \par}

It is necessary to understand how the concepts of \emph{eventually in} and \emph{frequently
in} of Appendix A2 apply in examples (a) and (b) of Fig. {\small \ref{Fig: Example2_1}.}
In these two examples we have two subsequences one each for the even indices
and the other for the odd. For a point-to-point functional relation, this would
mean that the sequence frequents the adherence set $\textrm{adh}(x)$ of the
sequence $(x_{n})$ but does not converge anywhere as it is not eventually in
every neighbourhood of any point. For a multifunctional limit however it is
possible, as demonstrated by these examples, for the subsequences to be eventually
in every neighbourhood of certain \emph{subsets} common to the eventual limiting
sets of the subsequences; this intersection of the subsequential limits is now
\emph{defined to be the limit of the original sequence.} A similar situation
obtains, for example, in the solution of simultaneous equations: The solution
of the equation $a_{11}x_{1}+a_{12}x_{2}=b_{1}$ for one of the variables $x_{2}$
say with $a_{12}\neq0$, is the set represented by the straight line $x_{2}=m_{1}x_{1}+c_{1}$
for all $x_{1}$ in its domain, while for a different set of constants $a_{21}$,
$a_{22}$ and $b_{2}$ the solution is the entirely different set $x_{2}=m_{2}x_{1}+c_{2}$,
under the assumption that $m_{1}\neq m_{2}$ and $c_{1}\neq c_{2}$. Thus even
though the individual equations (subsequences) of the simultaneous set of equations
(sequence) may have distinct solutions (limits), the solution of the equations
is their common point of intersection. 

Considered as sets in $X\times Y$, the discussion of convergence of a sequence
of graphs $f_{n}\!:X\rightarrow Y$ would be incomplete without a mention of
the convergence of a sequence of sets under the Hausdorff metric that is so
basic in the study of fractals. In this case, one talks about the convergence
of a sequence of compact subsets of the metric space $\mathbb{R}^{n}$ so that
the sequences, as also the limit points that are the fractals, are compact subsets
of $\mathbb{R}^{n}$. Let $\mathcal{K}$ denote the collection of all nonempty
compact subsets of $\mathbb{R}^{n}$. Then the \emph{Hausdorff metric} $d_{\textrm{H}}$
between two sets on $\mathcal{K}$ is defined to be \[
d_{\textrm{H}}(E,F)=\max\{\delta(E,F),\delta(F,E)\}\qquad E,F\in\mathcal{K},\]

\noindent where \[
\delta(E,F)=\max_{x\in E}\textrm{ }\min_{y\in F}\Vert\mathbf{x-y}\Vert_{2}\]

\noindent is $\delta(E,F)$ is the non-symmetric $2$-norm in $\mathbb{R}^{n}$.
The power and utility of the Hausdorff distance is best understood in terms
of the dilations $E+\varepsilon:=\bigcup_{x\in E}D_{\varepsilon}(x)$ of a subset
$E$ of $\mathbb{R}^{n}$ by $\varepsilon$ where $D_{\varepsilon}(x)$ is a
closed ball of radius $\varepsilon$ at $x$; physically a dilation of $E$
by $\varepsilon$ is a closed $\varepsilon$-neighbourhood of $E$. Then a fundamental
property of $d_{\textrm{H}}$ is that $d_{\textrm{H}}(E,F)\leq\varepsilon$
iff both $E\subseteq F+\varepsilon$ and $F\subseteq E+\varepsilon$ hold simultaneously
which leads \cite{Falconer1990} to the interesting consequence that 

\medskip{}
\textsl{If $(F_{n})_{n=1}^{\infty}$ and $F$ are nonempty compact sets, then
$\lim_{n\rightarrow\infty}F_{n}=F$ in the Hausdorff metric iff $F_{n}\subseteq F+\varepsilon$
and $F\subseteq F_{n}+\varepsilon$ eventually. Furthermore if $(F_{n})_{n=1}^{\infty}$
is a decreasing sequence of elements of a filter-base in $\mathbb{R}^{n}$,
then the nonempty and compact limit set $F$ is given by \[
\lim_{n\rightarrow\infty}F_{n}=F=\bigcap_{n=1}^{\infty}F_{n}.\]
} Note that since $\mathbb{R}^{n}$ is Hausdorff, the assumed compactness of
$F_{n}$ ensures that they are also closed in \textsl{}$\mathbb{R}^{n}$; $F$,
therefore, is just the adherent set of the filter-base. In the deterministic
algorithm for the generation of fractals by the so-called iterated function
system (IFS) approach, $F_{n}$ is the inverse image by the $n^{\textrm{th}}\textrm{ }$
iterate of a non-injective function $f$ having a finite number of injective
branches and converging graphically to a multifunction. Under the conditions
stated above, the Hausdorff metric ensures convergence of any class of compact
subsets in $\mathbb{R}^{n}$. It appears eminently plausible that our multifunctional
graphical convergence on $\textrm{Map}(\mathbb{R}^{n})$ implies Hausdorff convergence
on $\mathbb{R}^{n}$: in fact pointwise biconvergence involves simultaneous
convergence of image and preimage nets on $Y$ and $X$ respectively. Thus confining
ourselves to the simpler case of pointwise convergence, if $(f_{\alpha})_{\alpha\in\mathbb{D}}$
is a net of functions in $\textrm{Map}(X,Y)$, then the following theorem expresses
the link between convergence in $\textrm{Map}(X,Y)$ and in $Y$. 

\medskip{}
\noindent \textbf{Theorem 3.1.} \textsl{A net of functions} $(f_{\alpha})_{\alpha\in\mathbb{D}}$
\textsl{converges to a function} $f$ \textsl{in} $(\textrm{Map}(X,Y),\mathcal{T})$
\textsl{in the topology of pointwise convergence iff} $(f_{\alpha})$ \textsl{converges
pointwise to $f$ in the sense that $f_{\alpha}(x)\rightarrow f(x)$ in $Y$
for every $x$ in $X$.$\qquad\square$}

\noindent \textbf{Proof.} \emph{Necessity.} First consider $f_{\alpha}\rightarrow f$
in $(\textrm{Map}(X,Y),\mathcal{T})$. For an open neighbourhood $V$ of $f(x)$
in $Y$ with $x\in X$, let $B(x;V)$ be a local neighbourhood of $f$ in $(\textrm{Map}(X,Y),\mathcal{T})$,
see Eq. (\ref{Eqn: point}) in Appendix A1. By assumption of convergence, $(f_{\alpha})$
must eventually be in $B(x;V)$ implying that $f_{\alpha}(x)$ is eventually
in $V$. Hence $f_{\alpha}(x)\rightarrow f(x)$ in $Y$. 

\emph{Sufficiency.} Conversely, if $f_{\alpha}(x)\rightarrow f(x)$ in $Y$
for every $x\in X$, then for a \emph{finite} collection of points $(x_{i})_{i=1}^{I}$
of $X$ ($X$ may itself be uncountable) and corresponding open sets $(V_{i})_{i=1}^{I}$
in $Y$ with $f(x_{i})\in V_{i}$, let $B((x_{i})_{i=1}^{I};(V_{i})_{i=1}^{I})$
be an open neighbourhood of $f$. From the assumed pointwise convergence $f_{\alpha}(x_{i})\rightarrow f(x_{i})$
in $Y$ for $i=1.2.\cdots.I$, it follows that $(f_{\alpha}(x_{i}))$ is eventually
in $V_{i}$ for every $(x_{i})_{i=1}^{I}$. Because $\mathbb{D}$ is a directed
set, the existence of a residual applicable globally for all $i=1,2,\cdots,I$
is assured leading to the conclusion that $f_{\alpha}(x_{i})\in V_{i}$ eventually
for every $i=1,2,\cdots,I$. Hence $f_{\alpha}\in B((x_{i})_{i=1}^{I};(V_{i})_{i=1}^{I})$
eventually; this completes the demonstration that $f_{\alpha}\rightarrow f$
in $(\textrm{Map}(X,Y),\mathcal{T})$, and thus of the proof.$\qquad\blacksquare$

\vspace{-0.15cm}
\noindent \begin{flushright}\textbf{\textit{End Tutorial6}}\end{flushright}

\medskip{}
\noindent \begin{flushleft}\textbf{\emph{\large 3.2. The Extension}} \textbf{\large Multi$_{\mid}$(}\textbf{\emph{\large X,Y}}\textbf{\large )}
\textbf{\emph{\large of}} \textbf{\large Map(}\textbf{\emph{\large X,Y}}\textbf{\large )}\end{flushleft}{\large \par}

\noindent In this Section we show how the topological treatment of pointwise
convergence of functions to functions given in Example A1.1 of Appendix 1 can
be generalized to generate the boundary $\textrm{Multi}_{\mid}(X,Y)$ between
$\textrm{Map}(X,Y)$ and $\textrm{Multi}(X,Y)$; here $X$ and $Y$ are Hausdorff
spaces and $\textrm{Map}(X,Y)$ and $\textrm{Multi}(X,Y)$ are respectively
the sets of all functional and non-functional relations between $X$ and $Y$.
The generalization we seek defines neighbourhoods of $f\in\textrm{Map}(X,Y)$
to consist of those functional relations in $\textrm{Multi}(X,Y)$ whose images
at any point $x\in X$ lies not only arbitrarily close to $f(x)$ (this generates
the usual topology of pointwise convergence $\mathcal{T}_{Y}$ of Example A1.1)
but whose inverse images at $y=f(x)\in Y$ contain points arbitrarily close
to $x$. Thus the graph of $f$ must not only lie close enough to $f(x)$ at
$x$ in $V$, but must additionally be such that $f^{-}(y)$ has at least branch
in $U$ about $x$; thus $f$ is constrained to cling to $f$ as the number
of points on the graph of $f$ increases with convergence and, unlike in the
situation of simple pointwise convergence, no gaps in the graph of the limit
object is permitted not only, as in Example A1.1 on the domain of $f$, but
simultaneously on it range too. We call the resulting generated topology the
\emph{topology of pointwise biconvergence on} $\textrm{Map}(X,Y)$, to be denoted
by $\mathcal{T}$. Thus for any given integer $I\geq1$, the generalization
of Eq. (\ref{Eqn: point}) gives for $i=1,2,\cdots,I$, the open sets of $(\textrm{Map}(X,Y),\mathcal{T})$
to be \begin{multline}
B((x_{i}),(V_{i});(y_{i}),(U_{i}))=\{ g\in\mathrm{Map}(X,Y)\!:\\
(g(x_{i})\in V_{i})\wedge(g^{-}(y_{i})\bigcap U_{i}\neq\emptyset)\textrm{ },i=1,2,\cdots,I\},\label{Eqn: func_bi}\end{multline}

\noindent where $(x_{i})_{i=1}^{I},(V_{i})_{i=1}^{I}$ are as in that example,
$(y_{i})_{i=1}^{I}\in Y$, and the corresponding open sets $(U_{i})_{i=1}^{I}$
in $X$ are chosen arbitrarily%
\footnote{{\small \label{Foot: point_inter}Equation (\ref{Eqn: func_bi}) is essentially
the intersection of the pointwise topologies (\ref{Eqn: point}) due to $f$
and $f^{-}$. }%
}. A local base at $f$, for $(x_{i},y_{i})\in\mathbf{G}_{f}$, is the set of
functions of (\ref{Eqn: func_bi}) with $y_{i}=f(x_{i})$ and the collection
of all local bases \begin{equation}
B_{\alpha}=B((x_{i})_{i=1}^{I_{\alpha}},(V_{i})_{i=1}^{I_{\alpha}};(y_{i})_{i=1}^{I_{\alpha}},(U_{i})_{i=1}^{I_{\alpha}}),\label{Eqn: local_base}\end{equation}
 for every choice of $\alpha\in\mathbb{D}$, is a base $_{\textrm{T}}\mathcal{B}$
of $(\textrm{Map}(X,Y),\mathcal{T})$. Here the directed set $\mathbb{D}$ is
used as an indexing tool because, as pointed out in Example A1.1, the topology
of pointwise convergence is not first countable. 

In a manner similar to Eq. (\ref{Eqn: func_bi}), the open sets of $(\mathrm{Multi}(X,Y),\widehat{\mathcal{T}})$,
where $\textrm{Multi}(X,Y)$ are multifunctions with only countably many values
in $Y$ for every point of $X$ (so that we exclude continuous regions from
our discussion except for the {}``vertical lines'' of $\textrm{Multi}_{\mid}(X,Y)$),
can be defined by \begin{multline}
\widehat{B}((x_{i}),(V_{i});(y_{i}),(U_{i}))=\{\mathscr{G}\in\mathrm{Multi}(X,Y)\!:(\mathscr{G}(x_{i})\bigcap V_{i}\neq\emptyset)\wedge(\mathscr G^{-}(y_{i})\bigcap U_{i}\neq\emptyset)\}\label{Eqn: multi_bi}\end{multline}

\noindent where \[
\mathscr G^{-}(y)=\{ x\in X\!:y\in\mathscr{G}(x)\}.\]

\noindent and $(x_{i})_{i=1}^{I}\in\mathcal{D}(\mathscr{G}),(V_{i})_{i=1}^{I};(y_{i})_{i=1}^{I}\in\mathcal{R}(\mathscr{G}),(U_{i})_{i=1}^{I}$
are chosen as in the above. The topology $\widehat{\mathcal{T}}$ of $\textrm{Multi}(X,Y)$
is generated by the collection of all local bases $\widehat{B_{\alpha}}$ for
every choice of $\alpha\in\mathbb{D}$, and it is not difficult to see from
Eqs. (\ref{Eqn: func_bi}) and (\ref{Eqn: multi_bi}), that the restriction
\textbf{}\textsf{\textbf{$\widehat{\mathcal{T}}\mid_{\mathrm{Map}(X,Y)}$}}
of $\widehat{\mathcal{T}}$ to $\textrm{Map}(X,Y)$ is just $\mathcal{T}$. 

\medskip{}
Henceforth $\widehat{\mathcal{T}}$ and $\mathcal{T}$ will be denoted by the
same symbol $\mathcal{T}$, and convergence in the topology of pointwise biconvergence
in $(\textrm{Multi}(X,Y),\mathcal{T})$ will be denoted by $\rightrightarrows$,
with the notation being derived from Theorem 3.1.

\medskip{}
\noindent \textbf{Definition 3.2.} \textbf{\textit{Functionization of a multifunction.}}
\textsl{A net of functions} $(f_{\alpha})_{\alpha\in\mathbb{D}}$ \textsl{in}
$\textrm{Map}(X,Y)$ \textsl{converges in} $(\textrm{Multi}(X,Y),\mathcal{T})$,
$f_{\alpha}\rightrightarrows\mathscr{M}$, \textsl{if it biconverges pointwise
in} $(\textrm{Map}(X,Y),\mathcal{T}^{*})$. \textsl{Such a net of functions
will be said to be a} \emph{functionization of} $\mathscr{M}$\emph{.$\qquad\square$}

\medskip{}
\noindent \textbf{Theorem 3.2.} \textsl{Let $(f_{\alpha})_{\alpha\in\mathbb{D}}$
be a net of functions in $\textrm{Map}(X,Y)$. Then \[
f_{\alpha}\overset{\mathbf{G}}\longrightarrow\mathscr{M}\Longleftrightarrow f_{\alpha}\rightrightarrows\mathcal{M}.\qquad\square\]
} \textbf{Proof.} If $(f_{\alpha})$ converges graphically to $\mathscr{M}$
then either $\mathcal{D}_{-}$ or $\mathcal{R}_{-}$ is nonempty; let us assume
both of them to be so. Then the sequence of functions $(f_{\alpha})$ converges
pointwise to a function $F$ on $\mathcal{D}_{-}$ and to functions $G$ on
$\mathcal{R}_{-}$, and the local basic neighbourhoods of $F$ and $G$ generate
the topology of pointwise biconvergence. 

Conversely, for pointwise biconvergence on $X$ and $Y$, $\mathcal{R}_{-}$
and $\mathcal{D}_{-}$ must be non-empty.$\qquad\blacksquare$

{\small }%
\begin{figure}[htbp]
\noindent \begin{center}{\small \input{biconv.pstex_t}}\end{center}{\small \par}

\begin{singlespace}

\caption{{\footnotesize \label{Fig: biconv}The power of graphical convergence, illustrated
for Example 3.1 (1), shows a local neighbourhood of the functions $x$ and $2x$
in figures (a) and (b) at the four points $(x_{i})_{i=1}^{4}$ with corresponding
neighbourhoods $(U_{i})_{i=1}^{4}$ and $(V_{i})_{i=1}^{4}$ at $(x_{i},f(x_{i}))$
in $\mathbb{R}$ in the $X$ and $Y$ directions respectively, see Eqs. (\ref{Eqn: func_bi})
and (\ref{Eqn: point}) for the notations. In (a) is shown a function $g$ in
a pointwise neighbourhood of $f$ determined by the open sets $V_{i}$, while
(b) shows $g$ in a graphical neighbourhood of $f$ due to both $U_{i}$ and
$V_{i}$. A comparison of these figures demonstrates how the graphical neighbourhood
forces functions close to $f$ to remain closer to it than if they were in its
pointwise neighbourhood. This property is clearly visible in (a) where $g$,
if it were to be in a graphical neighbourhood of $f$, would be more faithful
to it by having to be also in $U_{2}$ and $U_{4}$. Thus in this case not only
the must the images $f(x_{ij})\overset{j}\rightarrow f(x_{i})$ as the $V_{i}$
decrease, but so also must the preimages $x_{ij}\overset{j}\rightarrow x_{i}$
with shrinking $U_{i}$. It is this simultaneous convergence of both images
and preimages at every $x$ that makes graphical convergence a natural candidate
for multifunctional convergence of functions. }}\end{singlespace}

\end{figure}
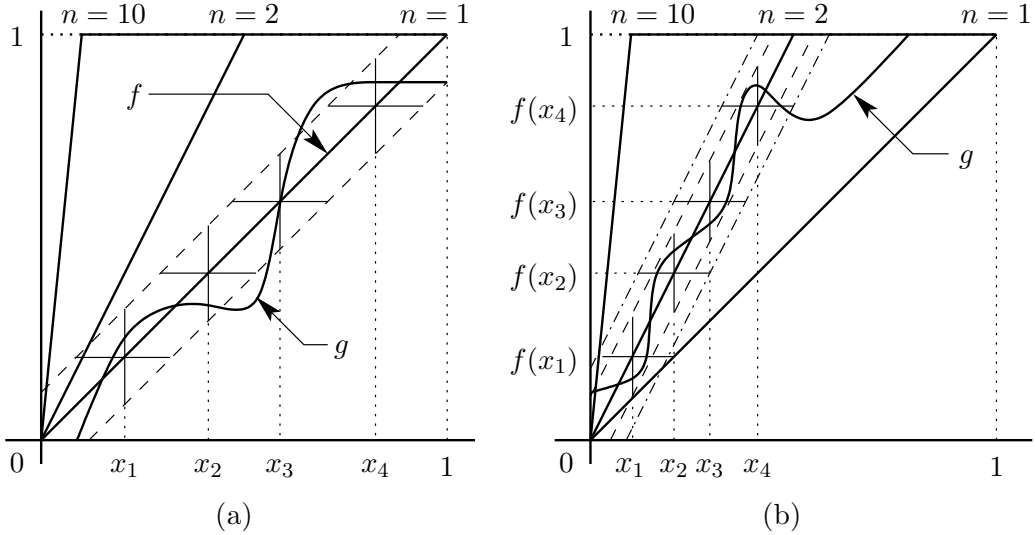
{\small \par}

\medskip{}
Observe that the boundary of $\textrm{Map}(X,Y)$ in the topology of pointwise
biconvergence is a {}``line parallel to the $Y$-axis''. We denote this closure
of $\textrm{Map}(X,Y)$ as 

\medskip{}
\noindent \textbf{Definition 3.3.} $\textrm{Multi}_{\mid}((X,Y),\mathcal{T})=\mathrm{Cl}(\mathrm{Map}((X,Y),\mathcal{T})).$$\qquad\square$
\medskip{}

The sense in which $\textrm{Multi}_{\mid}(X,Y)$ is the smallest closed topological
extension of $M=\textrm{Map}(X,Y)$ is the following, refer Thm. A1.4 and its
proof. Let $(M,\mathcal{T}_{0})$ be a topological space and suppose that\[
{\textstyle \widehat{M}=M\bigcup\{\widehat{m}\}}\]

\noindent is obtained by adjoining an extra point to $M$; here $M=\textrm{Map}(X,Y)$
and $\widehat{m}\in\textrm{Cl}(M)$ is the multifunctional limit in $\widehat{M}=\textrm{Multi}_{\mid}(X,Y)$.
Treat all open sets of $M$ generated by local bases of the type (\ref{Eqn: local_base})
with finite intersection property as a filter-base $_{\textrm{F}}\mathcal{B}$
on $X$ that induces a filter $\mathcal{F}$ on $M$ (by forming supersets of
all elements of $_{\textrm{F}}\mathcal{B}$; see Appendix A1) and thereby the
filter-base \[
{\textstyle \widehat{_{\textrm{F}}\mathcal{B}}=\{\widehat{B}=B\bigcup\{\widehat{m}\}\!:B\in\,_{\textrm{F}}\mathcal{B}\}}\]

\noindent on $\widehat{M}$; this filter-base at $m$ can also be obtained independently
from Eq. (\ref{Eqn: multi_bi}). Obviously $\widehat{_{\textrm{F}}\mathcal{B}}$
is an extension of $_{\textrm{F}}\mathcal{B}$ on $\widehat{M}$ and $_{\textrm{F}}\mathcal{B}$
is the filter induced on $M$ by $\widehat{_{\textrm{F}}\mathcal{B}}$. We may
also consider the filter-base to be a topological base on $M$ that defines
a coarser topology $\mathcal{T}$ on $M$ (through all unions of members of
$_{\textrm{F}}\mathcal{B}$) and hence the topology\[
{\textstyle \widehat{\mathcal{T}}=\{\widehat{G}=G\bigcup\{\widehat{m}\}\!:G\in\mathcal{T}\}}\]

\noindent on $\widehat{M}$ to be the topology associated with $\widehat{\mathcal{F}}$.
A finer topology on $\widehat{M}$ may be obtained by adding to $\widehat{\mathcal{T}}$
all the discarded elements of $\mathcal{T}_{0}$ that do not satisfy FIP. It
is clear that $\widehat{m}$ is on the boundary of $M$ because every neighbourhood
of $\widehat{m}$ intersects $M$ by construction; thus $(M,\mathcal{T})$ is
dense in $(\widehat{M,}\widehat{\mathcal{T}})$ which is the required topological
extension of $(M,\mathcal{T}).$ 

In the present case, a filter-base at $f\in\mathrm{Map}(X,Y)$ is the neighbourhood
system $_{\textrm{F}}\mathcal{B}_{f}$ at $f$ given by decreasing sequences
of neighbourhoods $(V_{k})$ and $(U_{k})$ of $f(x)$ and $x$ respectively,
and the filter $\widehat{\mathcal{F}}$ is the neighbourhood filter $\mathcal{N}_{f}\bigcup G$
where $G\in$$\textrm{Multi}_{\mid}(X,Y)$. We shall present an alternate, and
perhaps more intuitively appealing, description of graphical convergence based
on the adherence set of a filter on Sec. 4.1. 

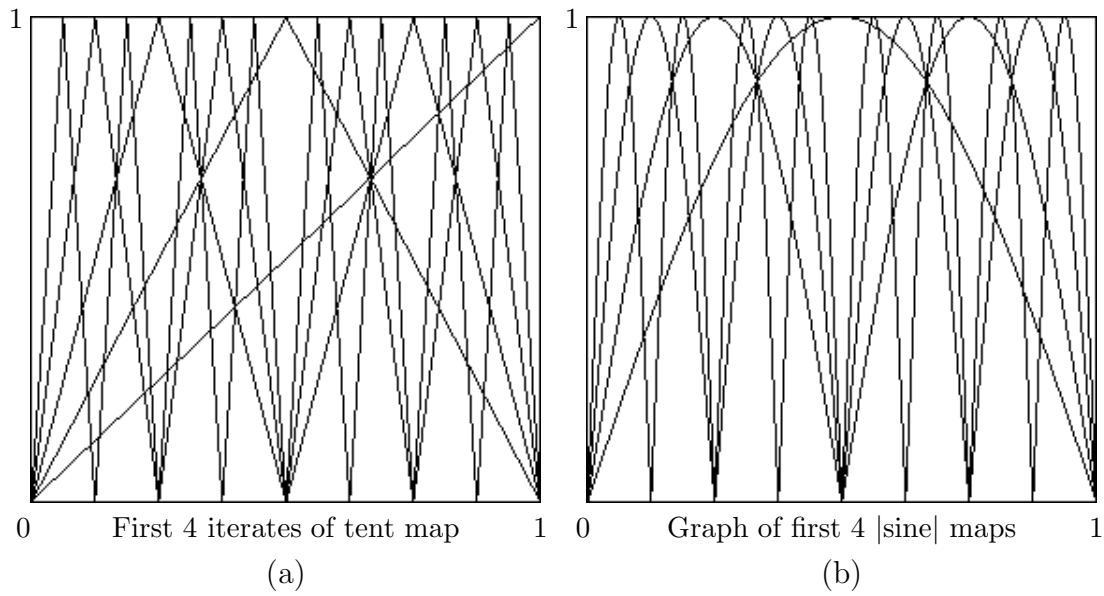
\begin{figure}[htbp]
\noindent \begin{center}{\small \input{tent4.pstex_t}}\end{center}{\small \par}

\begin{singlespace}

\caption{{\footnotesize \label{Fig: tent4}The first four iterates of (a) tent and (b)
$|\sin(2^{n-1}\pi x)|$ maps show the formal similarity of the dynamics of these
functions. It should be noted, as shown in Fig. \ref{Fig: Example2_1}, that
although $\sin(n\pi x)_{n=1}^{\infty}$ fails to converge at any point other
than $0$ and $1$, the subsequence $\sin(2^{n-1}\pi x)_{n=1}^{\infty}$ does
converge graphically on a set dense in $[0,1]$. }}\end{singlespace}

\end{figure}

As more serious examples of the graphical convergence of a net of functions
to multifunction than those considered above, Fig. \ref{Fig: tent4} shows the
first four iterates of the tent map \renewcommand{\arraystretch}{1.2}\[
t(x)=\left\{ \begin{array}{lc}
2x & 0\leq x<1/2\\
2(1-x) & 1/2\leq x\leq1\end{array}\right.\qquad\begin{array}{c}
(t^{1}=t).\end{array}\]
\renewcommand{\arraystretch}{1}

\noindent defined on $[0.1]$ and the sine map $f_{n}=|\sin(2^{n-1}\pi x)|,\; n=1,\cdots,4$
with domain $[0,1]$. 

{\small }%
{\small \par}

These examples illustrate the important generalization that \emph{periodic points
may be replaced by the more general equivalence classes} where a sequence of
functions converges graphically; this generalization based on the ill-posed
interpretation of dynamical systems is significant for non-iterative systems
as in second example above. The equivalence classes of the tent map for its
two fixed points $0$ and $2/3$ generated by the first 4 iterates are \[
[0]_{4}=\left\{ 0,\frac{1}{8},\frac{1}{4},\frac{3}{8},\frac{1}{2},\frac{5}{8},\frac{3}{4},\frac{7}{8},1\right\} \]
\[
\left[\frac{2}{3}\right]_{4}=\left\{ c,\frac{1}{8}\mp c,\frac{1}{4}\mp c,\frac{3}{8}\mp c,\frac{1}{2}\mp c,\frac{5}{8}\mp c,\frac{3}{4}\mp c,\frac{7}{8}\mp c,1-c\right\} \]

\noindent where $c=1/24$. If the moduli of the slopes of the graphs passing
through these equivalent fixed points are greater than 1 then the graphs converge
to multifunctions and when these slopes are less than 1 the corresponding graphs
converge to constant functions. It is to be noted that the number of equivalent
fixed points in a class increases with the number of iterations $k$ as $2^{k-1}+1;$
this \emph{increase in the degree of ill-posedness is typical of discrete chaotic
systems and can be regarded as a paradigm of chaos generated by} \emph{the convergence
of a family of functions. }

The $m^{\textrm{th}}$ iterate $t^{m}$ of the tent map has $2^{m}$ fixed points
corresponding to the $2^{m}$ injective branches of $t^{m}$ 

\renewcommand{\arraystretch}{1.75}\[
x_{mj}=\left\{ \begin{array}{ll}
{\displaystyle \frac{j-1}{2^{m}-1}}, & j=1,3,\cdots,(2^{m}-1),\\
{\displaystyle \frac{j}{2^{m}+1}}, & j=2,4,\cdots,2^{m},\end{array}\right.t^{m}(x_{mj})=x_{mj},\textrm{ }j=1,2,\cdots,2^{m}.\]

\noindent Let $X_{m}$ be the collection of these $2^{m}$ fixed points (thus
$X_{1}=\{0,2/3\}$), and denote by $[X_{m}]$ the set of the equivalent points,
one coming from each of the injective branches, for each of the fixed points:
thus

\noindent \renewcommand{\arraystretch}{1.2}\[
\begin{array}{crcl}
\mathcal{D}_{-}= & [X_{1}] & = & \{[0],[2/3]\}\\
 & [X_{2}] & = & \{[0],[2/5],[2/3],[4/5]\}\end{array}\]
\renewcommand{\arraystretch}{1} and $\mathcal{D}_{+}=\bigcap_{m=1}^{\infty}[X_{m}]$
is a nonempty countable set dense in $X$ at each of which the graphs of the
sequence $(t^{m})$ converge to a multifunction. New sets $[X_{n}]$ will be
formed by subsequences of the higher iterates $t^{n}$ for $m=in$ with $i=1,2,\cdots$
where these subsequences remain fixed. For example, the fixed points $2/5$
and $4/5$ produced respectively by the second and fourth injective branches
of $t^{2}$, are also fixed for the seventh and thirteenth branches of $t^{4}.$
For the shift map $2x\;\textrm{mod}(1)$ on $[0,1]$, $\mathcal{D}_{-}=\{[0],[1]\}$
where $[0]=\bigcap_{m=1}^{\infty}\{(i-1)/2^{m}\!:i=1,2,\cdots,2^{m}\}$ and
$[1]=\bigcap_{m=1}^{\infty}\{ i/2^{m}\!:i=1,2,\cdots,2^{m}\}$. 

It is useful to compare the graphical convergence of $(\sin(\pi nx))_{n=1}^{\infty}$
to $[0,1]$ at $0$ and to $0$ at $1$ with the usual integral and test-functional
convergences to $0$; note that the point $1/2$, for example, belongs to $\mathcal{D}_{+}$and
not to $\mathcal{D}_{-}=\{0,1\}$ because it is frequented by even $n$ only.
However for the subsequence $(f_{2^{m-1}})_{m\in\mathbb{Z}_{+}}$, $1/2$ is
in $\mathcal{D}_{-}$ because if the graph of $f_{2^{m-1}}$ passes through
$(1/2,0)$ for some $m$, then so do the graphs for all higher values . Therefore
$[0]=\bigcap_{m=1}^{\infty}\{ i/2^{m-1}\!:i=0,1,\cdots,2^{m-1}\}$ is the equivalence
class of $(f_{2^{m-1}})_{m=1}^{\infty}$ and this sequence converges to $[-1,1]$
on this set. Thus our extension $\textrm{Multi}(X)$ is distinct from the distributional
extension of function spaces with respect to test functions, and is able to
correctly generate the pathological behaviour of the limits that are so crucially
vital in producing chaos. 

\vspace{1cm}
\noindent \begin{flushleft}\textbf{\large 4. Discrete chaotic systems are maximally
ill-posed}\end{flushleft}{\large \par}

\noindent The above ideas apply to the development of a criterion for chaos
in discrete dynamical systems that is based on the limiting behaviour of the
graphs of a sequence of functions $(f_{n})$ on $X,$ rather than on the values
that the sequence generates as is customary. For the development of the maximality
of ill-posedness criterion of chaos, we need to refresh ourselves with the following
preliminaries. 

\medskip{}
\noindent \begin{flushright}\textbf{\textit{Resume Tutorial5: Axiom of Choice
and Zorn's Lemma}}\end{flushright}
\vspace{-0.15cm}

\noindent Let us recall from the first part of this Tutorial that for nonempty
subsets $(A_{\alpha})_{\alpha\in\mathbb{D}}$ of a nonempty set $X$, the Axiom
of Choice ensures the existence of a set $A$ such that $A\bigcap A_{\alpha}$
consists of a single element for every $\alpha$. The choice axiom has far reaching
consequences and a few equivalent statements, one of which the Zorn's lemma
that will be used immediately in the following, is the topic of this resumed
Tutorial. The beauty of the Axiom, and of its equivalents, is that they assert
the existence of mathematical objects that, in general, cannot be demonstrated
and it is often believed that Zorn's lemma is one of the most powerful tools
that a mathematician has available to him that is {}``almost indispensable
in many parts of modern pure mathematics'' with significant applications in
nearly all branches of contemporary mathematics. This {}``lemma'' talks about
maximal (as distinct from {}``maximum'') elements of a partially ordered set,
a set in which some notion of $x_{1}$ {}``preceding'' $x_{2}$ for two elements
of the set has been defined. 

A relation $\preceq$ on a set $X$ is said to be a \emph{partial order} (or
simply an \emph{order}) if it is (compare with the properties (ER1)--(ER3) of
an equivalence relation, Tutorial1)

\smallskip{}
(OR1) Reflexive, that is $(\forall x\in X)(x\preceq x)$. 

(OR2) Antisymmetric: $(\forall x,y\in X)(x\preceq y\wedge y\preceq x\Longrightarrow x=y)$.

(OR3) Transitive, that is $(\forall x,y,z\in X)(x\preceq y\wedge y\preceq z\Longrightarrow x\preceq z)$.
Any notion of order on a set $X$ in the sense of one element of $X$ preceding
another should possess at least this property. 
\smallskip{}

\noindent The \emph{}relation is a \emph{preorder} $\precsim$ if it is only
reflexive and transitive, that is if only (OR1) and (OR3) are true. If the hypothesis
of (OR2) is also satisfied by a preorder, then this $\precsim$ induces an equivalence
relation $\sim$ on $X$ according to $(x\precsim y)\wedge(y\precsim x)\Leftrightarrow x\sim y$
that evidently is actually a partial order iff $x\sim y\Leftrightarrow x=y$.
For any element $[x]\in X/\sim$ of the induced quotient space, let $\leq$
denote the generated order in $X/\sim$ so that \[
x\precsim y\Longleftrightarrow[x]\leq[y];\]
 then $\leq$ is a partial order on $X/\sim$. If every two element of $X$
are \emph{comparable}, in the sense that either $x_{1}\preceq x_{2}$ or $x_{2}\preceq x_{1}$
for all $x_{1},x_{2}\in X$, then $X$ is said to be a \emph{totally ordered
set} or a \emph{chain.} A totally ordered subset $(C,\preceq)$ of a partially
ordered set $(X,\preceq)$ with the ordering induced from $X$, is known as
a \emph{chain in $X$} if \begin{equation}
C=\{ x\in X\!:(\forall c\in X)(c\preceq x\vee x\preceq c)\}.\label{Eqn: chain}\end{equation}
The most important class of chains that we are concerned with in this work is
that on the subsets $\mathcal{P}(X)$ of a set $(X,\subseteq)$ under the inclusion
order; Eq. (\ref{Eqn: chain}), as we shall see in what follows, defines a family
of chains of nested subsets in $\mathcal{P}(X)$. Thus while the relation $\precsim$
in $\mathbb{Z}$ defined by $n_{1}\precsim n_{2}\Leftrightarrow\mid n_{1}\mid\,\leq\,\mid n_{2}\mid$
with $n_{1},n_{2}\in\mathbb{Z}$ preorders $\mathbb{Z}$, it is not a partial
order because although $-n\precsim n\textrm{ and }n\precsim-n$ for any $n\in\mathbb{Z}$,
it is does not follow that $-n=n$. A common example of partial order on a set
of sets, for example on the power set $\mathcal{P}(X)$ of a set $X$ (see footnote
\ref{Foot: notation}), is the inclusion relation $\subseteq$: the ordered
set $\mathcal{X}=(\mathcal{P}(\{ x,y,z\}),\subseteq)$ is partially ordered
but not totally ordered because, for example, $\{ x,y\}\not\subseteq\{ y,x\}$,
or $\{ x\}$ is not comparable to $\{ y\}$ unless $x=y$; however $C=\{\{\emptyset,\{ x\},\{ x,y\}\}$
does represent one of the many possible chains of $\mathcal{X}$. Another useful
example of partial order is the following: Let $X$ and $(Y,\leq)$ be sets
with $\leq$ ordering $Y$, and consider $f,g\in\textrm{Map}(X,Y)$ with $\mathcal{D}(f),\mathcal{D}(g)\subseteq X$.
Then 

\noindent \begin{eqnarray}
(\mathcal{D}(f)\subseteq\mathcal{D}(g))(f=g|_{\mathcal{D}(f)}) & \Longleftrightarrow & f\preceq g\nonumber \\
(\mathcal{D}(f)=\mathcal{D}(g))(\mathcal{R}(f)\subseteq\mathcal{R}(g)) & \Longleftrightarrow & f\preceq g\label{Eqn: FunctionOrder}\\
(\forall x\in\mathcal{D}(f)=\mathcal{D}(g))\textrm{ }(f(x)\leq g(x)) & \Longleftrightarrow & f\preceq g\nonumber \end{eqnarray}

\noindent define partial orders on $\textrm{Map}(X,Y)$. In the last case, the
order is not total because any two functions whose graphs cross at some point
in their common domain cannot be ordered by the given relation, while in the
first any $f$ whose graph does not coincide with that of $g$ on the common
domain is not comparable to it by this relation. 

Let $(X,\preceq)$ be a partially ordered set and let $A$ be a subset of $X$.
An element $a_{+}\in(A,\preceq)$ is said to be a \emph{maximal} element of
$A$ with respect to $\preceq$ if \begin{equation}
(\forall a\in(A,\preceq))(a_{+}\preceq a)\Longrightarrow\textrm{ }a=a_{+},\label{Eqn: maximal}\end{equation}
 that is iff there is no $a\in A$ with $a\neq a_{+}$ and $a\succ a_{+}$%
\footnote{{\small \label{Foot: strict reln}If $\preceq$ is an order relation in $X$
then the} \emph{\small strict relation $\prec$ in $X$} {\small corresponding
to $\preceq$, given by $x\prec y\Leftrightarrow(x\preceq y)\wedge(x\neq y)$,}
\emph{\small is not an order relation} {\small because unlike $\preceq$, $\prec$
is not reflexive even though it is both transitive and asymmetric.} \emph{\small }%
}. Expressed otherwise, this implies that an element $a_{+}$ of a subset $A\subseteq(X,\preceq)$
is maximal in $(A,\preceq)$ iff it is true that \begin{equation}
(a\preceq a_{+}\in A)\textrm{ }(\textrm{for every }a\in(A,\preceq)\textrm{ comparable to }a_{+});\label{Eqn: maximal1}\end{equation}
 thus $a_{+}$ in $A$ is a maximal element of $A$ iff it is strictly greater
than every \emph{other comparable} element of $A$. This of course does not
mean that each element $a$ of $A$ satisfies $a\preceq a_{+}$ because every
pair of elements of a partially ordered set need not be comparable: in a totally
ordered set there can be at most one maximal element. In comparison, an element
$a_{\infty}$ of a subset $A\subseteq(X,\preceq)$ is \emph{the} unique \emph{maximum}
(\emph{largest, greatest, last}) element of $A$ iff \begin{equation}
(a\preceq a_{\infty}\in A)\textrm{ }(\textrm{for every }a\in(A,\preceq)),\label{Eqn: maximum}\end{equation}

\noindent implying that $a_{\infty}$ is \emph{the} element of $A$ that is
strictly larger than every other element of $A$. As in the case of the maximal,
although this also does not require all elements of $A$ to be comparable to
each other, it does require $a_{\infty}$ to be larger than every element of
$A$. The dual concepts of minimal and minimum can be similarly defined by essentially
reversing the roles of $a$ and $b$ in relational expressions like $a\preceq b$. 

The last concept needed to formalize Zorn's lemma is that of an upper bound:
For a subset $(A,\preceq)$ of a partially ordered set $(X,\preceq)$, an element
$u$ of $X$ is an \emph{upper bound of} $A$ \emph{in} $X$ iff \begin{equation}
(a\preceq u\in(X,\preceq))\textrm{ }(\textrm{for every }a\in(A,\preceq))\label{Eqn: upper bound}\end{equation}
 which requires the upper bound $u$ to be larger than all members of $A$,
with the corresponding lower bounds of $A$ being defined in a similar manner.
Of course, it is again not necessary that the elements of $A$ be comparable
to each other, and it should be clear from Eqs. (\ref{Eqn: maximum}) and (\ref{Eqn: upper bound})
that when an upper bound of a set is in the set itself, then it is the maximum
element of the set. If the upper (lower) bounds of a subset $(A,\preceq)$ of
a set $(X,\preceq)$ has a least (greatest) element, then this smallest upper
bound (largest lower bound) is called \emph{the} \emph{least upper bound} (\emph{greatest
lower} \emph{bound}) or \emph{supremum} (\emph{infimum}) \emph{of $A$ in $X$}.
Combining Eqs. (\ref{Eqn: maximum}) and (\ref{Eqn: upper bound}) then yields
\begin{equation}
\begin{array}{rcl}
{\displaystyle \sup_{X}A} & = & \{ a_{\leftarrow}\in\Omega_{A}\!:a_{\leftarrow}\preceq u\textrm{ }\forall\textrm{ }u\in(\Omega_{A},\preceq)\}\\
{\displaystyle \inf_{X}A} & = & \{_{\rightarrow}a\in\Lambda_{A}\!:l\preceq\,_{\rightarrow}a\textrm{ }\forall\textrm{ }l\in(\Lambda_{A},\preceq)\}\end{array}\label{Eqn: supinf1}\end{equation}
 where \emph{}$\Omega_{A}=\{\textrm{ }u\in X\!:(\forall\textrm{ }a\in A)(a\preceq u)\}$
\emph{}and \emph{}$\Lambda_{A}=\{ l\in X\!:(\forall\textrm{ }a\in A)(l\preceq a)\}$
\emph{}are the sets of all upper and lower bounds of $A$ in $X$\emph{.} Equation
(\ref{Eqn: supinf1}) may be expressed in the equivalent but more transparent
form as \begin{equation}
\begin{array}{c}
{\displaystyle a_{\leftarrow}={\displaystyle \sup_{X}A}\Longleftrightarrow(a\in A\Rightarrow a\preceq a_{\leftarrow})\wedge(a_{0}\prec a_{\leftarrow}\Rightarrow a_{0}\prec b\preceq a_{\leftarrow}\textrm{ for some }b\in A)}\\
_{\rightarrow}a={\displaystyle \inf_{X}A}\Longleftrightarrow(a\in A\Rightarrow\,_{\rightarrow}a\preceq a)\wedge(_{\rightarrow}a\prec a_{1}\Rightarrow\,_{\rightarrow}a\preceq b\prec a_{1}\textrm{ for some }b\in A)\end{array}\label{Eqn: supinf2}\end{equation}
 to imply that \emph{$a_{\leftarrow}$} ($_{\rightarrow}a$) is \emph{the} upper
(lower) bound of $A$ in $X$ which precedes (succeeds) every other upper (lower)
bound of $A$ \emph{}in $X$. Notice that uniqueness in the definitions above
is a direct consequence of the uniqueness of greatest and least elements of
a set. \emph{}It must be noted that whereas maximal and maximum are properties
of the particular subset and have nothing to do with anything outside it, upper
and lower bounds of a set are defined only with respect to a superset that may
contain it. 

The following example, beside being useful in Zorn's lemma, is also of great
significance in fixing some of the basic ideas needed in our future arguments
involving classes of sets ordered by the inclusion relation. 

\medskip{}
\noindent \textbf{Example 4.1.} Let $\mathcal{X}=\mathcal{P}(\{ a,b,c\})$ be
ordered by the inclusion relation $\subseteq$. The subset $\mathcal{A}=\mathcal{P}(\{ a,b,c\})-\{ a,b,c\}$
has three maximals $\{ a,b\}$, $\{ b,c\}$ and $\{ c,a\}$ but no maximum as
there is no $A_{\infty}\in\mathcal{A}$ satisfying $A\preceq A_{\infty}$ for
every $A\in\mathcal{A}$, while $\mathcal{P}(\{ a,b,c\})-\emptyset$ the three
minimals $\{ a\}$, $\{ b\}$, and $\{ c\}$ but no minimum. This shows that
a subset of a partially ordered set may have many maximals (minimals) without
possessing a maximum (minimum), but a subset has a maximum (minimum) iff this
is its unique maximal (minimal). If $\mathcal{A}=\{\{ a,b\},\{ a,c\}\}$, then
every subset of the intersection of the elements of $\mathcal{A}$, namely $\{ a\}$
and $\emptyset$, are lower bounds of $\mathcal{A}$, and all supersets in $\mathcal{X}$
of the union of its elements --- which in this case is just $\{ a,b,c\}$ ---
are its upper bounds. Notice that while the maximal (minimal) and maximum (minimum)
are elements of $\mathcal{A}$, upper and lower bounds need not be contained
in their sets. In this class $(\mathcal{X},\subseteq)$ of subsets of a set
$X$, $X_{+}$ is a maximal element of $\mathcal{X}$ iff $X_{+}$ is not contained
in any other subset of $X$, while $X_{\infty}$ is a maximum of $\mathcal{X}$
iff $X_{\infty}$ contains every other subset of $X$. 

Let $\mathcal{A}:=\{ A_{\alpha}\in\mathcal{X}\}_{\alpha\in\mathbb{D}}$ be a
nonempty subclass of $(\mathcal{X},\subseteq)$, and suppose that both $\bigcup A_{\alpha}$
and $\bigcap A_{\alpha}$ are elements of $\mathcal{X}$. Since each $A_{\alpha}$
is $\subseteq$-less than $\bigcup A_{\alpha}$, it follows that $\bigcup A_{\alpha}$
is an upper bound of $\mathcal{A}$; this is also be the smallest of all such
bounds because if $U$ is any other upper bound then every $A_{\alpha}$ must
precede $U$ by Eq. (\ref{Eqn: upper bound}) and therefore so must $\bigcup A_{\alpha}$
(because the union of a class of subsets of a set is the smallest that contain
each member of the class: $A_{\alpha}\subseteq U\Rightarrow\bigcup A_{\alpha}\subseteq U$
for subsets $(A_{\alpha})$ and $U$ of $X$). Analogously, since $\bigcap A_{\alpha}$
is $\subseteq$-less than each $A_{\alpha}$ it is a lower bound of $\mathcal{A}$;
that it is the greatest of all the lower bounds $L$ in $\mathcal{X}$ follows
because the intersection of a class of subsets is the largest that is contained
in each of the subsets: $L\subseteq A_{\alpha}\Rightarrow L\subseteq\bigcap A_{\alpha}$
for subsets $L$ and $(A_{\alpha})$ of $X$. Hence the supremum and infimum
of $\mathcal{A}$ in $(\mathcal{X},\subseteq)$ given by \begin{equation}
A_{\leftarrow}=\sup_{(\mathcal{X},\subseteq)}\mathcal{A}=\bigcup_{A\in\mathcal{A}}A\qquad\textrm{and}\qquad_{\rightarrow}A=\inf_{(\mathcal{X},\subseteq)}\mathcal{A}=\bigcap_{A\in\mathcal{A}}A\label{Eqn: supinf3}\end{equation}
 are both elements of $(\mathcal{X},\subseteq)$. Intuitively, an upper (respectively,
lower) bound of $\mathcal{A}$ in $\mathcal{X}$ is any subset of $\mathcal{X}$
that contains (respectively, is contained in) every member of $\mathcal{A}$.$\qquad\blacksquare$
\medskip{}

The statement of Zorn's lemma and its proof can now be completed in three stages
as follows. For Theorem 4.1 below that constitutes the most significant technical
first stage, let $g$ be a function on $(X,\preceq)$ that assigns to every
$x\in X$ an \emph{immediate successor} $y\in X$ such that \[
{\textstyle \mathscr{M}(x)=\{\textrm{ }y\succ x\!:\not\exists\textrm{ }x_{*}\in X\textrm{ satisfying }x\prec x_{*}\prec y\}}\]
 are all the successors of $x$ in $X$ with no element of $X$ lying strictly
between $x$ and $y$. Select a representative of $\mathscr{M}(x)$ by a choice
function $f_{\textrm{C}}$ such that \[
g(x)=f_{\textrm{C}}(\mathscr{M}(x))\in\mathscr{M}(x)\]
 is an immediate successor of $x$ chosen from the many possible in the set
$\mathscr{M}(x)$. The basic idea in the proof of the first of the three-parts
is to express the existence of a maximal element of a partially ordered set
$X$ in terms of the existence of a fixed point in the set, which follows as
a contradiction of the assumed hypothesis that every point in $X$ has an immediate
successor. Our basic application of immediate successors in the following will
be to classes $\mathcal{X}\subseteq(\mathcal{P}(X),\subseteq)$ of subsets of
a set $X$ ordered by inclusion. In this case for any $A\in\mathcal{X}$, the
function $g$ can be taken to be the superset \begin{equation}
{\textstyle g(A)=A\bigcup f_{\textrm{C}}(\mathscr{G}(A)),\quad\textrm{where }\mathscr{G}(A)=\{ x\in X-A\!:A\bigcup\{ x\}\in\mathcal{X}\}}\label{Eqn: FilterTower}\end{equation}
 of $A$. Repeated application of $g$ to $A$ then generates a principal filter,
and hence an associated sequence, based at $A$. 

\medskip{}
\noindent \textbf{Theorem 4.1.} \textsl{Let $(X,\preceq)$ be a partially ordered
set that satisfies }

\smallskip{}
(ST1) \textsl{There is a smallest element $x_{0}$ of $X$ which has no immediate
predecessor in $X$.}

(ST2) \textsl{If $C\subseteq X$ is a totally ordered subset in $X$, then $c_{*}=\sup_{X}C$
is in $X$. }
\smallskip{}

\noindent \textsl{Then there exists a maximal element $x_{+}$ of $X$ which
has no immediate successor in $X$.}$\qquad\square$

\noindent \textbf{Proof.} Let $T\subseteq(X,\preceq)$ be a subset of $X$.
If the conclusion of the theorem is false then the alternative 

\smallskip{}
(ST3) \textsl{Every element $x\in T$ has an immediate successor $g(x)$ in
$T$}%
\footnote{{\small \label{Foot: infinite}This makes $T$, and hence $X$, inductively
defined infinite sets. It should be realized that (ST3)} \emph{\small does not
mean} {\small that every member of $T$ is obtained from $g$, but only ensures
that the immediate successor of any element of $T$ is also in $T.$ The infimum
$_{\rightarrow}T$ of these towers satisfies the additional property of being
totally ordered (and is therefore essentially a sequence or net) in $(X,\preceq)$
to which (ST2) can be applied. }%
} 
\smallskip{}

\noindent leads, as shown below, to a contradiction that can be resolved only
by the conclusion of the theorem. A subset $T$ of $(X,\preceq)$ satisfying
conditions (ST1)$-$(ST3) is sometimes known as an $g$\emph{-tower} or an $g$\emph{-sequence:}
an obvious example of a tower is $(X,\preceq)$ itself. If \[
{\textstyle _{\rightarrow}T=\bigcap\{ T\in\mathcal{T}\!:T\textrm{ is an }x_{0}-\textrm{tower}\}}\]
 is the $(\mathcal{P}(X),\subseteq)$-infimum of the class $\mathcal{T}$ of
all sequential towers of $(X,\preceq)$, we show that this smallest sequential
\emph{}tower is infact a \emph{sequential totally ordered chain} in $(X,\preceq)$
built from $x_{0}$ by the $g$-function. Let the subset \begin{equation}
C_{\textrm{T}}=\{ c\in X\!:(\forall t\in\,_{\rightarrow}T)(t\preceq c\vee c\preceq t)\}\subseteq X\label{Eqn: tower-chain}\end{equation}
 of $X$ be an $g$-chain in $_{\rightarrow}T$ in the sense that (cf. Eq. (\ref{Eqn: chain}))
it is that subset of $X$ each of whose elements is comparable with some element
of $_{\rightarrow}T$. The conditions (ST1)$-$(ST3) for $C_{\textrm{T}}$ can
be verified as follows to demonstrate that $C_{\textrm{T}}$ is an $g$-tower.

\smallskip{}
(1) $x_{0}\in C_{\textrm{T}}$, because it is less than each $x\in\,_{\rightarrow}T$. 

(2) Let $c_{\leftarrow}=\sup_{X}C_{\textrm{T}}$ be the supremum of the chain
$C_{\textrm{T}}$ in $X$ so that by (ST2), $c_{\leftarrow}\in X$. Let $t\in\,_{\rightarrow}T$.
If there is \emph{some} $c\in C_{\textrm{T}}$ such that $t\preceq c$, then
surely $t\preceq c_{\leftarrow}$. Else, $c\preceq t$ for \emph{every} $c\in C_{\textrm{T}}$
shows that $c_{\leftarrow}\preceq t$ because $c_{\leftarrow}$ is the smallest
of all the upper bounds $t$ of $C_{\textrm{T}}$. Therefore $c_{\leftarrow}\in C_{\textrm{T}}$.

(3) In order to show that $g(c)\in C$ whenever $c\in C$ it needs to verified
that for all $t\in\,_{\rightarrow}T$, either $t\preceq c\Rightarrow t\preceq g(c)$
or $c\preceq t\Rightarrow g(c)\preceq t$. As the former is clearly obvious,
we investigate the later as follows; note that $g(t)\in\,_{\rightarrow}T$ by
(ST3). The first step is to show that the subset \begin{equation}
C_{g}=\{ t\in\,_{\rightarrow}T\!:(\forall c\in C_{\textrm{T}})(t\preceq c\vee g(c)\preceq t)\}\label{Eqn: chain_g}\end{equation}
 of $_{\rightarrow}T$, which is a chain in $X$ (observe the inverse roles
of $t$ and $c$ here as compared to that in Eq. (\ref{Eqn: tower-chain})),
is a tower: Let $t_{\leftarrow}$ be the supremum of $C_{g}$ and take $c\in C$.
If there is \emph{some} $t\in C_{g}$ for which $g(c)\preceq t$, then clearly
$g(c)\preceq t_{\leftarrow}$. Else, $t\preceq x$ for \emph{each} $t\in C_{g}$
shows that $t_{\leftarrow}\preceq c$ because $t_{\leftarrow}$ is the smallest
of all the upper bounds $c$ of $C_{g}$. Hence $t_{\leftarrow}\in C_{g}$. 

Property (ST3) for $C_{g}$ follows from a small yet significant modification
of the above arguments in which the immediate successors $g(t)$ of $t\in C_{g}$
formally replaces the supremum $t_{\leftarrow}$ of $C_{g}$. Thus given a $c\in C$,
if there is \emph{some} $t\in C_{g}$ for which $g(c)\preceq t$ then $g(c)\prec g(t)$;
this combined with $(c=t)\Rightarrow(g(c)=g(t))$ yields $g(c)\preceq g(t)$.
On the other hand, $t\prec c$ for \emph{every} $t\in C_{g}$ requires $g(t)\preceq c$
as otherwise $(t\prec c)\Rightarrow(c\prec g(t))$ would, from the resulting
consequence $t\prec c\prec g(t)$, contradict the assumed hypothesis that $g(t)$
is the immediate successor of $t$. Hence, $C_{g}$ is a $g$-tower in $X$. 

To complete the proof that $g(c)\in C_{\textrm{T}}$, and thereby the argument
that $C_{\textrm{T}}$ is a tower, we first note that as $_{\rightarrow}T$
is the smallest tower and $C_{g}$ is built from it, $C_{g}=\,_{\rightarrow}T$
must infact be $_{\rightarrow}T$ itself. From Eq. (\ref{Eqn: chain_g}) therefore,
for every $t\in\,_{\rightarrow}T$ either $t\preceq g(c)$ or $g(c)\preceq t$,
so that $g(c)\in C_{\textrm{T}}$ whenever $c\in C_{\textrm{T}}$. This concludes
the proof that $C_{\textrm{T}}$ is actually the tower $_{\rightarrow}T$ in
$X$. 

From (ST2), the implication of the chain $C_{\textrm{T}}$ \begin{equation}
C_{\textrm{T}}=\,_{\rightarrow}T=C_{g}\label{Eqn: ChainedTower}\end{equation}
 being the minimal tower $_{\rightarrow}T$ is that the supremum $t_{\leftarrow}$
of the totally ordered $_{\rightarrow}T$ \emph{in its own tower} (as distinct
from in the tower $X$: recall that $_{\rightarrow}T$ is a subset of $X$)
must be contained in itself, that is \begin{equation}
\sup_{C_{\textrm{T}}}(C_{\textrm{T}})=t_{\leftarrow}\in\,_{\rightarrow}T\subseteq X.\label{Eqn: sup chain}\end{equation}
 This however leads to the contradiction from (ST3) that $g(t_{\leftarrow})$
be an element of $_{\rightarrow}T$, unless of course 

\begin{equation}
g(t_{\leftarrow})=t_{\leftarrow},\label{Eqn: fixed point}\end{equation}
 which because of (\ref{Eqn: ChainedTower}) may also be expressed equivalently
as $g(c_{\leftarrow})=c_{\leftarrow}\in C_{\textrm{T}}$. As the sequential
totally ordered set $_{\rightarrow}T$ is a subset of $X$, Eq. (48) implies
that $t_{\leftarrow}$ is a maximal element of $X$ which allows (ST3) to be
replaced by the remarkable inverse criterion that 

\smallskip{}
$(\textrm{ST}3^{\prime})$ If $x\in X$ and $w$ precedes $x,$ $w\prec x$,
then $w\in X$ 
\smallskip{}

\noindent that is obviously false for a general tower $T$. In fact, it follows
directly from Eq. (\ref{Eqn: maximal}) that under $(\textrm{ST}3^{\prime})$
\emph{any $x_{+}\in X$ is a maximal element of $X$ iff it is a fixed point
of $g$} as given by Eq. (\ref{Eqn: fixed point}). This proves the theorem
and also demonstrates how, starting from a minimum element of a partially ordered
set $X$, (ST3) can be used to generate inductively a totally ordered sequential
subset of $X$ \emph{leading to a maximal $x_{+}=c_{\leftarrow}\in(X,\preceq)$
that is a fixed point of the generating function $g$} \emph{whenever the supremum}
$t_{\leftarrow}$ \emph{of the chain $_{\rightarrow}T$ is in} $X$.$\qquad\blacksquare$

\medskip{}
\noindent \textbf{Remarks.} The proof of this theorem, despite its apparent
length and technically involved character, carries the highly significant underlying
message that 

\begin{LyXParagraphLeftIndent}{0.1in}
\smallskip{}
\noindent \emph{Any inductive sequential $g$-construction of an infinite chained
tower} $C_{\textrm{T}}$ \emph{starting with a smallest element $x_{0}\in(X,\preceq)$
such that a supremum $c_{\leftarrow}$ of the $g$-generated sequential chain}
$C_{\textrm{T}}$ \emph{in its own tower is contained in itself, must necessarily
terminate with a fixed point relation of the type} (\ref{Eqn: fixed point})
\emph{with respect to the supremum. Note from Eqs. (\ref{Eqn: sup chain}) and
(\ref{Eqn: fixed point}) that the role of} (ST2) \emph{applied to a fully ordered
tower is the identification of the maximal of the tower --- which depends only
the tower and has nothing to do with anything outside it --- with its supremum
that depends both on the tower and its complement. }
\smallskip{}

\end{LyXParagraphLeftIndent}
\noindent Thus although purely set-theoretic in nature, the filter-base associated
with a sequentially totally ordered set may be interpreted to lead to the usual
notions of adherence and convergence of filters and thereby of a generated topology
for $(X,\preceq)$, see Appendix A1 and Example A1.3. This very significant
apparent inter-relation between topologies, filters and orderings will form
the basis of our approach to the condition of maximal ill-posedness for chaos. 
\medskip{}

In the second stage of the three stage programme leading to Zorn's lemma, the
tower Theorem 4.1 and the comments of the preceding paragraph are applied at
one higher level to a very special class of the power set of a set, the class
of all the chains of a partially ordered set, to directly lead to the physically
significant 

\medskip{}
\noindent \textbf{Theorem 4.2.} \textbf{Hausdorff Maximal Principle.} \textsl{Every
partially ordered set $(X,\preceq)$ has a maximal totally ordered subset}.%
\footnote{{\small \label{Foot: Hausdorff}Recall that this means that if there is a totally
ordered chain $C$ in $(X,\preceq)$ that succeeds $C_{+}$, then $C$ must
be $C_{+}$ so that no chain in $X$ can be strictly larger than $C_{+}$. The
notation adopted here and below is the following: If $X=\{ x,y\}$ is a non-empty
set, then $\mathcal{X}:=\mathcal{P}(X)=\{ A\!:A\subseteq X\}=\{\emptyset,\{ x\},\{ y\},\{ x,y\}\}$
is the set of subsets of $X$, and $\mathfrak{X}:=\mathcal{P}^{2}(X)=\{\mathcal{A}:\mathcal{A}\subseteq\mathcal{X}\}$,
the set of all subsets of $\mathcal{X}$, consists of the $16$ elements $\emptyset$,
$\{\emptyset\}$, $\{\{ x\}\}$, $\{\{ y\}\}$, $\{\{ x,y\}\}$, $\{\{\emptyset\},\{ x\}\}$,
$\{\{\emptyset\},\{ y\}\}$, $\{\{\emptyset\},\{ x,y\}\}$, $\{\{ x\},\{ y\}\}$,
$\{\{ x\},\{ x,y\}\}$, $\{\{ y\},\{ x,y\}\}$, $\{\{\emptyset\},\{ x\},\{ y\}\}$,
$\{\{\emptyset\},\{ x\},\{ x,y\}\}$, $\{\{\emptyset\},\{ y\},\{ x,y\}\}$,
$\{\{ x\},\{ y\},\{ x,y\}\}$, and $\mathcal{X}$: an element of $\mathcal{P}^{2}(X)$
is a subset of $\mathcal{P}(X)$, any element of which is a subset of $X$.
Thus if $C=\{0,1,2\}$ is a chain in $(X=\{0,1,2\},\leq)$, then $\mathcal{C}=\{\{0\},\{0,1\},\{0,1,2\}\}\subseteq\mathcal{P}(X)$
and $\mathfrak{C}=\{\{\{0\}\},\{\{0\},\{0,1\}\},\{\{0\},\{0,1\},\{0,1,2\}\}\}\subseteq\mathcal{P}^{2}(X)$
represent chains in $(\mathcal{P}(X),\subseteq)$ and $(\mathcal{P}^{2}(X),\subseteq)$
respectively . }%
}$\qquad\square$

\noindent \textbf{Proof.} Here the base level is \begin{equation}
\mathcal{X}=\{ C\in\mathcal{P}(X)\!:C\textrm{ is a chain in }(X,\preceq)\}\subseteq\mathcal{P}(X)\label{Eqn: Hausdorff}\end{equation}
 be the set of all the totally ordered subsets of $(X,\preceq)$. Since $\mathcal{X}$
is a collection of (sub)sets of $X$, we order it by the inclusion relation
on $\mathcal{X}$ and use the tower Theorem to demonstrate that $(\mathcal{X},\subseteq)$
has a maximal element $C_{\leftarrow}$, which by the definition of $\mathcal{X}$,
is the required maximal chain in $(X,\preceq)$. 

Let $\mathcal{C}$ be a chain in $\mathcal{X}$ of the chains in $(X,\preceq)$.
In order to apply the tower Theorem to $(\mathcal{X},\subseteq)$ we need to
verify hypothesis (ST2) that the smallest \begin{equation}
C_{*}=\sup_{\mathcal{X}}\mathcal{C}=\bigcup_{C\in\mathcal{C}}C\label{Eqn: HausdorffChain}\end{equation}
 of the possible upper bounds of $\mathcal{C}$ (see Eq. (\ref{Eqn: supinf3}))
is a chain of $(X,\preceq)$. Indeed, if $x_{1},x_{2}\in X$ are two points
of $C_{\textrm{sup}}$ with $x_{1}\in C_{1}$ and $x_{2}\in C_{2}$, then from
the $\subseteq$-comparability of $C_{1}$ and $C_{2}$ we may choose $x_{1},x_{2}\in C_{1}\supseteq C_{2}$,
say. Thus $x_{1}$ and $x_{2}$ are $\preceq$-comparable as $C_{1}$ is a chain
in $(X,\preceq)$; $C_{*}\in\mathcal{X}$ is therefore a chain in $(X,\preceq)$
which establishes that the supremum of a chain of $(\mathcal{X},\subseteq)$
is a chain in $(X,\preceq)$. 

The tower Theorem 4.1 can now applied to $(\mathcal{X},\subseteq)$ with $C_{0}$
as its smallest element to construct a $g$-sequentially towered fully ordered
subset of $\mathcal{X}$ consisting of chains in $X$ \[
\mathcal{C}_{\textrm{T}}=\{ C_{i}\in\mathcal{P}(X)\!:C_{i}\subseteq C_{j}\textrm{ for }i\leq j\in\mathbb{N}\}=\,_{\rightarrow}\mathcal{T}\subseteq\mathcal{P}(X)\]
 of $(\mathcal{X},\subseteq)$ --- consisting of the common elements of all
$g$-sequential towers $\mathcal{T}\in\mathfrak{T}$ of $(\mathcal{X},\subseteq)$
--- that infact is a principal filter base of chained subsets of $(X,\preceq)$
at $C_{0}$. The supremum (chain in $X$) $C_{\leftarrow}$ of $\mathcal{C}_{\textrm{T}}$
in $\mathcal{C}_{\textrm{T}}$ must now satisfy, by Thm. 4.1, the fixed point
$g$-chain of $X$ \[
\sup_{\mathcal{C}_{\textrm{T}}}(\mathcal{C}_{\textrm{T}})=C_{\leftarrow}=g(C_{\leftarrow})\in\mathcal{C}_{\textrm{T}}\subseteq\mathcal{P}(X),\]
 where the chain $g(C)=C\bigcup f_{\textrm{C}}(\mathscr{G}(C)-C)$ with $\mathscr{G}(C)=\{ x\in X-C\!:C\bigcup\{ x\}\in\mathcal{X}\}$,
is an immediate successor of $C$ obtained by choosing one point $x=f_{\textrm{C}}(\mathscr{G}(C)-C)$
from the many possible in $\mathscr{G}(C)-C$ such that the resulting $g(C)=C\bigcup\{ x\}$
is a strict successor of the chain $C$ with no others lying between it and
$C$. Note that $C_{\leftarrow}\in(\mathcal{X},\subseteq)$ is only one of the
many maximal fully ordered subsets possible in $(X,\preceq)$.$\qquad\blacksquare$ 
\medskip{}

With the assurance of the existence of a maximal chain $C_{\leftarrow}$ among
all fully ordered subsets of a partially ordered set $(X,\preceq)$, the arguments
are completed by returning to the basic level of $X$. 

\medskip{}
\noindent \textbf{Theorem 4.3. Zorn's Lemma.} \textsl{Let $(X,\preceq)$ be
a partially ordered set such that every totally ordered subset of $X$ has an
upper bound in $X$. Then $X$ has at least one maximal element with respect
to its order.$\qquad\square$}

\noindent \textbf{Proof.} The proof of this final part is a mere application
of the Hausdorff Maximal Principle on the existence of a maximal chain $C_{\leftarrow}$
in $X$ to the hypothesis of this theorem that $C_{\leftarrow}$ has an upper
bound $u$ in $X$ that quickly leads to the identification of this bound as
a maximal element $x_{+}$ of $X$. Indeed, if there is an element $v\in X$
that is comparable to $u$ and $v\not\preceq u$, then $v$ cannot be in $C_{\leftarrow}$
as it is necessary for every $x\in C_{\leftarrow}$ to satisfy $x\preceq u$.
Clearly then $C_{\leftarrow}\bigcup\{ v\}$ is a chain in $(X,\preceq)$ bigger
than $C_{\leftarrow}$ which contradicts the assumed maximality of $C_{\leftarrow}$
among  the chains of $X$.$\qquad\blacksquare$
\medskip{}

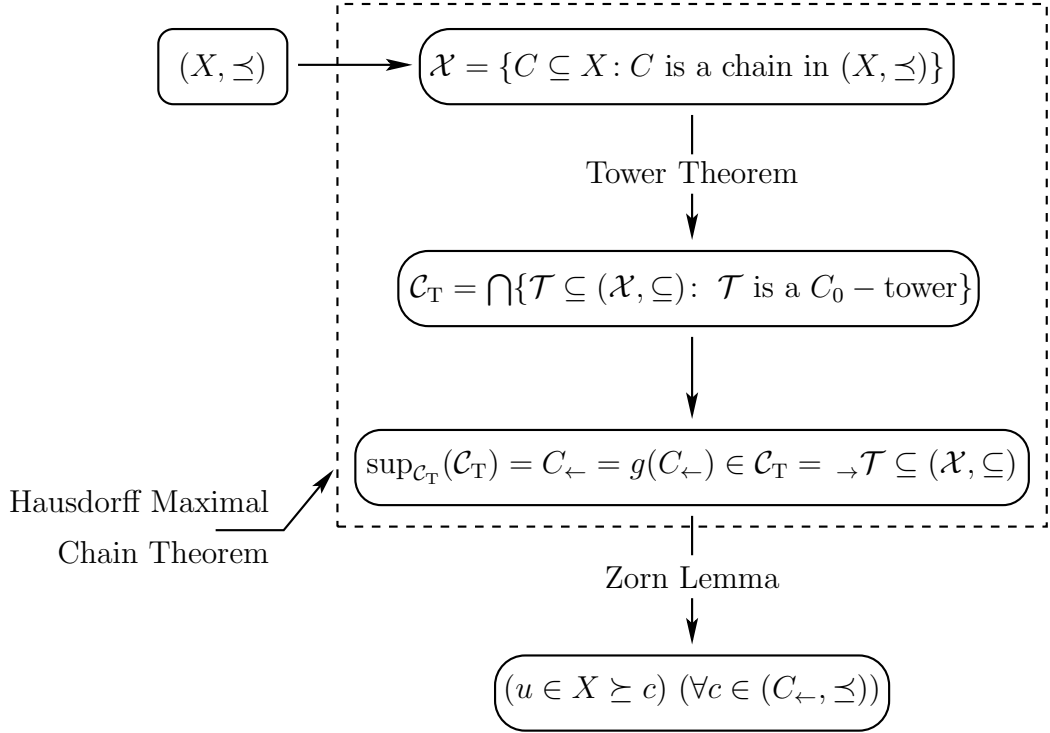
\begin{figure}[htbp]
\noindent \begin{center}{\small \input{Zorn.pstex_t}}\end{center}{\small \par}

\begin{singlespace}

\caption{{\footnotesize \label{Fig: Zorn}Application of Zorn's Lemma to $(X,\preceq)$.
Starting with a partially ordered set $(X,\preceq)$, construct: }}
\end{singlespace}

\begin{singlespace}
{\footnotesize (a) The one-level higher subset $\mathcal{X}=\{ C\in\mathcal{P}(X)\!:C\textrm{ is a chain in }(X,\preceq)\}$
of $\mathcal{P}(X)$ consisting of all the totally ordered subsets of $(X,\preceq)$, }{\footnotesize \par}

{\footnotesize (b) The smallest common $g$-sequential totally ordered towered
chain $\mathcal{C}_{\textrm{T}}=\{ C_{i}\in\mathcal{P}(X)\!:C_{i}\subseteq C_{j}\textrm{ for }i\leq j\}\subseteq\mathcal{P}(X)$
of all sequential $g$-towers of $\mathcal{X}$ by Thm. 4.1, which infact is
a principal filter base of totally ordered subsets of $(X,\preceq)$ at the
smallest element $C_{0}$. }{\footnotesize \par}

{\footnotesize (c) Apply Hausdorff Maximal Principle to $(\mathcal{X},\subseteq)$
to get the subset $\sup_{\mathcal{C}_{\textrm{T}}}(\mathcal{C}_{\textrm{T}})=C_{\leftarrow}=g(C_{\leftarrow})\in\mathcal{C}_{\textrm{T}}\subseteq\mathcal{P}(X)$
of $(X,\preceq)$ as the supremum of $(\mathcal{X},\subseteq)$ in $\mathcal{C}_{\textrm{T}}$.
The identification of this supremum as a maximal element of $(\mathcal{X},\subseteq)$
is a consequence of (ST2) and Eqs. (\ref{Eqn: sup chain}), (\ref{Eqn: fixed point})
that actually puts the supremum into $\mathcal{X}$ itself. }{\footnotesize \par}

{\footnotesize By returning to the original level $(X,\preceq)$ }{\footnotesize \par}

{\footnotesize (d) Zorn's Lemma finally yields the required maximal element
$u\in X$ as an upper bound of the maximal totally ordered subset $(C_{\leftarrow},\preceq)$
of $(X,\preceq)$. }{\footnotesize \par}
\end{singlespace}

\begin{singlespace}
\begin{center}{\footnotesize The dashed segment denotes the higher Hausdorff
$(\mathcal{X},\subseteq)$ level leading to the base $(X,\preceq)$ Zorn level. }\end{center}\end{singlespace}

\end{figure}

The sequence of steps leading to Zorn's Lemma, and thence to the maximal of
a partially ordered set, is summarised in Fig. \ref{Fig: Zorn}.

{\small }%

The three examples below of the application of Zorn's Lemma clearly reflect
the increasing complexity of the problem considered, with the maximals a point,
a subset, and a set of subsets of $X$, so that these are elements of $X$,
$\mathcal{P}(X)$, and $\mathcal{P}^{2}(X)$ respectively. 

\medskip{}
\noindent \textbf{Example 4.2.} (1) Let $X=(\{ a,b,c\},\preceq)$ be a three-point
base-level ground set ordered lexicographically, that is $a\prec b\prec c$.
A chain $\mathcal{C}$ of the partially ordered Hausdorff-level set $\mathcal{X}$
consisting of subsets of $X$ given by Eq. (\ref{Eqn: Hausdorff}) is, for example,
$\{\{ a\},\{ a,b\}\}$ and the six $g$-sequential chained towers

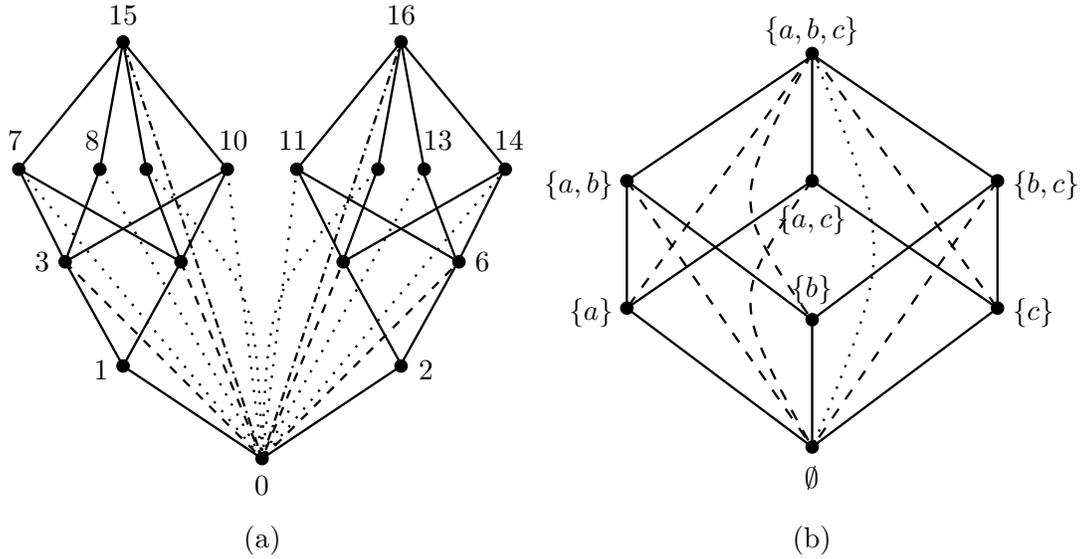
\begin{figure}[htbp]
\noindent \begin{center}{\small \input{order.pstex_t}}\end{center}{\small \par}

\begin{singlespace}

\caption{{\footnotesize \label{Fig: order}Tree diagrams of two partially ordered sets
where two points are connected by a line iff they are comparable to each other,
with the solid lines linking immediate neighbours and the dashed, dotted and
dashed-dotted lines denoting second, third and fourth generation orderings according
to the principle of transitivity of the order relation. There are $8\times2$
chains of (a) and 7 chains of (b) starting from respective smallest elements
with the immediate successor chains shown in solid lines. The 17 point set $X=\{0,1,2,\cdots,15,16\}$
in (a) has two maximals but no maximum, while in (b) there is a single maximum
of $\mathcal{P}(\{ a,b,c\})$, and three maximals without any maximum for $\mathcal{P}(\{ a,b,c\})-\{ a,b,c\}$.
In (a), let $A=\{1,3,4,7,9,10,15\}$, $B=\{1,3,4,6,7,13,15\}$, $C=\{1,3,4,10,11,16\}$
and $D=\{1,3,4\}$. The upper bounds of $D$ in $A$ are $7$, $10$ and $15$
without any supremum (as there is no smallest element of $\{7,10,15\}$), and
the upper bounds of $D$ in $B$ are $7\textrm{ and }15$ with $\sup_{B}(D)=7$,
while $\sup_{C}(D)=10$. Finally the maximal, maximum, and the supremum in $A$
of $\{1,3,4,7\}$ are all the same illustrating how the supremum of a set can
belong to itself. Observe how the supremum and upper bound of a set are with
reference to its complement in contrast with the maximum and maximal that have
nothing to do with anything outside the set.}}\end{singlespace}

\end{figure}

\noindent \renewcommand{\arraystretch}{1.2}\[
\begin{array}{c}
\mathcal{C}_{1}=\{\emptyset,\{ a\},\{ a,b\},\{ a,b,c\}\},\qquad\mathcal{C}_{2}=\{\emptyset,\{ a\},\{ a,c\},\{ a,b,c\}\}\\
\mathcal{C}_{3}=\{\emptyset,\{ b\},\{ a,b\},\{ a,b,c\}\},\qquad\mathcal{C}_{4}=\{\emptyset,\{ b\},\{ b,c\},\{ a,b,c\}\}\\
\mathcal{C}_{5}=\{\emptyset,\{ c\},\{ a,c\},\{ a,b,c\}\},\qquad\mathcal{C}_{6}=\{\emptyset,\{ c\},\{ b,c\},\{ a,b,c\}\}\end{array}\]
\renewcommand{\arraystretch}{1}

\noindent built from the smallest element $\emptyset$ corresponding to the
six distinct ways of reaching $\{ a,b,c\}$ from $\emptyset$ along the sides
of the cube marked on the figure with solid lines, all belong to $\mathcal{X}$;
see Fig. \ref{Fig: order}(b). An example of a tower in $(\mathcal{X},\subseteq)$
which is not a chain is \[
\mathcal{T}=\{\emptyset,\{ a\},\{ b\},\{ c\},\{ a,b\},\{ a,c\},\{ b,c\},\{ a,b,c\}\}.\]
 Hence the common infimum towered chained subset \[
\mathcal{C}_{\textrm{T}}=\{\emptyset,\{ a,b,c\}\}=\,_{\rightarrow}\mathcal{T}\subseteq\mathcal{P}(X)\]
 of $\mathcal{X}$, with \[
\sup_{\mathcal{C}_{\textrm{T}}}(\mathcal{C}_{\textrm{T}})=C_{\leftarrow}=\{ a,b,c\}=g(C_{\leftarrow})\in\mathcal{C}_{\textrm{T}}\subseteq\mathcal{P}(X)\]
 the only maximal element of $\mathcal{P}(X)$. Zorn's Lemma now assures the
existence of a maximal element of $c\in X$. Observe how the maximal element
of $(X,\preceq)$ is obtained by going one level higher to $\mathcal{X}$ at
the Hausdorff stage and returning to the base level $X$ at Zorn, see Fig. \ref{Fig: Zorn}
for a schematic summary of this sequence of steps. 

{\small }%
{\small \par}

(2) \emph{Basis of a vector space.} A linearly independent set of vectors in
a vector space $X$ that spans the space is known as the Hamel basis of $X$.
To prove the existence of a Hamel basis in a vector space, Zorn's lemma is invoked
as follows. 

The ground base level of the linearly independent subsets of $X$ \[
\mathcal{X}=\{\{ x_{i_{j}}\}_{j=1}^{J}\in\mathcal{P}(X)\!:\textrm{Span}(\{ x_{i_{j}}\}_{j=1}^{J})=0\Rightarrow(\alpha_{j})_{j=1}^{J}=0\,\forall J\geq1\}\subseteq\mathcal{P}(X)),\]
 with $\textrm{Span}(\{ x_{i_{j}}\}_{j=1}^{J}):=\sum_{j=1}^{J}\alpha_{j}x_{i_{j}}$,
is such that no $x\in\mathcal{X}$ can be expressed as a linear combination
of the elements of $\mathcal{X}-\{ x\}$. $\mathcal{X}$ clearly has a smallest
element, say $\{ x_{i_{1}}\}$, for some non-zero $x_{i_{1}}\in X$. Let the
higher Hausdorff level \[
\mathfrak{X}=\{\mathcal{C}\in\mathcal{P}^{2}(X)\!:\mathcal{C}\textrm{ is a chain in }(\mathcal{X},\subseteq)\}\subseteq\mathcal{P}^{2}(X)\]
 collection of the chains \[
\mathcal{C}_{i_{K}}=\{\{ x_{i_{1}}\},\{ x_{i_{1}},x_{i_{2}}\},\cdots,\{ x_{i_{1}},x_{i_{2}},\cdots,x_{i_{K}}\}\}\textrm{ }\in\mathcal{P}^{2}(X)\]
 of $\mathcal{X}$ comprising linearly independent subsets of $X$ be $g$-built
from the smallest $\{ x_{i_{1}}\}$. Any chain $\mathfrak{C}$ of $\mathfrak{X}$
is bounded above by the union $\mathcal{C}_{*}=\sup_{\mathfrak{X}}\mathfrak{C}=\bigcup_{\mathcal{C}\in\mathfrak{C}}\mathcal{C}$
which is a chain in $\mathcal{X}$ containing $\{ x_{i_{1}}\}$, thereby verifying
(ST2) for $\mathfrak{X}$. Application of the tower theorem to $\mathfrak{X}$
implies that the chain \[
\mathfrak{C}_{\textrm{T}}=\{\mathcal{C}_{i_{1}},\mathcal{C}_{i_{2}},\cdots,\mathcal{C}_{i_{n}},\cdots\}=\,_{\rightarrow}\mathfrak{T}\subseteq\mathcal{P}^{2}(X)\]
 in $\mathfrak{X}$ of chains of $\mathcal{X}$ is a $g$-sequential fully ordered
towered subset of $(\mathfrak{X},\subseteq)$ consisting of the common elements
of all $g$-sequential towers of $(\mathfrak{X},\subseteq)$, that infact is
a \emph{chained} \emph{principal ultrafilter on $(\mathcal{P}(X),\subseteq)$
generated by the filter-base $\{\{\{ x_{i_{1}}\}\}\}$} \emph{at $\{ x_{i_{1}}\}$},
where \[
\mathfrak{T}=\{\mathcal{C}_{i_{1}},\mathcal{C}_{i_{2}},\cdots,\mathcal{C}_{j_{n}},\mathcal{C}_{j_{n+1}},\cdots\}\]
 for some $n\in\mathbb{N}$ is an example of non-chained $g$-tower whenever
$(\mathcal{C}_{j_{k}})_{k=n}^{\infty}$ is neither contained in nor contains
any member of the $(\mathcal{C}_{i_{k}})_{k=1}^{\infty}$ chain. Hausdorff's
chain theorem now yields the fixed-point $g$-chain $\mathcal{C}_{\leftarrow}\,\in\mathfrak{X}$
of $\mathcal{X}$ \[
\sup_{\mathfrak{C}_{\textrm{T}}}(\mathfrak{C}_{\textrm{T}})=\mathcal{C}_{\leftarrow}=\{\{ x_{i_{1}}\},\{ x_{i_{1}},x_{i_{2}}\},\{ x_{i_{1}},x_{i_{2}},x_{i_{3}}\},\cdots\}=g(\mathcal{C}_{\leftarrow})\in\mathfrak{C}_{\textrm{T}}\subseteq\mathcal{P}^{2}(X)\]
 as a maximal \emph{totally ordered} \emph{principal filter on $X$ that is
generated by the filter-base $\{\{ x_{i_{1}}\}\}$} \emph{at $x_{i_{1}}$},
whose supremum $B=\{ x_{i_{1}},x_{i_{2}},\cdots\}\in\mathcal{P}(X)$ is, by
Zorn's lemma, a maximal element of the base level $\mathcal{X}$. This maximal
linearly independent subset of $X$ is the required Hamel basis for $X$: Indeed,
if the span of $B$ is not the whole of $X$, then $\textrm{Span}(B)\bigcup x$,
with $x\notin\textrm{Span}(B)$ would, by definition, be a linearly independent
set of $X$ strictly larger than $B$, contradicting the assumed maximality
of the later. It needs to be understood that since the infinite basis cannot
be classified as being linearly independent, we have here an important example
of the supremum of the maximal chained set not belonging to the set even though
this criterion was explicitly used in the construction process according to
(ST2) and (ST3).

Compared to this purely algebraic concept of basis in a vector space, is the
Schauder basis in a normed space which combines topological structure with the
linear in the form of convergence: If a normed vector space contains a sequence
$(e_{i})_{i\in\mathbb{Z}_{+}}$ with the property that for every $x\in X$ there
is an unique sequence of scalars $(\alpha_{i})_{i\in\mathbb{Z}_{+}}$ such that
the remainder $\parallel x-(\alpha_{1}e_{1}+\alpha_{2}e_{2}+\cdots+\alpha_{I}e_{I})\parallel$
approaches $0$ as $I\rightarrow\infty$, then the collection $(e_{i})$ is
known as a Schauder basis for $X$. 

(3) \emph{Ultrafilter.} Let $X$ be a set. The set \[
{\textstyle _{\textrm{F}}\mathcal{S}=\{ S_{\alpha}\in\mathcal{P}(X)\!:S_{\alpha}\bigcap S_{\beta}\neq\emptyset,\textrm{ }\forall\alpha\neq\beta\}\subseteq\mathcal{P}(X)}\]
 of all nonempty subsets of $X$ with finite intersection property is known
as a \emph{filter subbase on} $X$ and $_{\textrm{F}}\mathcal{B}=\{ B\subseteq X\!:B=\bigcap_{i\in I\subset\mathbb{D}}S_{i}\}$,
for $I\subset\mathbb{D}$ a finite subset of a directed set $\mathbb{D}$, is
a \emph{filter-base on $X$} \emph{associated with the subbase} $_{\textrm{F}}\mathcal{S}$;
cf. Appendix A1. Then the \emph{filter generated by} $_{\textrm{F}}\mathcal{S}$
consisting of every superset of the finite intersections $B\in\,_{\textrm{F}}\mathcal{B}$
of sets of $_{\textrm{F}}\mathcal{S}$ is the smallest filter that contain the
subbase $_{\textrm{F}}\mathcal{S}$ and base $_{\textrm{F}}\mathcal{B}$. For
notational simplicity, we will denote the subbase $_{\textrm{F}}\mathcal{S}$
in the rest of this example simply by $\mathcal{S}$. 

Consider the base-level ground set of all filter subbases on $X$ \[
\mathfrak{S}=\{\mathcal{S}\in\mathcal{P}^{2}(X)\!:\bigcap_{\emptyset\neq\mathcal{R}\subseteq\mathcal{S}}\mathcal{R}\neq\emptyset\textrm{ for every finite subset of }\mathcal{S}\}\subseteq\mathcal{P}^{2}(X),\]
 ordered by inclusion in the sense that $\mathcal{S}_{\alpha}\subseteq\mathcal{S}_{\beta}\textrm{ for all }\alpha\preceq\beta\in\mathbb{D}$,
and let the higher Hausdorff-level \[
\widetilde{\mathfrak{X}}=\{\mathfrak{C}\in\mathcal{P}^{3}(X)\!:\mathfrak{C}\textrm{ is a chain in }(\mathfrak{S},\subseteq)\}\subseteq\mathcal{P}^{3}(X)\]
 comprising the collection of the totally ordered chains \[
\mathfrak{C}_{\kappa}=\{\{ S_{\alpha}\},\{ S_{\alpha},S_{\beta}\},\cdots,\{ S_{\alpha},S_{\beta},\cdots,S_{\kappa}\}\}\in\mathcal{P}^{3}(X)\]
 of $\mathfrak{S}$ be $g$-built from the smallest $\{ S_{\alpha}\}$ then
an \emph{ultrafilter} on $X$ is a maximal member $\mathcal{S}_{+}$ of $(\mathfrak{S},\subseteq)$
in the usual sense that any subbase $\mathcal{S}$ on $X$ must necessarily
be contained in $\mathcal{S}_{+}$ so that $\mathcal{S}_{+}\subseteq\mathcal{S}\Rightarrow\mathcal{S}=\mathcal{S}_{+}$
for any $\mathcal{S}\subseteq\mathcal{P}(X)$ with FIP. The tower theorem now
implies that the element \[
\widetilde{\mathfrak{C}_{\textrm{T}}}=\{\mathfrak{C}_{\alpha},\mathfrak{C}_{\beta},\cdots,\mathfrak{C}_{\nu},\cdots\}=\,\widetilde{_{\rightarrow}\mathfrak{T}}\subseteq\mathcal{P}^{3}(X)\]
 of $\mathcal{P}^{4}(X)$, which is a chain in $\widetilde{\mathfrak{X}}$ of
the chains of $\mathfrak{S}$, is a $g$-sequential fully ordered towered subset
of the common elements of all sequential towers of $(\widetilde{\mathfrak{X}},\subseteq)$
and a \emph{chained} \emph{principal ultrafilter on $(\mathcal{P}^{2}(X),\subseteq)$
generated by the filter-base $\{\{\{ S_{\alpha}\}\}\}$} \emph{at} $\{ S_{\alpha}\}$;
here \[
\widetilde{\mathfrak{T}}=\{\mathfrak{C}_{\alpha},\mathfrak{C}_{\beta},\cdots,\mathfrak{C}_{\sigma},\mathfrak{C}_{\varsigma},\cdots\},\]
 is an obvious example of non-chained $g$-tower whenever $(\mathfrak{C}_{\sigma})$
is neither contained in, nor contains, any member of the $\mathfrak{C}_{\alpha}$-chain.
Hausdorff's chain theorem now yields the fixed-point $\widetilde{\mathfrak{C}_{\leftarrow}}\,\in\widetilde{\mathfrak{X}}$
\[
\sup_{\widetilde{\mathfrak{C}_{\textrm{T}}}}(\widetilde{\mathfrak{C}_{\textrm{T}}})=\widetilde{\mathfrak{C}_{\leftarrow}}=\{\{ S_{\alpha}\},\{ S_{\alpha},S_{\beta}\},\{ S_{\alpha},S_{\beta},S_{\gamma}\},\cdots\}=g(\widetilde{\mathfrak{C}_{\leftarrow}})\in\widetilde{\mathfrak{C}_{\textrm{T}}}\subseteq\mathcal{P}^{3}(X)\]
 as a maximal \emph{totally ordered} $g$-chained towered subset of $X$ that
is, by Zorn's lemma, a maximal element of the base level subset $\mathfrak{S}$
of $\mathcal{P}^{2}(X)$. $\widetilde{\mathfrak{C}_{\leftarrow}}$ is a \emph{chained
principal ultrafilter on} $(\mathcal{P}(X),\subseteq)$ \emph{generated by the
filter-base $\{\{ S_{\alpha}\}\}$} \emph{at $S_{\alpha}$}, while $\mathcal{S}_{+}=\{ S_{\alpha},S_{\beta},S_{\gamma},\cdots\}\in\mathcal{P}^{2}(X)$
is an (non-principal) \emph{ultrafilter on} $X$ --- characterized by the property
that any collection of subsets on $X$ with FIP (that is any filter subbase
on $X$) must be contained the maximal set $\mathcal{S}_{+}$ having FIP ---
that is not a principal filter unless $\mathcal{S}_{\alpha}$ is a singleton
set $\{ x_{\alpha}\}$. $\qquad\blacksquare$

\begin{spacing}{1.4}
\medskip{}
What emerges from these application of Zorn's Lemma is the remarkable fact that
\emph{infinities (the dot-dot-dots) can be formally introduced as {}``limiting
cases'' of finite systems in a purely set-theoretic context} \emph{without
the need for topologies, metrics or convergences.} The significance of this
observation will become clear from our discussions on filters and topology leading
to Sec. 4.2 below. Also, the observation on the successive iterates of the power
sets $\mathcal{P}(X)$ in the examples above was to suggest their anticipated
role in the complex evolution of a dynamical system that is expected to play
a significant part in our future interpretation and understanding of this adaptive
and self-organizing phenomenon of nature. {\small }{\small \par}
\end{spacing}

\vspace{-0.15cm}
\noindent \begin{flushright}\textbf{\textit{End Tutorial5}}\end{flushright}
\medskip{}

\noindent From the examples in Tutorial5, it should be clear that the sequential
steps summarized in Fig. \ref{Fig: Zorn} are involved in an application of
Zorn's lemma to show that a partially ordered set has a maximal element with
respect to its order. Thus for a partially ordered set $(X,\preceq)$, form
the set $\mathcal{X}$ of all chains $C$ in $X$. If $C_{+}$ is a maximal
chain of $X$ obtained by the Hausdorff Maximal Principle from the chain $\mathcal{C}$
of all chains of $X$, then its supremum $u$ is a maximal element of $(X,\preceq)$.
This sequence is now applied, paralleling Example 4.2(1), to the set of arbitrary
relations $\textrm{Multi}(X)$ on an infinite set $X$ in order to formulate
our definition of chaos that follows. {\large }{\large \par}

Let $f$ be a \emph{noninjective map} in $\textrm{Multi}(X)$ and $P(f)$ the
number of injective branches of $f$. Denote by \[
F=\{ f\in\textrm{Multi}(X)\!:f\textrm{ is a noninjective function on }X\}\subseteq\textrm{Multi}(X)\]
 the resulting basic collection of noninjective functions in $\textrm{Multi}(X)$. 

\smallskip{}
(i) For every $\alpha$ in some directed set $\mathbb{D}$, let $F$ have the
extension property \[
(\forall f_{\alpha}\in F)(\exists f_{\beta}\in F)\!:P(f_{\alpha})\leq P(f_{\beta})\]

(ii) Let a partial order $\preceq$ on $\textrm{Multi}(X)$ be defined, for
$f_{\alpha},f_{\beta}\in\textrm{Map}(X)\subseteq\textrm{Multi}(X)$ by \begin{equation}
P(f_{\alpha})\leq P(f_{\beta})\Longleftrightarrow f_{\alpha}\preceq f_{\beta},\label{Eqn: chaos1}\end{equation}

\noindent with $P(f):=1$ for the smallest $f$, define a partially ordered
subset $(F,\preceq)$ of $\textrm{Multi}(X)$. This is actually a preorder on
$\textrm{Multi}(X)$ in which functions with the same number of injective branches
are equivalent to each other. 

(iii) Let \[
C_{\nu}=\{ f_{\alpha}\in\textrm{Multi}(X)\!:f_{\alpha}\preceq f_{\nu}\}\in\mathcal{P}(F),\qquad\nu\in\mathbb{D},\]
 be the $g$-chains of non-injective functions of $\textrm{Multi}(X)$ and \[
\mathcal{X}=\{ C\in\mathcal{P}(F)\!:C\textrm{ is a chain in }(F,\preceq)\}\subseteq\mathcal{P}(F)\]
 denote the corresponding Hausdorff level of the chains of $F$, with \[
\mathcal{C}_{\textrm{T}}=\{ C_{\alpha},C_{\beta},\cdots,C_{\nu},\cdots\}=\,_{\rightarrow}\mathcal{T}\subseteq\mathcal{P}(F)\]
 being a $g$-sequential chain in $\mathcal{X}$ . \emph{}By Hausdorff Maximal
Principle, there is a maximal fixed-point $g$-towered chain $C_{\leftarrow}\in\mathcal{X}$
of $F$ \[
\sup_{\mathcal{C}_{\textrm{T}}}(\mathcal{C}_{\textrm{T}})=C_{\leftarrow}=\{ f_{\alpha},f_{\beta},f_{\gamma},\cdots\}=g(C_{\leftarrow})\in\mathcal{C}_{\textrm{T}}\subseteq\mathcal{P}(F).\]

\noindent Zorn's Lemma now applied to this maximal chain yields its supremum
as the maximal element of $C_{\leftarrow}$, and thereby of $F$. It needs to
be appreciated, as in the case of the algebraic Hamel basis, that the existence
of this maximal non-functional element was obtained purely set theoretically
as the {}``limit'' of a net of functions with increasing non-linearity, without
resorting to any topological arguments. Because it is not a function, this supremum
does not belong to the functional $g$-towered chain having it as a fixed point,
and this maximal chain does not possess a largest, or even a maximal, element,
although it does have a supremum.%
\footnote{{\small \label{Foot: supremum}A similar situation arises in the following more
intuitive example. Although the subset $A=\{1/n\}_{n\in Z_{+}}$ of the interval
$I=[-1,1]$ has no a smallest or minimal elements, it does have the infimum
0. Likewise, although $A$ is bounded below by any element of $[-1,0)$, it
has no greatest lower bound in $[-1,0)\bigcup(0,1]$. }%
} The supremum is a contribution of the inverse functional relations $(f_{\alpha}^{-})$
in the following sense. From Eq. (\ref{Eqn: func-multi}), the net of increasingly
non-injective functions of Eq. (\ref{Eqn: chaos1}) implies a corresponding
net of increasingly multivalued functions ordered inversely by the inverse relation
$f_{\alpha}\preceq f_{\beta}\Leftrightarrow f_{\beta}^{-}\preceq f_{\alpha}^{-}$.
Thus the inverse relations which are as much an integral part of graphical convergence
as are the direct relations, have a smallest element belonging to the multifunctional
class. Clearly, this smallest element as the required supremum of the increasingly
non-injective tower of functions defined by Eq. (\ref{Eqn: chaos1}), serves
to complete the significance of the tower by capping it with a {}``boundary''
element that can be taken to bridge the classes of functional and non-functional
relations on $X$.

We are now ready to define a \emph{maximally ill-posed problem $f(x)=y$} for
\emph{$x,y\in X$} in terms of a \emph{maximally non-injective map $f$} as
follows. 

\smallskip{}
\noindent \textbf{Definition 4.1.} \textbf{\emph{Chaotic map.}} \textsl{Let
$A$ be a non-empty closed set of a compact Hausdorff space $X.$ A function}
$f\in\textrm{Multi}(X)$ \textsl{}\textsl{\emph{(}}\textsl{equivalently the
sequence of functions $(f_{i})$}\textsl{\emph{)}} \textsl{is} \emph{maximally
non-injective} \textsl{or} \emph{chaotic on} \textsl{\emph{$A$}} \textsl{with
respect to the order relation} \textsl{\emph{(\ref{Eqn: chaos1})}} \textsl{if }

\textsl{(a) for any $f_{i}$ on $A$ there exists an $f_{j}$ on $A$ satisfying
$f_{i}\preceq f_{j}$ for every $j>i\in\mathbb{N}$.}

\emph{(b) the set $\mathcal{D}_{+}$ consists of a countable collection of isolated
singletons.$\qquad\square$}

\smallskip{}
\noindent \textbf{Definition 4.2.} \textbf{\emph{Maximally ill-posed problem.}}
\textsl{\noun{L}}\textsl{et $A$ be a non-empty closed set of a compact Hausdorff
space $X$ and let $f$ be a functional relation in} $\textrm{Multi}(X)$\textsl{.
The problem $f(x)=y$ is} \emph{maximally ill-posed at} \textsl{\emph{$y$}}
\textsl{if $f$ is chaotic on $A$}.$\qquad\square$

\smallskip{}
As an example of the application of these definitions, on the dense set $\mathcal{D}_{+}$,
the tent map satisfies both the conditions of sensitive dependence on initial
conditions and topological transitivity \cite{Devaney1989} and is also maximally
non-injective; the tent map is therefore chaotic on $\mathcal{D}_{+}.$ In contrast,
the examples of Secs. 1 and 2 are not chaotic as the maps are not topologically
transitive, although the Liapunov exponents, as in the case of the tent map,
are positive. Here the $(f_{n})$ are identified with the iterates of $f,$
and the {}``fixed point'' as one through which graphs of all the functions
on residual index subsets pass. When the set of points $\mathcal{D}_{+}$ is
dense in $[0,1]$ and both $\mathcal{D}_{+}$ and $[0,1]-\mathcal{D}_{+}=[0,1]-\bigcup_{i=0}^{\infty}f^{-i}(\textrm{Per}(f))$
(where $\textrm{Per}(f)$ denotes the set of periodic points of $f$) are totally
disconnected, it is expected that at any point on this complement the behaviour
of the limit will be similar to that on $\mathcal{D}_{+}$: these points are
special as they tie up the iterates on $\textrm{Per}(f)$ to yield the multifunctions.
Therefore in any neighbourhood $U$ of a $\mathcal{D}_{+}$-point, there is
an $x_{0}$ at which the \emph{forward orbit $\{ f^{i}(x_{0})\}_{i\geq0}$ is
chaotic} in the sense that 

(a) the sequence neither diverges nor does it converge in the image space of
$f$ to a periodic orbit of any period, and 

(b) the Liapunov exponent given by 

\begin{eqnarray*}
\lambda(x_{0}) & = & \lim_{n\rightarrow\infty}\ln\left|\frac{df^{n}(x_{0})}{dx}\right|^{1/n}\\
 & = & {\displaystyle \lim_{n\rightarrow\infty}\frac{1}{n}\sum_{i=0}^{n-1}\ln\left|\frac{df(x_{i})}{dx}\right|,\, x_{i}=f^{i}(x_{0}),}\end{eqnarray*}

\noindent which is a measure of the average slope of an orbit at $x_{0}$ or
equivalently of the average loss of information of the position of a point after
one iteration, is positive. Thus \emph{an orbit with positive Liapunov exponent
is chaotic if it is not} \emph{asymptotic} (that is neither convergent nor adherent,
having no convergent suborbit in the sense of Appendix A1) \emph{to an unstable
periodic orbit} \emph{or to any other limit set on which the dynamics is simple.}
A basic example of a chaotic orbit is that of an irrational in $[0,1]$ under
the shift map and that of the chaotic set its closure, the full unit interval. 

{\small }%
\begin{figure}[htbp]
\noindent \begin{center}{\small \input{logcob357_a.pstex_t}}\end{center}{\small \par}

\begin{singlespace}

\caption{{\footnotesize \label{Fig: logcob357}Multifunctional and cobweb plots of $3.569x(1-x)$.
Comparison of the graphs for the three values of $\lambda$ shown in figures
(a)--(f) illustrates how the dramatic changes in the character of the former
are conspicuously absent in the conventional plots that display no perceptible
distinction between the three cases. }}\end{singlespace}

\end{figure}
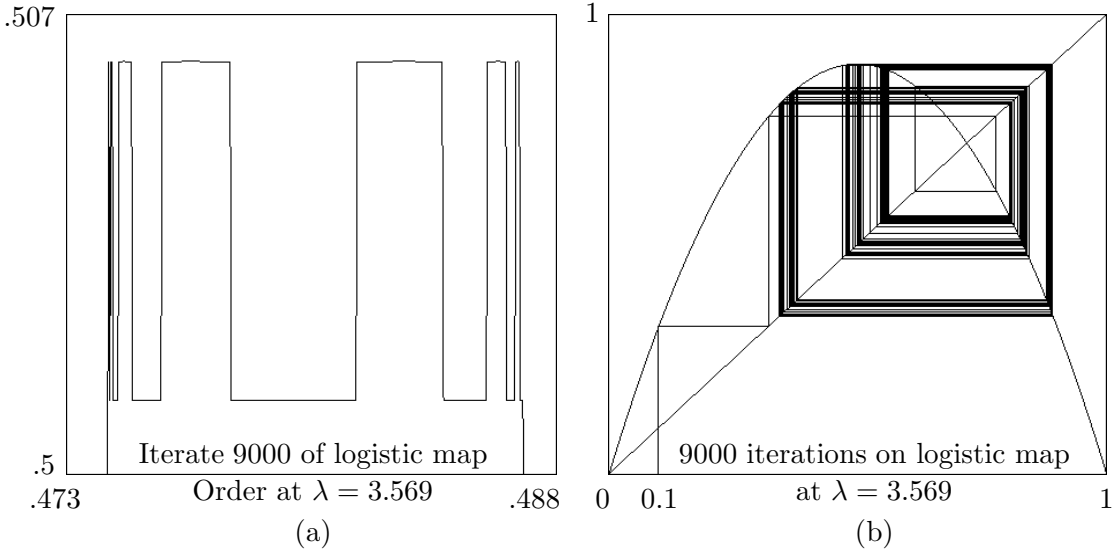
{\small \par}

Let $f\in\textrm{Map}((X,\mathcal{U}))$ and suppose that $A=\{ f^{j}(x_{0})\}_{j\in\mathbb{N}}$
is a sequential set corresponding to the orbit $\textrm{Orb}(x_{0})=(f^{j}(x_{0}))_{j\in\mathbb{N}}$,
and let $f_{\mathbb{R}_{i}}(x_{0})=\bigcup_{j\geq i}f^{j}(x_{0})$ be the $i$-residual
of the sequence $(f^{j}(x_{0}))_{j\in\mathbb{N}}$, with $_{\textrm{F}}\mathcal{B}_{x_{0}}=\{ f_{\mathbb{R}_{i}}(x_{0})\!:\textrm{Res}(\mathbb{N})\rightarrow X\textrm{ for all }i\in\mathbb{N}\}$
being the decreasingly nested filter-base associated with $\textrm{Orb}(x_{0})$.
The so-called \emph{$\omega$-limit set of} $x_{0}$ given by \begin{equation}
\begin{array}{ccl}
\omega(x_{0}) & \overset{\textrm{def}}= & \{ x\in X\!:(\exists n_{k}\in\mathbb{N})(n_{k}\rightarrow\infty)\textrm{ }(f^{n_{k}}(x_{0})\rightarrow x)\}\\
 & = & \{ x\in X\!:(\forall N\in\mathcal{N}_{x})(\forall f_{\mathbb{R}_{i}}\in\,_{\textrm{F}}\mathcal{B}_{x_{0}})\textrm{ }(f_{\mathbb{R}_{i}}(x_{0})\bigcap N\neq\emptyset)\}\end{array}\label{Eqn: Def: omega(x)}\end{equation}
 is simply the adherence set $\textrm{adh}(f^{j}(x_{0}))$ of the sequence $(f^{j}(x_{0}))_{j\in\mathbb{N}}$,
see Eq. (\ref{Eqn: net adh}); hence Def. A1.11 of the filter-base associated
with a sequence and Eqs. (\ref{Eqn: adh net2}), (\ref{Eqn: adh filter}), (\ref{Eqn: filter adh*})
and (\ref{Eqn: net-fil}) allow us to express $\omega(x_{0})$ more meaningfully
as \begin{equation}
\omega(x_{0})=\bigcap_{i\in\mathbb{N}}\textrm{Cl}(f_{\mathbb{R}_{i}}(x_{0})).\label{Eqn: adh_omega_x}\end{equation}
 It is clear from the second of Eqs. (\ref{Eqn: Def: omega(x)}) that for a
continuous $f$ and any $x\in X$, $x\in\omega(x_{0})$ implies $f(x)\in\omega(x_{0})$
so that the entire orbit of $x$ lies in $\omega(x_{0})$ whenever $x$ does
implying that the $\omega$-limit set is positively invariant; it is also closed
because the adherent set is a closed set according to Theorem A1.3. Hence $x_{0}\in\omega(x_{0})\Rightarrow A\subseteq\omega(x_{0})$
reduces the $\omega$-limit set to the closure of $A$ without any isolated
points, $A\subseteq\textrm{Der}(A)$. In terms of Eq. (\ref{Eqn: PrinFil_Cl(A)})
involving principal filters, Eq. (\ref{Eqn: adh_omega_x}) in this case may
be expressed in the more transparent form $\omega(x_{0})=\bigcap\textrm{Cl}(\,_{\textrm{F}}\mathcal{P}(\{ f^{j}(x_{0})\}_{j=0}^{\infty}))$
where the principal filter $_{\textrm{F}}\mathcal{P}(\{ f^{j}(x_{0})\}_{j=0}^{\infty})$
at $A$ consists of all supersets of $A=\{ f^{j}(x_{0})\}_{j=0}^{\infty}$,
and $\omega(x_{0})$ represents the adherence set of the principal filter at
$A$, see the discussion following Theorem A1.3. If $A$ represents a chaotic
orbit under this condition, then $\omega(x_{0})$ is sometimes known as a \emph{chaotic
set} \cite{Alligood1997}; thus the chaotic orbit infinitely often visits every
member of its chaotic set%
\footnote{{\small \label{Foot: omega-limit}How does this happen for $A=\{ f^{i}(x_{0})\}_{i\in\mathbb{N}}$
that is not the constant sequence $(x_{0})$ at a fixed point? As $i\in\mathbb{N}$
increases, points are added to $\{ x_{0},f(x_{0}),\cdots,f^{I}(x_{0})\}$ not,
as would be the case in a normal sequence, as a piled-up Cauchy tail, but as
points generally lying between those already present; recall a typical graph
as of Fig. \ref{Fig: tent4} for example.}%
} which is simply the $\omega$-limit set of a chaotic orbit that is itself contained
in its own limit set. Clearly the chaotic set if positive invariant, and from
Thm. A1.3 and its corollary it is also compact. Furthermore, if all (sub)sequences
emanating from points $x_{0}$ in some neighbourhood of the set converge to
it, then $\omega(x_{0})$ is called a \emph{chaotic attractor,} see \citet*{Alligood1997}.
As common examples of chaotic sets that are not attractors mention may be made
of the tent map with a peak value larger than $1$ at $0.5$, and the logistic
map with $\lambda\geq4$ again with a peak value at $0.5$ exceeding $1$.

{\small }%
\begin{figure}[htbp]
\noindent \begin{center}{\small \input{logcob357_b.pstex_t}}\end{center}{\small \par}

\begin{singlespace}
{\small Figure \ref{Fig: logcob357},} {\footnotesize contd: Multifunctional
and cobweb plots of $\lambda_{*}x(1-x)$ where $\lambda_{*}=3.5699456$}\end{singlespace}

\end{figure}
{\small \par}

\begin{figure}[htbp]
\noindent \begin{center}{\small \input{log357_a.pstex_t}}\end{center}{\small \par}

\begin{singlespace}

\caption{{\footnotesize \label{Fig: log357}Fixed points and cycles of logistic map.}
{\small }{\footnotesize The isolated fixed point of figure (b) yields two non-fixed
points to which the iterates converge} \emph{\footnotesize simultaneously} {\footnotesize in
the sense that the generated sequence converges to one iff it converges to the
other. This suggests that nonlinear dynamics of a system can lead to a situation
in which sequences in a Hausdorff space may converge to more than one point.
Since convergence depends on the topology (Corollary to Theorem A1.5), this
may be interpreted to mean that nonlinearity tends to modify the basic structure
of a space. }}\end{singlespace}

\end{figure}
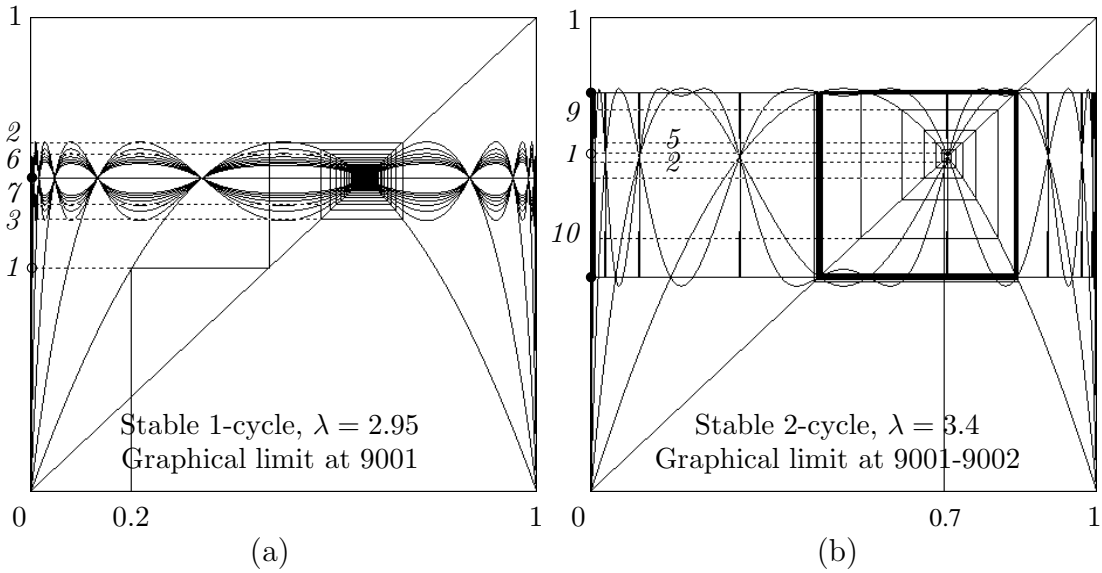

\begin{figure}[htbp]
\noindent \begin{center}{\small \input{attractor_a.pstex_t}}\end{center}{\small \par}

\begin{singlespace}

\caption{{\small \label{Fig: attractor}}{\footnotesize Chaotic attractors for $\lambda=\lambda_{*}$
and $\lambda=3.575$.} {\small }{\footnotesize For the logistic map the usual
bifurcation diagram (e) shows the chaotic attractors for $\lambda>\lambda_{*}=3.5699456$,
while (a)$-$(d) display the graphical limits for four values of $\lambda$
chosen for the Cantor set and 4,- 2-, and 1-piece attractors respectively. In
(f) the attractor $[0,1]$ (where the dotted lines represent odd iterates and
the solid lines even iterates of $f$) disappear if $f$ is reflected about
the $x$-axis. The function $f_{\textrm{f}}(x)$ is given by }}
\end{singlespace}

\begin{singlespace}
{\footnotesize \[
f_{\textrm{f}}(x)=\left\{ \begin{array}{ccl}
2(1+x)/3, &  & 0\leq x<1/2\\
2(1-x), &  & 1/2\leq x\leq1\end{array}\right.\]
}\end{singlespace}

\end{figure}
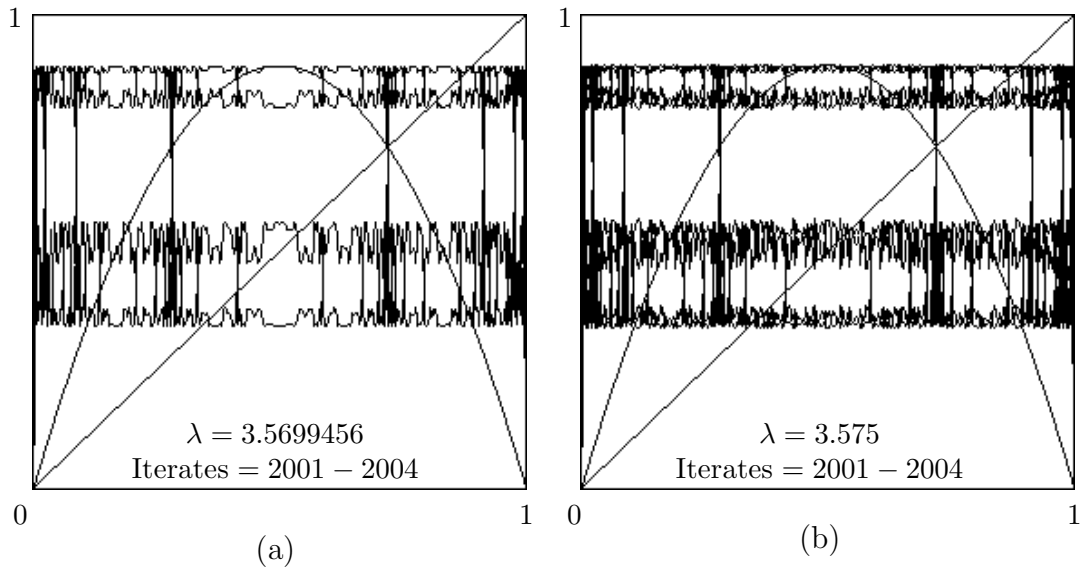

\medskip{}
It is important that the difference in the dynamical behaviour of the system
on $\mathcal{D}_{+}$ and its complement be appreciated. At any fixed point
$x$ of $f^{i}$ in $\mathcal{D}_{+}$ (or at its equivalent images in $[x]$)
the dynamics eventually gets attached to the (equivalent) fixed point, and the
sequence of iterates converges graphically in $\textrm{Multi}(X)$ to $x$ (or
its equivalent points). {\small }%
\begin{figure}[htbp]
\noindent \begin{center}{\small \input{logcob357_c.pstex_t}}\end{center}{\small \par}

\begin{singlespace}
{\small Figure \ref{Fig: logcob357},} {\footnotesize contd: Mulnctional and
cobweb plots of $3.57x(1-x)$. }\end{singlespace}

\end{figure}
 When $x\notin\mathcal{D}_{+}$, however, the orbit $A=\{ f^{i}(x)\}_{i\in\mathbb{N}}$
is chaotic in the sense that $(f^{i}(x))$ is not asymptotically periodic and
not being attached to any particular point they wander about in the closed chaotic
set $\omega(x)=\textrm{Der}(A)$ containing $A$ such that for any given point
in the set, some subsequence of the chaotic orbit gets arbitrarily close to
it. Such sequences do not converge anywhere but only frequent every point of
$\textrm{Der}(A)$. Thus although in the limit of progressively larger iterations
there is complete uncertainty of the outcome of an experiment conducted at either
of these two categories of initial points, whereas on $\mathcal{D}_{+}$ this
is due to a random choice from a multifunctional set of equally probable outputs
as dictated by the specific conditions under which the experiment was conducted
at that instant, on its complement the uncertainty is due to the chaotic behaviour
of the functional iterates themselves. Nevertheless it must be clearly understood
\emph{that this later behaviour is} \emph{entirely due to the multifunctional
limits at the $\mathcal{D}_{+}$ points which completely determine the behaviour
of the system on its complement.} As an explicit illustration of this situation,
recall that for the shift map $2x\textrm{ mod}(1)$ the $\mathcal{D}_{+}$ points
are the rationals on $[0,1]$, and any irrational is represented by a non-terminating
and non-repeating decimal so that almost all decimals in $[0,1]$ in any base
contain all possible sequences of any number of digits. For the logistic map,
the situation is more complex, however. Here the onset of chaos marking the
end of the period doubling sequence at $\lambda_{*}=3.5699456$ is signaled
by the disappearance of all stable fixed points, Fig. \ref{Fig: logcob357}(c),
with Fig. \ref{Fig: logcob357}(a) being a demonstration of the stable limits
for $\lambda=3.569$ that show up as convergence of the iterates to constant
valued functions (rather than as constant valued inverse functions) at stable
fixed points, shown more emphatically in Fig\ref{Fig: log357}(a). What actually
happens at $\lambda_{*}$ is shown in Fig. \ref{Fig: attractor}(a) in the next
subsection: the almost vertical lines produced at a large, but finite, iterations
$i$ (the multifunctions are generated only in the limiting sense of $i\rightarrow\infty$
and represent a boundary between functional and non-functional relations on
a set), decrease in magnitude with increasing iterations until they reduce to
points. This gives rise to a (totally disconnected) Cantor set on the $y$-axis
in contrast with the connected intervals that the multifunctional limits at
$\lambda>\lambda_{*}$ of Figs. \ref{Fig: attractor}(b)--(d) produce. By our
characterization Definition 4.1 of chaos therefore, $\lambda x(1-x)$ is chaotic
for the values of $\lambda>\lambda_{*}$ that are shown in Fig. \ref{Fig: attractor}.
We return to this case in the following subsection. 

\emph{}%
\begin{figure}[htbp]
\noindent \begin{center}{\small \input{log357_b.pstex_t}}\end{center}{\small \par}

\begin{singlespace}
{\small Figure} {\footnotesize \ref{Fig: log357}, contd: Isolated fixed points
of logistic map. The sequence of points generated by the iterates of the map
are marked on the $y$-axis of (a)--(c) in} \emph{\footnotesize italics}{\footnotesize .
The singletons $\{ x\}$ are $\omega$-limit sets of the respective fixed point
$x$ and is generated by the constant sequence $(x,x,\cdots)$. Whereas in (a)
this is the limit of every point in $(0,1)$, in the other cases these fixed
points are isolated in the sense of Def. 2.3. The isolated points, however,
give rise to sequences that converge to more than one point in the form of limit
cycles as shown in figures (b)--(d). }\end{singlespace}

\end{figure}

As an example of chaos \emph{in a noniterative system}, we investigate the following
question: While maximality of non-injectiveness produced by an increasing number
of injective branches is necessary for a family of functions to be chaotic,
is this also sufficient for the system to be chaotic? This is an important question
especially in the context of a non-iterative family of functions where fixed
points are of no longer relevant.

\begin{figure}[htbp]
\noindent \begin{center}{\small \input{tent17.pstex_t}}\end{center}{\small \par}

\begin{singlespace}

\caption{{\footnotesize \label{Fig: tent17}Similarity in the behaviour of the graphs
of the (a) tent and (b)$\mid\sin(2^{16}\pi x)\mid$ maps at 17 iterations demonstrate
chaoticity of the later.}}\end{singlespace}

\end{figure}
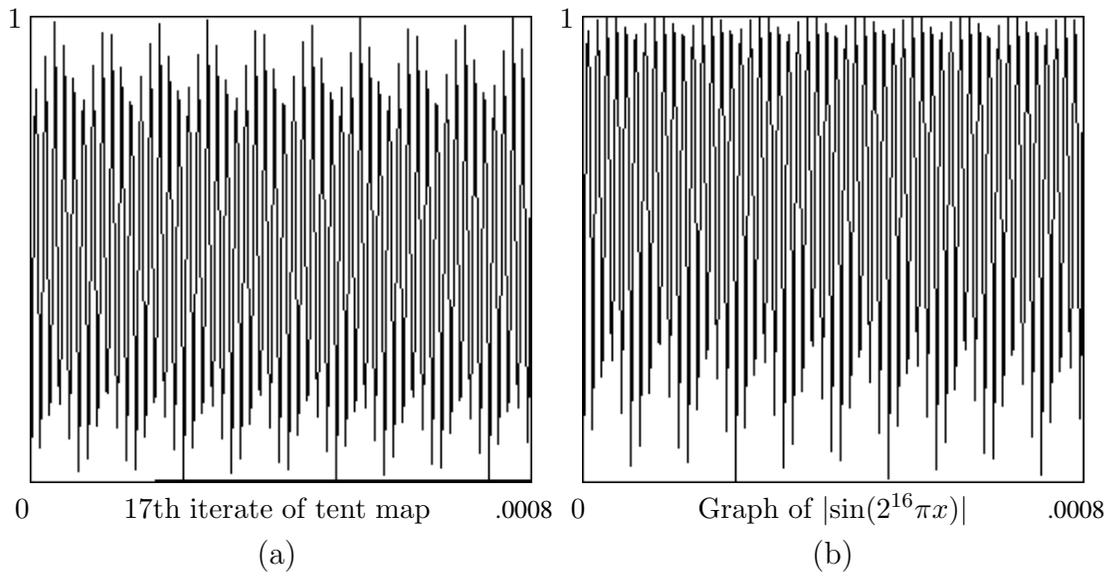

Consider the sequence of functions $|\sin(\pi nx)|_{n=1}^{\infty}.$ The graphs
of the subsequence $|\sin(2^{n-1}\pi x)|$ and of the sequence $(t^{n}(x))$
on {[}0,1{]} are qualitatively similar in that they both contain $2^{n-1}$
of their functional graphs each on a base of $1/2^{n-1}.$ Thus both $|\sin(2^{n-1}\pi x)|_{n=1}^{\infty}$
and $(t^{n}(x))_{n=1}^{\infty}$ converge graphically to the multifunction {[}0,1{]}
on the same set of points equivalent to 0. This is sufficient for us to conclude
that $|\sin(2^{n-1}\pi x)|_{n=1}^{\infty}$, and hence $|\sin(\pi nx)|_{n=1}^{\infty}$,
is chaotic on the infinite equivalent set {[}0{]}. While Fig. \ref{Fig: tent4}
was a comparison of the first four iterates of the tent and absolute sine maps,
Fig. {\small \ref{Fig: tent17}} following shows the {}``converged'' graphical
limits for after 17 iterations.

{\small }%
{\small \par}

\noindent \begin{flushleft}\textbf{\emph{\large 4.1. The chaotic attractor}}\end{flushleft}{\large \par}

\noindent One of the most fascinating characteristics of chaos in dynamical
systems is the appearance of attractors the dynamics on which are chaotic. \emph{}For
a subset $A$ of a topological space $(X,\mathcal{U})$ such that $\mathcal{R}(f(A))$
is contained in $A$ --- in this section, unless otherwise stated to the contrary,
$f(A)$ \emph{will} \emph{denote the} \emph{graph and not the range (image)}
\emph{of} $f$ --- which ensures that the iteration process can be carried out
in $A$, let \begin{equation}
\begin{array}{ccl}
{\displaystyle f_{\mathbb{R}_{i}}(A)} & = & {\displaystyle \bigcup_{j\geq i\in\mathbb{N}}f^{j}(A)}\\
 & = & {\displaystyle \bigcup_{j\geq i\in\mathbb{N}}\left(\bigcup_{x\in A}f^{j}(x)\right)}\end{array}\label{Eqn: absorbing set}\end{equation}
 generate the filter-base $_{\textrm{F}}\mathcal{B}$ with $A_{i}:=f_{\mathbb{R}_{i}}(A)\in\,_{\textrm{F}}\mathcal{B}$
being decreasingly nested, $A_{i+1}\subseteq A_{i}$ for all $i\in\mathbb{N}$,
in accordance with Def. A1.1. The existence of a maximal chain with a corresponding
maximal element as asssured by the Hausdorff Maximal Principle and Zorn's Lemma
respectively implies a nonempty core of $_{\textrm{F}}\mathcal{B}$. As in Sec.
3 following Def. 3.3, we now identify the filterbase with the neighbourhood
base at $f^{\infty}$ which allows us to define \begin{equation}
\begin{array}{ccl}
{\displaystyle \textrm{Atr}(A_{1})} & \overset{\textrm{def}}= & \textrm{adh}(\,_{\textrm{F}}\mathcal{B})\\
 & = & {\displaystyle \bigcap_{A_{i}\in\,_{\textrm{F}}\mathcal{B}}\textrm{Cl}(A_{i})}\end{array}\label{Eqn: attractor_adherence}\end{equation}
 as the attractor of the set $A_{1}$, where the last equality follows from
Eqs.(\ref{Eqn: Def: omega(A)}) and (\ref{Eqn: Def: Closure}) and the closure
is with respect to the topology induced by the neighbourhood filter base $_{\textrm{F}}\mathcal{B}$.
Clearly the attractor as defined here is the graphical limit of the sequence
of functions $(f^{i})_{i\in\mathbb{N}}$ which may be verified by reference
to Def. A1.8, Thm. A1.3 and the proofs of Thms. A1.4 and A1.5, together with
the directed set Eq. (\ref{Eqn: DirectedIndexed}) with direction (\ref{Eqn: DirectionIndexed}).
The \emph{basin of attraction} of the attractor is $A_{1}$ because the graphical
limit $(\mathcal{D}_{+},F(\mathcal{D}_{+}))\bigcup(G(\mathcal{R}_{+}),\mathcal{R}_{+})$
of Def. 3.1 may be obtained, as indicated above, by a proper choice of sequences
associated with $\mathcal{A}$. Note that in the context of iterations of functions,
the graphical limit $(\mathcal{D}_{+},y_{0})$ of the sequence $(f^{n}(x))$
denotes a stable fixed point $x_{*}$ with image $x_{*}=f(x_{*})=y_{0}$ to
which iterations starting at any point $x\in\mathcal{D}_{+}$ converge. The
graphical limits $(x_{i0},\mathcal{R}_{+})$ are generated with respect to the
class $\{ x_{i*}\}$ of points satisfying $f(x_{i0})=x_{i*}$, $i=0,1,2,\cdots$
equivalent to unstable fixed point $x_{*}:=x_{0*}$ to which inverse iterations
starting at any initial point in $\mathcal{R}_{+}$ must converge. Even though
only $x_{*}$ is inverse stable, an equivalent class of graphically converged
limit multis is produced at every member of the class $x_{i*}\in[x_{*}]$, resulting
in the far-reaching consequence \emph{that every member of the class is as significant
as the parent fixed point $x_{*}$ from which they were born in determining
the dynamics of the evolving system.} The point to remember about infinite intersections
of a collection of sets having finite intersection property, as in Eq. (\ref{Eqn: attractor_adherence}),
is that this may very well be empty; recall, however, that in a compact space
this is guaranteed not to be so. In the general case, if $\textrm{core}(\mathcal{A})\neq\emptyset$
then $\mathcal{A}$ is the principal filter at this core, and $\textrm{Atr}(A_{1})$
by Eqs. (\ref{Eqn: attractor_adherence}) and (\ref{Eqn: PrinFil_Cl(A)}) is
the closure of this core, which in this case of the topology being induced by
the filterbase, is just the core itself. $A_{1}$ by its very definition, is
a positively invariant set as any sequence of graphs converging to \emph{}$\textrm{Atr}(A_{1})$
must be eventually in $A_{1}$: the entire sequence therefore lies in $A_{1}$.
Clearly, from Thm. A3.1 and its corollary, the attractor is a positively invariant
compact set. A typical attractor is illustrated by the derived sets in the second
column of Fig. \ref{Fig: DerSets} which also illustrates that the set of functional
relations are open in $\textrm{Multi}(X)$; specifically functional-nonfunctional
correspondences are neutral-selfish related as in Fig. \ref{Fig: DerSets},
3-2, with the attracting graphical limit of Eq. (\ref{Eqn: attractor_adherence})
forming the boundary of (finitely)many-to-one functions and the one-to-(finitely)many
multifunctions. 

Equation (\ref{Eqn: attractor_adherence}) is to be compared with the \emph{image
definition of an attractor} \cite{Stuart1996} where $f(A)$ denotes the range
and not the graph of $f$. Then Eq. (\ref{Eqn: attractor_adherence}) can be
used to define a sequence of points $x_{k}\in A_{n_{k}}$ and hence the subset
\begin{eqnarray}
\omega(A) & \overset{\textrm{def}}= & \{ x\in X\!:(\exists n_{k}\in\mathbb{N})(n_{k}\rightarrow\infty)(\exists x_{k}\in A_{n_{k}})\textrm{ }(f^{n_{k}}(x_{k})\rightarrow x)\}\nonumber \\
 & = & \{ x\in X\!:(\forall N\in\mathcal{N}_{x})(\forall A_{i}\in\mathcal{A})(N\bigcap A_{i}\neq\emptyset)\}\label{Eqn: Def: omega(A)}\end{eqnarray}

\noindent as the corresponding attractor of $A$ that satisfies an equation
formally similar to (\ref{Eqn: attractor_adherence}) with the difference that
the filter-base $\mathcal{A}$ is now in terms of the image $f(A)$ of $A$,
which allows the adherence expression to take the particularly simple form \begin{equation}
\omega(A)=\bigcap_{i\in\mathbb{N}}\textrm{Cl}(f^{i}(A)).\label{Eqn: omega(A)_intersect}\end{equation}
 The complimentary subset excluded from this definition of $\omega(A)$, as
compared to $\textrm{Atr}(A_{1})$, that is required to complete the formalism
is given by Eq. (\ref{Eqn: basin}) below. Observe that the equation for $\omega(A)$
is essentially Eq. (\ref{Eqn: adh net1}), even though we prefer to use the
alternate form of Eq. (\ref{Eqn: adh net2}) as this brings out more clearly
the frequenting nature of the sequence. The basin of attraction \begin{equation}
\begin{array}{ccl}
B_{f}(A) & = & \{ x\in A\!:\omega(x)\subseteq\textrm{Atr}(A)\}\\
 & = & \{ x\in A\!:(\exists n_{k}\in\mathbb{N})(n_{k}\rightarrow\infty)\textrm{ }(f^{n_{k}}(x)\rightarrow x^{*}\in\omega(A)\textrm{ })\end{array}\label{Eqn: basin}\end{equation}
 of the attractor is the smallest subset of $X$ in which sequences generated
by $f$ must eventually lie in order to adhere at $\omega(A)$. Comparison of
Eqs. (\ref{Eqn: Attractor_R+}) with (\ref{Eqn: R+}) and (\ref{Eqn: basin})
with (\ref{Eqn: D+}) show that $\omega(A)$ can be identified with the subset
$\mathcal{R}_{+}$ on the $y$-axis on which the multifunctional limits $G\!:\mathcal{R}_{+}\rightarrow X$
of graphical convergence are generated, with its basin of attraction being contained
in the $\mathcal{D}_{+}$ associated with the injective branch of $f$ that
generates $\mathcal{R}_{+}$. In summary it may be concluded that since definitions
(\ref{Eqn: Def: omega(A)}) and (\ref{Eqn: basin}) involve both the domain
and range of $f$, a description of the attractor in terms of the graph of $f$,
like that of Eq. (\ref{Eqn: attractor_adherence}), is more pertinent and meaningful
as it combines the requirements of both these equations. Thus, for example,
as $\omega(A)$ is not the function $G(\mathcal{R}_{+})$, this attractor does
not include the equivalence class of inverse stable points that may be associated
with $x_{*}$, see for example Fig. \ref{Fig: omega}. 

From Eq. (\ref{Eqn: Def: omega(A)}), we may make the particularly simple choice
of $(x_{k})$ to satisfy $f^{n_{k}}(x_{-k})=x$ so that $x_{-k}=f_{\textrm{B}}^{-n_{k}}(x)$,
where $x_{-k}\in[x_{-k}]:=f^{-n_{k}}(x)$ is the element of the equivalence
class of the inverse image of $x$ corresponding to the injective branch $f_{\textrm{B}}$.
This choice is of special interest to us as it is the class that generates the
$G$-function on $\mathcal{R}_{+}$ in graphical convergence. This allows us
to express $\omega(A)$ as \begin{equation}
\omega(A)=\{ x\in X\!:(\exists n_{k}\in\mathbb{N})(n_{k}\rightarrow\infty)(f_{\textrm{B}}^{-n_{k}}(x)=x_{-k}\textrm{ converges in }(X,\mathcal{U}))\};\label{Eqn: Attractor_R+}\end{equation}
 note that the $x_{-k}$ of this equation and the $x_{k}$ of Eq. (\ref{Eqn: Def: omega(A)})
are, in general, quite different points. 

A simple illustrative example of the construction of $\omega(A)$ for the positive
injective branch of the homeomorphism $(4x^{2}-1)/3$, $-1\leq x\leq1$, is
shown in Fig. \ref{Fig: omega}, where the arrow-heads denote the converging
sequences $f^{n_{i}}(x_{i})\rightarrow x$ and $f^{n_{i}-m}(x_{i})\rightarrow x_{-m}$
which proves invariance of $\omega(A)$ for a homeomorphic $f$; here continuity
of the function and its inverse is explicitly required for invariance. Positive
invariance of a subset $A$ of $X$ implies that for any $n\in\mathbb{N}$ and
$x\in A$, $f^{n}(x)=y_{n}\in A$, while negative invariance assures that for
any $y\in A$, $f^{-n}(y)=x_{-n}\in A$. Invariance of $A$ in both the forward
and backward directions therefore means that for any $y\in A$ and $n\in\mathbb{N}$,
there exists a $x\in A$ such that $f^{n}(x)=y$. In interpreting this figure,
it may be useful to recall from Def. 4.1 that an increasing number of injective
branches of $f$ is a necessary, but not sufficient, condition for the occurrence
of chaos; thus in Figs. \ref{Fig: log357}(a) and \ref{Fig: omega}, increasing
noninjectivity of $f$ leads to constant valued limit functions over a connected
$\mathcal{D}_{+}$ in a manner similar to that associated with the classical
Gibb's phenomenon in the theory of Fourier series. 

{\small }%
\begin{figure}[htbp]
\noindent \begin{center}{\small \input{omega.pstex_t}}\end{center}{\small \par}

\begin{singlespace}

\caption{{\footnotesize \label{Fig: omega} The attractor for $f(x)=(4x^{2}-1)/3$,
for $-1\leq x\leq1$. The converging sequences are denoted by arrows on the
right, and $(x_{k})$ are chosen according to the construction shown. This example
demonstrates how although $A\subseteq f(A)$, where $A=[0,1]$ is the domain
of the positive injective branch of $f$, the succeeding images $(f^{i}(A))_{i\geq1}$
satisfy the required restriction for iteration, and $A$ in the discussion above
can be taken to be $f(A)$; this is permitted as only a finite number of iterates
is thereby discarded. It is straightforward to verify that $\textrm{Atr}(A_{1})=(-1,[-0.25,1])\bigcup((-1,1),-0.25)\bigcup(1,[-0.25,1])$
with $F(x)=-0.25$ on $\mathcal{D}_{-}=(-1,1)=\mathcal{D}_{+}$ and $G(y)=1,\textrm{ and }-1$
on $\mathcal{R}_{-}=[-0.25,1]=\mathcal{R}_{+}$. By comparison, $\omega(A)$
from either its definition Eq. (\ref{Eqn: Def: omega(A)}) or from the equivalent
intersection expression Eq. (\ref{Eqn: omega(A)_intersect}), is simply the
closed interval $\mathcal{R}_{+}=[-0.25,1]$. The italicized iterate numbers
on the graphs show how the oscillations die out with increasing iterations from
$x=\pm1$ and increasingly approach $-0.25$ in all neighbourhoods of $0$. }}\end{singlespace}

\end{figure}
{\small \par}

Graphical convergence of an increasingly nonlinear family of functions implied
by its increasing non-injectivity may now be combined with the requirements
of an attractor to lead to the concept of a chaotic attractor to be that on
which the dynamics is chaotic in the sense of Defs. 4.1. and 4.2. Hence 

\medskip{}
\noindent \textbf{Definition 4.3.} \textbf{\emph{Chaotic Attractor.}} \emph{Let
$A$ be a positively invariant subset of $X$. The attractor} $\textrm{Atr}(A)$
\emph{is chaotic on $A$ if there is sensitive dependence on initial conditions
for} all \emph{$x\in A$. The sensitive dependence manifests itself as multifunctional
graphical limits for all $x\in\mathcal{D}_{+}$ and as chaotic orbits when}
$x\not\in\mathcal{D}_{+}$\emph{.}$\qquad\square$
\medskip{}

{\small }%
{\small \par}

The picture of chaotic attractors that emerge from the foregoing discussions
and our characterization of chaos of Def. 4.1 is that it it is a subset of $X$
that is simultaneously {}``spiked'' multifunctional on the $y$-axis and consists
of a dense collection of singleton domains of attraction on the $x$-axis. This
is illustrated in Figure \ref{Fig: attractor} which shows some typical chaotic
attractors. The first four diagrams (a)$-$(d) are for the logistic map with
(b)$-$(d) showing the 4-, 2- and 1-piece attractors for $\lambda=3.575,\textrm{ }3.66,\textrm{ and }3.8$
respectively that are in qualitative agreement with the standard bifurcation
diagram reproduced in (e). Figs. (b)$-$(d) have the advantage of clearly demonstrating
how the attractors are formed by considering the graphically converged limit
as the object of study unlike in Fig. (e) which shows the values of the 501-1001th
iterates of $x_{0}=1/2$ as a function of $\lambda$. The difference in Figs.
(a) and (b) for a change of $\lambda$ from {\small $\lambda>\lambda_{*}=3.5699456$}
to 3.575 is significant as $\lambda=\lambda_{*}$ marks the boundary between
the nonchaotic region for $\lambda<\lambda_{*}$ and the chaotic for $\lambda>\lambda_{*}$
(this is to be understood as being suitably modified by the appearance of the
nonchaotic windows for some specific intervals in $\lambda>\lambda_{*}$). At
$\lambda_{*}$ the generated fractal Cantor set $\Lambda$ is an attractor as
it attracts almost every initial point $x_{0}$ so that the successive images
$x^{n}=f^{n}(x_{0})$ converge toward the Cantor set $\Lambda$. In Fig. (f)
the chaotic attractors for the piecewise continuous function on $[0,1]$

\noindent {\renewcommand{\arraystretch}{1.2}

\[
f_{\textrm{f}}(x)=\left\{ \begin{array}{ccc}
2(1+x)/3, &  & 0\leq x<1/2\\
2(1-x), &  & 1/2\leq x\leq1,\end{array}\right.\]
}

\noindent is $[0,1]$ where the dotted lines represent odd iterates and the
full lines even iterates of $f$; here the attractor disappears if the function
is reflected about the $x$-axis.

{\small }%
\begin{figure}[htbp]
\noindent \begin{center}{\small \input{attractor_b.pstex_t}}\end{center}{\small \par}

\begin{singlespace}
{\small Figure} {\footnotesize }{\small \ref{Fig: attractor}}{\footnotesize ,
contd.} {\small }{\footnotesize Chaotic attractors for $\lambda=3.66$ and $\lambda=3.8$.}\end{singlespace}

\end{figure}
{\small \par}

\vspace{0.75cm}
\noindent \begin{flushleft}\textbf{\emph{\large 4.2. Why Chaos? A Preliminary
Inquiry}}\end{flushleft}{\large \par}

\vspace{-0.2cm}
\noindent The question as to why a natural system should evolve chaotically
is both interesting and relevant, and this section attempts to advance a plausible
answer to this inquiry that is based on the connection between topology and
convergence contained in the Corollary to Theorem A1.5. Open sets are groupings
of elements that govern convergence of nets and filters, because the required
property of being either eventually of frequently in (open) neighbourhoods of
a point determines the eventual behaviour of the net; recall in this connection
the unusual convergence characteristics in cofinite and cocountable spaces.
Conversely for a given convergence characteristic of a class of nets, it is
possible to infer the topology of the space that is responsible for this convergence,
and it is this point of view that we adopt here to investigate the question
of this subsection: recall that our Definitions 4.1 and 4.2 were based on purely
algebraic set-theoretic arguments on ordered sets, just as the role of the choice
of an appropriate problem-dependent basis was highlighted at the end of Sec.
2.

\noindent {\small }%
\begin{figure}[htbp]
\noindent \begin{center}{\small \input{attractor_c.pstex_t}}\end{center}{\small \par}

\begin{singlespace}
{\small Figure} {\footnotesize \ref{Fig: attractor}, contd. Bifurcation diagram
and attractors for $f_{\textrm{f}}(x)$.}\end{singlespace}

\end{figure}
 Chaos as manifest in its attractors is a direct consequence of the increasing
nonlinearity of the map with increasing iteration; we reemphasize that this
is only a necessary condition so that the increasing nonlinearities of Figs.
\ref{Fig: log357} and \ref{Fig: omega} eventually lead to stable states and
not to chaotic instability. Under the right conditions as enunciated following
Fig. \ref{Fig: Zorn}, chaos appears to be the natural outcome of the difference
in the behaviour of a function $f$ and its inverse $f^{-}$ under their successive
applications. Thus $f=ff^{-}f$ allows $f$ to take advantage of its multi inverse
to generate all possible equivalence classes that is available to it, a feature
not accessible to $f^{-}=f^{-}ff^{-}$. As we have seen in the foregoing, equivalence
classes of fixed points, stable and unstable, are of defining significance in
determining the ultimate behaviour of an evolving dynamical system and as the
eventual (as also frequent) character of a filter or net in a set is dictated
by open neighbourhoods of points of the set, \emph{it is postulated that chaoticity
on a set $X$ leads to a reformulation of the open sets of $X$ to equivalence
classes generated by the evolving map $f$,} see Example 2.4(3). Such a redefinition
of open sets of equivalence classes allow the evolving system to temporally
access an ever increasing number of states even though the equivalent fixed
points are not fixed under iterations of $f$ except for the parent of the class,
and can be considered to be the governing criterion for the cooperative or collective
behaviour of the system. The predominance of the role of $f^{-}$ in $f=ff^{-}f$
in generating the equivalence classes (that is exploiting the many-to-one character)
of $f$ is reflected as limit multis for $f$ (that is constant $f^{-}$ on
$\mathcal{R}_{+}$) in $f^{-}=f^{-}ff^{-}$; this interpretation of the dynamics
of chaos is meaningful as graphical convergence leading to chaos is a result
of pointwise biconvergence of the sequence of iterates of the functions generated
by $f$. But as $f$ is a noninjective function \emph{on} $X$ \emph{possessing
the property of increasing nonlinearity in the form of increasing noninjectivity
with iteration,} various cycles of disjoint equivalence classes are generated
under iteration, see for example Fig. \ref{Fig: tent4}(a) for the tent map.
A reference to Fig. \ref{Fig: GenInv} shows that the basic set $X_{\textrm{B}}$,
for a finite number $n$ of iterations of $f$, contains the parent of each
of these open equivalent sets in the domain of $f$, with the topology on $X_{\textrm{B}}$
being the corresponding $p$-images of these disjoint saturated open sets of
the domain. In the limit of infinite iterations of $f$ leading to the multifunction
$\mathcal{M}$ (this is the $f^{\infty}$ of Sec. 4.1), the generated open sets
constitute a basis for a topology on $\mathcal{D}(f)$ and the basis for the
topology of $\mathcal{R}(f)$ are the corresponding $\mathcal{M}$-images of
these equivalent classes. \emph{It is our contention that the motive force behind
evolution toward a chaos, as defined by Def. 4.1, is the drive toward a state
of the dynamical system that supports ininality of the limit multi} $\mathcal{M}$\emph{;}
see Appendix A2 with the discussions on Fig. \ref{Fig: GenInv} and Eq. (\ref{Eqn: ininal})
in Sec. 2. In the limit of infinite iterations therefore, the open sets of the
range $\mathcal{R}(f)\subseteq X$ are the multi images that graphical convergence
generates at each of these inverse-stable fixed points. $X$ therefore has two
topologies imposed on it by the dynamics of $f$: the first of equivalence classes
generated by the limit multi $\mathcal{M}$ in the domain of $f$ and the second
as $\mathcal{M}$-images of these classes in the range of $f$. Quite clearly
these two topologies need not be the same; their intersection therefore can
be defined to be the \emph{chaotic topology} \emph{on} $X$ \emph{associated
with the chaotic map} $f$ on $X$. Neighbourhoods of points in this topology
cannot be arbitrarily small as they consist of all members of the equivalence
class to which any element belongs; hence a sequence converging to any of these
elements necessarily converges to all of them, and the eventual objective of
chaotic dynamics is to generate a topology in $X$ with respect to which elements
of the set can be grouped together in as large equivalence classes as possible
in the sense that if a net converges simultaneously to points $x\neq y\in X$
then $x\sim y$: $x$ is of course equivalent to itself while $x,y,z$ are equivalent
to each other iff they are simultaneously in every open set in which the net
may eventually belong. This hall-mark of chaos can be appreciated in terms of
a necessary obliteration of any separation property that the space might have
originally possessed, see property (H3) in Appendix A3. We reemphasize that
a set in this chaotic context is required to act in a dual capacity depending
on whether it carries the initial or final topology under $\mathcal{M}$. 

This preliminary inquiry into the nature of chaos is concluded in the final
section of this work. 

\vspace{1cm}
\noindent \begin{flushleft}\textbf{\large 5. Graphical convergence works}\end{flushleft}{\large \par}

\smallskip{}
\noindent We present in this section some real evidence in support of our hypothesis
of graphical convergence of functions in $\textrm{Multi}(X,Y)$. The example
is taken from neutron transport theory, and concerns the discretized spectral
approximation \cite{Sengupta1988,Sengupta1995} of Case's singular eigenfunction
solution of the monoenergetic neutron transport equation, \cite{Case1967}.
The neutron transport equation is a linear form of the Boltzmann equation that
is obtained as follows. Consider the neutron-moderator system as a mixture of
two species of gases each of which satisfies a Boltzmann equation of the type\begin{multline*}
\left(\frac{\partial}{\partial t}+v_{i}.\nabla\right)f_{i}(r,v,t)=\\
=\int dv^{\prime}\int dv_{1}\int dv_{1}^{\prime}\sum_{j}W_{ij}(v_{i}\rightarrow v^{\prime};v_{1}\rightarrow v_{1}^{\prime})\{ f_{i}(r,v^{\prime},t)f_{j}(r,v_{1}^{\prime},t)--f_{i}(r,v,t)f_{j}(r,v_{1},t)\})\end{multline*}

\noindent where\[
W_{ij}(v_{i}\rightarrow v^{\prime};v_{1}\rightarrow v_{1}^{\prime})=\mid v-v_{1}\mid\sigma_{ij}(v-v^{\prime},v_{1}-v_{1}^{\prime})\]

\noindent $\sigma_{ij}$ being the cross-section of interaction between species
$i$ and $j$. Denote neutrons by subscript 1 and the background moderator with
which the neutrons interact by 2, and make the assumptions that

(i) The neutron density $f_{1}$ is much less compared with that of the moderator
$f_{2}$ so that the terms $f_{1}f_{1}$ and $f_{1}f_{2}$ may be neglected
in the neutron and moderator equations respectively.

(ii) The moderator distribution $f_{2}$ is not affected by the neutrons. This
decouples the neutron and moderator equations and leads to an equilibrium Maxwellian
$f_{\textrm{M}}$ for the moderator while the neutrons are described by the
linear equation \begin{multline*}
\left(\frac{\partial}{\partial t}+v.\nabla\right)f(r,v,t)=\\
=\int dv^{\prime}\int dv_{1}\int dv_{1}^{\prime}W_{12}(v\rightarrow v^{\prime};v_{1}\rightarrow v_{1}^{\prime})\{ f(r,v^{\prime},t)f_{\textrm{M}}(v_{1}^{\prime})--f(r,v,t)f_{\textrm{M}}(v_{1})\})\end{multline*}
 This is now put in the standard form of the neutron transport equation \cite{Williams1967}\begin{multline*}
\left(\frac{1}{v}\frac{\partial}{\partial t}+\Omega.v+\mathcal{S}(E)\right)\Phi(r,E,\widehat{\Omega},t)=\int d\Omega^{\prime}\int dE^{\prime}\mathcal{S}(r,E^{\prime}\rightarrow E;\widehat{\Omega}^{\prime}\cdot\widehat{\Omega})\textrm{ }\Phi(r,E^{\prime},\widehat{\Omega}^{\prime},t).\end{multline*}
 where $E=mv^{2}/2$ is the energy and $\widehat{\Omega}$ the direction of
motion of the neutrons. The steady state, monoenergetic form of this equation
is Eq. (\ref{Eqn: NeutronTransport}) \[
\mu\frac{\partial\Phi(x,\mu)}{\partial x}+\Phi(x,\mu)=\frac{c}{2}\int_{-1}^{1}\Phi(x,\mu^{\prime})d\mu^{\prime},\qquad0<c<1,\,-1\leq\mu\leq1\]

\noindent and its singular eigenfunction solution for $x\in(-\infty,\infty)$
is given by Eq. (\ref{Eqn: CaseSolution_FR}) \begin{multline*}
\Phi(x,\mu)=a(\nu_{0})e^{-x/\nu_{0}}\phi(\mu,\nu_{0})+a(-\nu_{0})e^{x/\nu_{0}}\phi(-\nu_{0},\mu)+\int_{-1}^{1}a(\nu)e^{-x/\nu}\phi(\mu,\nu)d\nu;\end{multline*}

\noindent see Appendix A4 for an introductory review of Case's solution of the
one-speed neutron transport equation. \renewcommand{\arraystretch}{1.25} \renewcommand{\multirowsetup}{\centering}

\begin{table}[htbp]
\begin{center}\begin{tabular}{|c|c||c|c|c|}
\hline 
\multirow{2}{22mm}{$\mathscr L_{\lambda}$}&
\multirow{2}{22mm}{$\mathscr L_{\lambda}^{-1}$}&
\multicolumn{3}{c|}{$\mathcal{R}(\mathscr L_{\lambda})$}\tabularnewline
\cline{3-5} 
&
&
$\mathcal{R}=X$&
$\textrm{Cl}(\mathcal{R})=X$&
$\textrm{Cl}(\mathcal{R})\neq X$\tabularnewline
\hline
\hline 
Not injective&
$\cdots$&
$P\sigma(\mathscr{L})$&
$P\sigma(\mathscr{L})$&
$P\sigma(\mathscr{L})$\tabularnewline
\hline 
\multirow{2}{22mm}{Injective}&
 Not contiuous&
$C\sigma(\mathscr{L})$&
$C\sigma(\mathscr{L})$&
$R\sigma(\mathscr{L})$\tabularnewline
\cline{2-5}&
 Continuous&
$\rho(\mathscr{L})$&
$\rho(\mathscr{L})$&
$R\sigma(\mathscr{L})$\tabularnewline
\hline
\end{tabular}\end{center}

\begin{singlespace}

\caption{{\small \label{Table: spectrum}Spectrum of linear operator $\mathscr{L}\in\textrm{Map}(X)$.
Here $\mathscr L_{\lambda}:=\mathscr{L}-\lambda$ satisfies the equation $\mathscr L_{\lambda}(x)=0$,
with the resolvent set $\rho(\mathscr{L})$ of $\mathscr{L}$ consisting of
all those complex numbers $\lambda$ for which $\mathscr L_{\lambda}^{-1}$}
{\small exists as a continuous operator with dense domain. Any value of $\lambda$
for which this is not true is in the spectrum $\sigma(\mathscr{L})$ of $\mathscr{L}$,
that is further subdivided into three disjoint components of the point, continuous,
and residual spectra according to the criteria shown in the table. }}\end{singlespace}

\end{table}

The term {}``eigenfunction'' is motivated by the following considerations.
Consider the eigenvalue equation \begin{equation}
(\mu-\nu)\mathscr F_{\nu}(\mu)=0,\qquad\mu\in V(\mu),\textrm{ }\nu\in\mathbb{R}\label{Eqn: eigen}\end{equation}

\noindent in the space of multifunctions $\textrm{Multi}(V(\mu),(-\infty,\infty))$,
where $\mu$ is in either of the intervals $[-1,1]$ or $[0,1]$ depending on
whether the given boundary conditions for Eq. (\ref{Eqn: NeutronTransport})
is full-range or half range. If we are looking only for functional solutions
of Eq. (\ref{Eqn: eigen}), then the unique function $\mathcal{F}$ that satisfies
this equation for all possible $\mu\in V(\mu)$ and $\nu\in\mathbb{R}-V(\mu)$
is $\mathcal{F}_{\nu}(\mu)=0$ which means, according to Table \ref{Table: spectrum},
that the point spectrum of $\mu$ is empty and $(\mu-\nu)^{-1}$ exists for
all $\nu$. When $\nu\in V(\mu)$, however, this inverse is not continuous and
we show below that in $\textrm{Map}(V(\mu),0)$, $\nu\in V(\mu)$ belongs to
the continuous spectrum of $\mu$. This distinction between the nature of the
inverses depending on the relative values of $\mu$ and $\nu$ suggests a wider
{}``non-function'' space in which to look for the solutions of operator equations,
and in keeping with the philosophy embodied in Fig. \ref{Fig: GenInv} of treating
inverse problems in the space of multifunctions, we consider all $\mathscr F_{\nu}\in\textrm{Multi}(V(\mu),\mathbb{R}))$
satisfying Eq. (\ref{Eqn: eigen}) to be eigenfunctions of $\mu$ for the corresponding
eigenvalue $\nu$, leading to the following multifunctional solution of (\ref{Eqn: eigen})\begin{eqnarray*}
\mathscr F_{\nu}(\mu) & = & \left\{ \begin{array}{ccl}
(V(\mu),0), &  & \textrm{if }\nu\notin V(\mu)\\
(V(\mu)-\nu,0)\bigcup(\nu,\mathbb{R})), &  & \textrm{if }\nu\in V(\mu),\end{array}\right.\end{eqnarray*}

\noindent where $V(\mu)-\nu$ is used as a shorthand for the interval $V(\mu)$
with $\nu$ deleted. Rewriting the eigenvalue equation (\ref{Eqn: eigen}) as
$\mu_{\nu}(\mathscr F_{\nu}(\mu))=0$ and comparing this with Fig. \ref{Fig: GenInv},
allows us to draw the correspondences \begin{eqnarray}
f & \Longleftrightarrow & \mu_{\nu}\nonumber \\
X\textrm{ and }Y & \Longleftrightarrow & \{\mathscr F_{\nu}\in\textrm{Multi}(V(\mu),\mathbb{R})\!:\mathscr F_{\nu}\in\mathcal{D}(\mu_{\nu})\}\nonumber \\
f(X) & \Longleftrightarrow & \{0\!:0\in Y\}\label{Eqn: GenInv_Spectrum}\\
X_{\textrm{B}} & \Longleftrightarrow & \{0\!:0\in X\}\nonumber \\
f^{-} & \Longleftrightarrow & \mu_{\nu}^{-}.\nonumber \end{eqnarray}
 Thus a multifunction in $X$ is equivalent to $0$ in $X_{\textrm{B}}$ under
the linear map $\mu_{\nu}$, and we show below that this multifunction is infact
the Dirac delta {}``function'' $\delta_{\nu}(\mu)$, usually written as $\delta(\mu-\nu)$.
This suggests that in $\textrm{Multi}(V(\mu),\mathbb{R})$\emph{, every $\nu\in V(\mu)$
is in the point spectrum of $\mu$}, so that \emph{discontinuous functions that
are pointwise limits of functions in function space can be replaced by graphically
converged multifunctions in the space of multifunctions}. Completing the equivalence
class of $0$ in Fig. \ref{Fig: GenInv}, gives the multifunctional solution
of Eq. (\ref{Eqn: eigen}).

From a comparison of the definition of ill-posedness (Sec. 2) and the spectrum
(Table \ref{Table: spectrum}), it is clear that $\mathscr L_{\lambda}(x)=y$
is ill-posed iff 

\smallskip{}
(1) $\mathscr L_{\lambda}$ not injective $\Leftrightarrow$ $\lambda\in P\sigma(\mathscr L_{\lambda})$,
which corresponds to the first row of Table \ref{Table: spectrum}.

(2) $\mathscr L_{\lambda}$ not surjective $\Leftrightarrow$ the values of
$\lambda$ correspond to the second and third columns of Table \ref{Table: spectrum}. 

(3) $\mathscr L_{\lambda}$ is bijective but not open $\Leftrightarrow$ $\lambda\textrm{ is either in }C\sigma(\mathscr L_{\lambda})\textrm{ or }R\sigma(\mathscr L_{\lambda})$
corresponding to the second row of Table \ref{Table: spectrum}. 
\smallskip{}

We verify in the three steps below that $X=L_{1}[-1,1]$ of integrable functions,
$\nu\in V(\mu)=[-1,1]$ belongs to the continuous spectrum of $\mu$. 

(a) \emph{$\mathcal{R}(\mu_{\nu})$ is dense, but not equal to $L_{1}$}. The
set of functions $g(\mu)\in L_{1}$ such that $\mu_{\nu}^{-1}g\in L_{1}$ cannot
be the whole of $L_{1}$. Thus, for example, the piecewise constant function
$g=\textrm{const}\neq0$ on $\mid\mu-\nu\mid\leq\delta>0$ and $0$ otherwise
is in $L_{1}$ but not in \emph{$\mathcal{R}(\mu_{\nu})$} as $\mu_{\nu}^{-1}g\not\in L_{1}$.
Nevertheless for any $g\in L_{1}$, we may choose the sequence of functions
\[
g_{n}(\mu)=\left\{ \begin{array}{ccl}
0, &  & \textrm{if }\mid\mu-\nu\mid\leq1/n\\
g(\mu), &  & \textrm{otherwise}\end{array}\right.\]

\noindent in $\mathcal{R}(\mu_{\nu})$ to be eventually in every neighbourhood
of $g$ in the sense that $\lim_{n\rightarrow\infty}\int_{-1}^{1}\mid g-g_{n}\mid=0$. 

(b) \emph{The inverse $(\mu-\nu)^{-1}$ exists but is not continuous.} The inverse
exists because, as noted earlier, $0$ is the only functional solution of Eq.
(\ref{Eqn: eigen}). Nevertheless although the net of functions \[
\delta_{\nu\varepsilon}(\mu)=\frac{1}{\tan^{-1}(1+\nu)/\varepsilon+\tan^{-1}(1-\nu)/\varepsilon}\left(\frac{\varepsilon}{(\mu-\nu)^{2}+\varepsilon^{2}}\right),\qquad\varepsilon>0\]

\noindent is in the domain of $\mu_{\nu}$ because $\int_{-1}^{1}\delta_{\nu\varepsilon}(\mu)d\mu=1$
for all $\varepsilon>0$, \[
\lim_{\varepsilon\rightarrow0}\int_{-1}^{1}\mid\mu-\nu\mid\delta_{\nu\varepsilon}(\mu)d\mu=0\]

\noindent implying that $(\mu-\nu)^{-1}$ is unbounded. 

Taken together, (a) and (b) show that functional solutions of Eq. (\ref{Eqn: eigen})
lead to state 2-2 in Table \ref{Table: spectrum}; hence $\nu\in[-1,1]=C\sigma(\mu)$. 

\begin{figure}[htbp]
\noindent \begin{center}\input{Poison.pstex_t}\end{center}

\begin{singlespace}

\caption{\label{Fig: Poison}{\small Graphical convergence of: (a) Poisson kernel $\delta_{\varepsilon}(x)=\varepsilon/\pi(x^{2}+\varepsilon^{2})$
and (b) conjugate Poisson kernel $P_{\varepsilon}(x)=x/(x^{2}+\varepsilon^{2})$
to the Dirac delta and principal value respectively; the graphs, each for a
definite $\varepsilon$ value, converges to the respective limits as $\varepsilon\rightarrow0$.}}\end{singlespace}

\end{figure}
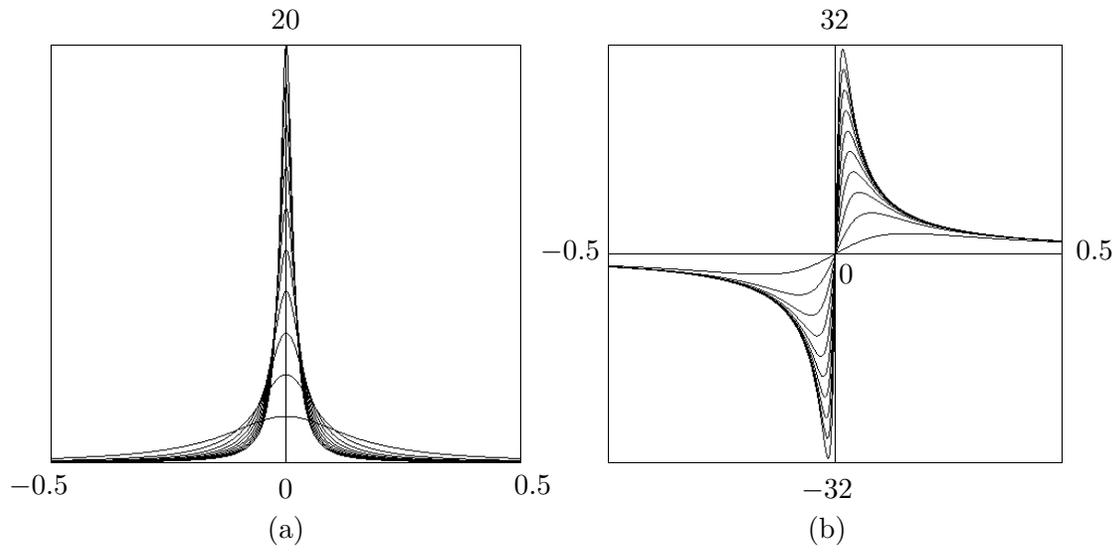

(c) The two integral constraints in (b) also mean that $\nu\in C\sigma(\mu)$
is a \emph{generalized eigenvalue} of $\mu$ which justifies calling the graphical
limit $\delta_{\nu\varepsilon}(\mu)\overset{\mathbf{G}}\rightarrow\delta_{\nu}(\mu)$
a \emph{generalized,} or singular, \emph{eigenfunction}, see Fig. \ref{Fig: Poison}
which clearly indicates the convergence of the net of functions%
\footnote{\label{Foot: gen_eigen}{\small The technical definition of a generalized eigenvalue
is as follows. Let $\mathcal{L}$ be a linear operator such that there exists
in the domain of $\mathcal{L}$ a sequence of elements $(x_{n})$ with $\Vert x_{n}\Vert=1$
for all $n$. If $\lim_{n\rightarrow\infty}\Vert(\mathcal{L}-\lambda)x_{n}\Vert=0$
for some $\lambda\in\mathbb{C}$, then this $\lambda$ is a} \emph{\small generalized
eigenvalue} {\small of $\mathcal{L}$, the corresponding eigenfunction $x_{\infty}$
being a} \emph{\small generalized eigenfunction. }%
}.%

From the fact that the solution Eq. (\ref{Eqn: CaseSolution_FR}) of the transport
equation contains an integral involving the multifunction $\phi(\mu,\nu)$,
we may draw an interesting physical interpretation. As the multi appears \emph{every
where} on $V(\mu)$ (that is there are no chaotic orbits but only the multifunctions
that produce them), we have here a situation typical of \emph{maximal ill-posedness}
characteristic of chaos: note that both the functions comprising $\phi_{\varepsilon}(\mu,\nu)$
are non-injective. As the solution (\ref{Eqn: CaseSolution_FR}) involves an
integral over all $\nu\in V(\mu)$, the singular eigenfunctions --- that collectively
may be regarded as representing a \emph{chaotic substate of} the system represented
by the solution of the neutron transport equation --- combine with the functional
components $\phi(\pm\nu_{0},\mu)$ to produce the well-defined, non-chaotic,
experimental end result of the neutron flux $\Phi(x,\mu)$.

The solution (\ref{Eqn: CaseSolution_FR}) is obtained by assuming $\Phi(x,\mu)=e^{-x/\nu}\phi(\mu,\nu)$
to get the equation for $\phi(\mu,\nu)$ to be $(\mu-\nu)\phi(\mu,\nu)=-c\nu/2$
with the normalization $\int_{-1}^{1}\phi(\mu,\nu)=1$. As $\mu_{\nu}^{-1}$
is not invertible in $\textrm{Multi}(V(\mu),\mathbb{R})$ and $\mu_{\nu\textrm{B}}\!:X_{\textrm{B}}\rightarrow f(X)$
does not exist, the alternate approach of regularization was adopted in \cite{Sengupta1988,Sengupta1995}
to rewrite $\mu_{\nu}\phi(\mu,\nu)=-c\nu/2$ as $\mu_{\nu\varepsilon}\phi_{\varepsilon}(\mu,\nu)=-c\nu/2$
with $\mu_{\nu\varepsilon}:=\mu-(\nu+i\varepsilon)$ being a net of bijective
functions for $\varepsilon>0$; this is a consequence of the fact that for the
multiplication operator every nonreal $\lambda$ belongs to the resolvent set
of the operator. The family of solutions of the later equation is given by \cite{Sengupta1988,Sengupta1995}
\begin{equation}
\phi_{\varepsilon}(\nu,\mu)=\frac{c\nu}{2}\frac{\nu-\mu}{(\mu-\nu)^{2}+\varepsilon^{2}}+\frac{\lambda_{\varepsilon}(\nu)}{\pi_{\varepsilon}}\frac{\varepsilon}{(\mu-\nu)^{2}+\varepsilon^{2}}\label{Eqn: phieps}\end{equation}

\noindent where the required normalization $\int_{-1}^{1}\phi_{\varepsilon}(\nu,\mu)=1$
gives 

\noindent {\renewcommand{\arraystretch}{1.85}\[
\begin{array}{ccl}
{\displaystyle \lambda_{\varepsilon}(\nu)} & = & {\displaystyle \frac{\pi_{\varepsilon}}{\tan^{-1}(1+\nu)/\varepsilon+\tan^{-1}(1-\nu)/\varepsilon}\left(1-\frac{c\nu}{4}\ln\frac{(1+\nu)^{2}+\varepsilon^{2}}{(1-\nu)^{2}+\varepsilon^{2}}\right)}\\
 & \overset{\varepsilon\rightarrow0}\longrightarrow & \pi\lambda(\nu)\end{array}\]
}

\noindent with \[
\pi_{\varepsilon}=\varepsilon\int_{-1}^{1}\frac{d\mu}{\mu^{2}+\varepsilon^{2}}=2\tan^{-1}\left(\frac{1}{\varepsilon}\right)\overset{\varepsilon\rightarrow0}\longrightarrow\pi.\]

\noindent These discretized equations should be compared with the corresponding
exact ones of Appendix A4. We shall see that the net of functions (\ref{Eqn: phieps})
converges graphically to the multifunction Eq. (\ref{Eqn: singular_eigen})
as $\varepsilon\rightarrow0$.

In the discretized spectral approximation., the singular eigenfunction $\phi(\mu,\nu)$
is replaced by $\phi_{\varepsilon}(\mu,\nu)$, $\varepsilon\rightarrow0$, with
the integral in $\nu$ being replaced by an appropriate sum. The solution Eq.
(\ref{Eqn: CaseSolution_HR}) of the physically interesting half-space $x\geq0$
problem then reduces to \cite{Sengupta1988,Sengupta1995} \begin{equation}
\Phi_{\varepsilon}(x,\mu)=a(\nu_{0})e^{-x/\nu_{0}}\phi(\mu,\nu_{0})+\sum_{i=1}^{N}a(\nu_{i})e^{-x/\nu_{i}}\phi_{\varepsilon}(\mu,\nu_{i})\qquad\mu\in[0,1]\label{Eqn: DiscSpect_HR}\end{equation}

\noindent where the nodes $\{\nu_{i}\}_{i=1}^{N}$ are chosen suitably. This
discretized spectral approximation to Case's solution has given surprisingly
accurate numerical results for a set of properly chosen nodes when compared
with exact calculations. Because of its involved nature \cite{Case1967}, the
exact calculations are basically numerical which leads to nonlinear integral
equations as part of the solution procedure. To appreciate the enormous complexity
of the exact treatment of the half-space problem, we recall that the complete
set of eigenfunctions $\{\phi(\mu,\nu_{0}),\{\phi(\mu,\nu)\}_{\nu\in[0,1]}\}$
are orthogonal with respect to the half-range weight function $W(\mu)$ of half-range
theory, Eq. (\ref{Eqn: W(mu)}), that is expressed only in terms of solution
of the nonlinear integral equation Eq. (\ref{Eqn: Omega(-mu)}). The solution
of a half-space problem then evaluates the coefficients $\{ a(\nu_{0}),a(\nu)_{\nu\in[0,1]}\}$
from the appropriate half range (that is $0\leq\mu\leq1$) orthogonality integrals
satisfied by the eigenfunctions $\{\phi(\mu,\nu_{0}),\{\phi(\mu,\nu)\}_{\nu\in[0,1]}\}$
with respect to the weight $W(\mu)$, see Appendix A4 for the necessary details
of the half-space problem in neutron transport theory.

As may be appreciated from this brief introduction, solutions to half-space
problems are not simple and actual numerical computations must rely a great
deal on tabulated values of the $X$-function. Self-consistent calculations
of sample benchmark problems performed by the discretized spectral approximation
in a full-range adaption of the half-range problem described below that generate
all necessary data, independent of numerical tables, with the quadrature nodes
$\{\nu_{i}\}_{i=1}^{N}$ taken at the zeros Legendre polynomials show that the
full range formulation of this approximation \cite{Sengupta1988,Sengupta1995}
can give very accurate results not only of integrated quantities like the flux
$\Phi$ and leakage of particles out of the half space, but of also basic \char`\"{}raw\char`\"{}
data like the extrapolated end point \begin{equation}
z_{0}=\frac{c\nu_{0}}{4}\int_{0}^{1}\frac{\nu}{N(\nu)}\left(1+\frac{c\nu^{2}}{1-\nu^{2}}\right)\ln\left(\frac{\nu_{0}+\nu}{\nu_{0}-\nu}\right)d\nu\label{Eqn: extrapolated}\end{equation}

\noindent and of the $X$-function itself. Given the involved nature of the
exact theory, it is our contention that the remarkable accuracy of these basic
data, some of which is reproduced in Table \ref{Table: extrapolated}, is due
to the graphical convergence of the net of functions \[
\phi_{\varepsilon}(\mu,\nu)\overset{\mathbf{G}}\longrightarrow\phi(\mu,\nu)\]

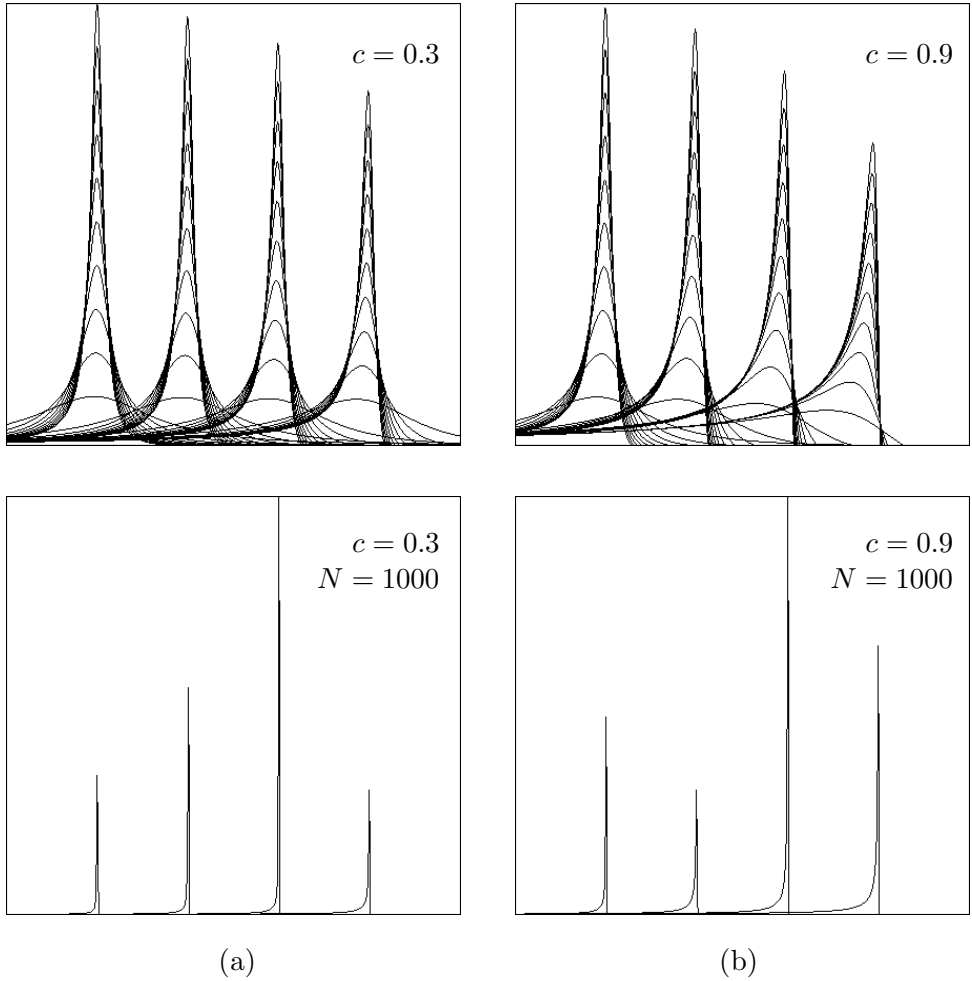
\begin{figure}[htbp]
\noindent \begin{center}\input{Case.pstex_t}\end{center}

\begin{singlespace}

\caption{\label{Fig: Case} {\small Rational function approximations $\phi_{\varepsilon}(\mu,\nu)$
of the singular eigenfunction $\phi(\mu,\nu)$ at four different values of $\nu$.
$N=1000$ denotes the {}``converged'' multifunction $\phi$, with the peaks
at the specific $\nu$-values chosen. }}\end{singlespace}

\end{figure}

\vspace{-0.2cm}
\noindent shown in Fig. \ref{Fig: Case}; here $\varepsilon=1/\pi N$ so that
$\varepsilon\rightarrow0$ as $N\rightarrow\infty$. By this convergence, the
delta function and principal values in $[-1,1]$ are the multifunctions $([-1,0),0)\bigcup(0,[0,\infty)\bigcup((0,1],0)$
and $\{1/x\}_{x\in[-1,0)}\bigcup(0,(-\infty,\infty))\bigcup\{1/x\}_{x\in(0,1]}$
respectively. %
 Tables \ref{Table: extrapolated} and \ref{Table: X-function}, taken from
\citet*{Sengupta1988} and \citet*{Sengupta1995}, show respectively the extrapolated
end point and $X$-function by the full-range adaption of the discretized spectral
approximation for two different half range problems denoted as Problems A and
B defined as 

\begin{align*}
Problem\textrm{ }A\quad & \textrm{Equation}\!:\textrm{ }{\textstyle {\mu\Phi_{x}+\Phi=(c/2)\int_{-1}^{1}\Phi(x,\mu^{\prime})d\mu^{\prime},\; x\geq0}}\\
 & \textrm{Boundary condition}:\textrm{ }\Phi(0,\mu)=0,\;\mu\geq0\\
 & \textrm{Asymptotic condition}:\textrm{ }\Phi\rightarrow e^{-x/\nu_{0}}\phi(\mu,\nu_{0}),\; x\rightarrow\infty.\\
Problem\textrm{ }B\quad & \textrm{Equation}\!:\textrm{ }{\textstyle {\mu\Phi_{x}+\Phi=(c/2)\int_{-1}^{1}\Phi(x,\mu^{\prime})d\mu^{\prime},\; x\geq0}}\\
 & \textrm{Boundary condition}:\textrm{ }\Phi(0,\mu)=1,\;\mu\geq0\\
 & \textrm{Asymptotic condition}:\textrm{ }\Phi\rightarrow0,\; x\rightarrow\infty.\end{align*}

\noindent The full $-1\leq\mu\leq1$ range form of the half $0\leq\mu\leq1$
range discretized spectral approximation replaces the exact integral boundary
condition at $x=0$ by a suitable quadrature sum over the values of $\nu$ taken
at the zeros of Legendre polynomials; thus the condition at $x=0$ can be expressed
as \begin{equation}
\psi(\mu)=a(\nu_{0})\phi(\mu,\nu_{0})+\sum_{i=1}^{N}a(\nu_{i})\phi_{\varepsilon}(\mu,\nu_{i}),\qquad\mu\in[0,1],\label{Eqn: BC}\end{equation}

\noindent where $\psi(\mu)=\Phi(0,\mu)$ is the specified incoming radiation
incident on the boundary from the left, and the half-range coefficients $a(\nu_{0})$,
$\{ a(\nu)\}_{\nu\in[0,1]}$ are to be evaluated using the $W$-function of
Appendix 4. We now exploit the relative simplicity of the full-range calculations
by replacing Eq. (\ref{Eqn: BC}) by Eq. (\ref{Eqn: HRFR_Discrete}) following,
where the coefficients $\{ b(\nu_{i})\}_{i=0}^{N}$ are used to distinguish
the full-range coefficients from the half-range ones. The significance of this
change lies in the overwhelming simplicity of the full-range weight function
$\mu$ as compared to the half-range function $W(\mu)$, and the resulting simplicity
of the orthogonality relations that follow, see Appendix A4. The basic data
of $z_{0}$ and $X(-\nu)$ are then completely generated self-consistently \cite{Sengupta1988,Sengupta1995}
by the discretized spectral approximation from the full-range adaption \begin{equation}
\sum_{i=0}^{N}b_{i}\phi_{\varepsilon}(\mu,\nu_{i})=\psi_{+}(\mu)+\psi_{-}(\mu),\qquad\mu\in[-1,1],\textrm{ }\nu_{i}\geq0\label{Eqn: HRFR_Discrete}\end{equation}

\noindent of the discretized boundary condition Eq. (\ref{Eqn: BC}), where
$\psi_{+}(\mu)$ is by definition the incident flux $\psi(\mu)$ for $\mu\in[0,1]$
and $0$ if $\mu\in[-1,0]$, while \[
\psi_{-}(\mu)=\left\{ \begin{array}{ccl}
{\displaystyle \sum_{i=0}^{N}b_{i}^{-}\phi_{\varepsilon}(\mu,\nu_{i})} &  & \textrm{if }\mu\in[-1,0],\textrm{ }\nu_{i}\geq0\textrm{ }\\
0 &  & \textrm{if }\mu\in[0,1]\end{array}\right.\]
 is the the emergent angular distribution out of the medium. Equation (\ref{Eqn: HRFR_Discrete})
corresponds to the full-range $\mu\in[-1,1],\textrm{ }\nu_{i}\geq0$ form \begin{equation}
b(\nu_{0})\phi(\mu,\nu_{0})+\int_{0}^{1}b(\nu)\phi(\mu,\nu)d\nu=\psi_{+}(\mu)+\left(b^{-}(\nu_{0})\phi(\mu,\nu_{0})+\int_{0}^{1}b^{-}(\nu)\phi(\mu,\nu)d\nu\right)\label{Eqn: HRFR}\end{equation}
 of boundary condition (\ref{Eqn: BC_HR}) with the first and second terms on
the right having the same interpretation as for Eq. (\ref{Eqn: HRFR_Discrete}).
This full-range simulation merely states that the solution (\ref{Eqn: CaseSolution_HR})
of Eq. (\ref{Eqn: NeutronTransport}) holds for all $\mu\in[-1,1]$, $x\geq0$,
although it was obtained, unlike in the regular full-range case, from the given
radiation $\psi(\mu)$ incident on the boundary at $x=0$ over only half the
interval $\mu\in[0,1]$. To obtain the simulated full-range coefficients $\{ b_{i}\}$
and $\{ b_{i}^{-}\}$ of the half-range problem, we observe that there are effectively
only half the number of coefficients as compared to a normal full-range problem
because $\nu$ is now only over half the full interval. This allows us to generate
two sets of equations from (\ref{Eqn: HRFR}) by integrating with respect to
$\mu\in[-1,1]$ with $\nu$ in the half intervals $[-1,0]$ and $[0,1]$ to
obtain the two sets of coefficients $b^{-}$ and $b$ respectively. Accordingly
we get from Eq. (\ref{Eqn: HRFR_Discrete}) with $\textrm{ }j=0,1,\cdots,N$
the sets of equations 

\noindent {\renewcommand{\arraystretch}{2}

\noindent \begin{equation}
\begin{array}{c}
{\displaystyle {\displaystyle (\psi,\phi_{j-})_{\mu}^{(+)}=-\sum_{i=0}^{N}b_{i}^{-}(\phi_{i+},\phi_{j-})_{\mu}^{(-)}}}\\
b_{j}={\displaystyle \left((\psi,\phi_{j+})_{\mu}^{(+)}+\sum_{i=0}^{N}b_{i}^{-}(\phi_{i+},\phi_{j+})_{\mu}^{(-)}\right)}\end{array}\label{Eqn: FRBC1}\end{equation}
}

\noindent where $(\phi_{j\pm})_{j=1}^{N}$ represents $(\phi_{\varepsilon}(\mu,\pm\nu_{j}))_{j=1}^{N}$,
$\phi_{0\pm}=\phi(\mu,\pm\nu_{0})$, the $(+)$ $(-)$ superscripts are used
to denote the integrations with respect to $\mu\in[0,1]$ and $\mu\in[-1,0]$
respectively, and $(f,g)_{\mu}$ denotes the usual inner product in $[-1,1]$
with respect to the full range weight $\mu$. While the first set of $N+1$
equations give $b_{i}^{-}$, the second set produces the required $b_{j}$ from
these \char`\"{}negative\char`\"{} coefficients. By equating these calculated
$b_{i}$ with the exact half-range expressions for $a(\nu)$ with respect to
$W(\mu)$ as outlined in Appendix A4, it is possible to find numerical values
of $z_{0}$ and $X(-\nu)$. Thus from the second of Eq. (\ref{Eqn: Constant_Coeff}),
$\{ X(-\nu_{i})\}_{i=1}^{N}$ is obtained with $b_{i\textrm{B}}\textrm{ }=a_{i\textrm{B}}$,
$i=1,\cdots,N$, which is then substituted in the second of Eq. (\ref{Eqn: Milne_Coeff})
with $X(-\nu_{0})$ obtained from $a_{\textrm{A}}(\nu_{0})$ according to Appendix
A4, to compare the respective $a_{i\textrm{A}}$ with the calculated $b_{i\textrm{A}}$
from (\ref{Eqn: FRBC1}). Finally the full-range coefficients of Problem A can
be used to obtain the $X(-\nu)$ values from the second of Eqs. (\ref{Eqn: Milne_Coeff})
and compared with the exact tabulated values as in Table \ref{Table: X-function}.
The tabulated values of $cz_{0}$ from Eq. (\ref{Eqn: extrapolated}) show a
consistent deviation from our calculations of Problem A according to $a_{\textrm{A}}(\nu_{0})=-\exp(-2z_{0}/\nu_{0})$.
Since the $X(-\nu)$ values of Problem A in Table \ref{Table: X-function} also
need the same $b_{0\textrm{A}}$ as input that was used in obtaining $z_{0}$,
it is reasonable to conclude that the \char`\"{}exact\char`\"{} numerical integration
of $z_{0}$ is inaccurate to the extent displayed in Table \ref{Table: extrapolated}.

\renewcommand{\arraystretch}{1.25}

\begin{table}[htbp]
\noindent \begin{center}\begin{tabular}{|c|c|c|c|c|}
\hline 
\multirow{2}{10mm}{$c$}&
\multicolumn{4}{c|}{$cz_{0}$}\tabularnewline
\cline{2-5} 
&
$N=2$&
$N=6$&
$N=10$&
Exact\tabularnewline
\hline
\hline 
0.2&
0.78478&
0.78478&
0.78478&
0.7851\tabularnewline
\hline 
0.4&
0.72996&
0.72996&
0.72996&
0.7305\tabularnewline
\hline 
0.6&
0.71535&
0.71536&
0.71536&
0.7155\tabularnewline
\hline 
0.8&
0.71124&
0.71124&
0.71124&
0.7113\tabularnewline
\hline
0.9&
0.71060&
0.71060&
0.71061&
0.7106\tabularnewline
\hline
\end{tabular}\end{center}

\begin{singlespace}

\caption{\label{Table: extrapolated}{\small Extrapolated end-point $z_{0}$.}}\end{singlespace}

\end{table}

\begin{table}[htbp]
\begin{center}\begin{tabular}{|c|c|c|c|c|c|}
\hline 
\multirow{2}{10mm}{$c$}&
\multirow{2}{10mm}{$N$}&
\multicolumn{4}{c|}{$X(-\nu)$}\tabularnewline
\cline{3-6} 
&
&
$\nu_{i}$&
Problem A&
Problem B&
Exact\tabularnewline
\hline
\hline 
\multirow{2}{10mm}{0.2}&
\multirow{2}{10mm}{2}&
0.2133&
0.8873091&
0.8873091&
0.887308\tabularnewline
\cline{3-6}&
&
0.7887&
0.5826001&
0.5826001&
0.582500\tabularnewline
\hline 
\multirow{6}{10mm}{0.6}&
\multirow{6}{10mm}{6}&
0.0338&
1.3370163&
1.3370163&
1.337015\tabularnewline
\cline{3-6}&
&
0.1694&
1.0999831&
1.0999831&
1.099983\tabularnewline
\cline{3-6}&
&
0.3807&
0.8792321&
0.8792321&
0.879232\tabularnewline
\cline{3-6}&
&
0.6193&
0.7215240&
0.7215240&
0.721524\tabularnewline
\cline{3-6}&
&
0.8306&
0.6239109&
0.6239109&
0.623911\tabularnewline
\cline{3-6}&
&
0.9662&
0.5743556&
0.5743556&
0.574355\tabularnewline
\hline 
\multirow{10}{10mm}{0.9}&
\multirow{10}{10mm}{10}&
0.0130&
1.5971784&
1.5971784&
1.597163\tabularnewline
\cline{3-6}&
&
0.0674&
1.4245314&
1.4245314&
1.424532\tabularnewline
\cline{3-6}&
&
0.1603&
1.2289940&
1.2289940&
1.228995\tabularnewline
\cline{3-6}&
&
0.2833&
1.0513750&
1.0513750&
1.051376\tabularnewline
\cline{3-6}&
&
0.4255&
0.9058140&
0.9058410&
0.905842\tabularnewline
\cline{3-6}&
&
0.5744&
0.7934295&
0.7934295&
0.793430\tabularnewline
\cline{3-6}&
&
0.7167&
0.7102823&
0.7102823&
0.710283\tabularnewline
\cline{3-6}&
&
0.8397&
0.6516836&
0.6516836&
0.651683\tabularnewline
\cline{3-6}&
&
0.9325&
0.6136514&
0.6136514&
0.613653\tabularnewline
\cline{3-6}&
&
0.9870&
0.5933988&
0.5933988&
0.593399\tabularnewline
\hline
\end{tabular}\end{center}

\begin{singlespace}

\caption{{\small \label{Table: X-function}$X(-\nu)$ by the full range method.}}\end{singlespace}

\end{table}

\renewcommand{\arraystretch}{1}

From these numerical experiments and Fig. \ref{Fig: Case} we may conclude that
the continuous spectrum $[-1,1]$ of the position operator $\mu$ acts as the
$\mathcal{D}_{+}$ points in generating the multifunctional Case singular eigenfunction
$\phi(\mu,\nu)$. Its rational approximation $\phi_{\varepsilon}(\mu,\nu)$
in the context of the simple simulated full-range computations of the complex
half-range exact theory of Appendix A4, clearly demonstrates the utility of
graphical convergence of sequence of functions to multifunction. The totality
of the multifunctions $\phi(\mu,\nu)$ for all $\nu$ in Fig. \ref{Fig: Case}(c)
and (d) endows the problem with the character of maximal ill-posedness that
is characteristic of chaos. This chaotic signature of the transport equation
is however latent as the experimental output $\Phi(x,\mu)$ is well-behaved
and regular. This important example shows how nature can use hidden and complex
chaotic substates to generate order through a process of superposition.

\vspace{1cm}
\noindent \begin{flushleft}\textbf{6. Does Nature support complexity?}\end{flushleft}

\noindent The question of this section is basic in the light of the theory of
chaos presented above as it may be reformulated to the inquiry of what makes
nature support chaoticity in the form of increasing non-injectivity of an input-output
system. It is the purpose of this Section to exploit the connection between
spectral theory and the dynamics of chaos that has been presented in the previous
section. Since linear operators on finite dimensional spaces do not possess
continuous or residual spectra, spectral theory on infinite dimensional spaces
essentially involves limiting behaviour to infinite dimensions of the familiar
matrix eigenvalue-eigenvector problem. As always this means extensions, dense
embeddings and completions of the finite dimensional problem that show up as
generalized eigenvalues and eigenvectors. In its usual form, the goal of nonlinear
spectral theory consists \cite{Appel2000} in the study of $T_{\lambda}^{-1}$
for nonlinear operators $T_{\lambda}$ that satisfy more general continuity
conditions, like differentiability and Lipschitz continuity, than simple boundedness
that is enough for linear operators. The following generalization of the concept
of the spectrum of a linear operator to the nonlinear case is suggestive. For
a nonlinear map, $\lambda$ need not appear only in a multiplying role, so that
an eigenvalue equation can be written more generally as a fixed-point equation
\[
f(\lambda;x)=x\]
 with a fixed point corresponding to the eigenfunction of a linear operator
and an {}``eigenvalue'' being the value of $\lambda$ for which this fixed
point appears. The correspondence of the residual and continuous parts of the
spectrum are, however, less trivial than for the point spectrum. This is seen
from the following two examples, \cite{Roman1975}. Let $Ae_{k}=\lambda_{k}e_{k},\textrm{ }k=1,2,\cdots$
be an eigenvalue equation with $e_{j}$ being the $j^{\textrm{th}}$ unit vector.
Then $(A-\lambda)e_{k}:=(\lambda_{k}-\lambda)e_{k}=0$ iff $\lambda=\lambda_{k}$
so that $\{\lambda_{k}\}_{k=1}^{\infty}\in P\sigma(A)$ are the only eigenvalues
of $A$. Consider now $(\lambda_{k})_{k=1}^{\infty}$ to be a sequence of real
numbers that tends to a finite $\lambda^{*}$; for example let $A$ be a diagonal
matrix having $1/k$ as its diagonal entries. Then $\lambda^{*}$ belongs to
the continuous spectrum of $A$ because $(A-\lambda^{*})e_{k}=(\lambda_{k}-\lambda^{*})e_{k}$
with $\lambda_{k}\rightarrow\lambda^{*}$ implies that $(A-\lambda^{*})^{-1}$
is an unbounded linear operator and $\lambda^{*}$ a generalized eigenvalue
of $A$. In the second example $Ae_{k}=e_{k+1}/(k+1)$, it is not difficult
to verify that: (a) The point spectrum of $A$ is empty, (b) The range of $A$
is not dense because it does not contain $e_{1}$, and (c) $A^{-1}$ is unbounded
because $Ae_{k}\rightarrow0$. Thus the generalized eigenvalue $\lambda^{*}=0$
in this case belongs to the residual spectrum of $A$. In either case, $\lim_{j\rightarrow\infty}e_{j}$
is the corresponding generalized eigenvector that enlarges the trivial null
space $\mathcal{N}(\mathscr L_{\lambda^{*}})$ of the generalized eigenvalue
$\lambda^{*}$. In fact in these two and the Dirac delta example of Sec. 5 of
continuous and residual spectra, the generalized eigenfunctions arise as the
limits of a sequence of functions whose images under the respective $\mathcal{L}_{\lambda}$
converge to $0$; recall the definition of footnote \ref{Foot: gen_eigen}.
This observation generalizes to the dense extension $\textrm{Multi}_{|}(X,Y)$
of $\textrm{Map}(X,Y)$ as follows. If $x\in\mathcal{D}_{+}$ is not a fixed
point of $f(\lambda;x)=x$, but there is some $n\in\mathbb{N}$ such that $f^{n}(\lambda;x)=x$,
then the limit $n\rightarrow\infty$ generates a multifunction at $x$ as was
the case with the delta function in the previous section and the various other
examples that we have seen so far in the earlier sections.

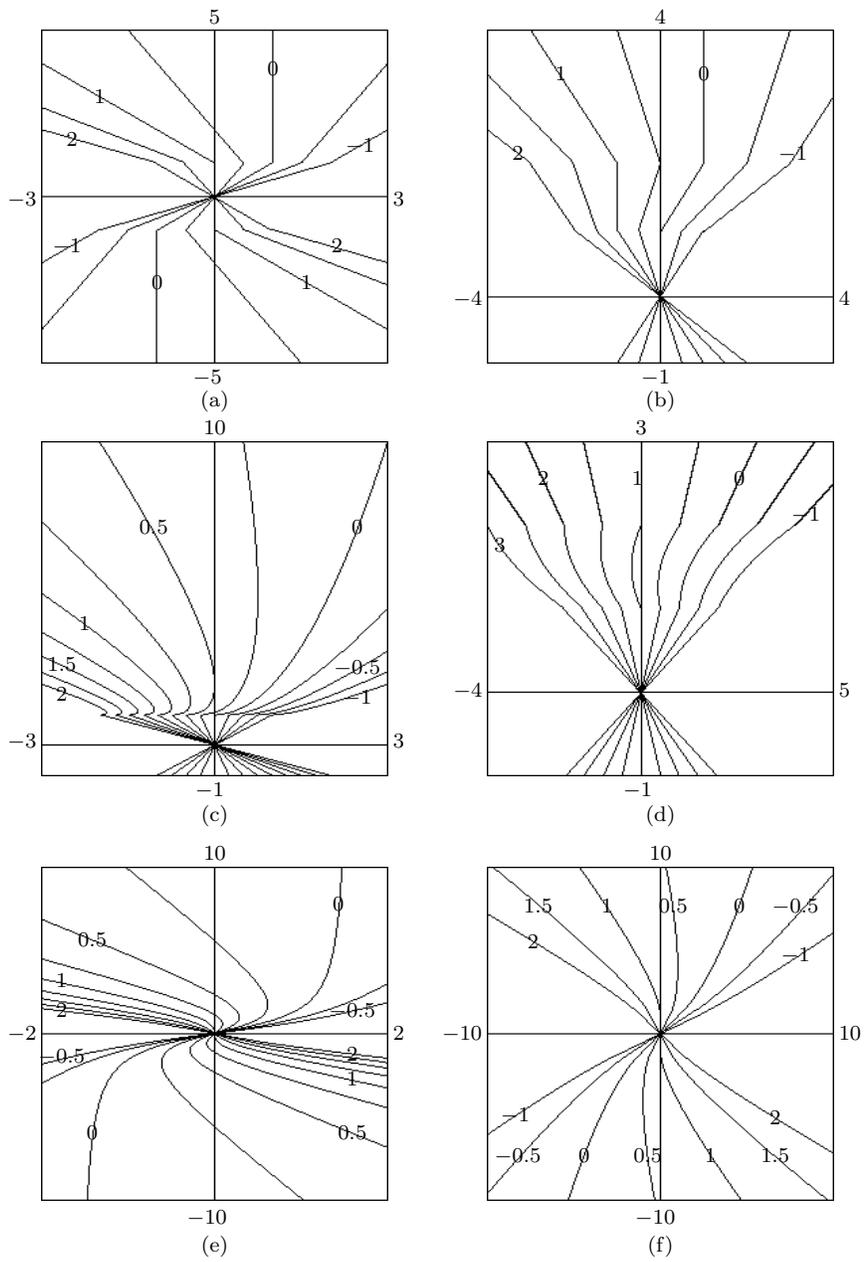
\begin{figure}[htbp]
\noindent \begin{center}\input{Appel.pstex_t}\end{center}

\begin{singlespace}

\caption{\label{Fig: Appel}{\small Inverses of $f-\lambda$. The $\lambda$-values
are shown on the graphs.}}\end{singlespace}

\end{figure}

One of the main goals of investigations on the spectrum of nonlinear operators
is to find a set in the complex plane that has the usual desirable properties
of the spectrum of a linear operator, \citet*{Appel2000}. In this case, the
focus has been to find a suitable class of operators $\mathcal{C}(X)$ with
$T\in\mathcal{C}(X)$, such that the resolvent set is expressed as\[
\rho(T)=\{\lambda\in\mathbb{C}\!:(T_{\lambda}\textrm{ is }1:1)(\textrm{Cl}(\mathcal{R}(T_{\lambda})=X)\textrm{ and }(T_{\lambda}^{-1}\in\mathcal{C}(X)\textrm{ on }\mathcal{R}(T_{\lambda}))\}\]

\noindent with the spectrum $\sigma(T)$ being defined as the complement of
this set. Among the classes $\mathcal{C}(X)$ that have been considered, beside
spaces of continuous functions $C(X)$, are linear boundedness $B(X)$, Frechet
differentiability $C^{1}(X)$, Lipschitz continuity $\textrm{Lip}(X)$, and
Granas quasiboundedness $Q(x)$, where $\textrm{Lip}(X)$ specifically takes
into account the nonlinearity of $T$ to define \begin{equation}
\Vert T\Vert_{\textrm{Lip}}=\sup_{x\neq y}\frac{\Vert T(x)-T(y)\Vert}{\Vert x-y\Vert},\qquad|T|_{\textrm{lip}}=\inf_{x\neq y}\frac{\Vert T(x)-T(y)\Vert}{\Vert x-y\Vert}\label{Eqn: LipNorm}\end{equation}

\noindent that are plainly generalizations of the corresponding norms of linear
operators.  Plots of $f_{\lambda}^{-}(y)=\{ x\in\mathcal{D}(f-\lambda)\!:(f-\lambda)x=y\}$
for the functions $f\!:\mathbb{R}\rightarrow\mathbb{R}$\[
\begin{array}{rcl}
f_{\lambda\textrm{a}}(x) & = & \left\{ \begin{array}{cll}
-1-\lambda x, &  & x<-1\\
(1-\lambda)x, &  & -1\leq x\leq1\\
1-\lambda x, &  & 1<x,\end{array}\right.\\
\\f_{\lambda\textrm{b}}(x) & = & \left\{ \begin{array}{cll}
-\lambda x, &  & x<1\\
(1-\lambda x)-1, &  & 1\leq x\leq2\\
1-\lambda x, &  & 2<x\end{array}\right.\\
\\f_{\lambda\textrm{c}}(x) & = & \left\{ \begin{array}{cll}
-\lambda x &  & x<1\\
\sqrt{x-1}-\lambda x &  & 1\leq x,\end{array}\right.\\
\\f_{\lambda\textrm{d}}(x) & = & \left\{ \begin{array}{cll}
(x-1)^{2}+1-\lambda x &  & 1\leq x\leq1\\
(1-\lambda)x &  & \textrm{otherwise}\end{array}\right.\\
\\f_{\lambda\textrm{e}}(x) & = & \tan^{-1}(x)-\lambda x,\\
\\f_{\lambda\textrm{f}}(x) & = & \left\{ \begin{array}{cll}
1-2\sqrt{-x}-\lambda x, &  & x<-1\\
(1-\lambda)x, &  & -1\leq x\leq1\\
2\sqrt{x}-1-\lambda x, &  & 1<x\end{array}\right.\end{array}\]
 taken from \citet*{Appel2000} are shown in Fig. \ref{Fig: Appel}. It is easy
to verify that the Lipschitz and linear upper and lower bounds of these maps
are as in Table \ref{Table: Appel_bnds}.

The point spectrum defined by \[
P\sigma(f)=\{\lambda\in\mathbb{C}\!:(f-\lambda)x=0\textrm{ for some }x\neq0\}\]

\noindent is the simplest to calculate. Because of the special role played by
the zero element $0$ in generating the point spectrum in the linear case, the
bounds $m\Vert x\Vert\leq\Vert\mathscr{L}x\Vert\leq M\Vert x\Vert$ together
with $\mathscr{L}x=\lambda x$ imply $\textrm{Cl}(P\sigma(\mathscr{L}))=[\Vert\mathscr{L}\Vert_{\textrm{b}},\Vert\mathscr{L}\Vert_{\textrm{B}}]$
--- where the subscripts denote the lower and upper bounds in Eq. (\ref{Eqn: LipNorm})
and which is sometimes taken to be a descriptor of the point spectrum of a nonlinear
operator --- as can be seen in Table \ref{Table: Appel_spectra} and verified
from Fig. \ref{Fig: Appel}. The remainder of the spectrum, as the complement
of the resolvent set, is more difficult to find. Here the convenient characterization
of the resolvent of a continuous linear operator as the set of all sufficiently
large $\lambda$ that satisfy $|\lambda|>M$ is of little significance as, unlike
for a linear operator, the non-existence of an inverse is not just due the set
$\{ f^{-1}(0)\}$ which happens to be the only way a linear map can fail to
be injective. Thus the map defined piecewise as $\alpha+2(1-\alpha)x$ for $0\leq x<1/2$
and $2(1-x)$ for $1/2\leq x\leq1$, with $0<\alpha<1$, is not invertible on
its range although $\{ f^{-}(0)\}=1$. Comparing Fig. \ref{Fig: Appel} and
Table \ref{Table: Appel_bnds}, it is seen that in cases (b), (c) and (d), the
intervals $[|f|_{\textrm{b}},\Vert f\Vert_{\textrm{B}}]$ are subsets of the
$\lambda$-values for which the respective maps are not injective; this is to
be compared with (a), (e) and (f) where the two sets are the same. Thus the
linear bounds are not good indicators of the uniqueness properties of solution
of nonlinear equations for which the Lipschitzian bounds are seen to be more
appropriate.

\noindent {\renewcommand{\arraystretch}{1.25}

\begin{table}[htbp]
\begin{center}\begin{tabular}{|c||c|c|c|c|}
\hline 
Function&
$|f|_{\textrm{b}}$&
$\Vert f\Vert_{\textrm{B}}$&
$|f|_{\textrm{lip}}$&
$\Vert f\Vert_{\textrm{Lip}}$\tabularnewline
\hline
\hline 
$f_{\textrm{a}}$&
$0$&
$1$&
$0$&
$1$\tabularnewline
\hline 
$f_{\textrm{b}}$&
$0$&
$1/2$&
$0$&
$1$\tabularnewline
\hline 
$f_{\textrm{c}}$&
$0$&
$1/2$&
$0$&
$\infty$\tabularnewline
\hline 
$f_{\textrm{d}}$&
$2(\sqrt{2}-1)$&
$\infty$&
$0$&
$2$\tabularnewline
\hline
$f_{\textrm{e}}$&
$0$&
$1$&
$0$&
$1$\tabularnewline
\hline 
$f_{\textrm{f}}$&
$0$&
$1$&
$0$&
$1$\tabularnewline
\hline
\end{tabular}\end{center}

\begin{singlespace}

\caption{\label{Table: Appel_bnds}{\small Bounds on the functions of Fig. \ref{Fig: Appel}.}}\end{singlespace}

\end{table}

\begin{table}[htbp]
\begin{center}\begin{tabular}{|c||c|c|}
\hline 
Functions&
$\sigma_{\textrm{Lip}}(f)$&
$P\sigma(f)$\tabularnewline
\hline
\hline 
$f_{\textrm{a}}$&
$[0,1]$&
$(0,1]$\tabularnewline
\hline 
$f_{\textrm{b}}$&
$[0,1]$&
$[0,1/2]$\tabularnewline
\hline 
$f_{\textrm{c}}$&
$[0,\infty)$&
$[0,1/2]$\tabularnewline
\hline 
$f_{\textrm{d}}$&
$[0,2]$&
$[2(\sqrt{2}-1),1]$\tabularnewline
\hline 
$f_{\textrm{e}}$&
$[0,1]$&
$(0,1)$\tabularnewline
\hline 
$f_{\textrm{f}}$&
$[0,1]$&
$(0,1)$\tabularnewline
\hline
\end{tabular}\end{center}

\begin{singlespace}

\caption{\label{Table: Appel_spectra}{\small Lipschitzian and point spectra of the
functions of Fig. \ref{Fig: Appel}.}}\end{singlespace}

\end{table}

}In view of the above, we may draw the following conclusions. If we choose
to work in the space of multifunctions $\textrm{Multi}(X,\mathcal{T})$, with
$\mathcal{T}$ the topology of pointwise biconvergence, when all functional
relations are (multi) invertible on their ranges, we may make the following
definition for the net of functions $f(\lambda;x)$ satisfying $f(\lambda;x)=x$.

\smallskip{}
\noindent \textbf{Definition 6.1.} \textsl{Let} $f(\lambda;\cdot)\in\textrm{Multi}(X,\mathcal{T})$
\textsl{be a function. The resolvent set of $f$ is given by} \[
\rho(f)=\{\lambda\!:(f(\lambda;\cdot)^{-1}\in\textrm{Map}(X,\mathcal{T}))\wedge(\textrm{Cl}(\mathcal{R}(f(\lambda;\cdot))=X)\},\]

\noindent \textsl{and any $\lambda$ not in $\rho$ is in the spectrum of $f$.$\qquad\square$}

\smallskip{}
Thus apart from multifunctions, $\lambda\in\sigma(f)$ also generates functions
on the boundary of functional and non-functional relations in $\textrm{Multi}(X,\mathcal{T})$.
While it is possible to classify the spectrum into point, continuous and residual
subsets, as in the linear case, it is more meaningful for nonlinear operators
to consider $\lambda$ as being either in the \emph{boundary spectrum} $\textrm{Bdy}(\sigma(f))$
or in the \emph{interior spectrum} $\textrm{Int}(\sigma(f))$, depending on
whether or not the multifunction $f(\lambda;\cdot)^{-}$ arises as the graphical
limit of a net of functions in either $\rho(f)$ or $R\sigma(f)$. This is suggested
by the spectra arising from the second row of Table \ref{Table: spectrum} (injective
$\mathcal{L}_{\lambda}$ and discontinuous $\mathcal{L}_{\lambda}^{-1}$) that
lies sandwiched in the $\lambda$-plane between the two components arising from
the first and third rows, see \citet*{Naylor1971} Sec. 6.6, for example. According
to this simple scheme, the spectral set is a closed set with its boundary and
interior belonging to $\textrm{Bdy}(\sigma(f))$ and $\textrm{Int}(\sigma(f))$
respectively. Table \ref{Table: Appel_multi} shows this division for the examples
in Fig. \ref{Fig: Appel}. Because $0$ is no more significant than any other
point in the domain of a nonlinear map in inducing non-injectivity, the division
of the spectrum into the traditional sets would be as shown in Table \ref{Table: Appel_multi};
compare also with the conventional linear point spectrum of Table \ref{Table: Appel_spectra}.
In this nonlinear classification, the point spectrum consists of any $\lambda$
for which the inverse $f(\lambda;\cdot)^{-}$ is set-valued, irrespective of
whether this is produced at $0$ or not, while the continuous and residual spectra
together comprise the boundary spectrum. Thus a $\lambda$ can be both in the
point and the continuous or residual spectra which need not be disjoint. The
continuous and residual spectra are included in the boundary spectrum which
may also contain parts of the point spectrum. 

\begin{table}[htbp]
\begin{center}\begin{tabular}{|c||c|c|c|c|c|}
\hline 
Function&
$\textrm{Int}(\sigma(f))$&
$\textrm{Bdy}(\sigma(f))$&
$P\sigma(f)$&
$C\sigma(f)$&
$R\sigma(f)$\tabularnewline
\hline
\hline 
$f_{\textrm{a}}$&
$(0,1)$&
$\{0,1\}$&
$[0,1]$&
$\{1\}$&
$\{0\}$\tabularnewline
\hline 
$f_{\textrm{b}}$&
$(0,1)$&
$\{0,1\}$&
$[0,1]$&
$\{1\}$&
$\{0\}$\tabularnewline
\hline 
$f_{\textrm{c}}$&
$(0,\infty)$&
$\{0\}$&
$[0,\infty)$&
$\{0\}$&
$\emptyset$\tabularnewline
\hline 
$f_{\textrm{d}}$&
$(0,2)$&
$\{0,2\}$&
$(0,2)$&
$\{0,2\}$&
$\emptyset$\tabularnewline
\hline 
$f_{\textrm{e}}$&
$(0,1)$&
$\{0,1\}$&
$(0,1)$&
$\{1\}$&
$\{0\}$\tabularnewline
\hline 
$f_{\textrm{f}}$&
$(0,1)$&
$\{0,1\}$&
$(0,1)$&
$\{0,1\}$&
$\emptyset$\tabularnewline
\hline
\end{tabular}\end{center}

\begin{singlespace}

\caption{\label{Table: Appel_multi}{\small Nonlinear spectra of functions of Fig. \ref{Fig: Appel}.
Compare the present point spectra with the usual linear spectra of Table \ref{Table: Appel_spectra}.}}\end{singlespace}

\end{table}

\noindent \textbf{Example 6.1.} To see how these concepts apply to linear mappings,
consider the equation $(D-\lambda)y(x)=r(x)$ where $D=d/dx$ is the differential
operator on $L^{2}[0,\infty)$, and let $\lambda$ be real. For $\lambda\neq0$,
the unique solution of this equation in $L^{2}[0,\infty)$, is 

\begin{figure}[htbp]
\noindent \begin{center}\input{spectrum.pstex_t}\end{center}

\begin{singlespace}

\caption{\label{Fig: spectrum}{\small Spectra of some linear operators in the $\lambda$-plane.
(a) Left shift $(\cdots,x_{-1},x_{0},x_{1},\cdots)\mapsto(\cdots x_{0},x_{1},x_{2},\cdots)$
on $l_{2}(-\infty,\infty)$, (b) Right shift $(x_{0},x_{1},x_{2},\cdots)\mapsto(0,x_{0},x_{1},\cdots)$
on $l_{2}[0,\infty)$, (c) Left shift $(x_{0},x_{1},x_{2},\cdots)\mapsto(x_{1},x_{2},x_{3},\cdots)$
on $l_{2}[0,\infty)$ of sequence spaces, and (d) $d/dx$ on $L_{2}(-\infty,\infty)$
(e) $d/dx$ on $L_{2}[0,\infty)$ with $y(0)=0$ and (f) $d/dx$ on $L_{2}[0,\infty)$.
The residual spectrum in (b) and (e) arise from block (3-3) in Table \ref{Table: spectrum},
i.e., $\mathcal{L}_{\lambda}$ is one-to-one and $\mathcal{L}_{\lambda}^{-1}$
is bounded on non-dense domains in $l_{2}[0,\infty)$ and $L_{2}[0,\infty)$
respectively. The continuous spectrum therefore marks the boundary between two
functional states, as in (a) and (e), now with dense and non-dense domains of
the inverse operator. }}\end{singlespace}

\end{figure}
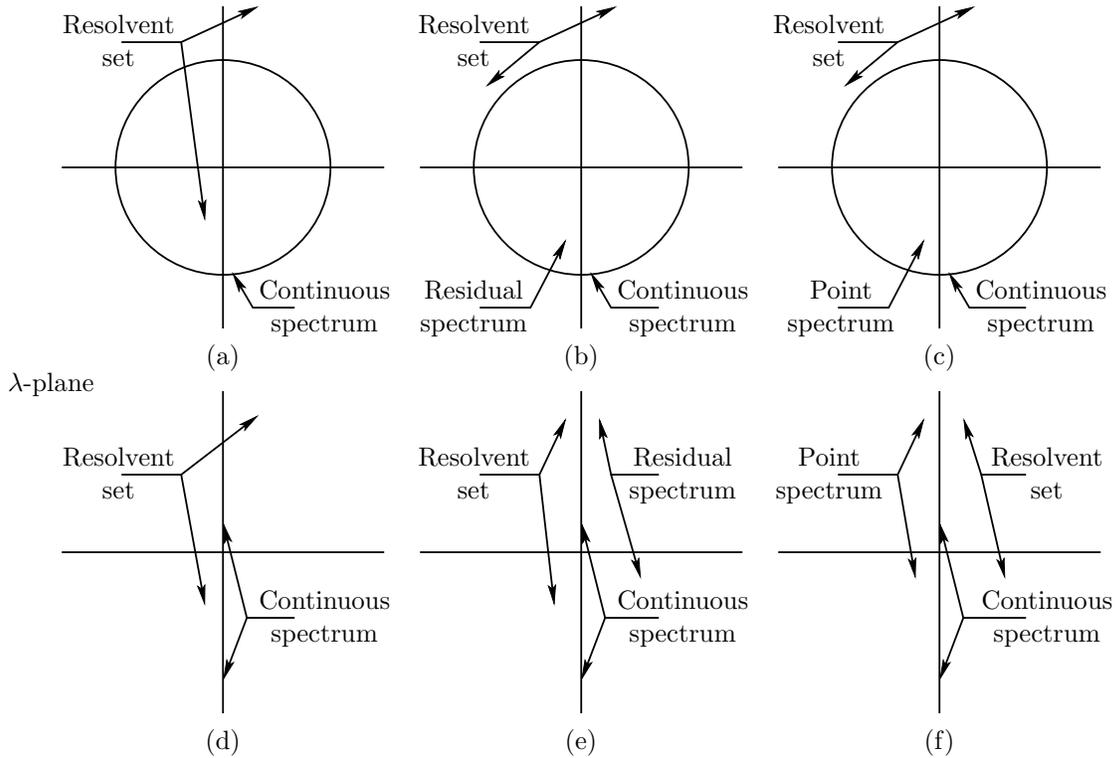

\noindent {\renewcommand{\arraystretch}{2}\begin{align*}
y(x)= & \left\{ \begin{array}{ll}
{\displaystyle e^{\lambda x}\left(y(0)+\int_{0}^{x}e^{-\lambda x^{\prime}}r(x^{\prime})dx^{\prime}\right)}, & \lambda<0\\
{\displaystyle e^{\lambda x}\left(y(0)-\int_{x}^{\infty}e^{-\lambda x^{\prime}}r(x^{\prime})dx^{\prime}\right),} & \lambda>0\end{array}\right.\end{align*}
}

\noindent showing that for $\lambda>0$ the inverse is functional so that $\lambda\in(0,\infty)$
belongs to the resolvent of $D$. However, when $\lambda<0$, apart from the
$y=0$ solution (since we are dealing a with linear problem, only $r=0$ is
to be considered), $e^{\lambda x}$ is also in $L^{2}[0,\infty)$ so that all
such $\lambda$ are in the point spectrum of $D$. For $\lambda=0$ and $r\neq0$,
the two solutions are not necessarily equal unless $\int_{0}^{\infty}r(x)=0$,
so that the range $\mathcal{R}(D-I)$ is a subspace of $L^{2}[0,\infty)$. To
complete the problem, it is possible to show \cite{Naylor1971} that $0\in C\sigma(D)$,
see Ex. 2.2; hence the continuous spectrum forms at the boundary of the functional
solution for the resolvent-$\lambda$ and the multifunctional solution for the
point spectrum. With a slight variation of problem to $y(0)=0$, all $\lambda<0$
are in the resolvent set, while $\lambda>0$ the inverse is bounded but must
satisfy $y(0)=\int_{0}^{\infty}e^{-\lambda x}r(x)dx=0$ so that $\textrm{Cl}(\mathcal{R}(D-\lambda))\neq L^{2}[0,\infty)$.
Hence $\lambda>0$ belong to the residual spectrum. The decomposition of the
complex $\lambda$-plane for these and some other linear spectral problems taken
from \citet*{Naylor1971} is shown in Fig. \ref{Fig: spectrum}. In all cases,
the spectrum due to the second row of Table \ref{Table: spectrum} acts as a
boundary between that arising from the first and third rows, which justifies
our division of the spectrum for a nonlinear operator into the interior and
boundary components. Compare Example 2.2.$\qquad\blacksquare$
\medskip{}

From the basic representation of the resolvent operator $(\mathbf{1}-f)^{-1}$
\[
\mathbf{1}+f+f^{2}+\cdots+f^{i}+\cdots\]

\noindent in $\textrm{Multi}(X)$, if the iterates of $f$ converge to a multifunction
for some $\lambda$, then that $\lambda$ must be in the spectrum of $f$, which
means that the control parameter of a chaotic dynamical system is in its spectrum.
Of course, the series can sum to a multi even otherwise: take $f_{\lambda}(x)$
to be identically $x$ with $\lambda=1$, for example, to get $1\in P\sigma(f)$.
A comparison of Tables \ref{Table: spectrum} and \ref{Table: Appel_spectra}
reveal that in case (d), for example, $0$ and $2$ belong to the Lipschtiz
spectrum because although $f_{\textrm{d}}^{-1}$ is not Lipschitz continuous,
$\Vert f\Vert_{\textrm{Lip}}=2$. It should also be noted that the boundary
between the functional resolvent and multifunctional spectral set is formed
by the graphical convergence of a net of resolvent functions while the multifunctions
in the interior of the spectral set evolve graphically independent of the functions
in the resolvent. The chaotic states forming the boundary of the functional
and multifunctional subsets of $\textrm{Multi}(X)$ marks the transition from
the less efficient functional state to the more efficient multifunctional one.

These arguments also suggest the following. The countably many outputs arising
from the non-injectivity of $f(\lambda;\cdot)$ corresponding a given input
can be interpreted to define \emph{complexity because} \emph{in a nonlinear
system each of these possibilities constitute a experimental result in itself
that may not be combined in any definite predetermined manner.} This is in sharp
contrast to linear systems where a linear combination, governed by the initial
conditions, always generate a unique end result; recall also the combination
offered by the singular generalized eigenfunctions of neutron transport theory.
This multiplicity of possibilities that have no definite combinatorial property
is the basis of the diversity of nature, and is possibly responsible for Feigenbaum's
{}``historical prejudice'', \cite{Feigenbaum1992}, see Prelude, 2. Thus \emph{order}
represented by the functional resolvent passes over to \emph{complexity} of
the countably multifunctional interior spectrum via the uncountably multifunctional
boundary that is a prerequisite for \emph{chaos.} We may now strengthen our
hypothesis offered at the end of the previous section in terms of the examples
of Figs. \ref{Fig: Appel} and \ref{Fig: spectrum}, that nature uses chaoticity
as an intermediate step to the attainment of states that would otherwise be
inaccessible to it. Well-posedness of a system is an extremely inefficient way
of expressing a multitude of possibilities as this requires a different input
for every possible output. Nature chooses to express its myriad manifestations
through the multifunctional route leading either to averaging as in the delta
function case or to a countable set of well-defined states, as in the examples
of Fig. \ref{Fig: Appel} corresponding to the interior spectrum. Of course
it is no distraction that the multifunctional states arise respectively from
$f_{\lambda}$ and $f_{\lambda}^{-}$ in these examples as $f$ is a function
on $X$ that is under the influence of both $f$ and its inverse. The functional
resolvent is, for all practical purposes, only a tool in this structure of nature.

The equation $f(x)=y$ is typically an input-output system in which the inverse
images at a functional value $y_{0}$ represents a set of input parameters leading
to the same experimental output $y_{0};$ this is stability characterized by
a complete insensitivity of the output to changes in input. On the other hand,
a continuous multifunction at $x_{0}$ is a signal for a hypersensitivity to
input because the output, which is a definite experimental quantity, is a choice
from the possibly infinite set $\{ f(x_{0})\}$ made by a choice function which
represents the experiment at that particular point in time. Since there will
always be finite differences in the experimental parameters when an experiment
is repeated, the choice function (that is the experimental output) will select
a point from $\{ f(x_{0})\}$ that is representative of that experiment and
which need not bear any definite relation to the previous values; this is instability
and signals sensitivity to initial conditions. Such a state is of high entropy
as the number of available states $f_{\textrm{C}}(\{ f(x_{0})\})$ --- where
$f_{\textrm{C}}$ is the choice function --- is larger than a functional state
represented by the singleton $\{ f(x_{0})\}.$ 

\vspace{1cm}
\noindent \begin{flushleft}\textbf{Epilogue}\end{flushleft}

\noindent \textsf{\textsl{The most passionate advocates of the new science go
so far as to say that twentieth-century science will be remembered for just
three things: relativity, quantum mechanics and chaos. Chaos, they contend,
has become the century's third great revolution in the physical sciences. Like
the first two revolutions, chaos cuts away at the tenets of Newton's physics.
As one physicist put it: {}``Relativity eliminated the Newtonian illusion of
absolute space and time; quantum theory eliminated the Newtonian dream of a
controllable measurement process; and chaos eliminates the Laplacian fantasy
of deterministic predictability.'' Of the three, the revolution in chaos applies
to the universe we see and touch, to objects at human scale. $\cdots$ There
has long been a feeling, not always expressed openly, that theoretical physics
has strayed far from human intuition about the world. Whether this will prove
to be fruitful heresy, or just plain heresy, no one knows. But some of those
who thought that physics might be working its way into a corner now look to
chaos as a way out.}}

\noindent \begin{flushright}\citet*{Gleick1987}\end{flushright}

\bigskip{}
\noindent \begin{center}\textbf{Appendix}\end{center}
\medskip{}

\noindent This Appendix gives a brief overview of some aspects of topology that
are necessary for a proper understanding of the concepts introduced in this
work. 

\vspace{0.75cm}
\noindent \begin{center}\textbf{A1. Convergence in Topological Spaces: Sequence,
Net and Filter.}\end{center}

\noindent In the theory of convergence in topological spaces, \emph{countability}
plays an important role. To understand the significance of this concept, some
preliminaries are needed. 

The notion of a basis, or base, is a familiar one in analysis: a base is a subcollection
of a set which may be used to construct, in a specified manner, any element
of the set. This simplifies the statement of a problem since a smaller number
of elements of the base can be used to generate the larger class of every element
of the set. This philosophy finds application in topological spaces as follows. 

\smallskip{}
Among the three properties $(\textrm{N}1)-(\textrm{N}3)$ of the neighbourhood
system $\mathcal{N}_{x}$ of Tutorial4, (N1) and (N2) are basic in the sense
that the resulting subcollection of $\mathcal{N}_{x}$ can be used to generate
the full system by applying $(\textrm{N}3)$; this \emph{basic neighbourhood}
\emph{system}, or \emph{neighbourhood (local) bas}e $\mathcal{B}_{x}$ \emph{at}
$x$, is characterized by 

\smallskip{}
(NB1) $x$ belongs to each member $B$ of $\mathcal{B}_{x}$\emph{. }

(NB2) The intersection of any two members of \emph{}$\mathcal{B}_{x}$ \emph{}contains
another member of $\mathcal{B}_{x}$: $B_{1},B_{2}\in\mathcal{B}_{x}\Rightarrow(\exists B\in\mathcal{B}_{x}\!:B\subseteq B_{1}\bigcap B_{2})$. \emph{}
\smallskip{}

Formally, compare Eq. (\ref{Eqn: nbd-topology}), 

\medskip{}
\noindent \textbf{Definition A1.1.} \textsl{A neighbourhood (local) base} $\mathcal{B}_{x}$
\textsl{at $x$ in a topological space $(X,\mathcal{U})$ is a subcollection
of the neighbourhood system $\mathcal{N}_{x}$ having the property that each
$N\in\mathcal{N}_{x}$ contains some member of} $\mathcal{B}_{x}$\textsl{.}
\textsl{Thus} \begin{equation}
\mathcal{B}_{x}\overset{\textrm{def}}=\{ B\in\mathcal{N}_{x}\!:x\in B\subseteq N\textrm{ for each }N\in\mathcal{N}_{x}\}\label{Eqn: TBx}\end{equation}
 \textsl{determines the full neighbourhood system} \begin{equation}
\mathcal{N}_{x}=\{ N\subseteq X\!:x\in B\subseteq N\textrm{ for some }B\textrm{ }\in\,\mathcal{B}_{x}\}\label{Eqn: TBx_nbd}\end{equation}

\noindent \textsl{reciprocally as all supersets of the basic elements.$\qquad\square$}

\medskip{}
\noindent The entire neighbourhood system $\mathcal{N}_{x}$, which is recovered
from the base by forming all supersets of the basic neighbourhoods, \textit{}is
trivially a local base at $x$; non-trivial examples are given below. 

The second example of a base, consisting as usual of a subcollection of a given
collection, is the topological base $_{\textrm{T}}\mathcal{B}$ that allows
the specification of the topology on a set $X$ in terms of a smaller collection
of open sets. 

\medskip{}
\noindent \textbf{Definition A1.2.} \textsl{A base} $_{\textrm{T}}\mathcal{B}$
\textsl{in a topological space $(X,\mathcal{U})$ is a subcollection of the
topology $\mathcal{U}$ having the property that each $U\in\mathcal{U}$ contains
some member of} $_{\textrm{T}}\mathcal{B}$\textsl{.} \textsl{Thus} \begin{equation}
_{\textrm{T}}\mathcal{B}\overset{\textrm{def}}=\{ B\in\mathcal{U}\!:B\subseteq U\textrm{ for each }U\in\mathcal{U}\}\label{Eqn: TB}\end{equation}
 \textsl{determines reciprocally the topology $\mathcal{U}$ as} \begin{equation}
\mathcal{U}=\left\{ U\subseteq X\!:U=\bigcup_{B\in\,\!_{\textrm{T}}\mathcal{B}\,}B\right\} \qquad\square\label{Eqn: TB_topo}\end{equation}

\noindent This means that the topology on $X$ can be reconstructed form the
base by taking all possible unions of members of the base, and a collection
of subsets of a set $X$ is a topological base iff Eq. (\ref{Eqn: TB_topo})
of arbitrary unions of elements of $_{\textrm{T}}\mathcal{B}$ generates a topology
on $X$. This topology, which is the coarsest (that is the smallest) that contains
$_{\textrm{T}}\mathcal{B}$, is obviously closed under finite intersections.
Since the open set $\textrm{Int}(N)$ is a neighbourhood of $x$ whenever $N$
is, Eq. (\ref{Eqn: TBx_nbd}) and the definition Eq. (\ref{Eqn: Def: nbd system})
of $\mathcal{N}_{x}$ implies that \emph{the open neighbourhood system of any
point in a topological space is an example of a neighbourhood base at that point,}
an observation that has often led, together with Eq. (\ref{Eqn: TB}), to the
use of the term {}``neighbourhood'' as a synonym for {}``non-empty open set''.
The distinction between the two however is significant as neighbourhoods need
not necessarily be open sets; thus while not necessary, it is clearly sufficient
for the local basic sets $B$ to be open in Eqs. (\ref{Eqn: TBx}) and (\ref{Eqn: TBx_nbd}).
If Eq. (\ref{Eqn: TBx_nbd}) holds for every $x\in N$, then the resulting $\mathcal{N}_{x}$
reduces to the topology induced by the open basic neighbourhood system $\mathcal{B}_{x}$
as given by Eq. (\ref{Eqn: nbd-topology}). 

In order to check if a collection of subsets $_{\textrm{T}}\mathcal{B}$ of
$X$ qualifies to be a basis, it is not necessary to verify properties $(\textrm{T}1)-(\textrm{T}3)$
of Tutorial4 for the class (\ref{Eqn: TB_topo}) generated by it because of
the properties (TB1) and (TB2) below whose strong affinity to (NB1) and (NB2)
is formalized in Theorem A1.1. 

\medskip{}
\noindent \textbf{Theorem A1.1.} \textsl{A collection} $_{\textrm{T}}\mathcal{B}$
\textsl{of subsets of $X$ is a} \textsl{topological basis on} $X$ \textsl{iff }

\smallskip{}
(TB1) \textsl{$X=\bigcup_{B\in\,_{\textrm{T}}\mathcal{B}}B$. Thus each $x\in X$
must belong to some} $B\in\,_{\textrm{T}}\mathcal{B}$ \textsl{which implies
the existence of a} \emph{}\textsl{local base} \emph{}\textsl{at each point}
\emph{$x\in X$. }

(TB2) \textsl{The intersection of any two members $B_{1}$ and $B_{2}$ of}
$_{\textrm{T}}\mathcal{B}$ \textsl{with $x\in B_{1}\bigcap B_{2}$} \textsl{\emph{}}\textsl{contains
another member of} $_{\textrm{T}}\mathcal{B}$: $(B_{1},B_{2}\in\,_{\textrm{T}}\mathcal{B})\wedge(x\in B_{1}\bigcap B_{2})\Rightarrow(\exists B\in\,_{\textrm{T}}\mathcal{B}\!:x\in B\subseteq B_{1}\bigcap B_{2})$.$\qquad\square$

\medskip{}
\noindent This theorem, together with Eq. (\ref{Eqn: TB_topo}) ensures that
a given collection of subsets of a set $X$ satisfying (TB1) and (TB2) induces
\emph{some} topology on $X$; compared to this is the result that \emph{any}
collection of subsets of a set $X$ is a \emph{subbasis} for some topology on
$X$. If $X$, however, already has a topology $\mathcal{U}$ imposed on it,
then Eq. \textsl{}(\ref{Eqn: TB}) must also be satisfied in order that the
topology generated by $_{\textrm{T}}\mathcal{B}$ is indeed $\mathcal{U}$.
The next theorem connects the two types of bases of Defs. A1.1 and A1.2 by asserting
that although a local base of a space need not consist of open sets and a topological
base need not have any reference to a point of $X$, any subcollection of the
base containing a point is a local base at that point. 

\medskip{}
\noindent \textbf{Theorem A1.2.} \textsl{A collection of open sets} $_{\textrm{T}}\mathcal{B}$
\textsl{is a base for a topological space $(X,\mathcal{U})$ iff for each $x\in X$,
the subcollection} \begin{equation}
\mathcal{B}_{x}=\{ B\in\mathcal{U}\!:x\in B\in\!\,_{\textrm{T}}\mathcal{B}\}\label{Eqn: base_local base}\end{equation}
 \textsl{of basic sets containing $x$ is a local base at} $x$.$\qquad\square$ 

\noindent \textbf{Proof.} \emph{Necessity.} Let $_{\textrm{T}}\mathcal{B}$
be a base of \textsl{$(X,\mathcal{U})$} and $N$ be a neighbourhood of $x$,
so that $x\in U\subseteq N$ for some open set $U=\bigcup_{B\in\!\,_{\textrm{T}}\mathcal{B}}B$
and basic open sets $B$. Hence $x\in B\subseteq N$ shows, from Eq. (\ref{Eqn: TBx}),
that $B\in\mathcal{B}_{x}$ is a local basic set at $x$. 

\emph{Sufficiency.} If $U$ is an open set of $X$ containing $x$, then the
definition of local base Eq. (\ref{Eqn: TBx}) requires $x\in B_{x}\subseteq U$
for some subcollection of basic sets $B_{x}$ in $\mathcal{B}_{x}$; hence $U=\bigcup_{x\in U}B_{x}$.
By Eq. (\ref{Eqn: TB_topo}) therefore, $_{\textrm{T}}\mathcal{B}$ is a topological
base for $X$.$\qquad\blacksquare$
\medskip{}

Because the basic sets are open, (TB2) of Theorem A1.1 leads to the following
physically appealing paraphrase of Thm. A1.2. 

\medskip{}
\noindent \textbf{Corollary.} \textsl{A collection} $_{\textrm{T}}\mathcal{B}$
\textsl{of open sets of} $(X,\mathcal{U})$ \textsl{is a topological base that
generates} $\mathcal{U}$ \textsl{iff for each open set $U$ of $X$ and each
$x\in U$ there is an open set} $B\in\!\,_{\textrm{T}}\mathcal{B}$ \textsl{such
that $x\in B\subseteq U$}; \textsl{that is iff} \[
x\in U\in\mathcal{U}\Longrightarrow(\exists B\in\,_{\textrm{T}}\mathcal{B}\!:x\in B\subseteq U).\qquad\square\]
 \emph{}\textbf{Example A1.1.} Some examples of local bases in $\mathbb{R}$
are intervals of the type $(x-\varepsilon,x+\varepsilon)$, $[x-\varepsilon,x+\varepsilon]$
for real $\varepsilon$, $(x-q,x+q)$ for rational $q$, $(x-1/n,x+1/n)$ for
$n\in\mathbb{Z}_{+}$, while for a metrizable space with the topology induced
by a metric $d$, each of the following is a local base at $x\in X$: $B_{\varepsilon}(x;d):=\{ y\in X:d(x,y)<\varepsilon\}$
and $D_{\varepsilon}(x;d):=\{ y\in X:d(x,y)\leq\varepsilon\}$ for $\varepsilon>0$,
$B_{q}(x;d)$ for $\mathbb{Q}\ni q>0$ and $B_{1/n}(x;d)$ for $n\in\mathbb{Z}_{+}$.
In $\mathbb{R}^{2}$, two neighbourhood bases at any $x\in\mathbb{R}^{2}$ are
the disks centered at $x$ and the set of all squares at $x$ with sides parallel
to the axes. Although these bases have no elements in common, they are nevertheless
equivalent in the sense that they both generate the same (usual) topology in
$\mathbb{R}^{2}$. Of course, the entire neighbourhood system at any point of
a topological space is itself a (less useful) local base at that point. By Theorem
A1.2, $B_{\varepsilon}(x;d)$, $D_{\varepsilon}(x;d)$, $\varepsilon>0$, $B_{q}(x;d)$,
$\mathbb{Q}\ni q>0$ and $B_{1/n}(x;d)$, $n\in\mathbb{Z}_{+}$, for all $x\in X$
are examples of bases in a metrizable space with topology induced by a metric
$d$.$\qquad\square$

\medskip{}
In terms of local bases and bases, it is now possible to formulate the notions
of first and second countability as follows. 

\medskip{}
\noindent \textbf{Definition A1.3.} \textsl{A topological space is} \emph{first
countable} \textsl{if each $x\in X$ has some countable neighbourhood base,
and is} \emph{second countable} \textsl{if it has a countable base.} $\qquad\square$
\medskip{}

Every metrizable space $(X,d)$ is first countable as both $\{ B(x,q)\}_{\mathbb{Q}\ni q>0}$
and $\{ B(x,1/n)\}_{n\in\mathbb{Z}_{+}}$ are examples of countable neighbourhood
bases at any $x\in(X,d)$; hence $\mathbb{R}^{n}$ is first countable. It should
be clear that although every second countable space is first countable, \emph{only
a countable first countable space can be second countable}, and a common example
of a uncountable first countable space that is also second countable is provided
by $\mathbb{R}^{n}$. Metrizable spaces need not be second countable: any uncountable
set having the discrete topology is as an example. 

\medskip{}
\noindent \textbf{Example A1.2.} The following is an important example of a
space that is not first countable as it is needed for our pointwise biconvergence
of Section 3. Let $\textrm{Map}(X,Y)$ be the set of all functions between the
uncountable spaces $(X,\mathcal{U})$ and $(Y,\mathcal{V})$. Given any integer
$I\geq1$, and any \emph{finite} collection of points $(x_{i})_{i=1}^{I}$ of
$X$ and of open sets $(V_{i})_{i=1}^{I}$ in $Y$, let \begin{equation}
B((x_{i})_{i=1}^{I};(V_{i})_{i=1}^{I})=\{ g\in\textrm{Map}(X,Y)\!:(g(x_{i})\in V_{i})(i=1,2,\cdots,I)\}\label{Eqn: point}\end{equation}

\noindent be the functions in $\textrm{Map}(X,Y)$ whose graphs pass through
each of the sets $(V_{i})_{i=1}^{I}$ at $(x_{i})_{i=1}^{I}$, and let $_{\textrm{T}}\mathcal{B}$
be the collection of all such subsets of $\textrm{Map}(X,Y)$ for every choice
of $I$, $(x_{i})_{i=1}^{I}$, and $(V_{i})_{i=1}^{I}$. The existence of a
unique topology $\mathcal{T}$ --- the \emph{topology of pointwise convergence}
on $\textrm{Map}(X,Y)$ --- that is generated by the open sets $B$ of the collection
$_{\textrm{T}}\mathcal{B}$ now follows because 

\smallskip{}
(TB1) is satisfied: For any $f\in\textrm{Map}(X,Y)$ there must be some $x\in X$
and a corresponding $V\subseteq Y$ such that $f(x)\in V$, and 

(TB2) is satisfied because \[
B((s_{i})_{i=1}^{I};(V_{i})_{i=1}^{I})\bigcap B((t_{j})_{j=1}^{J};(W_{j})_{j=1}^{J})=B((s_{i})_{i=1}^{I},(t_{j})_{j=1}^{J};(V_{i})_{i=1}^{I},(W_{j})_{j=1}^{J})\]

\noindent implies that a function simultaneously belonging to the two open sets
on the left must pass through each of the points defining the open set on the
right. 

We now demonstrate that $(\textrm{Map}(X,Y),\mathcal{T})$ is not first countable
by verifying that it is not possible to have a countable local base at any $f\in\textrm{Map}(X,Y)$.
If this is not indeed true, let $B_{f}^{I}((x_{i})_{i=1}^{I};(V_{i})_{i=1}^{I})=\{ g\in\textrm{Map}(X,Y)\!:(g(x_{i})\in V_{i})_{i=1}^{I}\}$,
which denotes those members of $_{\textrm{T}}\mathcal{B}$ that contain $f$
with $V_{i}$ an open neighbourhood of $f(x_{i})$ in $Y$, be a countable local
base at $f$, see Thm. A1.2. Since $X$ is uncountable, it is now possible to
choose some $x^{*}\in X$ different from any of the $(x_{i})_{i=1}^{I}\textrm{ }$
(for example, let $x^{*}\in\mathbb{R}$ be an irrational for rational $(x_{i})_{i}^{I}\textrm{ }$),
and let $f(x^{*})\in V^{*}$ where $V^{*}$ is an open neighbourhood of $f(x^{*})$.
Then $B(x^{*};V^{*})$ is an open set in $\textrm{Map}(X,Y)$ containing $f$;
hence from the definition of the local base, Eq. (\ref{Eqn: TBx}), or equivalently
from the Corollary to Theorem A1.2, there exists some (countable) $I\in\mathbb{N}$
such that $f\in B^{I}\subseteq B(x^{*};V^{*})$. However, \[
\begin{array}{ccc}
f^{*}(x) & = & \begin{cases}
y_{i}\in V_{i}, & \textrm{if }x=x_{i},\textrm{ and }1\leq i\leq I\\
y^{*}\in V^{*} & \textrm{if }x=x^{*}\\
\textrm{arbitrary}, & \textrm{otherwise}\end{cases}\end{array}\]

\noindent is a simple example of a function on $X$ that is in $B^{I}$ (as
it is immaterial as to what values the function takes at points other than those
defining $B^{I}$), but not in $B(x^{*};V^{*})$. From this it follows that
\emph{a sufficient condition for the topology of pointwise convergence to be
first countable is that $X$ be countable.}$\qquad\blacksquare$
\medskip{}

Even though it is not first countable, $(\textrm{Map}(X,Y),\mathcal{T})$ is
a Hausdorff space when $Y$ is Hausdorff. Indeed, if $f,g\in(\textrm{Map}(X,Y),\mathcal{T})$
with $f\neq g$, then $f(x)\neq g(x)$ for some $x\in X$. But then as $Y$
is Hausdorff, it is possible to choose disjoint open intervals $V_{f}$ and
$V_{g}$ at $f(x)$ and $g(x)$ respectively. 

With this background on first and second countability, it is now possible to
go back to the question of nets, filters and sequences. Technically, a sequence
on a set $X$ is a map $x\!:\mathbb{N}\rightarrow X$ from the set of natural
numbers to $X$; instead of denoting this is in the usual functional manner
of $x(i)\textrm{ with }i\in\mathbb{N}$, it is the standard practice to use
the notation $(x_{i})_{i\in\mathbb{N}}$ for the terms of a sequence. However,
if the space $(X,\mathcal{U})$ is not first countable (and as seen above this
is not a rare situation), it is not difficult to realize that sequences are
inadequate to describe convergence in $X$ simply because it can have only countably
many values whereas the space may require uncountably many neighbourhoods to
completely define the neighbourhood system at a point. The resulting uncountable
generalizations of a sequence in the form of \emph{nets} and \emph{filters}
is achieved through a corresponding generalization of the index set $\mathbb{N}$
to the directed set $\mathbb{D}$. 

\medskip{}
\noindent \textbf{Definition A1.4.} \textsl{A} \emph{directed set} \textsl{$\mathbb{D}$
is a preordered set for which the order $\preceq$, known as a} \emph{direction
of} $\mathbb{D}$, \textsl{satisfies}

(a) \textsl{$\alpha\in\mathbb{D}$ $\Rightarrow$ $\alpha\preceq\alpha$} (that
is $\preceq$ is reflexive)\textsl{. }

(b) \textsl{}$\alpha,\beta,\gamma\in\mathbb{D}\textrm{ such that }(\alpha\preceq\beta\wedge\beta\preceq\gamma)$
$\Rightarrow$ $\alpha\preceq\gamma$ (that is $\preceq$ is transitive).

(c) $\alpha,\beta\in\mathbb{D}$ $\Rightarrow$ $\exists\gamma\in\mathbb{D}\textrm{ such that }(\alpha\preceq\gamma)\wedge(\beta\preceq\gamma)$\textsl{.$\qquad\square$ }
\medskip{}

\noindent While the first two properties are obvious enough and constitutes
the preordering of $\mathbb{{D}}$, the third which replaces antisymmetry, ensures
that for any finite number of elements of the directed set (recall that a preordered
set need not be fully ordered), there is always a successor. Examples of directed
sets can be both straight forward, as any totally ordered set like $\mathbb{N}$,
$\mathbb{R}$, $\mathbb{Q}$, or $\mathbb{Z}$ and all subsets of a set $X$
under the superset or subset relation (that is $(\mathcal{P}(X),\supseteq)$
or $(\mathcal{P}(X),\subseteq)$ that are directed by their usual ordering,
and not quite so obvious as the following examples which are significantly useful
in dealing with convergence questions in topological spaces, amply illustrate. 

The neighbourhood system \[
_{\mathbb{D}}N=\{ N\!:N\in\mathcal{N}_{x}\}\]
 at a point $x\in X$, directed by the reverse inclusion direction $\preceq$
defined as \begin{equation}
M\preceq N\Longleftrightarrow N\subseteq M\qquad\textrm{for }M,N\in\mathcal{N}_{x},\label{Eqn: Direction1}\end{equation}

\noindent is a fundamental example of a \emph{natural direction of $\mathcal{N}_{x}$}.
In fact while reflexivity and transitivity are clearly obvious, (c) follows
because for any $M,N\in\mathcal{N}_{x}$, $M\preceq M\bigcap N$ and $N\preceq M\bigcap N$.
Of course, this direction is not a total ordering on $\mathcal{N}_{x}$. A more
naturally useful directed set in convergence theory is \begin{equation}
_{\mathbb{D}}N_{t}=\{(N,t)\!:(N\in\mathcal{N}_{x})(t\in N)\}\label{Eqn: Directed}\end{equation}

\noindent under its \emph{natural direction} \begin{equation}
(M,s)\preceq(N,t)\Longleftrightarrow N\subseteq M\qquad\textrm{for }M,N\in\mathcal{N}_{x};\label{Eqn: Direction2}\end{equation}
 \emph{}$_{\mathbb{D}}N_{t}$ is more useful than $_{\mathbb{D}}N$ because,
unlike the later, $_{\mathbb{D}}N_{t}$ does not require a simultaneous choice
of points from every $N\in\mathcal{N}_{x}$ that implicitly involves a simultaneous
application of the Axiom of Choice; see Examples A1.2(2) and (3) below. The
general indexed variation 

\begin{equation}
_{\mathbb{D}}N_{\beta}=\{(N,\beta)\!:(N\in\mathcal{N}_{x})(\beta\in\mathbb{D})(x_{\beta}\in N)\}\label{Eqn: DirectedIndexed}\end{equation}
 of Eq. (\ref{Eqn: Directed}), with natural direction \begin{equation}
(M,\alpha)\leq(N,\beta)\Longleftrightarrow(\alpha\preceq\beta)\wedge(N\subseteq M),\label{Eqn: DirectionIndexed}\end{equation}

\noindent often proves useful in applications as will be clear from the proofs
of Theorems A1.3 and A1.4.

\medskip{}
\noindent \textbf{Definition A1.5.} \textbf{\textit{Net.}} \textsl{Let $X$
be any set and $\mathbb{D}$ a directed set. A net $\chi\!:\mathbb{D}\rightarrow X$}
\emph{in $X$} \textsl{is a function} \emph{on the directed set $\mathbb{D}$
with values in $X$.$\qquad\square$}
\medskip{}

A net, to be denoted as $\chi(\alpha)$, $\alpha\in\mathbb{D}$, is therefore
a function indexed by a directed set. We adopt the convention of denoting nets
in the manner of functions and do not use the sequential notation $\chi_{\alpha}$
that can also be found in the literature. Thus, while every sequence is a special
type of net, $\chi:\!\mathbb{Z}\rightarrow X$ is an example of a net that is
not a sequence. 

Convergence of sequences and nets are described most conveniently in terms of
the notions of being \emph{eventually in} and \emph{frequently in} every neighbourhood
of points. We describe these concepts in terms of nets which apply to sequences
with obvious modifications. 

\medskip{}
\noindent \textbf{Definition A1.6.} \textsl{A net} $\chi\!:\mathbb{D}\rightarrow X$
\textsl{is said to be} 

(a) \emph{Eventually in} \textsl{a subset $A$} \emph{of} \textsl{$X$ if its
tail is eventually in $A$}: \textsl{$(\exists\beta\in\mathbb{D})\!:(\forall\gamma\succeq\beta)(\chi(\gamma)\in A).$} 

(b) \emph{Frequently in} \textsl{a subset $A$} \emph{of} \textsl{$X$ if for
any index $\beta\in\mathbb{D}$, there is a successor index $\gamma\in\mathbb{D}$
such that $\chi(\gamma)$} is in $A$: \textsl{$(\forall\beta\in\mathbb{D})(\exists\gamma\succeq\beta)\!:(\chi(\gamma)\in A).\qquad\square$}
\medskip{}

It is not difficult to appreciate that 

(i) A net eventually in a subset is also frequently in it but not conversely, 

(ii) A net eventually (respectively, frequently) in a subset cannot be frequently
(respectively, eventually) in its complement. 

With these notions of eventually in and frequently in, convergence characteristics
of a net may be expressed as follows. 

\medskip{}
\noindent \textbf{Definition A1.7.} \textit{A net} \textsl{$\chi\!:\mathbb{D}\rightarrow X$
converges to $x\in X$ if it is eventually in every neighbourhood of $x$, that
is} \[
(\forall N\in\mathcal{N}_{x})(\exists\mu\in\mathbb{D})(\chi(\nu\succeq\mu)\in N).\]
 \textsl{The point $x$ is known as the} \textit{limit} \textsl{of $\chi$ and
the collection of all limits of a net is the} \textit{limit set} \begin{equation}
\textrm{lim}(\chi)=\{ x\in X\!:(\forall N\in\mathcal{N}_{x})(\exists\mathbb{R}_{\beta}\in\textrm{Res}(\mathbb{D}))(\chi(\mathbb{R}_{\beta})\subseteq N)\}\label{Eqn: lim net}\end{equation}
\textsl{of $\chi$, with the set of} \textit{residuals} $\textrm{Res}(\mathbb{D})$
\textsl{in $\mathbb{D}$ given by} \begin{equation}
\textrm{Res}(\mathbb{D})=\{\mathbb{R}_{\alpha}\in\mathcal{P}(\mathbb{D})\!:\mathbb{R}_{\alpha}=\{\beta\in\mathbb{D}\textrm{ for all }\beta\succeq\alpha\in\mathbb{D}\}\}.\label{Eqn: residual}\end{equation}

\noindent \textsl{The net} \emph{adheres at} \textit{$x\in X$}%
\footnote{\label{Foot: cluster}This is also known as a \emph{cluster point}; we shall,
however, use this new term exclusively in the sense of the elements of a derived
set, see Definition 2.3. %
} \textsl{if it is frequently in every neighbourhood of $x$, that is} \[
((\forall N\in\mathcal{N}_{x})(\forall\mu\in\mathbb{D}))((\exists\nu\succeq\mu)\!:\chi(\nu)\in N).\]
 \textsl{The point $x$ is known as the} \emph{adherent} \textsl{of $\chi$
and the collection of all adherents of $\chi$ is the} \emph{adherent set of
the net, which} \textsl{may be expressed in terms of the} \emph{cofinal subset}
\textsl{of $\mathbb{D}$} \begin{equation}
\textrm{Cof}(\mathbb{D})=\{\mathbb{C}_{\alpha}\in\mathcal{P}(\mathbb{D})\!:\mathbb{C}_{\alpha}=\{\beta\in\mathbb{D}\textrm{ for some }\beta\succeq\alpha\in\mathbb{D}\}\}\label{Eqn: cofinal}\end{equation}
 (thus $\mathbb{D}_{\alpha}$ is cofinal in $\mathbb{D}$ iff it intersects
every residual in $\mathbb{D}$), \textsl{as} \begin{equation}
\textrm{adh}(\chi)=\{ x\in X\!:(\forall N\in\mathcal{N}_{x})(\exists\mathbb{C}_{\beta}\in\textrm{Cof}(\mathbb{D}))(\chi(\mathbb{C}_{\beta})\subseteq N)\}.\label{Eqn: adh net1}\end{equation}

\noindent \textsl{This recognizes, in keeping with the limit set, each subnet
of a net to be a net in its own right, and is equivalent to} \begin{equation}
{\textstyle \textrm{adh}(\chi)=\{ x\in X\!:(\forall N\in\mathcal{N}_{x})(\forall\mathbb{R}_{\alpha}\in\textrm{Res}(\mathbb{D}))(\chi(\mathbb{R}_{\alpha})\bigcap N\neq\emptyset)\}.\qquad\square}\label{Eqn: adh net2}\end{equation}

Intuitively, a sequence is eventually in a set $A$ if it is always in it after
a finite number of terms (of course, the concept of a \emph{finite number of
terms} is unavailable for nets; in this case the situation may be described
by saying that a net is eventually in $A$ if its \emph{tail is in} $A$) and
it is frequently in $A$ if it always returns to $A$ to leave it again. It
can be shown that a net is eventually (resp. frequently) in a set iff it is
not frequently (resp.eventually) in its complement. 

The following examples illustrate graphically the role of a proper choice of
the index set $\mathbb{D}$ in the description of convergence. 

\smallskip{}
\noindent \textbf{Example A1.3.} (1) Let $\gamma\in\mathbb{D}$. The eventually
constant net $\chi(\delta)=x$ for $\delta\succeq\gamma$ converges to $x$. 

(2) Let $\mathcal{N}_{x}$ be a neighbourhood system at a point $x$ in $X$
and suppose that the net $(\chi(N))_{N\in\mathcal{N}_{x}}$ is defined by \begin{equation}
\chi(M)\overset{\textrm{def}}=s\in M;\label{Eqn: Def: Net1}\end{equation}
 here the directed index set $_{\mathbb{D}}N$ is ordered by the natural direction
(\ref{Eqn: Direction1}) of $\mathcal{N}_{x}$. Then $\chi(N)\rightarrow x$
because given any $x$-neighbourhood $M\in\!\:_{\mathbb{D}}N$, it follows from
\begin{equation}
M\preceq N\in\,{}_{\mathbb{D}}N\Longrightarrow\chi(N)=t\in N\subseteq M\label{Eqn: DirectedNet1}\end{equation}

\noindent that a point in any subset of $M$ is also in $M$; $\chi(N)$ is
therefore eventually in every neighbourhood of $x$. 

(3) This slightly more general form of the previous example provides a link
between the complimentary concepts of nets and filters that is considered below.
For a point $x\in X$, and $M,N\in\mathcal{N}_{x}$ with the corresponding directed
set $_{\mathbb{D}}M_{s}$ of Eq. (\ref{Eqn: Directed}) ordered by its natural
order (\ref{Eqn: Direction2}), the net \begin{equation}
\chi(M,s)\overset{\textrm{def}}=s\label{Eqn: Def: Net2}\end{equation}
 converges to $x$ because, as in the previous example, for any given $(M,s)\in\:\!_{\mathbb{D}}N_{s}$,
it follows from \begin{equation}
(M,s)\preceq(N,t)\in\!\:_{\mathbb{D}}M_{s}\Longrightarrow\chi(N,t)=t\in N\subseteq M\label{Eqn: DirectedNet2}\end{equation}
 that $\chi(N,t)$ is eventually in every neighbourhood $M$ of $x$. The significance
of the directed set $_{\mathbb{D}}N_{t}$ of Eq. (\ref{Eqn: Directed}), as
compared to $_{\mathbb{D}}N$, is evident from the net that it induces \emph{without
using the Axiom of Choice}: For a subset $A$ of $X$, the net $\chi(N,t)=t\in A$
indexed by the directed set \begin{equation}
{\textstyle _{\mathbb{D}}N_{t}=\{(N,t)\!:(N\in\mathcal{N}_{x})(t\in N\bigcap A)\}}\label{Eqn: Closure_Directed}\end{equation}
under the direction of Eq. (\ref{Eqn: Direction2}), converges to $x\in X$
with all such $x$ defining the closure $\textrm{Cl}(A)$ of $A$. Furthermore
taking the directed set to be \begin{equation}
{\textstyle _{\mathbb{D}}N_{t}=\{(N,t)\!:(N\in\mathcal{N}_{x})(t\in N\bigcap A-\{ x\})\}}\label{Eqn: Der_Directed}\end{equation}
 which, unlike Eq. (\ref{Eqn: Closure_Directed}), excludes the point $x$ that
may or may not be in the subset $A$ of $X$, induces the net $\chi(N,t)=t\in A-\{ x\}$
converging to $x\in X$, with the set of all such $x$ yielding the derived
set $\textrm{Der}(A)$ of $A$. In contrast, Eq. (\ref{Eqn: Closure_Directed})
also includes the isolated points $t=x$ of $A$ so as to generate its closure.
Observe how neighbourhoods of a point, which define convergence of nets and
filters in a topological space $X$, double up here as index sets to yield a
self-consistent tool for the description of convergence.

As compared with sequences where, the index set is restricted to positive integers,
the considerable freedom in the choice of directed sets as is abundantly borne
out by the two preceding examples, is not without its associated drawbacks.
Thus as a trade-off, the wide range of choice of the directed sets may imply
that induction methods, so common in the analysis of sequences, need no longer
apply to arbitrary nets.
\smallskip{}

(4) The non-convergent nets (actually these are sequences) 

(a) $(1,-1,1,-1,\cdots)$ adheres at $1$ and $-1$ and 

(b) $\begin{array}{ccl}
x_{n} & = & {\displaystyle \left\{ \begin{array}{lcl}
n &  & \textrm{if }n\textrm{ is odd}\\
1-1/(1+n) &  & \textrm{if }n\textrm{ is even}\end{array},\right.}\end{array}$ adheres at $1$ for its even terms, but is unbounded in the odd terms.$\qquad\blacksquare$
\medskip{}

A converging sequence or net is also adhering but, as examples (4) show, the
converse is false. Nevertheless it is true, as again is evident from examples
(4), that in a first countable space where sequences suffice, a sequence $(x_{n})$
adheres at $x$ iff some subsequence $(x_{n_{m}})_{m\in\mathbb{N}}$ of $(x_{n})$
converges to $x$. If the space is not first countable this has a corresponding
equivalent formulation for nets with subnets replacing subsequences as follows. 

Let $(\chi(\alpha))_{\alpha\in\mathbb{D}}$ be a net. A \emph{subnet} of $\chi(\alpha)$
is the net $\zeta(\beta)=\chi(\sigma(\beta))$, $\beta\in\mathbb{E}$, where
$\sigma\!:(\mathbb{E},\leq)\rightarrow(\mathbb{D},\preceq)$ is a function that
captures the essence of the subsequential mapping $n\mapsto n_{m}$ in $\mathbb{N}$
by satisfying 

\smallskip{}
(SN1) $\sigma$ is an increasing order-preserving function: it respects the
order of $\mathbb{E}$: $\sigma(\beta)\preceq\sigma(\beta^{\prime})$ for every
$\beta\leq\beta^{\prime}\in\mathbb{E}$, and 

(SN2) For every $\alpha\in\mathbb{D}$ there exists a $\beta\in\mathbb{E}$
such that $\alpha\preceq\sigma(\beta)$.
\smallskip{}

\noindent These generalize the essential properties of a subsequence in the
sense that (1) Even though the index sets $\mathbb{D}$ and $\mathbb{E}$ may
be different, it is necessary that the values of $\mathbb{E}$ be contained
in $\mathbb{D}$, and (2) There are arbitrarily large $\alpha\in\mathbb{D}$
such that $\chi(\alpha=\sigma(\beta))$ is a value of the subnet $\zeta(\beta)$
for some $\beta\in\mathbb{E}$. Recalling the first of the order relations Eq.
(\ref{Eqn: FunctionOrder}) on $\textrm{Map}(X,Y)$, we will denote a subnet
$\zeta$ of $\chi$ by $\zeta\preceq\chi$. 

We now consider the concept of filter on a set $X$ that is very useful in visualizing
the behaviour of sequences and nets, and in fact filters constitute an alternate
way of looking at convergence questions in topological spaces. A filter $\mathcal{F}$
on a set $X$ is a collection of \emph{nonempty} subsets of $X$ satisfying
properties $(\textrm{F}1)-(\textrm{F}3)$ below that are simply those of a neighbourhood
system $\mathcal{N}_{x}$ without specification of the reference point $x$. 

\smallskip{}
(F1) The empty set $\emptyset$ does not belong to $\mathcal{F}$, 

(F2) The intersection of any two members of a filter is another member of the
filter: $F_{1},F_{2}\in\mathcal{F}\Rightarrow F_{1}\bigcap F_{2}\in\mathcal{F}$, 

(F3) Every superset of \emph{}a member of a filter belongs to the filter: $(F\in\mathcal{F})\wedge(F\subseteq G)\Rightarrow G\in\mathcal{F}$;
in particular $X\in\mathcal{F}$. 

\medskip{}
\noindent \textbf{Example A1.4.} (1) The \emph{indiscrete filter} is the smallest
filter on $X$. 

(2) The neighbourhood system $\mathcal{N}_{x}$ is the important \emph{neighbourhood
filter at $x$ on $X$,} and any local base at $x$ is also a filter-base for
$\mathcal{N}_{x}$. In general for any subset $A$ of $X$, $\{ N\subseteq X\!:A\subseteq\textrm{Int}(N)\}$
is a filter on $X$ at $A$. 

(3) All subsets of $X$ containing a point $x\in X$ is the \emph{principal
filter} $_{\textrm{F}}\mathcal{P}(x)$ \emph{on $X$ at $x$.} More generally,
if $\mathcal{F}$ consists of all supersets of a \emph{nonempty} subset $A$
of $X$, then $\mathcal{F}$ is the \emph{principal filter} $_{\textrm{F}}\mathcal{P}(A)=\{ N\subseteq X\!:A\subseteq\textrm{Int}(N)\}$
\emph{at $A$. By adjoining the empty set to this filter give the $p$-inclusion
and $A$-inclusion topologies on $X$ respectively.} The single element sets
$\{\{ x\}\}$ and $\{ A\}$ are particularly simple examples of filter-bases
that generate the principal filters at $x$ and $A$. 

(4) For an uncountable (resp. infinite) set $X$, all cocountable (resp. cofinite)
subsets of $X$ constitute the \emph{cocountable} (resp. \emph{cofinite} or
\emph{Frechet}) filter on $X$. Again, adding to these filters the empty set
gives the respective topologies.$\qquad\blacksquare$
\medskip{}

Like the topological and local bases $_{\textrm{T}}\mathcal{B}$ and $\mathcal{B}_{x}$
respectively, a subclass of $\mathcal{F}$ may be used to define a filter-base
$_{\textrm{F}}\mathcal{B}$ that in turn generate $\mathcal{F}$ on $X$, just
as it is possible to define the concepts of limit and adherence sets for a filter
to parallel those for nets that follow straightforwardly from Def. A1.7, taken
with Def. A1.11. 

\medskip{}
\noindent \textbf{Definition A1.8.} \textsl{Let $(X,\mathcal{T})$ be a topological
space and $\mathcal{F}$ a filter on $X$. Then}\begin{equation}
\textrm{lim}(\mathcal{F})=\{ x\in X\!:(\forall N\in\mathcal{N}_{x})(\exists F\in\mathcal{F})(F\subseteq N)\}\label{Eqn: lim filter}\end{equation}
 and \begin{equation}
{\textstyle \textrm{adh}(\mathcal{F})=\{ x\in X\!:(\forall N\in\mathcal{N}_{x})(\forall F\in\mathcal{F})(F\bigcap N\neq\emptyset)\}}\label{Eqn: adh filter}\end{equation}

\noindent \textsl{are respectively the sets of} \textit{limit points} \textsl{and}
\textit{adherent} \textsl{}\textit{points} \textsl{of $\mathcal{F}$}%
\footnote{\label{Foot: Filter_conv}{\small The restatement \begin{equation}
\mathcal{F}\rightarrow x\Longleftrightarrow\mathcal{N}_{x}\subseteq\mathcal{F}\label{Eqn: Def: LimFilter}\end{equation}
of Eq. (\ref{Eqn: lim filter}) that follows from (F3), and sometimes taken
as the definition of convergence of a filter, is significant as it ties up the
algebraic filter with the topological neighbourhood system to produce the filter
theory of convergence in topological spaces. From the defining properties of
$\mathcal{F}$ it follows that for each $x\in X$, $\mathcal{N}_{x}$ is the
coarsest (that is smallest) filter on $X$ that converges to $x$.}%
}\textsl{.$\qquad\square$}
\medskip{}

A comparison of Eqs. (\ref{Eqn: lim net}) and (\ref{Eqn: adh net2}) with Eqs.
(\ref{Eqn: lim filter}) and (\ref{Eqn: adh filter}) respectively demonstrate
their formal similarity; this inter-relation between filters and nets will be
made precise in Definitions A1.10 and A1.11 below. It should be clear from the
preceding two equations that \begin{equation}
\textrm{lim}(\mathcal{F})\subseteq\textrm{adh}(\mathcal{F}),\label{Eqn: lim/adh(fil)}\end{equation}
 with a similar result \begin{equation}
\textrm{lim}(\chi)\subseteq\textrm{adh}(\chi)\label{Eqn: lim/adh(net)}\end{equation}
 holding for nets because of the duality between nets and filters as displayed
by Defs. A1.9 and A1.10 below, with the equality in Eqs. (\ref{Eqn: lim/adh(fil)})
and (\ref{Eqn: lim/adh(net)}) being true (but not characterizing) for ultrafilters
and ultranets respectively, see Example 4.2(3) for an account of this notion
. It should be clear from the equations of Definition A1.8 that \begin{equation}
\textrm{adh}(\mathcal{F})=\{ x\in X\!:(\exists\textrm{ a finer filter }\mathcal{G}\supseteq\mathcal{F}\textrm{ on }X)\textrm{ }(\mathcal{G}\rightarrow x)\}\label{Eqn: filter adh}\end{equation}
 consists of all the points of $X$ to which some finer filter $\mathcal{G}$
(in the sense that $\mathcal{F}\subseteq\mathcal{G}$ implies every element
of $\mathcal{F}$ is also in $\mathcal{G}$) converges in $X$; thus \[
{\textstyle \textrm{adh}(\mathcal{F})=\bigcup\lim(\mathcal{G}\!:\mathcal{G}\supseteq\mathcal{F}),}\]
which corresponds to the net-result of Theorem A1.5 below, that a net \emph{$\chi$}
adheres at \emph{$x$} iff there is some subnet of \emph{$\chi$} that converges
to \emph{$x$} in \emph{$X$}. Thus if $\zeta\preceq\chi$ is a subnet of $\chi$
and $\mathcal{F}\subseteq\mathcal{G}$ is a filter coarser than $\mathcal{G}$
then \begin{eqnarray*}
\lim(\chi)\subseteq\lim(\zeta) &  & \lim(\mathcal{F})\subseteq\lim(\mathcal{G})\\
\textrm{adh}(\zeta)\subseteq\textrm{adh}(\chi) &  & \textrm{adh}(\mathcal{G})\subseteq\textrm{adh}(\mathcal{F});\end{eqnarray*}

\noindent a filter $\mathcal{G}$ finer than a given filter $\mathcal{F}$ corresponds
to a subnet $\zeta$ of a given net $\chi$. The implication of this correspondence
should be clear from the association between nets and filters contained in Definitions
A1.10 and A1.11. 

A filter-base in $X$ is a \emph{nonempty} family $(B_{\alpha})_{\alpha\in\mathbb{D}}=\!\,_{\textrm{F}}\mathcal{B}$
of subsets of $X$ characterized by

\smallskip{}
(FB1) There are no empty sets in the collection $_{\textrm{F}}\mathcal{B}$:
$(\forall\alpha\in\mathbb{D})(B_{\alpha}\neq\emptyset)$

(FB2) The intersection of any two members of \emph{}$_{\textrm{F}}\mathcal{B}$
\emph{}contains another member of $_{\textrm{F}}\mathcal{B}$: $B_{\alpha},B_{\beta}\in\,_{\textrm{F}}\mathcal{B}\Rightarrow(\exists B\in\,_{\textrm{F}}\mathcal{B}\!:B\subseteq B_{\alpha}\bigcap B_{\beta})$;
\smallskip{}

\noindent hence any class of subsets of $X$ that does not contain the empty
set and is closed under finite intersections is a base for a unique filter on
$X$; compare the properties (NB1) and (NB2) of a local basis given at the beginning
of this Appendix. Similar to Def. A1.1 for the local base, it is possible to
define 

\medskip{}
\noindent \textbf{Definition A1.9.} \textsl{A filter-base} $_{\textrm{F}}\mathcal{B}$
\textsl{in a set $X$ is a subcollection of the filter} $\mathcal{F}$ \textsl{on
$X$ having the property that each $F\in\mathcal{F}$ contains some member of}
$_{\textrm{F}}\mathcal{B}$\textsl{.} \textsl{Thus} \begin{equation}
_{\textrm{F}}\mathcal{B}\overset{\textrm{def}}=\{ B\in\mathcal{F}\!:B\subseteq F\textrm{ for each }F\in\mathcal{F}\}\label{Eqn: FB}\end{equation}
 \textsl{determines the filter} \begin{equation}
\mathcal{F}=\{ F\subseteq X\!:B\subseteq F\textrm{ for some }B\textrm{ }\in\!\,_{\textrm{F}}\mathcal{B}\}\label{Eqn: filter_base}\end{equation}

\noindent \textsl{reciprocally as all supersets of the basic elements.$\qquad\square$}

\medskip{}
\noindent This is the smallest filter on $X$ that contains $_{\textrm{F}}\mathcal{B}$
and is said to be \emph{the filter generated by its filter-base} $_{\textrm{F}}\mathcal{B}$;
alternatively $_{\textrm{F}}\mathcal{B}$ is the filter-base of $\mathcal{F}$.
The entire neighbourhood system $\mathcal{N}_{x}$, the local base $\mathcal{B}_{x}$,
$\mathcal{N}_{x}\bigcap A$ for $x\in\textrm{Cl}(A)$, and the set of all residuals
of a directed set $\mathbb{D}$ are among the most useful examples of filter-bases
on $X$, $A$ and $\mathbb{D}$ respectively. Of course, every filter is trivially
a filter-base of itself, and \emph{the singletons $\{\{ x\}\}$, $\{ A\}$ are
filter-bases that generate the principal filters $_{\textrm{F}}\mathcal{P}(x)$
and $_{\textrm{F}}\mathcal{P}(A)$ at $x$, and $A$ respectively}. 

\smallskip{}
Paralleling the case of topological subbase $_{\textrm{T}}\mathcal{S}$, a filter
subbase $_{\textrm{F}}\mathcal{S}$ can be defined on $X$ to be any collection
of subsets of $X$ \emph{with the finite intersection property} (as compared
with $_{\textrm{T}}\mathcal{S}$ where no such condition was necessary, this
represents the fundamental point of departure between topology and filter) and
it is not difficult to deduce that the filter generated by \emph{}$_{\textrm{F}}\mathcal{S}$
on $X$ is obtained by taking all finite intersections $_{\textrm{F}}\mathcal{S}_{\wedge}$
of members of $_{\textrm{F}}\mathcal{S}$ followed by their supersets $_{\textrm{F}}\mathcal{S}_{\Sigma\wedge}$.
$\mathcal{F}(_{\textrm{F}}\mathcal{S}):=\,_{\textrm{F}}\mathcal{S}_{\Sigma\wedge}$
is the smallest filter on $X$ that contains $_{\textrm{F}}\mathcal{S}$ and
is the filter \emph{generated by} $_{\textrm{F}}\mathcal{S}$. 

Equation (\ref{Eqn: adh filter}) can be put in the more useful and transparent
form given by 

\medskip{}
\noindent \textbf{Theorem A1.3.} \textsl{For a filter $\mathcal{F}$ in a space
$(X,\mathcal{T})$} \begin{eqnarray}
{\displaystyle \textrm{adh}(\mathcal{F})} & = & {\displaystyle \bigcap_{F\in\mathcal{F}}\textrm{Cl}(F)}\label{Eqn: filter adh*}\\
 & = & {\displaystyle \bigcap_{B\in\,_{\textrm{F}}\mathcal{B}}\textrm{Cl}(B)},\nonumber \end{eqnarray}

\noindent \textsl{and dually} $\textrm{adh}(\chi)$, \textsl{are closed set}s.$\qquad\square$

\noindent \textbf{Proof.} Follows immediately from the definitions for the closure
of a set Eq. (\ref{Eqn: Def: Closure}) and the adherence of a filter Eq. (\ref{Eqn: adh filter}).
As always, it is a matter of convenience in using the basic filters \textbf{$_{\textrm{F}}\mathcal{B}$}
instead of $\mathcal{F}$ to generate the adherence set.$\qquad\blacksquare$
\medskip{}

It is infact true that the limit sets $\lim(\mathcal{F})$ and $\lim(\chi)$
are also closed set of $X$; the arguments involving ultrafilters are omitted.

Similar to the notion of the adherence set of a filter is its \emph{core ---}
a concept that unlike the adherence, is purely set-theoretic being the infimum
of the filter and is not linked with any topological structure of the underlying
(infinite) set $X$ --- defined as \begin{equation}
{\displaystyle \textrm{core }(\mathcal{F})=\bigcap_{F\in\mathcal{F}}F.}\label{Eqn: core}\end{equation}
 From Theorem A1.3 and the fact that the closure of a set $A$ is the smallest
closed set that contains $A$, see Eq. (\ref{Eqn: closure}) at the end of Tutorial4,
it is clear that in terms of filters\begin{eqnarray}
A & = & \textrm{core}(\,_{\textrm{F}}\mathcal{P}(A))\nonumber \\
\textrm{Cl}(A) & = & \textrm{adh}(\,_{\textrm{F}}\mathcal{P}(A))\label{Eqn: PrinFil_Cl(A)}\\
 & = & \textrm{core}(\textrm{Cl}(\,_{\textrm{F}}\mathcal{P}(A)))\nonumber \end{eqnarray}
 where $_{\textrm{F}}\mathcal{P}(A)$ is the principal filter at $A$; thus
\emph{the core and adherence sets of the principal filter at $A$ are equal
respectively to $A$ and} $\textrm{Cl}(A)$ \emph{---} a classic example of
equality in the general relation $\textrm{Cl}(\bigcap A_{\alpha})\subseteq\bigcap\textrm{Cl}(A_{\alpha})$
--- but both are empty, for example, in the case of an infinitely decreasing
family of rationals centered at any irrational (leading to a principal filter-base
of rationals at the chosen irrational). This is an important example demonstrating
that \emph{the infinite intersection of a non-empty family of (closed) sets
with the finite intersection property may be empty,} \emph{a situation that
cannot arise on a finite set or an infinite compact set}. Filters on $X$ with
an empty core are said to be \emph{free,} and are \emph{fixed} otherwise: notice
that by its very definition filters cannot be free on a finite set, and a free
filter represents an additional feature that may arise in passing from finite
to infinite sets. Clearly $(\textrm{adh}(\mathcal{F})=\emptyset)\Rightarrow(\textrm{core}(\mathcal{F})=\emptyset)$,
but as the important example of the rational space in the reals illustrate,
the converse need not be true. Another example of a free filter of the same
type is provided by the filter-base $\{[a,\infty)\!:a\in\mathbb{R}\}$ in $\mathbb{R}$.
Both these examples illustrate the important property that \emph{a filter is
free iff it contains the cofinite filter,} and the cofinite filter is the smallest
possible free filter on an infinite set. The free cofinite filter, as these
examples illustrate, may be typically generated as follows. Let $A$ be a subset
of $X$, $x\in\textrm{Bdy}_{X-A}(A)$, and consider the directed set Eq. (\ref{Eqn: Closure_Directed})
to generate the corresponding net in $A$ given by $\chi(N\in\mathcal{N}_{x},t)=t\in A$.
Quite clearly, the core of any Frechet filter based on this net must be empty
as the point $x$ does not lie in $A$. In general, the intersection is empty
because if it were not so then the complement of the intersection --- which
is an element of the filter --- would be infinite in contravention of the hypothesis
that the filter is Frechet. It should be clear that every filter finer than
a free filter is also free, and any filter coarser than a fixed filter is fixed. 

Nets and filters are complimentary concepts and one may switch from one to the
other as follows. 

\medskip{}
\noindent \textbf{Definition A1.10.} \textsl{Let $\mathcal{F}$ be a filter
on $X$ and let $_{\mathbb{D}}F_{x}=\{(F,x)\!:(F\in\mathcal{F})(x\in F)\}$
be a directed set with its natural direction $(F,x)\preceq(G,y)\Rightarrow(G\subseteq F)$.
The net $\chi_{\mathcal{F}}\,\!:\,_{\mathbb{D}}F_{x}\rightarrow X$ defined
by} \[
\chi_{\mathcal{F}}(F,x)=x\]
 \textsl{is said to be} \textit{associated with} \emph{the filter} \textsl{$\mathcal{F}$,
see Eq. (\ref{Eqn: DirectedNet2}).$\qquad\square$}

\medskip{}
\noindent \textbf{Definition A1.11.} \textsl{Let $\chi\!:\mathbb{D}\rightarrow X$
be a net and $\mathbb{R}_{\alpha}=\{\beta\in\mathbb{D}\!:\beta\succeq\alpha\in\mathbb{D}\}$
a residual in $\mathbb{D}$. Then} \[
_{\textrm{F}}\mathcal{B}_{\chi}\overset{\textrm{def}}=\{\chi(\mathbb{R}_{\alpha})\!:\textrm{Res}(\mathbb{D})\rightarrow X\textrm{ for all }\alpha\in\mathbb{D}\}\]
 \textsl{is the} \emph{filter-base associated with} \textsl{$\chi$, and the
corresponding filter $\mathcal{F}_{\chi}$ obtained by taking all supersets
of the elements of} $_{\textrm{F}}\mathcal{B}_{\chi}$ \textsl{is the} \emph{filter}
\textsl{}\emph{associated with} \textsl{$\chi$.$\qquad\square$}

\medskip{}
$_{\textrm{F}}\mathcal{B}_{\chi}$ is a filter-base in $X$ because $\chi(\bigcap\mathbb{R}_{\alpha})\subseteq\bigcap\chi(\mathbb{R}_{\alpha})$,
that holds for any functional relation, proves (FB2). It is not difficult to
verify that 

(i) $\chi$ is eventually in $A\Longrightarrow A\in\mathcal{F}_{\chi}$, and 

(ii) $\chi$ is frequently in $A\Longrightarrow(\forall\mathbb{R}_{\alpha}\in\textrm{Res}(\mathbb{D}))(A\bigcap\chi(\mathbb{R}_{\alpha})\neq\emptyset)$
$\Longrightarrow A\bigcap\mathcal{F}_{\chi}\neq\emptyset$ . 

\noindent Limits and adherences are obviously preserved in switching between
nets (respectively, filters) and the filters (respectively, nets) that they
generate: \begin{eqnarray}
\lim(\chi)=\lim(\mathcal{F}_{\chi}), &  & \textrm{adh}(\chi)=\textrm{adh}(\mathcal{F}_{\chi})\label{Eqn: net-fil}\\
\lim(\mathcal{F})=\lim(\chi_{\mathcal{F}}), &  & \textrm{adh}(\mathcal{F})=\textrm{adh}(\chi_{\mathcal{F}}).\label{Eqn: fil-net}\end{eqnarray}

\noindent The proofs of the two parts of Eq. (\ref{Eqn: net-fil}), for example,
go respectively as follows. $x\in\lim(\chi)\Leftrightarrow\chi\textrm{ is eventually in }\mathcal{N}_{x}\Leftrightarrow(\forall N\in\mathcal{N}_{x})(\exists F\in\mathcal{F}_{\chi})\textrm{ such that }(F\subseteq N)\Leftrightarrow x\in\lim(\mathcal{F}_{\chi})$,
and $x\in\textrm{adh}(\chi)\Leftrightarrow\chi\textrm{ is frequently in }\mathcal{N}_{x}\Leftrightarrow(\forall N\in\mathcal{N}_{x})(\forall F\in\mathcal{F}_{\chi})\textrm{ }(N\bigcap F\neq\emptyset)\Leftrightarrow x\in\textrm{adh}(\mathcal{F}_{\chi})$;
here $F$ is a superset of $\chi(\mathbb{R}_{\alpha})$.

Some examples of convergence of filters are 

\smallskip{}
(1) Any filter on an indiscrete space $X$ converges to every point of $X$. 

(2) Any filter on a space that coincides with its topology (minus the empty
set, of course) converges to every point of the space. 

(3) For each $x\in X$, the neighbourhood filter $\mathcal{N}_{x}$ converges
to $x$; this is the smallest filter on $X$ that converges to $x$. 

(4) The \emph{indiscrete} filter $\mathcal{F}=\{ X\}$ converges to no point
in the space $(X,\{\emptyset,A,X-A,X\})$, but converges to every point of $X-A$
if $X$ has the topology $\{\emptyset,A,X\}$ because the only neighbourhood
of any point in $X-A$ is $X$ which is contained in the filter. 
\smallskip{}

One of the most significant consequences of convergence theory of sequences
and nets, as shown by the two theorems and the corollary following, is that
this can be used to describe the topology of a set. The proofs of the theorems
also illustrate the close inter-relationship between nets and filters. 

\medskip{}
\noindent \textbf{Theorem A1.4.} \textsl{For a subset $A$ of a topological
space $X$,} \begin{equation}
\textrm{Cl}(A)=\{ x\in X\!:(\exists\textrm{ a net }\chi\textrm{ in }A)\textrm{ }(\chi\rightarrow x)\}.\qquad\square\label{Eqn: net closure}\end{equation}
\textbf{Proof.} \emph{Necessity.} For \emph{}$x\in\textrm{Cl}(A)$, construct
a \emph{}net \emph{}$\chi\rightarrow x$ in \emph{$A$} as \emph{}follows. Let
$\mathcal{B}_{x}$ be a topological local base at $x$, which by definition
is the collection of all open sets of $X$ containing $x$. For each $\beta\in\mathbb{D}$,
the sets \[
N_{\beta}=\bigcap_{\alpha\preceq\beta}\{ B_{\alpha}\!:B_{\alpha}\in\mathcal{B}_{x}\}\]
 form a nested decreasing local neighbourhood filter base at $x$. With respect
to the directed set $_{\mathbb{D}}N_{\beta}=\{(N_{\beta},\beta)\!:(\beta\in\mathbb{D})(x_{\beta}\in N_{\beta})\}$
of Eq. (\ref{Eqn: DirectedIndexed}), define the desired net in $A$ by \[
{\textstyle \chi(N_{\beta},\beta)=x_{\beta}\in N_{\beta}\bigcap A}\]
 where the family of nonempty decreasing subsets $N_{\beta}\bigcap A$ of $X$
constitute the filter-base in $A$ as required by the directed set $_{\mathbb{D}}N_{\beta}$.
It now follows from Eq. (\ref{Eqn: DirectionIndexed}) and the arguments in
Example A1.3(3) that $x_{\beta}\rightarrow x$; compare the directed set of
Eq. (\ref{Eqn: Closure_Directed}) for a more compact, yet essentially identical,
argument. Carefully observe the dual roles of $\mathcal{N}_{x}$ as a neighbourhood
filter base at $x$. 

\emph{Sufficiency.} Let $\chi$ be a net in $A$ that converges to $x\in X$.
For any $N_{\alpha}\in\mathcal{N}_{x}$, there is a $\mathbb{R}_{\alpha}\in\textrm{Res}(\mathbb{D})$
of Eq. (\ref{Eqn: residual}) such that $\chi(\mathbb{R}_{\alpha})\subseteq N_{\alpha}$.
Hence the point $\chi(\alpha)=x_{\alpha}$ of $A$ belongs to $N_{\alpha}$
so that $A\bigcap N_{\alpha}\neq\emptyset$ which means, from Eq. (\ref{Eqn: Def: Closure}),
that $x\in\textrm{Cl}(A)$.$\qquad\blacksquare$
\medskip{}

\noindent \textbf{Corollary.} Together with Eqs. (\ref{Eqn: Def: Closure})
and (\ref{Eqn: Def: Derived}), is follows that \begin{equation}
\textrm{Der}(A)=\{ x\in X\!:(\exists\textrm{ a net }\zeta\textrm{ in }A-\{ x\})(\zeta\rightarrow x)\}\qquad\square\label{Eqn: net derived}\end{equation}

\noindent The filter forms of Eqs. (\ref{Eqn: net closure}) and (\ref{Eqn: net derived})
\begin{eqnarray}
\textrm{Cl}(A) & = & \{ x\in X\!:(\exists\textrm{ a filter }\mathcal{F}\textrm{ on }X)(A\in\mathcal{F})(\mathcal{F}\rightarrow x)\}\label{Eqn: filter cls_der}\\
\textrm{Der}(A) & = & \{ x\in X\!:(\exists\textrm{ a filter }\mathcal{F}\textrm{ on }X)(A-\{ x\}\in\mathcal{F})(\mathcal{F}\rightarrow x)\}\nonumber \end{eqnarray}
 then follows from Eq. (\ref{Eqn: Def: LimFilter}) and the finite intersection
property (F2) of $\mathcal{F}$ so that every neighbourhood of $x$ must intersect
$A$ (respectively $A-\{ x\}$) in Eq. (\ref{Eqn: filter cls_der}) to produce
the converging net needed in the proof of Theorem A1.3. 

We end this discussion of convergence in topological spaces with a proof of
the following theorem which demonstrates the relationship that {}``eventually
in'' and {}``frequently in'' bears with each other; Eq. (\ref{Eqn: net adh})
below is the net-counterpart of the filter equation (\ref{Eqn: filter adh}). 

\medskip{}
\noindent \textbf{Theorem A1.5.} \textsl{If $\chi$ is a net in a topological
space $X$, then} $x\in\textrm{adh}(\chi)$ \textsl{iff some subnet $\zeta(\beta)=\chi(\sigma(\beta))$
of $\chi(\alpha)$, with $\alpha\in\mathbb{D}$ and $\beta\in\mathbb{E}$ ,
converges in $X$ to $x$; thus} \begin{equation}
\textrm{adh}(\chi)=\{ x\in X\!:(\exists\textrm{ a subnet }\zeta\preceq\chi\textrm{ in }X)(\zeta\rightarrow x)\}.\qquad\square\label{Eqn: net adh}\end{equation}
 \textsl{}\textbf{Proof.} \emph{Necessity.} Let $x\in\textrm{adh}(\chi)$. Define
a subnet function $\sigma\!:\,_{\mathbb{D}}N_{\alpha}\rightarrow\mathbb{D}$
by $\sigma(N_{\alpha},\alpha)=\alpha$ where $_{\mathbb{D}}N_{\alpha}$ is the
directed set of Eq. (\ref{Eqn: DirectedIndexed}): (SN1) and (SN2) are quite
evidently satisfied according to Eq. (\ref{Eqn: DirectionIndexed}). Proceeding
as in the proof of the preceding theorem it follows that $x_{\beta}=\chi(\sigma(N_{\alpha},\alpha))=\zeta(N_{\alpha},\alpha)\rightarrow x$
is the required converging subnet that exists from Eq. (\ref{Eqn: adh net1})
and the fact that $\chi(\mathbb{R}_{\alpha})\bigcap N_{\alpha}\neq\emptyset$
for every $N_{\alpha}\in\mathcal{N}_{x}$, by hypothesis. 

\emph{Sufficiency.} Assume now that $\chi$ has a subnet $\zeta(N_{\alpha},\alpha)$
that converges to $x$. If $\chi$ does not adhere at $x$, there is a neighbourhood
$N_{\alpha}$ of $x$ not frequented by it, in which case $\chi$ must be eventually
in $X-N_{\alpha}$. Then $\zeta(N_{\alpha},\alpha)$ is also eventually in $X-N_{\alpha}$
so that $\zeta$ cannot be eventually in $N_{\alpha}$, a contradiction of the
hypothesis that $\zeta(N_{\alpha},\alpha)\rightarrow x$.%
\footnote{\label{Foot: adh_seq}{\small In a first countable space, while the corresponding
proof of the first part of the theorem for sequences is essentially the same
as in the present case, the more direct proof of the converse illustrates how
the convenience of nets and directed sets may require more general arguments.
Thus if a sequence $(x_{i})_{i\in\mathbb{N}}$ has a subsequence $(x_{i_{k}})_{k\in\mathbb{N}}$
converging to $x$, then a more direct line of reasoning proceeds as follows.
Since the subsequence converges to $x$, its tail $(x_{i_{k}})_{k\geq j}$ must
be in every neighbourhood $N$ of $x$. But as the number of such terms is infinite
whereas $\{ i_{k}\!:k<j\}$ is only finite, it is necessary that for any given
$n\in\mathbb{N}$, cofinitely many elements of the sequence $(x_{i_{k}})_{i_{k}\geq n}$
be in $N$. Hence $x\in\textrm{adh}((x_{i})_{i\in\mathbb{N}})$. }%
}$\qquad\blacksquare$ 

\medskip{}
Eqs. (\ref{Eqn: net closure}) and (\ref{Eqn: net adh}) imply that the closure
of a subset $A$ of $X$ is the class of $X$-adherences of all the (sub)nets
of $X$ that are eventually in $A$. This includes both the constant nets yielding
the isolated points of $A$ and the non-constant nets leading to the cluster
points of $A$, and implies the following physically useful relationship between
convergence and topology that can be used as defining criteria for open and
closed sets having a more appealing physical significance than the original
definitions of these terms. Clearly, the term {}``net'' is justifiably used
here to include the subnets too. 

The following corollary of Theorem A1.5 summarizes the basic topological properties
of sets in terms of nets (respectively, filters). 

\medskip{}
\noindent \textbf{Corollary.} Let $A$ be a subset of a topological space $X$.
Then 

(1) $A$ is closed in $X$ iff every convergent net of $X$ that is eventually
in $A$ actually converges to a point in $A$ (respectively, iff the adhering
points of each filter-base on $A$ all belong to $A$). Thus no $X$-convergent
net in a closed subset may converge to a point outside it. 

(2) $A$ is open in $X$ iff every convergent net of $X$ that converges to
a point in $A$ is eventually in $A$. Thus no $X$-convergent net outside an
open subset may converge to a point in the set.

(3) $A$ is closed-and-open (clopen) in $X$ iff every convergent net of $X$
that converges in $A$ is eventually in $A$ and conversely.

(4) $x\in\textrm{Der}(A)$ iff some net (respectively, filter-base) in $A-\{ x\}$
converges to $x$; this clearly eliminates the isolated points of $A$ and $x\in\textrm{Cl}(A)$
iff some net (respectively, filter-base) in $A$ converges to $x$.$\qquad\square$
\medskip{}

\noindent \textbf{Remark.} The differences in these characterizations should
be fully appreciated: If we consider the cluster points $\textrm{Der}(A)$ of
a net $\chi$ in $A$ as the \emph{resource generated by} $\chi$, then a closed
subset of $X$ can be considered to be \emph{selfish} as its keeps all its resource
to itself: $\textrm{Der}(A)\cap A=\textrm{Der}(A)$. The opposite of this is
a \emph{donor} set that donates all its generated resources to its neighbour:
$\textrm{Der}(A)\cap X-A=\textrm{Der}(A)$, while for a \emph{neutral} set,
both $\textrm{Der}(A)\cap A\neq\emptyset$ and $\textrm{Der}(A)\cap X-A\neq\emptyset$
implying that the convergence resources generated in $A$ and $X-A$ can be
deposited only in the respective sets. The clopen sets (see diagram 2-2 of Fig.
\ref{Fig: DerSets}) are of some special interest as they are boundary less
so that no net-resources can be generated in this case as any such limit are
required to be simultaneously in the set and its complement.

\medskip{}
\noindent \textbf{Example A1.1, Continued.} This continuation \textbf{}of Example
A1.2 illustrates how sequential convergence is inadequate in spaces that are
not first countable like the uncountable set with cocountable topology. In this
topology, a sequence can converge to a point $x$ in the space iff it has only
a finite number of distinct terms, and is therefore eventually constant. Indeed,
let the complement \[
G\overset{\textrm{def}}=X-F,\qquad F=\{ x_{i}\!:x_{i}\neq x,\textrm{ }i\in\mathbb{N}\}\]
 of the countably closed sequential set $F$ be an open neighbourhood of $x\in X$.
Because a sequence $(x_{i})_{i\in\mathbb{N}}$ in $X$ converges to a point
$x\in X$ iff it is eventually in \emph{every} neighbourhood (including $G$)
of $x$, the sequence represented by the set $F$ cannot converge to $x$ unless
it is of the uncountable type%
\footnote{\label{Foot: seq xxx}{\small This is uncountable because interchanging any
two eventual terms of the sequence does not alter the sequence. }%
}\begin{equation}
(x_{0},x_{1},\cdots,x_{I},x_{I+1},x_{I+1},\cdots)\label{Eqn: cocount}\end{equation}
 with only a finite number $I$ of distinct terms actually belonging to the
closed sequential set $F=X-G$, and $x_{I+1}=x$. Note that as we are concerned
only with the eventual behaviour of the sequence, we may discard all distinct
terms from $G$ by considering them to be in $F$, and retain only the constant
sequence $(x,x,\cdots)$ in $G$. In comparison with the cofinite case that
was considered in Sec. 4, the entire countably infinite sequence can now lie
outside a neighbourhood of $x$ thereby enforcing the eventual constancy of
the sequence. This leads to a generalization of our earlier cofinite result
in the sense that a cocountable filter on a cocountable space converges to every
point in the space. 

It is now straightforward to verify that for a point $x_{0}$ in an uncountable
cocountable space $X$

\smallskip{}
(a) Even though no sequence in the open set $G=X-\{ x_{0}\}$ can converge to
$x_{0}$, yet $x_{0}\in\textrm{Cl}(G)$ since the intersection of any (uncountable)
open neighbourhood $U$ of $x_{0}$ with $G$, being an uncountable set, is
not empty. 

(b) By corollary (1) of Theorem A1.5, the uncountable open set $G=X-\{ x_{0}\}$
is also closed in $X$ because if any sequence $(x_{1},x_{2},\cdots)$ in $G$
converges to some $x\in X$, then $x$ must be in $G$ as the sequence must
be eventually constant in order for it to converge. But this is a contradiction
as $G$ cannot be closed since it is not countable.%
\footnote{{\small Note that $\{ x\}$ is a $1$-point set but $(x)$ is an uncountable
sequence.}%
} By the same reckoning, although $\{ x_{0}\}$ is not an open set because its
complement is not countable, nevertheless it follows from Eq. (\ref{Eqn: cocount})
that should any sequence converge to the only point $x_{0}$ of this set, then
it must eventually be in $\{ x_{0}\}$ so by corollary (2) of the same theorem,
$\{ x_{0}\}$ becomes an open set. 

(c) The identity map $\mathbf{1}\!:X\rightarrow X_{\textrm{d}}$, where $X_{\textrm{d}}$
is $X$ with discrete topology, is not continuous because the inverse image
of any singleton of $X_{\textrm{d}}$ is not open in $X$. Yet if a sequence
converges in $X$ to $x$, then its image $(\mathbf{1}(x))=(x)$ must actually
converge to $x$ in $X_{\textrm{d}}$ because a sequence converges in a discrete
space, as in the cofinite or cocountable spaces, iff it is eventually constant;
this is so because each element of a discrete space being clopen is boundaryless.
\smallskip{}

This pathological behaviour of sequences in a non Hausdorff, non first countable
space does not arise if the discrete indexing set of sequences is replaced by
a continuous, uncountable directed set like $\mathbb{R}$ for example, leading
to nets in place of sequences. In this case the net can be in an open set without
having to be constant-valued in order to converge to a point in it as the open
set can be defined as the complement of a closed countable part of the uncountable
net. The careful reader could not have failed to notice that the burden of the
above arguments, as also of that in the example following Theorem 4.6, is to
formalize the fact that since \emph{a closed set is already defined as a countable
(respectively finite) set,} the closure operation cannot add further points
to it from its complement, and any sequence that converges in an open set in
these topologies must necessarily be eventually constant at its point of convergence,
a restriction that no longer applies to a net. The cocountable topology thus
has the very interesting property of filtering out a countable part from an
uncountable set, as for example the rationals in $\mathbb{R}$.$\qquad\blacksquare$
\medskip{}

This example serves to illustrate the hard truth that in a space that is not
first countable, the simplicity of sequences is not enough to describe its topological
character, and in fact {}``sequential convergence will be able to describe
only those topologies in which the number of (basic) neighbourhoods around each
point is no greater than the number of terms in the sequences'', \citet*{Willard1970}.
It is important to appreciate the significance of this interplay of convergence
of sequences and nets (and of continuity of functions of Appendix A1) and the
topology of the underlying spaces. 

A comparison of the defining properties (T1), (T2), (T3) of topology $\mathcal{T}$
with (F1), (F2), (F3) of that of the filter $\mathcal{F}$, shows that a filter
is very close to a topology with the main difference being with regard to the
empty set which must always be in $\mathcal{T}$ but never in $\mathcal{F}$.
Addition of the empty set to a filter yields a topology, but removal of the
empty set from a topology need not produce the corresponding filter as the topology
may contain nonintersecting sets. 

\medskip{}
The distinction between the topological and filter-bases should be carefully
noted. Thus 

(a) While the topological base may contain the empty set, a filter-base cannot. 

(b) From a given topology, form a common base by dropping all basic open sets
that do not intersect. Then a (coarser) topology can be generated from this
base by taking all unions, and a filter by taking all supersets according to
Eq. (\ref{Eqn: filter_base}). For any given filter this expression may be used
to extract a subclass $_{\textrm{F}}\mathcal{B}$ as a base for $\mathcal{F}$.

\vspace{0.75cm}
\noindent \begin{center}\textbf{A2. Initial and Final topology}\end{center}

\noindent The commutative diagram of Fig. \ref{Fig: GenInv} contains four sub-diagrams
$X-X_{\textrm{B}}-f(X)$, $Y-X_{\textrm{B}}-f(X)$, $X-X_{\textrm{B}}-Y$ and
$X-f(X)-Y$. Of these, the first two are especially significant as they can
be used to conveniently define the topologies on $X_{\textrm{B}}$ and $f(X)$
from those of $X$ and $Y$, so that $f_{\textrm{B}}$, $f_{\textrm{B}}^{-1}$
and $G$ have some desirable continuity properties; we recall that a function
$f\!:X\rightarrow Y$ is continuous if inverse images of open sets of $Y$ are
open in $X$. This simple notion of continuity needs refinement in order that
topologies on $X_{\textrm{B}}$ and $f(X)$ be unambiguously defined from those
of $X$ and $Y$, a requirement that leads to the concepts of the so-called
\emph{final} and \emph{initial topologies.} To appreciate the significance of
these new constructs, note that if $f\!:(X,\mathcal{U})\rightarrow(Y,\mathcal{V})$
is a continuous function, there may be open sets in $X$ that are not inverse
images of open --- or for that matter of any --- subset of $Y$, just as it
is possible for non-open subsets of $Y$ to contribute to $\mathcal{U}$. When
the triple $\{\mathcal{U},f,\mathcal{V}\}$ are tuned in such a manner that
these are impossible, the topologies so generated on $X$ and $Y$ are the initial
and final topologies respectively; they are the smallest (coarsest) and largest
(finest) topologies on $X$ and $Y$ that make $f\!:X\rightarrow Y$ continuous.
It should be clear that every image and preimage continuous function is continuous,
but the converse is not true. 

Let $\textrm{sat}(U):=f^{-}f(U)\subseteq X$ be the saturation of an open set
$U$ of $X$ and $\textrm{comp}(V):=ff^{-}(V)=V\bigcap f(X)\in Y$ be the component
of an open set $V$ of $Y$ on the range $f(X)$ of $f$. Let $\mathcal{U}_{\textrm{sat}}$,
$\mathcal{V}_{\textrm{comp}}$ denote respectively the saturations $U_{\textrm{sat}}=\{\textrm{sat}(U)\!:U\in\mathcal{U}\}$
of the open sets of $X$ and the components $V_{\textrm{comp}}=\{\textrm{comp}(V)\!:V\in\mathcal{V}\}$
of the open sets of $Y$ whenever these are also open in $X$ and $Y$ respectively.
Plainly, $\mathcal{U}_{\textrm{sat}}\subseteq\mathcal{U}$ and $\mathcal{V}_{\textrm{comp}}\subseteq\mathcal{V}$. 

\smallskip{}
\noindent \textbf{Definition A2.1.} \textsl{For a function} $e\!:X\rightarrow(Y,\mathcal{V})$,
\textsl{the} \emph{preimage} \textsl{or} \emph{initial topology of} $X$ \emph{based
on (generated by)} \textsl{\emph{$e$}} \emph{and $\mathcal{V}$} \textsl{is}
\begin{equation}
\textrm{IT}\{ e;\mathcal{V}\}\overset{\textrm{def}}=\{ U\subseteq X\!:U=e^{-}(V)\textrm{ if }V\in\mathcal{V}_{\textrm{comp}}\},\label{Eqn: IT}\end{equation}

\noindent \textsl{while for $q\!:(X,\mathcal{U})\rightarrow Y$, the} \emph{image}
\textsl{or} \emph{final topology of} $Y$ \emph{based on (generated by) $\mathcal{U}$
and} \textsl{\emph{$q$}} \textsl{is} \begin{equation}
\textrm{FT}\{\mathcal{U};q\}\overset{\textrm{def}}=\{ V\subseteq Y\!:q^{-}(V)=U\textrm{ if }U\in\mathcal{U}_{\textrm{sat}}\}.\qquad\square\label{Eqn: FT'}\end{equation}

\noindent Thus, the topology of $(X,\textrm{IT}\{ e;\mathcal{V}\})$ consists
of, and only of, the $e$-saturations of all the open sets of $e(X)$, while
the open sets of $(Y,\textrm{FT}\{\mathcal{U};q\})$ are the $q$-images \emph{in}
$Y$ (and not just in $q(X)$) of all the $q$-saturated open sets of $X$.%
\footnote{\label{Foot: e&q}{\small We adopt the convention of denoting arbitrary preimage
and image continuous functions by $e$ and $q$ respectively even though they
are not be injective or surjective; recall that the embedding $e\!:X\supseteq A\rightarrow Y$
and the association $q\!:X\rightarrow f(X)$ are $1:1$ and onto respectively. }%
} The need for defining (\ref{Eqn: IT}) in terms of $\mathcal{V}_{\textrm{comp}}$
rather than $\mathcal{V}$ will become clear in the following. The subspace
topology $\textrm{IT}\{ i;\mathcal{U}\}$ of a subset $A\subseteq(X,\mathcal{U})$
is a basic example of the initial topology \emph{}by the inclusion map $i\!:X\supseteq A\rightarrow(X,\mathcal{U})$,
and we take its generalization $e\!:(A,\textrm{IT}\{ e;\mathcal{V}\})\rightarrow(Y,\mathcal{V})$
that embeds a subset $A$ of $X$ into $Y$ as the prototype of a preimage continuous
map. Clearly the topology of $Y$ may also contain open sets not in $e(X)$,
and any subset in $Y-e(X)$ may be added to the topology of $Y$ without altering
the preimage topology of $X$: \emph{open sets of $Y$ not in $e(X)$ may be
neglected in obtaining the preimage topology} as $e^{-}(Y-e(X))=\emptyset$.
The final topology on a quotient set by the quotient map $Q\!:(X,\mathcal{U})\rightarrow X/\sim$,
which is just the collection of $Q$-images of the $Q$-saturated open sets
of $X$, known as the \emph{quotient topology of $X/\sim$,} is the basic example
of the image topology and the resulting space $(X/\sim,\textrm{FT}\{\mathcal{U};Q\})$
is called the \emph{quotient space.} We take the generalization $q\!:(X,\mathcal{U})\rightarrow(Y,\textrm{FT}\{\mathcal{U};q\})$
of $Q$ as the prototype of a image continuous function.

The following results are specifically useful in dealing with initial and final
topologies; compare the corresponding results for open maps given later. 

\smallskip{}
\noindent \textbf{Theorem A2.1.} \textsl{Let $(X,\mathcal{U})$ and $(Y_{1},\mathcal{V}_{1})$
be topological spaces and let $X_{1}$ be a set. If $f\!:X_{1}\rightarrow(Y_{1},\mathcal{V}_{1})$,
$q\!:(X,\mathcal{U})\rightarrow X_{1}$, and $h=f\circ q\!:(X,\mathcal{U})\rightarrow(Y_{1},\mathcal{V}_{1})$
are functions with the topology $\mathcal{U}_{1}$ of $X_{1}$ given by} $\textrm{FT}\{\mathcal{U};q\}$,
\textsl{then }

(a) \textsl{$f$ is continuous iff $h$ is continuous. }

(b) \textsl{$f$ is image continuous iff} $\mathcal{V}_{1}=\textrm{FT}\{\mathcal{U};h\}$.$\qquad\square$
\smallskip{}

\vspace{-0.2cm}

\noindent \textbf{Theorem A2.2.} \textsl{Let $(Y,\mathcal{V})$ and $(X_{1},\mathcal{U}_{1})$
be topological spaces and let $Y_{1}$ be a set. If $f\!:(X_{1},\mathcal{U}_{1})\rightarrow Y_{1}$,
$e\!:Y_{1}\rightarrow(Y,\mathcal{V})$ and $g=e\circ f\!:(X_{1},\mathcal{U}_{1})\rightarrow(Y,\mathcal{V})$
are function with the topology $\mathcal{V}_{1}$ of $Y_{1}$ given by} $\textrm{IT}\{ e;\mathcal{V}\}$\textsl{,
then }

(a) \textsl{$f$ is continuous iff $g$ is continuous. }

(b) \textsl{$f$ is preimage continuous iff} $\mathcal{U}_{1}=\textrm{IT}\{ g;\mathcal{V}\}$.$\qquad\square$

\smallskip{}
\noindent As we need the second part of these theorems in our applications,
their proofs are indicated below. The special significance of the first parts
is that they ensure the converse of the usual result that the composition of
two continuous functions is continuous, namely that one of the components of
a composition is continuous whenever the composition is so. 

\smallskip{}
\noindent \textbf{Proof of Theorem A2.1.} If $f$ be image continuous, $\mathcal{V}_{1}=\{ V_{1}\subseteq Y_{1}\!:f^{-}(V_{1})\in\mathcal{U}_{1}\}$
and $\mathcal{U}_{1}=\{ U_{1}\subseteq X_{1}\!:q^{-}(U_{1})\in\mathcal{U}\}$
are the final topologies of $Y_{1}$ and $X_{1}$ based on the topologies of
$X_{1}$ and $X$ respectively. Then $\mathcal{V}_{1}=\{ V_{1}\subseteq Y_{1}\!:q^{-}f^{-}(V_{1})\in\mathcal{U}\}$
shows that $h$ is image continuous. 

Conversely, when $h$ is image continuous, $\mathcal{V}_{1}=\{ V_{1}\subseteq Y_{1}\!:h^{-}(V_{1})\}\in\mathcal{U}\}=\{ V_{1}\subseteq Y_{1}\!:q^{-}f^{-}(V_{1})\}\in\mathcal{U}\}$,
with $\mathcal{U}_{1}=\{ U_{1}\subseteq X_{1}\!:q^{-}(U_{1})\in\mathcal{U}\}$,
proves $f^{-}(V_{1})$ to be open in $X_{1}$ and thereby $f$ to be image continuous.

\smallskip{}
\noindent \textbf{Proof of Theorem A2.2.} If $f$ be preimage continuous, $\mathcal{V}_{1}=\{ V_{1}\subseteq Y_{1}\!:V_{1}=e^{-}(V)\textrm{ if }V\in\mathcal{V}\}$
and $\mathcal{U}_{1}=\{ U_{1}\subseteq X_{1}\!:U_{1}=f^{-}(V_{1})\textrm{ if }V_{1}\in\mathcal{V}_{1}\}$
are the initial topologies of $Y_{1}$ and $X_{1}$ respectively. Hence from
$\mathcal{U}_{1}=\{ U_{1}\subseteq X_{1}\!:U_{1}=f^{-}e^{-}(V)\textrm{ if }V\in\mathcal{V}\}$
it follows that $g$ is preimage continuous. 

Conversely, when $g$ is preimage continuous, $\mathcal{U}_{1}=\{ U_{1}\subseteq X_{1}\!:U_{1}=g^{-}(V)\textrm{ if }V\in\mathcal{V}\textrm{ }\}=\{ U_{1}\subseteq X_{1}\!:U_{1}=f^{-}e^{-}(V)\textrm{ if }V\in\mathcal{V}\}$
and $\mathcal{V}_{1}=\{ V_{1}\subseteq Y_{1}\!:V_{1}=e^{-}(V)\textrm{ if }V\in\mathcal{V}\}$
shows that $f$ is preimage continuous.$\qquad\blacksquare$

\smallskip{}
Since both Eqs. (\ref{Eqn: IT}) and (\ref{Eqn: FT'}) are in terms of inverse
images (the first of which constitutes a direct, and the second an inverse,
problem) the image $f(U)=\textrm{comp}(V)$ for $V\in\mathcal{V}$ is of interest
as it indicates the relationship of the openness of $f$ with its continuity.
This, and other related concepts are examined below, where the range space $f(X)$
is always taken to be a subspace of $Y$. Openness of a function \textsl{$f\!:(X,\mathcal{U})\rightarrow(Y,\mathcal{V})$}
is the {}``inverse'' of continuity, when images of open sets of $X$ are required
to be open in $Y$; such a function is said to be \emph{open.} Following are
two of the important properties of open functions.

\smallskip{}
(1) \textsl{If $f\!:(X,\mathcal{U})\rightarrow(Y,f(\mathcal{U}))$ is an open
function, then so is} $f_{<}\!:(X,\mathcal{U})\rightarrow(f(X),\textrm{IT}\{ i;f(\mathcal{U})\})$\textsl{.
The converse is true if $f(X)$ is an open set of $Y$; thus openness of} $f_{<}\!:(X,\mathcal{U})\rightarrow(f(X),f_{<}(\mathcal{U}))$
\textsl{implies tha}t \textsl{of $f\!:(X,\mathcal{U})\rightarrow(Y,\mathcal{V})$
whenever $f(X)$ is open in $Y$ such that $f_{<}(U)\in\mathcal{V}$ for $U\in\mathcal{U}$.}
The truth of this last assertion follows easily from the fact that if $f_{<}(U)$
is an open set of $f(X)\subset Y$, then necessarily $f_{<}(U)=V\bigcap f(X)$
for some $V\in\mathcal{V}$, and the intersection of two open sets of $Y$ is
again an open set of $Y$. 

(2) \textsl{If $f\!:(X,\mathcal{U})\rightarrow(Y,\mathcal{V})$ and $g\!:(Y,\mathcal{V})\rightarrow(Z,\mathcal{W})$
are open functions then $g\circ f\!:(X,\mathcal{U})\rightarrow(Z,\mathcal{W})$}
\textsl{is also open.} It follows that the condition in (1) on $f(X)$ can be
replaced by the requirement that the inclusion $i\!:(f(X),\textrm{IT}\{ i;\mathcal{V}\})\rightarrow(Y,\mathcal{V})$
be an open map. This interchange of $f(X)$ with its inclusion $i\!:f(X)\rightarrow Y$
into $Y$ is a basic result that finds application in many situations. 

\smallskip{}
Collected below are some useful properties of the initial and final topologies
that we need in this work. 

\medskip{}
\noindent \textbf{\emph{Initial Topology.}} In Fig. \ref{Fig: Initial-Final}(b),
consider $Y_{1}=h(X_{1})$, $e\rightarrow i$ and $f\rightarrow h_{<}\!:X_{1}\rightarrow(h(X_{1}),\textrm{IT}\{ i;\mathcal{V}\})$.
From $h^{-}(B)=h^{-}(B\bigcap h(X_{1}))$ for any $B\subseteq Y$, it follows
that for an open set $V$ of $Y$, $h^{-}(V_{\textrm{comp}})=h^{-}(V)$ is an
open set of $X_{1}$ which, if the topology of $X_{1}$ is $\textrm{IT}\{ h;\mathcal{V}\}$,
are the only open sets of $X_{1}$. Because $V_{\textrm{comp}}$ is an open
set of $h(X_{1})$ in its subspace topology, this implies that \emph{the preimage
topologies} $\textrm{IT}\{ h;\mathcal{V}\}$ and $\textrm{IT}\{ h_{<};\textrm{IT}\{ i;\mathcal{V}\}\}$
\emph{of $X_{1}$ generated by $h$ and} $h_{<}$ \textsl{are the same.} Thus
the preimage topology of $X_{1}$ is not affected if $Y$ is replaced by the
subspace $h(X_{1})$, the part $Y-h(X_{1})$ contributing nothing to $\textrm{IT}\{ h;\mathcal{V}\}$. 

\emph{A preimage continuous function} $e\!:X\rightarrow(Y,\mathcal{V})$ \emph{is
not necessarily an open function.} Indeed, if $U=e^{-}(V)\in\textrm{IT}\{ e;\mathcal{V}\}$,
it is almost trivial to verify along the lines of the restriction of open maps
to its range, that $e(U)=ee^{-}(V)=e(X)\bigcap V$, $V\in\mathcal{V}$, is open
in $Y$ (implying that $e$ is an open map) iff $e(X)$ is an open subset of
$Y$ (because finite intersections of open sets are open). A special case of
this is the important consequence that \textsl{the restriction} $e_{<}\!:(X,\textrm{IT}\{ e;\mathcal{V}\})\rightarrow(e(X),\textrm{IT}\{ i;\mathcal{V}\})$
\textsl{of} $e\!:(X,\textrm{IT}\{ h;\mathcal{V}\})\rightarrow(Y,\mathcal{V})$
\textsl{to its range is an open map.} Even though a preimage continuous map
need not be open, it is true that \textsl{an injective, continuous and open
map $f\!:X\rightarrow(Y,\mathcal{V})$ is preimage continuous.} Indeed, from
its injectivity and continuity, inverse images of all open subsets of $Y$ are
saturated-open in $X$, and openness of $f$ ensures that these are the only
open sets of $X$ the condition of injectivity being required to exclude non-saturated
sets from the preimage topology. It is therefore possible to rewrite Eq. (\ref{Eqn: IT})
as 

\begin{equation}
U\in\textrm{IT}\{ e;\mathcal{V}\}\Longleftrightarrow e(U)=V\textrm{ if }V\in\mathcal{V}_{\textrm{comp}},\label{Eqn: IT'}\end{equation}

\noindent and to compare it with the following criterion for an \emph{injective,
open-continuous} \emph{map} $f\!:(X,\mathcal{U})\rightarrow(Y,\mathcal{V})$
that necessarily satisfies $\textrm{sat}(A)=A$ for all $A\subseteq X$ \begin{equation}
U\in\mathcal{U}\Longleftrightarrow(\{\{ f(U)\}_{U\in\mathcal{U}}=\mathcal{V}_{\textrm{comp}})\wedge(f^{-1}(V)|_{V\in\mathcal{V}}\in\mathcal{U}).\label{Eqn: OCINJ}\end{equation}

\smallskip{}
\noindent \textbf{\emph{Final Topology.}} Since it is necessarily produced on
the range $\mathcal{R}(q)$ of $q$, the final topology is often considered
in terms of a surjection. This however is not necessary as, much in the spirit
of the initial topology, $Y-q(X)\neq\emptyset$ inherits the discrete topology
without altering anything, thereby allowing condition (\ref{Eqn: FT'}) to be
restated in the following more transparent form \begin{equation}
V\in\textrm{FT}\{\mathcal{U};q\}\Longleftrightarrow V=q(U)\textrm{ if }U\in\mathcal{U}_{\textrm{sat}},\label{Eqn: FT}\end{equation}

\noindent and to compare it with the following criterion for a \emph{surjective,
open-continuous} \emph{map} $f\!:(X,\mathcal{U})\rightarrow(Y,\mathcal{V})$
that necessarily satisfies $_{f}B=B$ for all $B\subseteq Y$ \begin{equation}
V\in\mathcal{V}\Longleftrightarrow(\mathcal{U}_{\textrm{sat}}=\{ f^{-}(V)\}_{V\in\mathcal{V}})\wedge(f(U)|_{U\in\mathcal{U}}\in\mathcal{V}).\label{Eqn: OCSUR}\end{equation}

\noindent As may be anticipated from Fig. \ref{Fig: Initial-Final}, the final
topology does not behave as well for subspaces as the initial topology does.
This is so because in Fig. \ref{Fig: Initial-Final}(a) the two image continuous
functions $h$ and $q$ are connected by a preimage continuous inclusion $f$,
whereas in Fig. \ref{Fig: Initial-Final}(b) all the three functions are preimage
continuous. Thus quite like open functions, although image continuity of $h\!:(X,\mathcal{U})\rightarrow(Y_{1},\textrm{FT}\{\mathcal{U};h\})$
implies that of $h_{<}\!:(X,\mathcal{U})\rightarrow(h(X),\textrm{IT}\{ i;\textrm{FT}\{\mathcal{U};h\}))$
for a subspace $h(X)$ of $Y_{1}$, the converse need not be true unless ---
entirely like open functions again --- either $h(X)$ is an open set of $Y_{1}$
or $i\!:(h(X),\textrm{IT}\{ i;\textrm{FT}\{\mathcal{U};h\}))\rightarrow(X,\textrm{FT}\{\mathcal{U};h\})$
is an open map. Since an open preimage continuous map is image continuous, this
makes $i\!:h(X)\rightarrow Y_{1}$ an ininal function and hence all the three
legs of the commutative diagram image continuous.

Like preimage continuity, \emph{an image continuous function $q\!:(X,\mathcal{U})\rightarrow Y$
need not be open.} However, although \emph{the restriction of an image continuous
function to the saturated open sets of its domain is an open function}, $q$
is unrestrictedly open iff the saturation of every open set of $X$ is also
open in $X$. Infact it can be verified without much effort that a continuous,
open surjection is image continuous. 

Combining Eqs. (\ref{Eqn: IT'}) and (\ref{Eqn: FT}) gives the following criterion
for ininality \begin{equation}
U\textrm{ and }V\in\textrm{IFT}\{\mathcal{U}_{\textrm{sat}};f;\mathcal{V}\}\Longleftrightarrow(\{ f(U)\}_{U\in\mathcal{U}_{\textrm{sat}}}=\mathcal{V})(\mathcal{U}_{\textrm{sat}}=\{ f^{-}(V)\}_{V\in\mathcal{V}}),\label{Eqn: INI}\end{equation}

\noindent which reduces to the following for a homeomorphism $f$ that satisfies
both $\textrm{sat}(A)=A$ for $A\subseteq X$ and $_{f}B=B$ for $B\subseteq Y$
\begin{equation}
U\textrm{ and }V\in\textrm{HOM}\{\mathcal{U};f;\mathcal{V}\}\Longleftrightarrow(\mathcal{U}=\{ f^{-1}(V)\}_{V\in\mathcal{V}})(\{ f(U)\}_{U\in\mathcal{U}}=\mathcal{V})\label{Eqn: HOM}\end{equation}

\noindent and compares with \begin{multline}
U\textrm{ and }V\in\textrm{OC}\{\mathcal{U};f;\mathcal{V}\}\Longleftrightarrow(\textrm{sat}(U)\in\mathcal{U}\!:\{ f(U)\}_{U\in\mathcal{U}}=\mathcal{V}_{\textrm{comp}})\wedge\\
\wedge(\textrm{comp}(V)\in\mathcal{V}\!:\{ f^{-}(V)\}_{V\in\mathcal{V}}=\mathcal{U}_{\textrm{sat}})\label{Eqn: OC}\end{multline}

\noindent for an open-continuous $f$.

The following is a slightly more general form of the restriction on the inclusion
that is needed for image continuity to behave well for subspaces of $Y$.

\smallskip{}
\noindent \textbf{Theorem A2.3.} \textsl{Let} $q\!:(X,\mathcal{U})\rightarrow(Y,\textrm{FT}\{\mathcal{U};q\})$
\textsl{be an image continuous} \textsl{function. For a subspace} $B$ of $(Y,\textrm{FT}\{\mathcal{U};q\})$,\[
\textrm{FT}\{\textrm{IT}\{ j;\mathcal{U}\};q_{<}\}=\textrm{IT}\{ i;\textrm{FT}\{\mathcal{U};q\}\}\]

\noindent \textsl{where} $q_{<}\!:(q^{-}(B),\textrm{IT}\{ j;\mathcal{U}\})\rightarrow(B,\textrm{FT}\{\textrm{IT}\{ j;\mathcal{U}\};q_{<}\})$,
\textsl{if either $q$ is an} \textsl{open map or $B$ is an open set of} $Y$.$\qquad\square$

In summary we have the useful result that an open preimage continuous function
is image continuous and an open image continuous function is preimage continuous,
where the second assertion follows on neglecting non-saturated open sets in
$X$; this is permitted in as far as the generation of the final topology is
concerned, as these sets produce the same images as their saturations. Hence
\emph{an image continuous function} $q\!:X\rightarrow Y$ \emph{is preimage
continuous iff every open set in $X$ is saturated with respect to $q$,} and
\emph{a preimage continuous function} $e\!:X\rightarrow Y$ \emph{is image continuous
iff the $e$-image of every open set of $X$ is open in $Y$. }

\vspace{0.75cm}
\noindent \begin{center}\textbf{A3. More on Topological Spaces}\end{center}

\noindent This Appendix --- which completes the review of those concepts of
topological spaces begun in Tutorial4 that are needed for a proper understanding
of this work --- begins with the following summary of the different possibilities
in the distribution of $\textrm{Der}(A)$ and $\textrm{Bdy}(A)$ between sets
$A\subseteq X$ and its complement $X-A$, and follows it up with a few other
important topological concepts that have been used, explicitly or otherwise,
in this work. 

\begin{figure}[htbp]
\noindent \begin{center}\input{DerSets.pstex_t}\end{center}

\begin{singlespace}

\caption{\label{Fig: DerSets}{\small Classification of a subset $A$ of $X$ relative
to the topology of $X$. The derived set of $A$ may intersect both $A$ and
$X-A$ (row 3), may be entirely in $A$ (row 2), or may be wholly in $X-A$
(row 1). $A$ is closed iff $\textrm{Bdy}(A):=\textrm{$\textrm{Bdy}_{X-A}$}(A)\cup\textrm{Bdy}_{A}(X-A)=(\textrm{Cl}(A)\cap(X-A))\cup(\textrm{Cl}(X-A)\cap A)\subseteq A$
(row 2), open iff $\textrm{Bdy}(A)\subseteq X-A$ (column 2), and clopen iff
$\textrm{Bdy}(A)=\emptyset$ when the derived sets of both $A$ and $X-A$ are
contained in the respective sets. An open set, beside being closed, may also
be neutral or donor.}}\end{singlespace}

\end{figure}

\medskip{}
\noindent \textbf{Definition A3.1.} \textbf{\textit{Separation, Connected Space}}\textbf{.}
\textsl{A} \emph{separation} \textit{}\emph{(disconnection)} \textsl{of $X$
is a pair of mutually disjoint nonempty open (and therefore closed) subsets
$H_{1}$ and $H_{2}$ such that $X=H_{1}\cup H_{2}$}\textsf{\textsl{.}} \textsl{A
space $X$ is said to be} \emph{connected} \textsl{if it has no separation,
that is if it cannot be partitioned into two open or two closed nonempty subsets.
$X$ is} \emph{separated (disconnected)} \textsl{if it is not connected.}\emph{$\qquad\square$}
\medskip{}

It follows from the definition, that for a disconnected space $X$ the following
are equivalent statements. 

\smallskip{}
(a) There exist a pair of disjoint nonempty open subsets of $X$ that cover
$X$. 

(b) There exist a pair of disjoint nonempty closed subsets of $X$ that cover
$X.$ 

(c) There exist a pair of disjoint nonempty clopen subsets of $X$ that cover
$X.$ 

(d) There exists a nonempty, proper, clopen subset of $X$. 

\smallskip{}
\noindent By a \emph{connected subset} is meant a subset of $X$ that is connected
\emph{when provided with its relative topology making it a subspace of $X$.}
Thus any connected subset of a topological space must necessarily be contained
in any clopen set that might intersect it: if $C$ and $H$ are respectively
connected and clopen subsets of $X$ such that $C\bigcap H\neq\emptyset$, then
$C\subset H$ because $C\bigcap H$ is a nonempty clopen set in $C$ which must
contain $C$ because $C$ is connected. 

For testing whether a subset of a topological space is connected, the following
relativized form of (a)$-$(d) is often useful. 

\medskip{}
\noindent \textbf{Lemma A3.1.} \textsl{A subset $A$ of $X$ is disconnected
iff there are disjoint open sets $U$ and $V$ of $X$ satisfying} \begin{equation}
{\textstyle U\bigcap A\neq\emptyset\neq V\bigcap A\textrm{ such that }A\subseteq U\bigcup V,\;\textrm{with }U\bigcap V\bigcap A=\emptyset}\label{Eqn: SubDisconnect1}\end{equation}
\textsl{or there are disjoint closed sets $E$ and $F$ of $X$ satisfying}
\begin{equation}
{\textstyle E\bigcap A\neq\emptyset\neq F\bigcap A\textrm{ such that }A\subseteq E\bigcup F,\;\textrm{with }E\bigcap F\bigcap A=\emptyset.}\label{Eqn: SybDisconnect2}\end{equation}
 \emph{Thus $A$ is disconnected iff there are disjoint clopen subsets in the
relative topology of $A$ that cover $A$.$\qquad\square$}
\medskip{}

\noindent \textbf{Lemma A3.2.} \textsl{If $A$ is a subspace of $X$, a} \textit{separation
of} \textsl{$A$ is a pair of disjoint nonempty subsets $H_{1}$ and $H_{2}$
of $A$ whose union is $A$ neither of which contains a cluster point of the
other. $A$ is connected iff there is no separation of $A.$} \emph{$\qquad\square$}

\noindent \textbf{Proof.} Let $H_{1}$ and $H_{2}$ be a separation of $A$
so that they are clopen subsets of $A$ whose union is $A$. As $H_{1}$ is
a closed subset of $A$ it follows that $H_{1}=\textrm{Cl}_{X}(H_{1})\bigcap A$,
where $\textrm{Cl}_{X}(H_{1})\bigcap A$ is the closure of $H_{1}$ in $A$;
hence $\textrm{Cl}_{X}(H_{1})\bigcap H_{2}=\emptyset$. But as the closure of
a subset is the union of the set and its adherents, an empty intersection signifies
that $H_{2}$ cannot contain any of the cluster points of $H_{1}$. A similar
argument shows that $H_{1}$ does not contain any adherent of $H_{2}$. 

Conversely suppose that neither $H_{1}$ nor $H_{2}$ contain an adherent of
the other: $\textrm{Cl}_{X}(H_{1})\bigcap H_{2}=\emptyset$ and $\textrm{Cl}_{X}(H_{2})\bigcap H_{1}=\emptyset$.
Hence $\textrm{Cl}_{X}(H_{1})\bigcap A=H_{1}$ and $\textrm{Cl}_{X}(H_{2})\bigcap A=H_{2}$
so that both $H_{1}$ and $H_{2}$ are closed in $A.$ But since $H_{1}=A-H_{2}$
and $H_{2}=A-H_{1}$, they must also be open in the relative topology of $A$.
\emph{$\qquad\blacksquare$}
\medskip{}

Following are some useful properties of connected spaces. 

\smallskip{}
(c1) \textsf{\textbf{}}The closure of any connected subspace of a space is connected.
In general, every $B$ satisfying \[
A\subseteq B\subseteq\textrm{Cl}(A)\]
 \textsf{\textsl{}}is connected. \textsf{\textsl{}}Thus any subset of $X$ formed
from $A$ by adjoining to it some or all of its adherents is connected so that
\textsl{a topological space with a dense connected subset is connected. }

(c2) \textsf{\textbf{}}The union of any class of connected subspaces of $X$
with nonempty intersection is a connected subspace of $X$. \textsf{\textsl{}}

(c3) A topological space is connected iff there is a covering of the space consisting
of connected sets with nonempty intersection. Connectedness is a topological
property: Any space homeomorphic to a connected space is itself connected. 

(c4) If $H_{1}$ and $H_{2}$ is a separation of $X$ and $A$ is any connected
subset $A$ of $X$, then either $A\subseteq H_{1}$ or $A\subseteq H_{2}$\textsl{. }
\smallskip{}

While the real line $\mathbb{R}$ is connected, a subspace of $\mathbb{R}$
is connected iff it is an interval in $\mathbb{R}$. 

The important concept of total disconnectedness introduced below needs the following 

\medskip{}
\noindent \textbf{Definition A3.2.} \textbf{\textit{Component}}\textbf{.} \textsl{A}
\textit{component $C^{*}$} \textsl{of a space $X$ is a maximally} (with respect
to inclusion) \textsl{connected subset of $X$.} \emph{$\qquad\square$}
\medskip{}

\noindent Thus a component is a connected subspace which is not properly contained
in any larger connected subspace of $X$. The maximal element need not be unique
as there can be more than one component of a given space and a {}``maximal''
criterion rather than {}``maximum'' is used as the component need not contain
every connected subsets of $X$; it simply must not be contained in any other
connected subset of $X$. Components can be constructively defined as follows:
Let $x\in X$ be any point. Consider the collection of all connected subsets
of $X$ to which $x$ belongs Since $\{ x\}$ is one such set, the collection
is nonempty. As the intersection of the collection is nonempty, its union is
a nonempty connected set $C$. This the largest connected set containing $x$
and is therefore a component containing $x$ and we have 

\smallskip{}
(C1) Let $x\in X$. The unique component of $X$ containing $x$ is the union
of all the connected subsets of $X$ that contain \emph{$x$.} Conversely \emph{}any
nonempty connected subset $A$ of $X$ \emph{}is contained in that unique component
of $X$ to which each of the points of $A$ belong\emph{.} Hence \emph{a} \textsl{topological
space is connected iff it is the unique component of itself. }

(C2) Each component $C^{*}$ of $X$ is a closed set of $X$: By property (c1)
above, $\textrm{Cl}(C^{*})$ is also connected and from $C^{*}\subseteq\textrm{Cl}(C^{*})$
it follows that $C^{*}=\textrm{Cl}(C^{*})$. Components need not be open sets
of $X$: an example of this is the space of rationals $\mathbb{Q}$ in reals
in which the components are the individual points which cannot be open in $\mathbb{R}$;
see Example (2) below. 

(C3) Components of $X$ are equivalence classes of $(X,\sim)$ with $x\sim y$
iff they are in the same component: while reflexivity and symmetry are obvious
enough, transitivity follows because if $x,y\in C_{1}$ and $y,z\in C_{2}$
with $C_{1}$, $C_{2}$ connected subsets of $X$, then $x$ and $z$ are in
the set $C_{1}\bigcup C_{2}$ which is connected by property c(2) above as they
have the point $y$ in common. Components are connected disjoint subsets of
$X$ whose union is $X$ (that is they form a partition of $X$ with each point
of $X$ contained in exactly one component of $X$) such that any connected
subset of $X$ can be contained in only one of them. Because a connected subspace
cannot contain in it any clopen subset of $X$, it follows that \emph{every
clopen connected subspace must be a component of $X$. }

Even when a space is disconnected, it is always possible to decompose it into
pairwise disjoint connected subsets. If $X$ is a discrete space this is the
only way in which $X$ may be decomposed into connected pieces. If $X$ is not
discrete, there may be other ways of doing this. For example, the space \[
X=\{ x\in\mathbb{R}\!:(0\leq x\leq1)\vee(2<x<3)\}\]

\noindent has the following three distinct decomposition into connected subsets:

\renewcommand{\arraystretch}{1.5}

\noindent \[
\begin{array}{rcl}
{\displaystyle X} & = & [0,1/2)\bigcup[1/2,1]\bigcup(2,7/3]\bigcup(7/3,3)\\
X & = & \{0\}\bigcup{\displaystyle \left(\bigcup_{n=1}^{\infty}\left(\frac{1}{n+1},\frac{1}{n}\right]\right)}\bigcup(2,3)\\
X & = & [0,1]\bigcup(2,3).\end{array}\]

\noindent Intuition tells us that only in the third of these decompositions
have we really broken up $X$ into its connected pieces. What distinguishes
the third from the other two is that neither of the pieces $[0,1]$ or $(2,3)$
can be enlarged into bigger connected subsets of $X$. 

As connected spaces, the empty set and the singleton are considered to be \emph{degenerate}
and any connected subspace with more than one point is \emph{nondegenerate.}
At the opposite extreme of the largest possible component of a space $X$ which
is $X$ itself, are the singletons $\{ x\}$ for every $x\in X$. This leads
to the extremely important notion of a 

\medskip{}
\noindent \textbf{Definition A3.3.} \textbf{\textit{Totally disconnected space.}}
\textsl{A space $X$ is} \textit{totally disconnected} \textsl{if every pair
of distinct points in it can be separated by a disconnection of $X$.}\emph{$\qquad\square$}
\medskip{}

\noindent $X$ is totally disconnected iff the components in $X$ are single
points with the only nonempty connected subsets of $X$ being the one-point
sets: If $x\neq y\in A\subseteq X$ are distinct points of a subset $A$ of
$X$ then $A=(A\bigcap H_{1})\bigcup(A\bigcap H_{2})$, where $X=H_{1}\bigcup H_{2}$
with $x\in H_{1}$ and $y\in H_{2}$ is a disconnection of $X$ (it is possible
to choose $H_{1}$ and $H_{2}$ in this manner because $X$ is assumed to be
totally disconnected), is a separation of $A$ that demonstrates that any subspace
of a totally disconnected space with more than one point is disconnected. 

A totally disconnected space has interesting physically appealing separation
properties in terms of the (separated) Hausdorff spaces; here a topological
space $X$ is \emph{Hausdorff, or $T_{2}$,} iff each two distinct points of
$X$ can be \emph{separated} by disjoint neighbourhoods, so that for every $x\neq y\in X$,
there are neighbourhoods $M\in\mathcal{N}_{x}$ and $N\in\mathcal{N}_{y}$ such
that $M\bigcap N=\emptyset$. This means that for any two distinct points $x\neq y\in X$,
it is impossible to find points that are arbitrarily close to both of them.
Among the properties of Hausdorff spaces, the following need to be mentioned. 

\smallskip{}
(H1) $X$ is Hausdorff iff for each $x\in X$ and any point $y\neq x$, there
is a neighbourhood $N$ of $x$ such that $y\not\in\textrm{Cl}(N)$. This leads
to the significant result that for any $x\in X$ the closed singleton \[
\{ x\}=\bigcap_{N\in\mathcal{N}_{x}}\textrm{Cl}(N)\]
 \emph{is the intersection of the closures of any local base at that point,}
which in the language of nets and filters (Appendix A1) means that a net in
a Hausdorff space cannot converge to more than one point in the space and the
adherent set $\textrm{adh}(\mathcal{N}_{x})$ of the neighbourhood filter at
$x$ is the singleton $\{ x\}$. 

(H2) Since each singleton is a closed set, each finite set in a Hausdorff space
is also closed in $X$. Unlike a cofinite space, however, there can clearly
be infinite closed sets in a Hausdorff space. 

(H3) Any point $x$ in a Hausdorff space $X$ is a cluster point of $A\subseteq X$
iff every neighbourhood of $x$ contains infinitely many points of $A$, a fact
that has led to our mental conditioning of the points of a (Cauchy) sequence
piling up in neighbourhoods of the limit. Thus suppose for the sake of argument
that although some neighbourhood of $x$ contains only a finite number of points,
$x$ is nonetheless a cluster point of $A$. Then there is an open neighbourhood
$U$ of $x$ such that $U\bigcap(A-\{ x\})=\{ x_{1},\cdots,x_{n}\}$ is a finite
closed set of $X$ not containing $x$, and $U\bigcap(X-\{ x_{1},\cdots,x_{n}\})$
being the intersection of two open sets, is an open neighbourhood of $x$ not
intersecting $A-\{ x\}$ implying thereby that $x\not\in\textrm{Der}(A)$; infact
$U\bigcap(X-\{ x_{1},\cdots,x_{n}\})$ is simply $\{ x\}$ if $x\in A$ or belongs
to $\textrm{Bdy}_{X-A}(A)$ when $x\in X-A$. Conversely if every neighbourhood
of a point of $X$ intersects $A$ in infinitely many points, that point must
belong to $\textrm{Der}(A)$ by definition. 

Weaker separation axioms than Hausdorffness are those of $T_{0}$, respectively
$T_{1}$, spaces in which for every pair of distinct points \emph{at least one,}
respectively \emph{each one,} has some neighbourhood not containing the other;
the following table is a listing of the separation properties of some useful
spaces. 

\renewcommand{\arraystretch}{1.25}

\begin{table}[htbp]
\noindent \begin{center}\begin{tabular}{|c||c|c|c|}
\hline 
Space&
$T_{0}$&
$T_{1}$&
$T_{2}$\tabularnewline
\hline
\hline 
Discrete&
$\checkmark$&
$\checkmark$&
$\checkmark$\tabularnewline
Indiscrete&
$\times$&
$\times$&
$\times$\tabularnewline
\hline 
$\mathbb{R}$, standard&
$\checkmark$&
$\checkmark$&
$\checkmark$\tabularnewline
left/right ray&
$\checkmark$&
$\times$&
$\times$\tabularnewline
\hline 
Infinite cofinite&
$\checkmark$&
$\checkmark$&
$\times$\tabularnewline
Uncountable cocountable&
$\checkmark$&
$\checkmark$&
$\times$\tabularnewline
\hline 
$x$-inclusion/exclusion&
$\checkmark$&
$\times$&
$\times$\tabularnewline
$A$-inclusion/exclusion&
$\times$&
$\times$&
$\times$\tabularnewline
\hline
\end{tabular}\end{center}

\caption{\label{Table: separation}{\small Separation properties of some useful spaces.}}
\end{table}

It should be noted that that as none of the properties (H1)--(H3) need neighbourhoods
of both the points simultaneously, it is sufficient for $X$ to be $T_{1}$
for the conclusions to remain valid. 

From its definition it follows that any totally disconnected space is a Hausdorff
space and is therefore both $T_{1}$ and $T_{0}$ spaces as well. However, if
a Hausdorff space has a base of clopen sets then it is totally disconnected;
this is so because if $x$ and $y$ are distinct points of $X$, then the assumed
property of $x\in H\subseteq M$ for every $M\in\mathcal{N}_{x}$ and some clopen
set $M$ yields $X=H\bigcup(X-H)$ as a disconnection of $X$ that separates
$x$ and $y\in X-H$; note that the assumed Hausdorffness of $X$ allows $M$
to be chosen so as not to contain $y$. 
\smallskip{}

\smallskip{}
\noindent \textbf{Example A3.1.} (1) Every indiscrete space is connected; every
subset of an indiscrete space is connected. Hence if $X$ is empty or a singleton,
it is connected. A discrete space is connected iff it is either empty or is
a singleton; the only connected subsets in a discrete space are the degenerate
ones. This is an extreme case of lack of connectedness, and a discrete space
is the simplest example of a total disconnected space. 

(2) $\mathbb{Q}$, the set of rationals considered as a subspace of the real
line, is (totally) disconnected because all rationals larger than a given irrational
$r$ is a clopen set in $\mathbb{Q}$, and \[
{\textstyle \mathbb{Q}=((-\infty,r)\bigcap\mathbb{Q})\bigcup(\mathbb{Q}\bigcap(r,\infty))\qquad r\textrm{ is an irrational}}\]

\noindent is the union of two disjoint clopen sets in the relative topology
of $\mathbb{Q}$. The sets \emph{$(-\infty,r)\cap\mathbb{Q}$} and $\mathbb{Q}\cap(r,\infty)$
are clopen in $\mathbb{Q}$ because neither contains a cluster point of the
other. Thus for example, any neighbourhood of the second must contain the irrational
$r$ in order to be able to cut the first which means that any neighbourhood
of a point in either of the relatively open sets cannot be wholly contained
in the other. The only connected sets of $\mathbb{Q}$ are one point subsets
consisting of the individual rationals. In fact, a connected piece of $\mathbb{Q}$,
being a connected subset of $\mathbb{R}$, is an interval in $\mathbb{R}$,
and a nonempty interval cannot be contained in $\mathbb{Q}$ unless it is a
singleton. It needs to be noted that the individual points of the rational line
are not (cl)open because any open subset of $\mathbb{R}$ that contains a rational
must also contain others different from it. This example shows that a space
need not be discrete for each of its points to be a component and thereby for
the space to be totally disconnected. 

In a similar fashion, the set of irrationals is (totally) disconnected because
all the irrationals larger than a given rational is an example of a clopen set
in $\mathbb{R}-\mathbb{Q}$. 

(3) The $p$-inclusion ($A$-inclusion) topology is connected; a subset in this
topology is connected iff it is degenerate or contains $p$. For, a subset inherits
the discrete topology if it does not contain $p$, and $p$-inclusion topology
if it contains $p$. 

(4) The cofinite (cocountable) topology on an infinite (uncountable) space is
connected; a subset in a cofinite (cocountable) space is connected iff it is
degenerate or infinite (countable). 

(5) Removal of a single point may render a connected space disconnected and
even totally disconnected. In the former case, the point removed is called a
\emph{cut point} and in the second, it is a \emph{dispersion point.} Any real
number is a cut point of $\mathbb{R}$ and it does not have any dispersion point
only. 

(6) Let $X$ be a topological space. Considering components of $X$ as equivalence
classes by the equivalence relation $\sim$ with $Q\!:X\rightarrow X/\sim$
denoting the quotient map, $X/\sim$ is totally disconnected: As $Q^{-}([x])$
is connected for each $[x]\in X/\sim$ in a component class of $X$, and as
any open or closed subset $A\subseteq X/\sim$ is connected iff $Q^{-}(A)$
is open or closed, it must follow that $A$ can only be a singleton.$\qquad\blacksquare$
\medskip{}

The next notion of compactness in topological spaces provides an insight of
the role of nonempty adherent sets of filters that lead in a natural fashion
to the concept of attractors in the dynamical systems theory that we take up
next. 

\medskip{}
\noindent \textbf{Definition A3.4.} \textbf{\textit{Compactness.}} \textsl{A
topological space $X$ is} \textit{compact} \textsl{iff every open cover of
$X$ contains a finite subcover of $X$}. \textsf{\textsl{}}\emph{$\qquad\square$}
\medskip{}

\noindent This definition of compactness has an useful equivalent contrapositive
reformulation: \textsl{For any given collection of open sets of $X$ if none
of its finite subcollections cover $X$, then the entire collection also cannot
cover $X$.} The following theorem is a statement of the fundamental property
of compact spaces in terms of adherences of filters in such spaces, the proof
of which uses this contrapositive characterization of compactness. 

\smallskip{}
\noindent \textbf{Theorem A3.1.} \textsl{A topological space $X$ is compact
iff each class of closed subsets of $X$ with finite intersection property has
nonempty intersection.} \emph{$\qquad\square$ }

\noindent \textbf{Proof.} \emph{Necessity.} Let $X$ be a compact space. Let
$\mathcal{F}=\{ F_{\alpha}\}_{\alpha\in\mathbb{D}}$ be a collection of closed
subsets of $X$ with finite FIP, and let $\mathcal{G}=\{ X-F_{\alpha}\}_{\alpha\in\mathbb{D}}$
be the corresponding open sets of $X$. If $\{ G_{i}\}_{i=1}^{N}$ is a nonempty
finite subcollection from $\mathcal{G}$, then $\{ X-G_{i}\}_{i=1}^{N}$ is
the corresponding nonempty finite subcollection of $\mathcal{F}$. Hence from
the assumed finite intersection property of $\mathcal{F}$, it must be true
that \[
\begin{array}{ccl}
{\displaystyle X-\bigcup_{i=1}^{N}G_{i}} & = & {\displaystyle \bigcap_{i=1}^{N}(X-G_{i})}\qquad(\textrm{DeMorgan}'\textrm{s Law})\\
 & \neq & \emptyset,\end{array}\]

\noindent so that no finite subcollection of $\mathcal{G}$ can cover $X$.
Compactness of $X$ now implies that $\mathcal{G}$ too cannot cover $X$ and
therefore \[
\bigcap_{\alpha}F_{\alpha}=\bigcap_{\alpha}(X-G_{\alpha})=X-\bigcup_{\alpha}G_{\alpha}\neq\emptyset.\]
 The proof of the converse is a simple exercise of reversing the arguments involving
the two equations in the proof above.$\qquad\blacksquare$
\medskip{}

Our interest in this theorem and its proof lies in the following corollary ---
\emph{which essentially means that for every filter $\mathcal{F}$ on a compact
space the adherent set} $\textrm{adh}(\mathcal{F})$ \emph{is not empty ---}
from which follows that every net in a compact space must have a convergent
subnet. 

\medskip{}
\noindent \textbf{Corollary.} \textsl{A space $X$ is compact iff for every
class $\mathcal{A}=(A_{\alpha})$ of nonempty subsets of $X$ with} FIP\textsl{,}
$\textrm{adh}(\mathcal{A})=\bigcap_{A_{\alpha}\in\mathcal{A}}\textrm{Cl}(A_{\alpha})\neq\emptyset$\textsl{.}$\qquad\square$
\medskip{}

The proof of this result for nets given by the next theorem illustrates the
general approach in such cases which is all that is basically needed in dealing
with attractors of dynamical systems; compare Theorem A1.3. 

\medskip{}
\noindent \textbf{Theorem A3.2.} \textsl{A topological space $X$ is compact
iff each net in $X$ adheres in $X$}.$\qquad\square$

\noindent \textbf{Proof.} \emph{Necessity.} Let $X$ be a compact space, $\chi\!:\mathbb{D}\rightarrow X$
a net in $X$, and $\mathbb{R}_{\alpha}$ the residual of $\alpha$ in the directed
set $\mathbb{D}$. For the filter-base $(_{\textrm{F}}\mathcal{B}_{\chi(\mathbb{R}_{\alpha})})_{\alpha\in\mathbb{D}}$
of nonempty, decreasing, nested subsets of $X$ associated with the net $\chi$,
compactness of $X$ requires from $\bigcap_{\alpha\preceq\delta}\textrm{Cl}(\chi(\mathbb{R}_{\alpha})\supseteq\chi(\mathbb{R}_{\delta})\neq\emptyset$,
that the uncountably intersecting subset \[
\textrm{adh}(_{\textrm{F}}\mathcal{B}_{\chi}):=\bigcap_{\alpha\in\mathbb{D}}\textrm{Cl}(\chi(\mathbb{R}_{\alpha}))\]

\noindent of $X$ be non-empty. If $x\in\textrm{adh}(_{\textrm{F}}\mathcal{B}_{\chi})$
then because $x$ is in the closure of $\chi(\mathbb{R}_{\beta})$, it follows
from Eq. (\ref{Eqn: Def: Closure}) that $N\bigcap\chi(\mathbb{R}_{\beta})\neq\emptyset$%
\footnote{\label{Foot: fil-nbd}{\small This is of course a triviality if we identify
each $\chi(\mathbb{R}_{\beta})$ (or $F$ in the proof of the converse that
follows) with a neighbourhood $N$ of $X$ that generates a topology on $X$.}%
} for every $N\in\mathcal{N}_{x}$, $\beta\in\mathbb{D}$. Hence $\chi(\gamma)\in N$
for some $\gamma\succeq\beta$ so that $x\in\textrm{adh}(\chi)$; see Eq. (\ref{Eqn: adh net2}). 

\emph{Sufficiency.} Let \emph{$\chi$} be a net in $X$ that adheres at $x\in X$.
From any class $\mathcal{F}$ of closed subsets of $X$ with FIP, construct
as in the proof of Thm. A1.4, a decreasing nested sequence of closed subsets
$C_{\beta}=\bigcap_{\alpha\preceq\beta\in\mathbb{D}}\{ F_{\alpha}\!:F_{\alpha}\in\mathcal{F}\}$
and consider the directed set $_{\mathbb{D}}C_{\beta}=\{(C_{\beta},\beta)\!:(\beta\in\mathbb{D})(x_{\beta}\in C_{\beta})\}$
with its natural direction (\ref{Eqn: DirectionIndexed}) to define the net
$\chi(C_{\beta},\beta)=x_{\beta}$ in $X$; see Def. A1.10. From the assumed
adherence of $\chi$ at some $x\in X$, it follows that $N\bigcap F\neq\emptyset$
for every $N\in\mathcal{N}_{x}$ and $F\in\mathcal{F}$. Hence $x$ belongs
to the closed set $F$ so that $x\in\textrm{adh}(\mathcal{F})$; see Eq. (\ref{Eqn: adh filter}).
Hence $X$ is compact.$\qquad\blacksquare$
\medskip{}

Using Theorem A1.5 that specifies a definite criterion for the adherence of
a net, this theorem reduces to the useful formulation that \textsl{a space is
compact iff each net in it has some convergent subnet.} An important application
is the following: Since every decreasing sequence $(F_{m})$ of nonempty sets
has FIP (because $\bigcap_{m=1}^{M}F_{m}=F_{M}$ for every finite $M$), \emph{every
decreasing sequence of nonempty} closed \emph{subsets} \emph{of a compact spac}e
\emph{has nonempty intersection.} For a complete metric space this is known
as the \emph{Nested Set Theorem,} and for $[0,1]$ and other compact subspaces
of $\mathbb{R}$ as the \emph{Cantor Intersection Theorem.}%
\footnote{\noindent \textbf{Nested-set theorem.} \textsl{\small If $(E_{n})$ is a decreasing
sequence of nonempty, closed, subsets of a complete metric space $(X,d)$ such
that} {\small $\lim_{n\rightarrow\infty}\textrm{dia}(E_{n})=0$}\textsl{\small ,
then there is a unique point} {\small \[
x\in\bigcap_{n=0}^{\infty}E_{n}.\]
 The uniqueness arises because the limiting condition on the diameters of $E_{n}$
imply, from property (H1), that $(X,d)$ is a Hausdorff space. }%
}

For subspaces $A$ of $X$, it is the relative topology that determines as usual
compactness of $A$; however the following criterion renders this test in terms
of the relative topology unnecessary and shows that the topology of $X$ itself
is sufficient to determine compactness of subspaces: \textsl{A subspace $K$
of a topological space $X$ is compact iff each open cover of $K$ in $X$ contains
a finite cover of $K$. }

A proper understanding of the distinction between compactness and closedness
of subspaces --- which often causes much confusion to the non-specialist ---
is expressed in the next two theorems. As a motivation for the first that establishes
that not every subset of a compact space need be compact, mention may be made
of the subset $(a,b)$ of the compact closed interval $[a,b]$ in $\mathbb{R}$. 

\medskip{}
\noindent \textbf{Theorem A3.3.} \textsl{A closed subset $F$ of a compact space
$X$ is compact.} \emph{$\qquad\square$}

\noindent \textbf{Proof.} Let $\mathcal{G}$ be an open cover of $F$ so that
an open cover of $X$ is $\mathcal{G}\bigcup(X-F)$, which because of compactness
of $X$ contains a finite subcover $\mathcal{U}$. Then $\mathcal{U}-(X-F)$
is a finite collection of $\mathcal{G}$ that covers $F$.\emph{$\qquad\blacksquare$}

\medskip{}
It is not true in general that a compact subset of a space is necessarily closed.
For example, in an infinite set $X$ with the cofinite topology, let $F$ be
an infinite subset of $X$ with $X-F$ also infinite. Then although $F$ is
not closed in $X$, it is nevertheless compact because $X$ is compact. Indeed,
let $\mathcal{G}$ be an open cover of $X$ and choose any nonempty $G_{0}\in\mathcal{G}$.
If $G_{0}=X$ then $\{ G_{0}\}$ is the required finite cover of $X$. If this
is not the case, then because $X-G_{0}=\{ x_{i}\}_{i=1}^{n}$ is a finite set,
there is a $G_{i}\in\mathcal{G}$ with $x_{i}\in G_{i}$ for each $1\leq i\leq n$,
and therefore $\{ G_{i}\}_{i=0}^{n}$ is the finite cover that demonstrates
the compactness of the cofinite space $X$. Compactness of $F$ now follows
because the subspace topology on $F$ is the induced cofinite topology from
$X$. The distinguishing feature of this topology is that it, like the cocountable,
is not Hausdorff: If $U$ and $V$ are any two nonempty open sets of $X$, then
they cannot be disjoint as the complements of the open sets can only be finite
and if $U\bigcap V$ were to be indeed empty, then \[
{\textstyle X=X-\emptyset={X-(U\bigcap V)=(X-U)\bigcup(X-V)}}\]

\noindent would be a finite set. An immediate fallout of this is that in an
infinite cofinite space, a sequence $(x_{i})_{i\in\mathbb{N}}$ (and even a
net) with $x_{i}\neq x_{j}$ for $i\neq j$ behaves in an extremely unusual
way: \emph{It converges,} as in the indiscrete space, \emph{to} \emph{every
point of the space.} Indeed if $x\in X$, where $X$ is an infinite set provided
with its cofinite topology, and $U$ is any neighbourhood of $x$, any infinite
sequence $(x_{i})_{i\in\mathbb{N}}$ in $X$ must be eventually in $U$ because
$X-U$ is finite, and ignoring of the initial set of its values lying in $X-U$
in no way alters the ultimate behaviour of the sequence (note that this implies
that the filter induced on $X$ by the sequence agrees with its topology). Thus
$x_{i}\rightarrow x$ for any $x\in X$ is a reflection of the fact that there
are no small neighbourhoods of any point of $X$ with every neighbourhood being
almost the whole of $X$, except for a null set consisting of only a finite
number of points. This is in sharp contrast with Hausdorff spaces where, although
every finite set is also closed, every point has arbitrarily small neighbourhoods
that lead to unique limits of sequences. A corresponding result for cocountable
spaces can be found in Example A1.2 Continued. 

This example of the cofinite topology motivates the following {}``converse''
of the previous theorem.

\medskip{}
\noindent \textbf{Theorem A3.4.} \textsl{Every compact subspace of a Hausdorff
space is closed.}\emph{$\qquad\square$}

\noindent \begin{center}%
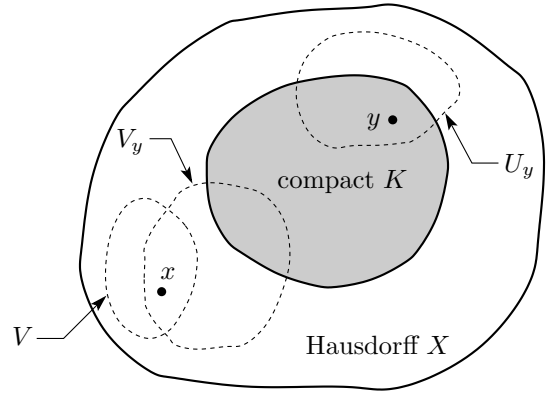
\begin{figure}[htbp]
\noindent \begin{center}\input{cmpct_clsd.pstex_t}\end{center}

\caption{\label{Fig: cmpct_clsd}{\small Closedness of compact subsets of a Hausdorff
space. }}
\end{figure}
\end{center}

\noindent \textbf{Proof.} Let $K$ be a nonempty compact subset of $X$, Fig.
\ref{Fig: cmpct_clsd}, and let $x\in X-K$. Because of the separation of $X$,
for every $y\in K$ there are disjoint open subsets $U_{y}$ and $V_{y}$ of
$X$ with $y\in U_{y}$, and $x\in V_{y}$. Hence $\{ U_{y}\}_{y\in K}$ is
an open cover for $K$, and from its compactness there is a finite subset $A$
of $K$ such that $K\subseteq\bigcup_{y\in A}U_{y}$ with $V=\bigcap_{y\in A}V_{y}$
an open neighbourhood of $x$; $V$ is open because each $V_{y}$ is a neighbourhood
of $x$ and the intersection is over finitely many points $y$ of $A$. To prove
that $K$ is closed in $X$ it is enough to show that $V$ is disjoint from
$K$: If there is indeed some $z\in V\bigcap K$ then $z$ must be in some $U_{y}$
for $y\in A$. But as $z\in V$ it is also in $V_{y}$ which is impossible as
$U_{y}$ and $V_{y}$ are to be disjoint. \emph{}This last part of the argument
infact shows that \textsl{if $K$ is a compact subspace of a Hausdorff space
$X$ and $x\notin K$, then there are disjoint open sets $U$ and $V$ of $X$
containing $x$ and $K$.}\emph{$\qquad\blacksquare$} 
\medskip{}

The last two theorems may be combined to give the obviously important

\medskip{}
\noindent \textbf{Corollary.} \textsl{In a compact Hausdorff space, closedness
and compactness of its subsets are equivalent concepts.}\emph{$\qquad\square$}
\medskip{}

In the absence of Hausdorffness, it is not possible to conclude from the assumed
compactness of the space that every point to which the net may converge actually
belongs to the subspace. 

\medskip{}
\noindent \textbf{Definition A3.5.} \textsl{A subset $D$ of a topological space}
\emph{$(X,\mathcal{U})$} \textsl{is} \emph{dense in $X$ if} $\textrm{Cl}(D)=X$\emph{.
Thus the closure of $D$ is the largest open subset of $X$, and every neighbourhood
of any point of $X$ contains a point of $D$ not necessarily distinct from
it; refer to the distinction between Eqs. (\ref{Eqn: Def: Closure}) and (\ref{Eqn: Def: Derived}).$\qquad\square$}
\medskip{}

Loosely, $D$ is dense in $X$ iff every point of $X$ has points of $D$ arbitrarily
close to it. A \emph{self-dense} (\emph{dense in itself}) set is a set without
any isolated points; hence $A$ is self-dense iff $A\subseteq\textrm{Der}(A)$.
A closed self-dense set is called a \emph{perfect set} so that a closed set
$A$ is perfect iff it has no isolated points. Accordingly \emph{}\[
A\textrm{ is perfect}\Longleftrightarrow A=\textrm{Der}(A),\]

\noindent means that the closure of a set without any isolated points is a perfect
set. 

\medskip{}
\noindent \textbf{Theorem A3.5.} \textsl{The following are equivalent statements. }

(1) \textsl{$D$ is dense in $X$}.

(2) \textsl{If $F$ is any closed set of $X$ with $D\subseteq F$, then $F=X$};
\textsl{thus the only closed superset of $D$ is $X$. }

(3) \textsl{Every nonempty (basic) open set of $X$ cuts $D;$ thus the only
open set disjoint from $D$ is the empty set $\emptyset$. }

(4) \textsl{The exterior of $D$ is empty.$\qquad\square$} 

\noindent \textbf{Proof.} (3) If $U$ indeed is a nonempty open set of $X$
with $U\bigcap D=\emptyset$, then $D\subseteq X-U\neq X$ leads to the contradiction
$X=\textrm{Cl}(D)\subseteq\textrm{Cl}(X-U)=X-U\neq X$, which also incidentally
proves (2). From (3) it follows that for any open set $U$ of $X$, $\textrm{Cl}(U)=\textrm{Cl}(U\bigcap D)$
because if $V$ is any open neighbourhood of $x\in\textrm{Cl}(U)$ then $V\bigcap U$
is a nonempty open set of $X$ that must cut $D$ so that $V\bigcap(U\bigcap D)\neq\emptyset$
implies $x\in\textrm{Cl}(U\bigcap D)$. Finally, $\textrm{Cl}(U\bigcap D)\subseteq\textrm{Cl}(U)$
completes the proof.$\qquad\blacksquare$

\medskip{}
\noindent \textbf{Definition A3.6.} (a) \textsl{A set $A\subseteq X$ is said
to be} \textit{nowhere dense} \textsl{in} \textsl{\emph{}}\textsl{$X$ if} $\textrm{Int}(\textrm{Cl}(A))=\emptyset$
\textsl{and} \textit{residual} \textsl{in} \textsl{\emph{}}\textsl{$X$ if}
$\textrm{Int}(A)=\emptyset$\textsl{.$\qquad\square$}
\medskip{}

$A$ is nowhere dense in $X$ iff \[
\textrm{Bdy}(X-\textrm{Cl}(A))=\textrm{Bdy}(\textrm{Cl}(A))=\textrm{Cl}(A)\]

\noindent so that \[
{\textstyle \textrm{Cl}(X-\textrm{Cl}(A))={(X-\textrm{Cl}(A))\bigcup\textrm{Cl}(A)=X}}\]

\noindent from which it follows that \[
A\textrm{ is nwd in }X\Longleftrightarrow X-\textrm{Cl}(A)\textrm{ is dense in }X\]

\noindent and \[
A\textrm{ is residual in }X\Longleftrightarrow X-A\textrm{ is dense in }X.\]

Thus $A$ is nowhere dense iff $\textrm{Ext}(A):=X-\textrm{Cl}(A)$ \textsl{}is
dense in \textsl{$X$,} and in particular a closed set is nowhere dense in $X$
iff its complement is open dense in $X$ with open-denseness being complimentarily
dual to closed-nowhere denseness. The rationals in reals is an example of a
set that is residual but not nowhere dense. The following are readily verifiable
properties of subsets of $X$.

\smallskip{}
(1) A set $A\subseteq X$ is nowhere dense in $X$ iff it is contained in its
own boundary, iff it is contained in the closure of the complement of its closure,
that is $A\subseteq\textrm{Cl}(X-\textrm{Cl}(A))$. In particular a closed subset
$A$ is nowhere dense in $X$ iff $A=\textrm{Bdy}(A)$, that is iff it contains
no open set. 

(2) From $M\subseteq N\Rightarrow\textrm{Cl}(M)\subseteq\textrm{Cl}(N)$ it
follows, with $M=X-\textrm{Cl}(A)$ and $N=X-A$, that a nowhere dense set is
residual, but a residual set need not be nowhere dense unless it is also closed
in $X$.

(3) Since $\textrm{Cl}(\textrm{Cl}(A))=\textrm{Cl}(A)$, $\textrm{Cl}(A)$ is
nowhere dense in $X$ iff $A$ is. 

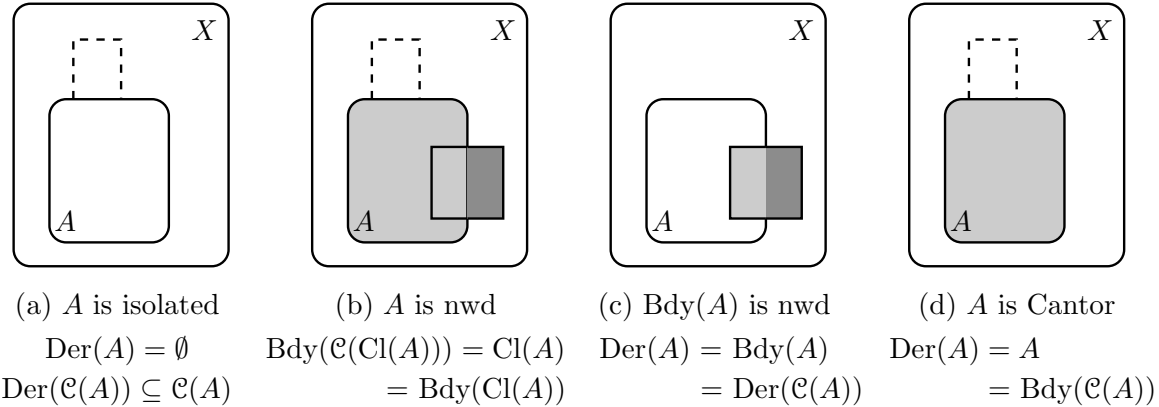
\begin{figure}[htbp]
\noindent \begin{center}\input{DerSets1.pstex_t}\end{center}

\begin{singlespace}

\caption{\label{Fig: DerSets1}{\small Shows the distinction between isolated, nowhere
dense and Cantor sets. Topologically, the Cantor set can be described as a perfect,
nowhere dense, totally disconnected and compact subset of a space. In (b), the
closed nowhere dense set $\textrm{Cl}(A)$ is the boundary of its open complement.
Here downward and upward inclined hatching denote respectively $\textrm{Bdy}_{A}(X-A)$
and $\textrm{Bdy}_{X-A}(A)$.}}\end{singlespace}

\end{figure}

(4) For any $A\subseteq X$, both $\textrm{Bdy}_{A}(X-A):=\textrm{Cl}(X-A)\bigcap A$
and $\textrm{Bdy}_{X-A}(A):=\textrm{Cl}(A)\bigcap(X-A)$ are residual sets and
as Fig. \ref{Fig: DerSets} shows \[
\textrm{Bdy}_{X}(A)=\textrm{Bdy}_{X-A}(A)\bigcup\textrm{Bdy}_{A}(X-A)\]
 is the union of these two residual sets. When $A$ is closed (or open) in $X$,
its boundary consisting of the only component $\textrm{Bdy}_{A}(X-A)$ (or $\textrm{Bdy}_{X-A}(A)$)
as shown by the second row (or column) of the figure, being a closed set of
$X$ is also nowhere dense in $X$; infact \emph{a closed nowhere dense set
is always the boundary of some open set.} Otherwise, the boundary components
of the two residual parts --- as in the donor-donor, donor-neutral, neutral-donor
and neutral-neutral cases --- need not be individually closed in $X$ (although
their union is) and their union is a residual set that need not be nowhere dense
in $X$: the union of two nowhere dense sets is nowhere dense but the union
of a residual and a nowhere dense set is a residual set. One way in which a
two-component boundary can be nowhere dense is by having $\textrm{Bdy}_{A}(X-A)\supseteq\textrm{Der}(A)$
or $\textrm{Bdy}_{X-A}(A)\supseteq\textrm{Der}(X-A)$, so that it is effectively
in one piece rather than in two, as shown in Fig. \ref{Fig: DerSets1}(b). 

\noindent \textbf{Theorem A3.6.} \textsl{$A$ is nowhere dense in $X$ iff each
non-empty open set of $X$ has a non-empty open subset disjoint from} Cl\textsf{\textsl{$(A)$.}}
\emph{$\qquad\square$}

\noindent \textbf{Proof.} If $U$ is a nonempty open set of $X$, then $U_{0}=U\cap\textrm{Ext}(A)\neq\emptyset$
as $\textrm{Ext}(A)$ is dense in $X$; $U_{0}$ is the open subset that is
disjoint from $\textrm{Cl}(A)$. It clearly follows from this that each non-empty
open set of $X$ has a non-empty open subset disjoint from a nowhere dense set
$A$.$\qquad\blacksquare$

\medskip{}
What this result (which follows just from the definition of nowhere dense sets)
actually means is that no point in $\textrm{Bdy}_{X-A}(A)$ can be isolated
in it. 

\medskip{}
\noindent \textbf{Corollary.} $A$ \textsl{is nowhere dense in $X$ iff} Cl$(A)$
\textsf{\textsl{}}\textsl{does not contain any nonempty open set of $X$} \textsf{\textsl{iff}}
\textsl{any nonempty open set that contains $A$ also contains its closure.}
\emph{$\qquad\square$}

\medskip{}
\noindent \textbf{Example A3.2.} Each finite subset of $\mathbb{R}^{n}$ is
nowhere dense in $\mathbb{R}^{n}$; the set $\{1/n\}_{n=1}^{\infty}$ is nowhere
dense in $\mathbb{R}$. The Cantor set $\mathcal{C}$ is nowhere dense in $[0,1]$
because every neighbourhood of any point in $\mathcal{C}$ must contain, by
its very construction, a point with $1$ in its ternary representation. That
the interior and the interior of the closure of a set are not necessarily the
same is seen in the example of the rationals in reals: The set of rational numbers
$\mathbb{Q}$ has empty interior because any neighbourhood of a rational number
contains irrational numbers (so also is the case for irrational numbers) and
$\mathbb{R}=\textrm{Int}(\textrm{Cl}(\mathbb{Q}))\supseteq\textrm{Int}(\mathbb{Q})=\emptyset$
justifies the notion of a nowhere dense set.$\qquad\blacksquare$
\medskip{}

The following properties of $\mathcal{C}$ can be taken to define any subset
of a topological space as a Cantor set; set-theoretically it should be clear
from its classical middle-third construction that the Cantor set consists of
all points of the closed interval $[0,1]$ whose infinite triadic (base 3) representation,
expressed so as not to terminate with an infinite string of $1$'s, does not
contain the digit $1$. Accordingly, any end-point of the infinite set of closed
intervals whose intersection yields the Cantor set, is represented by a repeating
string of either $0$ or $2$ while a non end-point has every other arbitrary
collection of these two digits. Recalling that any number in $[0,1]$ is a rational
iff its representation in any base is terminating or recurring --- thus any
decimal that neither repeats or terminates but consists of all possible sequences
of all possible digits represents an irrational number --- it follows that both
rationals and irrationals belong to the Cantor set.

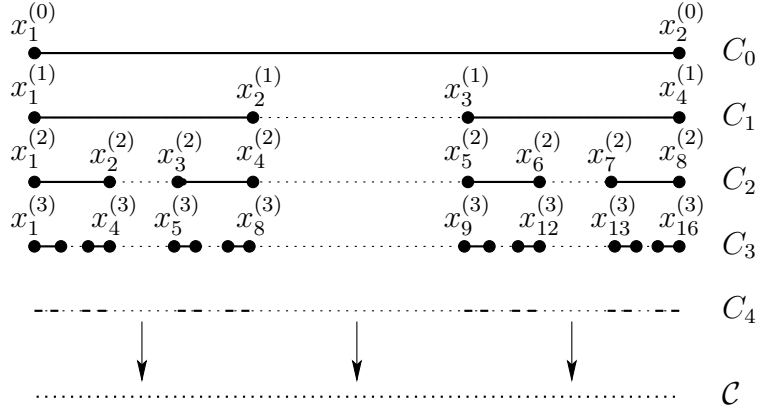
\begin{figure}[htbp]
\noindent \begin{center}\input{Cantor.pstex_t}\end{center}

\begin{singlespace}

\caption{{\small \label{Fig: Cantor}Construction of the classical $1/3$-Cantor set.
The endpoints of $C_{3}$, for example, in increasing order are: $\left|0,\frac{1}{27}\right|;\left|\frac{2}{27},\frac{1}{9}\right|;\left|\frac{2}{9},\frac{7}{27}\right|;\left|\frac{8}{27},\frac{1}{3}\right|;\left|\frac{2}{3},\frac{19}{27}\right|;\left|\frac{20}{27},\frac{7}{9}\right|;\left|\frac{8}{9},\frac{25}{27}\right|;\left|\frac{26}{27},1\right|$.
$C_{i}$ is the union of $2^{i}$ pairwise disjoint closed intervals each of
length $3^{-i}$ and the nonempty infinite intersection $\mathcal{C}=\cap_{i=0}^{\infty}C_{i}$
is the adherent Cantor set of the filter-base of closed sets $\{ C_{0},C_{1},C_{2},\cdots\}$. }}\end{singlespace}

\end{figure}

($\mathcal{C}1$) \textbf{\textit{$\mathcal{C}$ is}} \textbf{\textit{\emph{}}}\textbf{\textit{totally
disconnected.}} If possible, let $\mathcal{C}$ have a component containing
points $a$ and $b$ with $a<b$. Then $[a,b]\subseteq\mathcal{C}\Rightarrow[a,b]\subseteq C_{i}$
for all $i$. But this is impossible because we may choose $i$ large enough
to have $3^{-i}<b-a$ so that $a$ and $b$ must belong to two different members
of the pairwise disjoint closed $2^{i}$ subintervals each of length $3^{-i}$
that constitutes $C_{i}$. Hence \[
[a,b]\textrm{ is not a subset of any }C_{i}\Longrightarrow[a,b]\textrm{ is not a subset of }\mathcal{C}.\]

($\mathcal{C}2$) \textbf{\textit{$\mathcal{C}$ is perfect}} so that for any
$x\in\mathcal{C}$ every neighbourhood of $x$ must contain some other point
of $\mathcal{C}$. Supposing to the contrary that the singleton $\{ x\}$ is
an open set of $\mathcal{C}$, there must be an $\varepsilon>0$ such that in
the usual topology of $\mathbb{R}$\begin{equation}
{\textstyle \{ x\}=\mathcal{C}\bigcap(x-\varepsilon,x+\varepsilon).}\label{Eqn: Cantor_Perfect}\end{equation}

\noindent Choose a positive integer $i$ large enough to satisfy $3^{-i}<\varepsilon$.
Since $x$ is in every $C_{i}$, it must be in one of the $2^{i}$ pairwise
disjoint closed intervals $[a,b]\subset(x-\varepsilon,x+\varepsilon)$ each
of length $3^{-i}$ whose union is $C_{i}$. As $[a,b]$ is an interval, at
least one of the endpoints of $[a,b]$ is different from $x$, and since an
endpoint belongs to $\mathcal{C}$, $\mathcal{C}\cap(x-\varepsilon,x+\varepsilon)$
must also contain this point thereby violating Eq. (\ref{Eqn: Cantor_Perfect}). 

($\mathcal{C}3$) \textbf{\textit{$\mathcal{C}$ is nowhere dense}} because
each neighbourhood of any point of $\mathcal{C}$ intersects $\textrm{Ext}(\mathcal{C})$;
see Thm. A3.6. 

($\mathcal{C}4$) \textbf{\textit{$\mathcal{C}$ is compact}} because it is
a closed subset contained in the compact subspace $[0,1]$ of $\mathbb{R}$,
see Thm. A3.3. The compactness of $[0,1]$ follows from the Heine-Borel Theorem
which states that any subset of the real line is compact iff it is both closed
and bounded with respect to the Euclidean metric on $\mathbb{R}$. 
\smallskip{}

Compare ($\mathcal{C}1$) and ($\mathcal{C}2$) with the essentially similar
arguments of Example A3.1(2) for the subspace of rationals in $\mathbb{R}$. 

\vspace{0.75cm}
\noindent \begin{center}\textbf{A4. Neutron Transport Theory}\end{center}

\noindent This section introduces the reader to the basics of the \emph{linear}
neutron transport theory where graphical convergence approximations to the singular
distributions, interpreted here as multifunctions, led to the present study
of this work. The one-speed (that is mono-energetic) neutron transport equation
in one dimension and plane geometry, is \begin{equation}
\mu\frac{\partial\Phi(x,\mu)}{\partial x}+\Phi(x,\mu)=\frac{c}{2}\int_{-1}^{1}\Phi(x,\mu^{\prime})d\mu^{\prime},\,0<c<1,\,-1\leq\mu\leq1\label{Eqn: NeutronTransport}\end{equation}

\noindent where $x$ is a non-dimensional physical space variable that denotes
the location of the neutron moving in a direction $\theta=\cos^{-1}(\mu)$,
$\Phi(x,\mu)$ is a neutron density distribution function such that $\Phi(x,\mu)dxd\mu$
is the expected number of neutrons in a distance $dx$ about the point $x$
moving at constant speed with their direction cosines of motion in $d\mu$ about
$\mu$, and $c$ is a physical constant that will be taken to satisfy the restriction
shown above. Case's method starts by assuming the solution to be of the form
$\Phi_{\nu}(x,\mu)=e^{-x/\mu}\phi(\mu,\nu)$ with a normalization integral constraint
of $\int_{-1}^{1}\phi(\mu,\nu)d\mu=1$ to lead to the simple equation \begin{equation}
(\nu-\mu)\phi(\mu,\nu)=\frac{c\nu}{2}\label{Eqn: case_eigen}\end{equation}
 for the unknown function $\phi(\nu,\mu)$. Case then suggested, see \citet*{Case1967},
the non-simple complete solution of this equation to be \begin{equation}
\phi(\mu,\nu)=\frac{c\nu}{2}\mathcal{P}\frac{1}{\nu-\mu}+\lambda(v)\delta(v-\mu),\label{Eqn: singular_eigen}\end{equation}

\noindent where $\lambda(\nu)$ is the usual combination coefficient of the
solutions of the homogeneous and non-homogeneous parts of a linear equation,
$\mathcal{P}(\cdot)$ is a principal value and $\delta(x)$ the Dirac delta,
to lead to the full-range $-1\leq\mu\leq1$ solution valid for $-\infty<x<\infty$
\begin{equation}
\Phi(x,\mu)=a(\nu_{0})e^{-x/\nu_{0}}\phi(\mu,\nu_{0})+a(-\nu_{0})e^{x/\nu_{0}}\phi(-\nu_{0},\mu)+\int_{-1}^{1}a(\nu)e^{-x/\nu}\phi(\mu,\nu)d\nu\label{Eqn: CaseSolution_FR}\end{equation}

\noindent of the one-speed neutron transport equation (\ref{Eqn: NeutronTransport}).
Here the real $\nu_{0}$ and $\nu$ satisfy respectively the integral constraints
\[
\frac{c\nu_{0}}{2}\ln\frac{\nu_{0}+1}{\nu_{0}-1}=1,\qquad\mid\nu_{0}\mid>1\]
 \[
\lambda(\nu)=1-\frac{c\nu}{2}\ln\frac{1+\nu}{1-\nu},\qquad\nu\in[-1,1],\]
 with \[
\phi(\mu,\nu_{0})=\frac{c\nu_{0}}{2}\frac{1}{\nu_{0}-\mu}\]

\noindent following from Eq. (\ref{Eqn: singular_eigen}).

It can be shown \cite{Case1967} that the eigenfunctions \textbf{$\phi(\nu,\mu)$}
satisfy the full-range orthogonality condition \[
\int_{-1}^{1}\mu\phi(\nu,\mu)\phi(\nu^{\prime},\mu)d\mu=N(\nu)\delta(\nu-\nu^{\prime}),\]

\noindent where the odd normalization constants $N$ are given by

\noindent \renewcommand{\arraystretch}{2}\[
\begin{array}{ccl}
{\displaystyle N(\pm\nu_{0})} & = & {\displaystyle \int_{-1}^{1}\mu\phi^{2}(\pm\nu_{0},\mu)d\mu}\qquad\textrm{for }\mid\nu_{0}\mid>1\\
 & = & {\displaystyle \pm\frac{c\nu_{0}^{3}}{2}\left(\frac{c}{\nu_{0}^{2}-1}-\frac{1}{\nu_{0}^{2}}\right)},\end{array}\]

\noindent and\[
N(\nu)=\nu\left(\lambda^{2}(\nu)+\left(\frac{\pi c\nu}{2}\right)^{2}\right)\qquad\textrm{for }\nu\in[-1,1].\]

\noindent With a source of particles $\psi(x_{0},\mu)$ located at $x=x_{0}$
in an infinite medium, Eq. (\ref{Eqn: CaseSolution_FR}) reduces to the boundary
condition, with $\mu,\textrm{ }\nu\in[-1,1]$, \begin{equation}
\psi(x_{0},\mu)=a(\nu_{0})e^{-x_{0}/\nu_{0}}\phi(\mu,\nu_{0})+a(-\nu_{0})e^{x_{0}/\nu_{0}}\phi(-\nu_{0},\mu)+\int_{-1}^{1}a(\nu)e^{-x_{0}/\nu}\phi(\mu,\nu)d\nu\label{Eqn: BC_FR}\end{equation}

\noindent for the determination of the expansion coefficients $a(\pm\nu_{0}),\textrm{ }\{ a(\nu)\}_{\nu\in[-1,1]}$.
Use of the above orthogonality integrals then lead to the complete solution
of the problem to be \[
a(\nu)=\frac{e^{x_{0}/\nu}}{N(\nu)}\int_{-1}^{1}\mu\psi(x_{0},\mu)\phi(\mu,\nu)d\mu,\qquad\nu=\pm\nu_{0}\textrm{ or }\nu\in[-1,1].\]

\noindent For example, in the infinite-medium Greens function problem with $x_{0}=0$
and $\psi(x_{0},\mu)=\delta(\mu-\mu_{0})/\mu$, the coefficients are $a(\pm\nu_{0})=\phi(\mu_{0},\pm\nu_{0})/N(\pm\nu_{0})$
when $\nu=\pm\nu_{0}$, and $a(\nu)=\phi(\mu_{0},\nu)/N(\nu)$ for $\nu\in[-1,1]$. 

For a half-space $0\leq x<\infty$, the obvious reduction of Eq. (\ref{Eqn: CaseSolution_FR})
to 

\noindent \begin{equation}
\Phi(x,\mu)=a(\nu_{0})e^{-x/\nu_{0}}\phi(\mu,\nu_{0})+\int_{0}^{1}a(\nu)e^{-x/\nu}\phi(\mu,\nu)d\nu\label{Eqn: CaseSolution_HR}\end{equation}

\noindent with boundary condition, $\mu,\textrm{ }\nu\in[0,1]$, \begin{equation}
\psi(x_{0},\mu)=a(\nu_{0})e^{-x_{0}/\nu_{0}}\phi(\mu,\nu_{0})+\int_{0}^{1}a(\nu)e^{-x_{0}/\nu}\phi(\mu,\nu)d\nu,\label{Eqn: BC_HR}\end{equation}

\noindent leads to an infinitely more difficult determination of the expansion
coefficients due to the more involved nature of the orthogonality relations
of the eigenfunctions in the half-interval $[0,1]$ that now reads for $\nu,\textrm{ }\nu^{\prime}\in[0,1]$
\cite{Case1967} 

\noindent \begin{eqnarray}
\int_{0}^{1}W(\mu)\phi(\mu,\nu^{\prime})\phi(\mu,\nu)d\mu & = & \frac{W(\nu)N(\nu)}{\nu}\delta(\nu-\nu^{\prime})\nonumber \\
\int_{0}^{1}W(\mu)\phi(\mu,\nu_{0})\phi(\mu,\nu)d\mu & = & 0\nonumber \\
\int_{0}^{1}W(\mu)\phi(\mu,-\nu_{0})\phi(\mu,\nu)d\mu & = & c\nu\nu_{0}X(-\nu_{0})\phi(\nu,-\nu_{0})\nonumber \\
\int_{0}^{1}W(\mu)\phi(\mu,\pm\nu_{0})\phi(\mu,\nu_{0})d\mu & = & \mp\left(\frac{c\nu_{0}}{2}\right)^{2}X(\pm\nu_{0})\label{Eqn: HR Ortho}\\
\int_{0}^{1}W(\mu)\phi(\mu,\nu_{0})\phi(\mu,-\nu)d\mu & = & \frac{c^{2}\nu\nu_{0}}{4}X(-\nu)\nonumber \\
\int_{0}^{1}W(\mu)\phi(\mu,\nu^{\prime})\phi(\mu,-\nu)d\mu & = & \frac{c\nu^{\prime}}{2}(\nu_{0}+\nu)\phi(\nu^{\prime},-\nu)X(-\nu)\nonumber \end{eqnarray}

\noindent where the half-range weight function $W(\mu)$ is defined as 

\begin{equation}
W(\mu)=\frac{c\mu}{2(1-c)(\nu_{0}+\mu)X(-\mu)}\label{Eqn: W(mu)}\end{equation}
 in terms of the $X$-function \[
X(-\mu)=\textrm{exp}-\left\{ \frac{c}{2}\int_{0}^{1}\frac{\nu}{N(\nu)}\left[1+\frac{c\nu^{2}}{1-\nu^{2}}\right]\ln(\nu+\mu)d\nu\right\} ,\qquad0\leq\mu\leq1,\]

\noindent that is conveniently obtained from a numerical solution of the nonlinear
integral equation \begin{equation}
\Omega(-\mu)=1-\frac{c\mu}{2(1-c)}\int_{0}^{1}\frac{\nu_{0}^{2}(1-c)-\nu^{2}}{(\nu_{0}^{2}-\nu^{2})(\mu+\nu)\Omega(-\nu)}d\nu\label{Eqn: Omega(-mu)}\end{equation}

\noindent to yield \[
X(-\mu)=\frac{\Omega(-\mu)}{\mu+\nu_{0}\sqrt{1-c}},\]

\noindent and the $X(\pm\nu_{0})$ satisfy \[
X(\nu_{0})X(-\nu_{0})=\frac{\nu_{0}^{2}(1-c)-1}{2(1-c)v_{0}^{2}(\nu_{0}^{2}-1)}.\]
 Two other useful relations involving the $W$-function are given by $\int_{0}^{1}W(\mu)\phi(\mu,\nu_{0})d\mu=c\nu_{0}/2$
and $\int_{0}^{1}W(\mu)\phi(\mu,\nu)d\mu=c\nu/2$. 

The utility of these full and half range orthogonality relations lie in the
fact that a suitable class of functions of the type that is involved here can
always be expanded in terms of them, see \citet*{Case1967}. An example of this
for a full-range problem has been given above; we end this introduction to the
generalized --- traditionally known as singular in neutron transport theory
--- eigenfunction method with two examples of half-range orthogonality integrals
to the half-space problems A and B of Sec. 5. 

\medskip{}
\textbf{Problem A: The Milne Problem.} In this case there is no incident flux
of particles from outside the medium at $x=0$, but for large $x>0$ the neutron
distribution inside the medium behaves like $e^{x/\nu_{0}}\phi(-\nu_{0},\mu)$.
Hence the boundary condition (\ref{Eqn: BC_HR}) at $x=0$ reduces to \[
-\phi(\mu,-\nu_{0})=a_{\textrm{A}}(\nu_{0})\phi(\mu,\nu_{0})+\int_{0}^{1}a_{\textrm{A}}(\nu)\phi(\mu,\nu)d\nu\qquad\mu\geq0.\]

\noindent Use of the fourth and third equations of Eq. (\ref{Eqn: HR Ortho})
and the explicit relation Eq. (\ref{Eqn: W(mu)}) for $W(\mu)$ gives respectively
the coefficients \begin{eqnarray}
{\displaystyle a_{\textrm{A}}(\nu_{0})} & = & X(-\nu_{0})/X(v_{0})\nonumber \\
a_{\textrm{A}}(\nu) & = & -\frac{1}{N(\nu)}\textrm{ }c(1-c)\nu_{0}^{2}\nu X(-\nu_{0})X(-\nu)\label{Eqn: Milne_Coeff}\end{eqnarray}

\noindent The extrapolated end-point $z_{0}$ of Eq. (\ref{Eqn: extrapolated})
is related to $a_{\textrm{A}}(\nu_{0})$ of the Milne problem by $a_{\textrm{A}}(\nu_{0})=-\exp(-2z_{0}/\nu_{0})$. 

\smallskip{}
\textbf{Problem B: The Constant Source Problem.} Here \textbf{}the boundary
condition at $x=0$ is \[
1=a_{\textrm{B}}(\nu_{0})\phi(\mu,\nu_{0})+\int_{0}^{1}a_{\textrm{B}}(\nu)\phi(\mu,\nu)d\nu\qquad\mu\geq0\]

\noindent which leads, using the integral relations satisfied by $W$, to the
expansion coefficients 

\noindent \begin{eqnarray}
{\displaystyle a_{\textrm{B}}(\nu_{0})} & = & -2/c\nu_{0}X(v_{0})\label{Eqn: Constant_Coeff}\\
a_{\textrm{B}}(\nu) & = & \frac{1}{N(\nu)}\textrm{ }(1-c)\nu(\nu_{0}+\nu)X(-\nu)\nonumber \end{eqnarray}

\noindent where the $X(\pm\nu_{0})$ are related to Problem A as \begin{eqnarray*}
X(\nu_{0}) & = & \frac{1}{\nu_{0}}\sqrt{\frac{\nu_{0}^{2}(1-c)-1}{2a_{\textrm{A}}(\nu_{0})(1-c)(\nu_{0}^{2}-1)}}\\
X(-\nu_{0}) & = & \frac{1}{\nu_{0}}\sqrt{\frac{a_{A}(\nu_{0})\left(\nu_{0}^{2}(1-c)-1\right)}{2(1-c)(\nu_{0}^{2}-1)}}.\end{eqnarray*}

This brief introduction to the singular eigenfunction method should convince
the reader of the great difficulties associated with half-space, half-range
methods in particle transport theory; note that the $X$-functions in the coefficients
above must be obtained from numerically computed tables. In contrast, full-range
methods are more direct due to the simplicity of the weight function $\mu$,
which suggests the full-range formulation of half-range problems presented in
Sec. 5. Finally it should be mentioned that this singular eigenfunction method
is based on the theory of singular integral equations. 

\vspace{1.25cm}
\noindent \textbf{\large Acknowledgment}{\large \par}

\noindent It is a pleasure to thank the referees for recommending an enlarged
Tutorial and Review revision of the original submission \emph{Graphical Convergence,
Chaos and Complexity}, \emph{}and the Editor Professor Leon O Chua for suggesting
a pedagogically self-contained, jargonless no-page limit version accessible
to a wider audience for the present form of the paper. Financial assistance
during the initial stages of this work from the National Board for Higher Mathematics
is also acknowledged. 

\bibliographystyle{/mnt/win/MikTex/texmf/bibtex/bst/harvard/jmps}
\bibliography{/home/osegu/LyxDocs/osegu}

\end{document}

%% file: functions.pstex_t
\begin{picture}(0,0)%
\includegraphics{functions.pstex}%
\end{picture}%
\setlength{\unitlength}{3947sp}%
\begingroup\makeatletter\ifx\SetFigFont\undefined%
\gdef\SetFigFont#1#2#3#4#5{%
  \reset@font\fontsize{#1}{#2pt}%
  \fontfamily{#3}\fontseries{#4}\fontshape{#5}%
  \selectfont}%
\fi\endgroup%
\begin{picture}(6376,3773)(1028,-4226)
\put(5851,-4186){\makebox(0,0)[b]{\smash{{\SetFigFont{11}{13.2}{\familydefault}{\mddefault}{\updefault}{\color[rgb]{0,0,0}(d)}%
}}}}
\put(5851,-2236){\makebox(0,0)[b]{\smash{{\SetFigFont{11}{13.2}{\familydefault}{\mddefault}{\updefault}{\color[rgb]{0,0,0}(b)}%
}}}}
\put(2551,-2236){\makebox(0,0)[b]{\smash{{\SetFigFont{11}{13.2}{\familydefault}{\mddefault}{\updefault}{\color[rgb]{0,0,0}(a)}%
}}}}
\put(2551,-4186){\makebox(0,0)[b]{\smash{{\SetFigFont{11}{13.2}{\familydefault}{\mddefault}{\updefault}{\color[rgb]{0,0,0}(c)}%
}}}}
\put(5776,-561){\makebox(0,0)[b]{\smash{{\SetFigFont{11}{13.2}{\familydefault}{\mddefault}{\updefault}{\color[rgb]{0,0,0}$f$}%
}}}}
\put(4951,-3736){\makebox(0,0)[b]{\smash{{\SetFigFont{11}{13.2}{\familydefault}{\mddefault}{\updefault}{\color[rgb]{0,0,0}$X$}%
}}}}
\put(2626,-1861){\makebox(0,0)[b]{\smash{{\SetFigFont{11}{13.2}{\familydefault}{\mddefault}{\updefault}{\color[rgb]{0,0,0}$g$}%
}}}}
\put(2626,-3511){\makebox(0,0)[b]{\smash{{\SetFigFont{11}{13.2}{\familydefault}{\mddefault}{\updefault}{\color[rgb]{0,0,0}$g$}%
}}}}
\put(6676,-1861){\makebox(0,0)[b]{\smash{{\SetFigFont{11}{13.2}{\familydefault}{\mddefault}{\updefault}{\color[rgb]{0,0,0}$Y$}%
}}}}
\put(3901,-3286){\makebox(0,0)[b]{\smash{{\SetFigFont{11}{13.2}{\familydefault}{\mddefault}{\updefault}{\color[rgb]{0,0,0}$Y$}%
}}}}
\put(3901,-961){\makebox(0,0)[b]{\smash{{\SetFigFont{11}{13.2}{\familydefault}{\mddefault}{\updefault}{\color[rgb]{0,0,0}$Y$}%
}}}}
\put(4951,-1786){\makebox(0,0)[b]{\smash{{\SetFigFont{11}{13.2}{\familydefault}{\mddefault}{\updefault}{\color[rgb]{0,0,0}$X$}%
}}}}
\put(4751,-1111){\makebox(0,0)[b]{\smash{{\SetFigFont{11}{13.2}{\familydefault}{\mddefault}{\updefault}{\color[rgb]{0,0,0}$x_{1}$}%
}}}}
\put(4601,-1411){\makebox(0,0)[b]{\smash{{\SetFigFont{11}{13.2}{\familydefault}{\mddefault}{\updefault}{\color[rgb]{0,0,0}$x_{2}$}%
}}}}
\put(4826,-3236){\makebox(0,0)[b]{\smash{{\SetFigFont{11}{13.2}{\familydefault}{\mddefault}{\updefault}{\color[rgb]{0,0,0}$x$}%
}}}}
\put(3676,-1261){\makebox(0,0)[b]{\smash{{\SetFigFont{11}{13.2}{\familydefault}{\mddefault}{\updefault}{\color[rgb]{0,0,0}$y_{1}$}%
}}}}
\put(3651,-2961){\makebox(0,0)[b]{\smash{{\SetFigFont{11}{13.2}{\familydefault}{\mddefault}{\updefault}{\color[rgb]{0,0,0}$y_{1}$}%
}}}}
\put(6726,-3536){\makebox(0,0)[b]{\smash{{\SetFigFont{11}{13.2}{\familydefault}{\mddefault}{\updefault}{\color[rgb]{0,0,0}$y$}%
}}}}
\put(6951,-1061){\makebox(0,0)[b]{\smash{{\SetFigFont{11}{13.2}{\familydefault}{\mddefault}{\updefault}{\color[rgb]{0,0,0}$y_{1}$}%
}}}}
\put(3601,-1661){\makebox(0,0)[b]{\smash{{\SetFigFont{11}{13.2}{\familydefault}{\mddefault}{\updefault}{\color[rgb]{0,0,0}$y_{2}$}%
}}}}
\put(7151,-1601){\makebox(0,0)[b]{\smash{{\SetFigFont{11}{13.2}{\familydefault}{\mddefault}{\updefault}{\color[rgb]{0,0,0}$y_{2}$}%
}}}}
\put(3601,-3486){\makebox(0,0)[b]{\smash{{\SetFigFont{11}{13.2}{\familydefault}{\mddefault}{\updefault}{\color[rgb]{0,0,0}$y_{2}$}%
}}}}
\put(6751,-1336){\makebox(0,0)[b]{\smash{{\SetFigFont{11}{13.2}{\familydefault}{\mddefault}{\updefault}{\color[rgb]{0,0,0}$\mathcal {R}(f)$}%
}}}}
\put(6926,-3236){\makebox(0,0)[b]{\smash{{\SetFigFont{11}{13.2}{\familydefault}{\mddefault}{\updefault}{\color[rgb]{0,0,0}$B$}%
}}}}
\put(1391,-2986){\makebox(0,0)[b]{\smash{{\SetFigFont{11}{13.2}{\familydefault}{\mddefault}{\updefault}{\color[rgb]{0,0,0}$A$}%
}}}}
\put(2626,-2776){\makebox(0,0)[b]{\smash{{\SetFigFont{11}{13.2}{\familydefault}{\mddefault}{\updefault}{\color[rgb]{0,0,0}$f$}%
}}}}
\put(7201,-3436){\makebox(0,0)[b]{\smash{{\SetFigFont{11}{13.2}{\familydefault}{\mddefault}{\updefault}{\color[rgb]{0,0,0}$Y$}%
}}}}
\put(5776,-1761){\makebox(0,0)[b]{\smash{{\SetFigFont{11}{13.2}{\familydefault}{\mddefault}{\updefault}{\color[rgb]{0,0,0}$g$}%
}}}}
\put(1576,-1786){\makebox(0,0)[b]{\smash{{\SetFigFont{11}{13.2}{\familydefault}{\mddefault}{\updefault}{\color[rgb]{0,0,0}$X$}%
}}}}
\put(1576,-3736){\makebox(0,0)[b]{\smash{{\SetFigFont{11}{13.2}{\familydefault}{\mddefault}{\updefault}{\color[rgb]{0,0,0}$X$}%
}}}}
\put(2626,-561){\makebox(0,0)[b]{\smash{{\SetFigFont{11}{13.2}{\familydefault}{\mddefault}{\updefault}{\color[rgb]{0,0,0}$f$}%
}}}}
\put(1476,-1111){\makebox(0,0)[b]{\smash{{\SetFigFont{11}{13.2}{\familydefault}{\mddefault}{\updefault}{\color[rgb]{0,0,0}$x_{1}$}%
}}}}
\put(5901,-3196){\makebox(0,0)[b]{\smash{{\SetFigFont{11}{13.2}{\familydefault}{\mddefault}{\updefault}{\color[rgb]{0,0,0}$\mathscr{M}$}%
}}}}
\put(1591,-3436){\makebox(0,0)[b]{\smash{{\SetFigFont{11}{13.2}{\familydefault}{\mddefault}{\updefault}{\color[rgb]{0,0,0}$x$}%
}}}}
\put(1586,-1361){\makebox(0,0)[b]{\smash{{\SetFigFont{11}{13.2}{\familydefault}{\mddefault}{\updefault}{\color[rgb]{0,0,0}$x_{2}$}%
}}}}
\end{picture}%

%% file: quotient.pstex_t
\begin{picture}(0,0)%
\includegraphics{quotient.pstex}%
\end{picture}%
\setlength{\unitlength}{2368sp}%
\begingroup\makeatletter\ifx\SetFigFont\undefined%
\gdef\SetFigFont#1#2#3#4#5{%
  \reset@font\fontsize{#1}{#2pt}%
  \fontfamily{#3}\fontseries{#4}\fontshape{#5}%
  \selectfont}%
\fi\endgroup%
\begin{picture}(4819,3319)(-121,-2558)
\put(451,-911){\makebox(0,0)[b]{\smash{{\SetFigFont{10}{12.0}{\familydefault}{\mddefault}{\updefault}{\color[rgb]{0,0,0}$Q$}%
}}}}
\put(451,314){\makebox(0,0)[b]{\smash{{\SetFigFont{10}{12.0}{\familydefault}{\mddefault}{\updefault}{\color[rgb]{0,0,0}$X$}%
}}}}
\put(2301,-961){\makebox(0,0)[b]{\smash{{\SetFigFont{10}{12.0}{\familydefault}{\mddefault}{\updefault}{\color[rgb]{0,0,0}$q$}%
}}}}
\put(2201,-2301){\makebox(0,0)[b]{\smash{{\SetFigFont{10}{12.0}{\familydefault}{\mddefault}{\updefault}{\color[rgb]{0,0,0}$h$}%
}}}}
\put(4126,-2286){\makebox(0,0)[b]{\smash{{\SetFigFont{10}{12.0}{\familydefault}{\mddefault}{\updefault}{\color[rgb]{0,0,0}$S^1$}%
}}}}
\put(451,-2286){\makebox(0,0)[b]{\smash{{\SetFigFont{10}{12.0}{\familydefault}{\mddefault}{\updefault}{\color[rgb]{0,0,0}$X/\sim$}%
}}}}
\end{picture}%

%% file: FuncSpace.pstex_t
\begin{picture}(0,0)%
\includegraphics{FuncSpace.pstex}%
\end{picture}%
\setlength{\unitlength}{3947sp}%
\begingroup\makeatletter\ifx\SetFigFont\undefined%
\gdef\SetFigFont#1#2#3#4#5{%
  \reset@font\fontsize{#1}{#2pt}%
  \fontfamily{#3}\fontseries{#4}\fontshape{#5}%
  \selectfont}%
\fi\endgroup%
\begin{picture}(6702,2916)(1414,-3340)
\put(4801,-3286){\makebox(0,0)[b]{\smash{{\SetFigFont{11}{13.2}{\familydefault}{\mddefault}{\updefault}{\color[rgb]{0,0,0}(b)}%
}}}}
\put(2476,-3286){\makebox(0,0)[b]{\smash{{\SetFigFont{11}{13.2}{\familydefault}{\mddefault}{\updefault}{\color[rgb]{0,0,0}(a)}%
}}}}
\put(6976,-3286){\makebox(0,0)[b]{\smash{{\SetFigFont{12}{14.4}{\familydefault}{\mddefault}{\updefault}{\color[rgb]{0,0,0}(c)}%
}}}}
\put(2326,-2971){\makebox(0,0)[b]{\smash{{\SetFigFont{11}{13.2}{\familydefault}{\mddefault}{\updefault}{\color[rgb]{0,0,0}$.25$}%
}}}}
\put(2101,-2971){\makebox(0,0)[b]{\smash{{\SetFigFont{11}{13.2}{\familydefault}{\mddefault}{\updefault}{\color[rgb]{0,0,0}$.1$}%
}}}}
\put(3451,-2971){\makebox(0,0)[b]{\smash{{\SetFigFont{11}{13.2}{\familydefault}{\mddefault}{\updefault}{\color[rgb]{0,0,0}$1$}%
}}}}
\put(4351,-2971){\makebox(0,0)[b]{\smash{{\SetFigFont{11}{13.2}{\familydefault}{\mddefault}{\updefault}{\color[rgb]{0,0,0}0}%
}}}}
\put(6601,-2971){\makebox(0,0)[b]{\smash{{\SetFigFont{11}{13.2}{\familydefault}{\mddefault}{\updefault}{\color[rgb]{0,0,0}$0$}%
}}}}
\put(7201,-2971){\makebox(0,0)[b]{\smash{{\SetFigFont{11}{13.2}{\familydefault}{\mddefault}{\updefault}{\color[rgb]{0,0,0}$0.5$}%
}}}}
\put(5551,-2971){\makebox(0,0)[b]{\smash{{\SetFigFont{11}{13.2}{\familydefault}{\mddefault}{\updefault}{\color[rgb]{0,0,0}$1$}%
}}}}
\put(4276,-1861){\makebox(0,0)[rb]{\smash{{\SetFigFont{11}{13.2}{\familydefault}{\mddefault}{\updefault}{\color[rgb]{0,0,0}$1$}%
}}}}
\put(4951,-2971){\makebox(0,0)[b]{\smash{{\SetFigFont{11}{13.2}{\familydefault}{\mddefault}{\updefault}{\color[rgb]{0,0,0}$0.5$}%
}}}}
\put(3926,-2971){\makebox(0,0)[b]{\smash{{\SetFigFont{11}{13.2}{\familydefault}{\mddefault}{\updefault}{\color[rgb]{0,0,0}$a$}%
}}}}
\put(6176,-2971){\makebox(0,0)[b]{\smash{{\SetFigFont{11}{13.2}{\familydefault}{\mddefault}{\updefault}{\color[rgb]{0,0,0}$a$}%
}}}}
\put(5751,-2971){\makebox(0,0)[b]{\smash{{\SetFigFont{11}{13.2}{\familydefault}{\mddefault}{\updefault}{\color[rgb]{0,0,0}$b$}%
}}}}
\put(8001,-2971){\makebox(0,0)[b]{\smash{{\SetFigFont{11}{13.2}{\familydefault}{\mddefault}{\updefault}{\color[rgb]{0,0,0}$b$}%
}}}}
\put(7801,-2971){\makebox(0,0)[b]{\smash{{\SetFigFont{11}{13.2}{\familydefault}{\mddefault}{\updefault}{\color[rgb]{0,0,0}$1$}%
}}}}
\put(6526,-886){\makebox(0,0)[rb]{\smash{{\SetFigFont{11}{13.2}{\familydefault}{\mddefault}{\updefault}{\color[rgb]{0,0,0}$1$}%
}}}}
\put(3201,-1411){\makebox(0,0)[b]{\smash{{\SetFigFont{11}{13.2}{\familydefault}{\mddefault}{\updefault}{\color[rgb]{0,0,0}$f_1$}%
}}}}
\put(2476,-711){\makebox(0,0)[b]{\smash{{\SetFigFont{11}{13.2}{\familydefault}{\mddefault}{\updefault}{\color[rgb]{0,0,0}$f_4$}%
}}}}
\put(2176,-711){\makebox(0,0)[b]{\smash{{\SetFigFont{11}{13.2}{\familydefault}{\mddefault}{\updefault}{\color[rgb]{0,0,0}$f_{10}$}%
}}}}
\put(1826,-2971){\makebox(0,0)[b]{\smash{{\SetFigFont{11}{13.2}{\familydefault}{\mddefault}{\updefault}{\color[rgb]{0,0,0}$0$}%
}}}}
\put(4276,-886){\makebox(0,0)[rb]{\smash{{\SetFigFont{11}{13.2}{\familydefault}{\mddefault}{\updefault}{\color[rgb]{0,0,0}$2$}%
}}}}
\put(3076,-711){\makebox(0,0)[b]{\smash{{\SetFigFont{11}{13.2}{\familydefault}{\mddefault}{\updefault}{\color[rgb]{0,0,0}$f_2$}%
}}}}
\put(5251,-1686){\makebox(0,0)[b]{\smash{{\SetFigFont{11}{13.2}{\familydefault}{\mddefault}{\updefault}{\color[rgb]{0,0,0}$\delta_1$}%
}}}}
\put(7501,-1786){\makebox(0,0)[b]{\smash{{\SetFigFont{11}{13.2}{\familydefault}{\mddefault}{\updefault}{\color[rgb]{0,0,0}$\delta_1^{(-1)}$}%
}}}}
\put(7501,-711){\makebox(0,0)[b]{\smash{{\SetFigFont{11}{13.2}{\familydefault}{\mddefault}{\updefault}{\color[rgb]{0,0,0}$\delta_2^{(-1)}$}%
}}}}
\put(4651,-711){\makebox(0,0)[b]{\smash{{\SetFigFont{11}{13.2}{\familydefault}{\mddefault}{\updefault}{\color[rgb]{0,0,0}$\delta_2$}%
}}}}
\put(1876,-886){\makebox(0,0)[rb]{\smash{{\SetFigFont{11}{13.2}{\familydefault}{\mddefault}{\updefault}{\color[rgb]{0,0,0}$1$}%
}}}}
\put(2701,-2971){\makebox(0,0)[b]{\smash{{\SetFigFont{11}{13.2}{\familydefault}{\mddefault}{\updefault}{\color[rgb]{0,0,0}$.5$}%
}}}}
\put(1526,-2971){\makebox(0,0)[b]{\smash{{\SetFigFont{11}{13.2}{\familydefault}{\mddefault}{\updefault}{\color[rgb]{0,0,0}$a$}%
}}}}
\end{picture}%

%% file: MP_Inverse.pstex_t
\begin{picture}(0,0)%
\includegraphics{MP_Inverse.pstex}%
\end{picture}%
\setlength{\unitlength}{3631sp}%
\begingroup\makeatletter\ifx\SetFigFont\undefined%
\gdef\SetFigFont#1#2#3#4#5{%
  \reset@font\fontsize{#1}{#2pt}%
  \fontfamily{#3}\fontseries{#4}\fontshape{#5}%
  \selectfont}%
\fi\endgroup%
\begin{picture}(7158,3427)(654,-3251)
\put(2226,-3211){\makebox(0,0)[b]{\smash{{\SetFigFont{10}{12.0}{\familydefault}{\mddefault}{\updefault}{\color[rgb]{0,0,0}(a)}%
}}}}
\put(5851,-3211){\makebox(0,0)[b]{\smash{{\SetFigFont{10}{12.0}{\familydefault}{\mddefault}{\updefault}{\color[rgb]{0,0,0}(b)}%
}}}}
\put(2701,-1461){\makebox(0,0)[lb]{\smash{{\SetFigFont{11}{13.2}{\familydefault}{\mddefault}{\updefault}{\color[rgb]{0,0,0}$b$}%
}}}}
\put(3151,-2536){\makebox(0,0)[lb]{\smash{{\SetFigFont{11}{13.2}{\familydefault}{\mddefault}{\updefault}{\color[rgb]{0,0,0}$b_{\bot }$}%
}}}}
\put(2251,-2911){\makebox(0,0)[rb]{\smash{{\SetFigFont{11}{13.2}{\familydefault}{\mddefault}{\updefault}{\color[rgb]{0,0,0}$\mathcal{N}(A)$}%
}}}}
\put(2401,-1536){\makebox(0,0)[rb]{\smash{{\SetFigFont{11}{13.2}{\familydefault}{\mddefault}{\updefault}{\color[rgb]{0,0,0}$G_\mathrm{MP}(b)$}%
}}}}
\put(1876,-2011){\makebox(0,0)[rb]{\smash{{\SetFigFont{11}{13.2}{\familydefault}{\mddefault}{\updefault}{\color[rgb]{0,0,0}$G_\mathrm{MP}(b_{\bot })$}%
}}}}
\put(2826,-861){\makebox(0,0)[rb]{\smash{{\SetFigFont{11}{13.2}{\familydefault}{\mddefault}{\updefault}{\color[rgb]{0,0,0}$G_\mathrm{MP}(b_{\Vert })$}%
}}}}
\put(5651,-2561){\makebox(0,0)[b]{\smash{{\SetFigFont{11}{13.2}{\familydefault}{\mddefault}{\updefault}{\color[rgb]{0,0,0}$\mathcal{D}(f)$}%
}}}}
\put(2626,-511){\makebox(0,0)[b]{\smash{{\SetFigFont{11}{13.2}{\familydefault}{\mddefault}{\updefault}{\color[rgb]{0,0,0}$\mathbb{R}^m$}%
}}}}
\put(2201,-511){\makebox(0,0)[b]{\smash{{\SetFigFont{11}{13.2}{\familydefault}{\mddefault}{\updefault}{\color[rgb]{0,0,0}$\mathbb{R}^n$}%
}}}}
\put(6691,-2936){\makebox(0,0)[b]{\smash{{\SetFigFont{11}{13.2}{\familydefault}{\mddefault}{\updefault}{\color[rgb]{0,0,0}$x_2$}%
}}}}
\put(4636,-2936){\makebox(0,0)[b]{\smash{{\SetFigFont{10}{12.0}{\familydefault}{\mddefault}{\updefault}{\color[rgb]{0,0,0}$x_1$}%
}}}}
\put(4286,-1016){\makebox(0,0)[rb]{\smash{{\SetFigFont{10}{12.0}{\familydefault}{\mddefault}{\updefault}{\color[rgb]{0,0,0}$r_1$}%
}}}}
\put(3326,-961){\makebox(0,0)[lb]{\smash{{\SetFigFont{11}{13.2}{\familydefault}{\mddefault}{\updefault}{\color[rgb]{0,0,0}$b_{\Vert }$}%
}}}}
\put(7051,-311){\makebox(0,0)[lb]{\smash{{\SetFigFont{11}{13.2}{\familydefault}{\mddefault}{\updefault}{\color[rgb]{0,0,0}$g(x)$}%
}}}}
\put(7651,-2836){\makebox(0,0)[b]{\smash{{\SetFigFont{11}{13.2}{\familydefault}{\mddefault}{\updefault}{\color[rgb]{0,0,0}$X$}%
}}}}
\put(7276,-1861){\makebox(0,0)[b]{\smash{{\SetFigFont{11}{13.2}{\familydefault}{\mddefault}{\updefault}{\color[rgb]{0,0,0}$\mathcal{R}(f)$}%
}}}}
\put(4286,-536){\makebox(0,0)[rb]{\smash{{\SetFigFont{11}{13.2}{\familydefault}{\mddefault}{\updefault}{\color[rgb]{0,0,0}$r_2$}%
}}}}
\put(4276,-786){\makebox(0,0)[rb]{\smash{{\SetFigFont{11}{13.2}{\familydefault}{\mddefault}{\updefault}{\color[rgb]{0,0,0}$y_2$}%
}}}}
\put(4276,-1836){\makebox(0,0)[rb]{\smash{{\SetFigFont{11}{13.2}{\familydefault}{\mddefault}{\updefault}{\color[rgb]{0,0,0}$y_1$}%
}}}}
\put(5551,-1636){\makebox(0,0)[rb]{\smash{{\SetFigFont{11}{13.2}{\familydefault}{\mddefault}{\updefault}{\color[rgb]{0,0,0}$f(x)$}%
}}}}
\put(4351,-61){\makebox(0,0)[b]{\smash{{\SetFigFont{11}{13.2}{\familydefault}{\mddefault}{\updefault}{\color[rgb]{0,0,0}$Y$}%
}}}}
\put(826,-111){\makebox(0,0)[b]{\smash{{\SetFigFont{11}{13.2}{\familydefault}{\mddefault}{\updefault}{\color[rgb]{0,0,0}$\mathcal{R}(A^\mathrm{T})$}%
}}}}
\put(2776,-111){\makebox(0,0)[b]{\smash{{\SetFigFont{11}{13.2}{\familydefault}{\mddefault}{\updefault}{\color[rgb]{0,0,0}$\mathcal{R}(A)$}%
}}}}
\put(2526,-2911){\makebox(0,0)[lb]{\smash{{\SetFigFont{11}{13.2}{\familydefault}{\mddefault}{\updefault}{\color[rgb]{0,0,0}$\mathcal{N}(A^\mathrm{T})$}%
}}}}
\put(4286,-2956){\makebox(0,0)[rb]{\smash{{\SetFigFont{10}{12.0}{\familydefault}{\mddefault}{\updefault}{\color[rgb]{0,0,0}$0$}%
}}}}
\end{picture}%

%% file: GenInv.pstex_t
\begin{picture}(0,0)%
\includegraphics{GenInv.pstex}%
\end{picture}%
\setlength{\unitlength}{3158sp}%
\begingroup\makeatletter\ifx\SetFigFont\undefined%
\gdef\SetFigFont#1#2#3#4#5{%
  \reset@font\fontsize{#1}{#2pt}%
  \fontfamily{#3}\fontseries{#4}\fontshape{#5}%
  \selectfont}%
\fi\endgroup%
\begin{picture}(6194,3469)(-721,-2708)
\put(2326,164){\makebox(0,0)[b]{\smash{{\SetFigFont{12}{14.4}{\familydefault}{\mddefault}{\updefault}{\color[rgb]{0,0,0}$f^{-}$}%
}}}}
\put(2326,474){\makebox(0,0)[b]{\smash{{\SetFigFont{12}{14.4}{\familydefault}{\mddefault}{\updefault}{\color[rgb]{0,0,0}$f$}%
}}}}
\put(4351,314){\makebox(0,0)[b]{\smash{{\SetFigFont{12}{14.4}{\familydefault}{\mddefault}{\updefault}{\color[rgb]{0,0,0}$(Y,\mathcal {V})$}%
}}}}
\put(2076,-1461){\makebox(0,0)[b]{\smash{{\SetFigFont{12}{14.4}{\familydefault}{\mddefault}{\updefault}{\color[rgb]{0,0,0}$G$}%
}}}}
\put(326,-2486){\makebox(0,0)[b]{\smash{{\SetFigFont{12}{14.4}{\familydefault}{\mddefault}{\updefault}{\color[rgb]{0,0,0}$(X_{\textrm{B}},\textrm{FT}\{\mathcal{U};q\})$}%
}}}}
\put(4351,-1086){\makebox(0,0)[b]{\smash{{\SetFigFont{12}{14.4}{\familydefault}{\mddefault}{\updefault}{\color[rgb]{0,0,0}$e$}%
}}}}
\put(326,-1061){\makebox(0,0)[b]{\smash{{\SetFigFont{12}{14.4}{\familydefault}{\mddefault}{\updefault}{\color[rgb]{0,0,0}$q$}%
}}}}
\put(2766,-611){\makebox(0,0)[b]{\smash{{\SetFigFont{12}{14.4}{\familydefault}{\mddefault}{\updefault}{\color[rgb]{0,0,0}$g$}%
}}}}
\put(351,314){\makebox(0,0)[b]{\smash{{\SetFigFont{12}{14.4}{\familydefault}{\mddefault}{\updefault}{\color[rgb]{0,0,0}$(X,\mathcal {U})$}%
}}}}
\put(4401,-2486){\makebox(0,0)[b]{\smash{{\SetFigFont{12}{14.4}{\familydefault}{\mddefault}{\updefault}{\color[rgb]{0,0,0}$(f(X),\textrm{IT}\{e;\mathcal{V}\})$}%
}}}}
\put(1616,-636){\makebox(0,0)[b]{\smash{{\SetFigFont{12}{14.4}{\familydefault}{\mddefault}{\updefault}{\color[rgb]{0,0,0}$h$}%
}}}}
\put(2276,-2336){\makebox(0,0)[b]{\smash{{\SetFigFont{12}{14.4}{\familydefault}{\mddefault}{\updefault}{\color[rgb]{0,0,0}$f_{\textrm{B}}$}%
}}}}
\put(2326,-2626){\makebox(0,0)[b]{\smash{{\SetFigFont{12}{14.4}{\familydefault}{\mddefault}{\updefault}{\color[rgb]{0,0,0}$f_{\textrm{B}}^{-1}$}%
}}}}
\end{picture}%

%% file: gen-inv.pstex_t
\begin{picture}(0,0)%
\includegraphics{gen-inv.pstex}%
\end{picture}%
\setlength{\unitlength}{2368sp}%
\begingroup\makeatletter\ifx\SetFigFont\undefined%
\gdef\SetFigFont#1#2#3#4#5{%
  \reset@font\fontsize{#1}{#2pt}%
  \fontfamily{#3}\fontseries{#4}\fontshape{#5}%
  \selectfont}%
\fi\endgroup%
\begin{picture}(4347,4332)(-564,-3376)
\put(-199,-3161){\makebox(0,0)[b]{\smash{{\SetFigFont{12}{14.4}{\familydefault}{\mddefault}{\updefault}{\color[rgb]{0,0,0}$0$}%
}}}}
\put(1351,-3286){\makebox(0,0)[b]{\smash{{\SetFigFont{12}{14.4}{\familydefault}{\mddefault}{\updefault}{\color[rgb]{0,0,0}$\frac{3}{8}$}%
}}}}
\put(3576,-3161){\makebox(0,0)[b]{\smash{{\SetFigFont{12}{14.4}{\familydefault}{\mddefault}{\updefault}{\color[rgb]{0,0,0}$1$}%
}}}}
\put(-149,-161){\makebox(0,0)[rb]{\smash{{\SetFigFont{12}{14.4}{\familydefault}{\mddefault}{\updefault}{\color[rgb]{0,0,0}$\frac{3}{4}$}%
}}}}
\put(-149,-1036){\makebox(0,0)[rb]{\smash{{\SetFigFont{12}{14.4}{\familydefault}{\mddefault}{\updefault}{\color[rgb]{0,0,0}$\frac{1}{2}$}%
}}}}
\put(921,-3286){\makebox(0,0)[b]{\smash{{\SetFigFont{12}{14.4}{\familydefault}{\mddefault}{\updefault}{\color[rgb]{0,0,0}$\frac{1}{4}$}%
}}}}
\put(-149,764){\makebox(0,0)[rb]{\smash{{\SetFigFont{12}{14.4}{\familydefault}{\mddefault}{\updefault}{\color[rgb]{0,0,0}$1$}%
}}}}
\put(2101,-3286){\makebox(0,0)[b]{\smash{{\SetFigFont{12}{14.4}{\familydefault}{\mddefault}{\updefault}{\color[rgb]{0,0,0}$\frac{5}{8}$}%
}}}}
\end{picture}%

%% file: Example2_1.pstex_t
\begin{picture}(0,0)%
\includegraphics{Example2_1.pstex}%
\end{picture}%
\setlength{\unitlength}{3947sp}%
\begingroup\makeatletter\ifx\SetFigFont\undefined%
\gdef\SetFigFont#1#2#3#4#5{%
  \reset@font\fontsize{#1}{#2pt}%
  \fontfamily{#3}\fontseries{#4}\fontshape{#5}%
  \selectfont}%
\fi\endgroup%
\begin{picture}(6039,6201)(1112,-6690)
\put(1426,-2236){\makebox(0,0)[rb]{\smash{{\SetFigFont{11}{13.2}{\familydefault}{\mddefault}{\updefault}{\color[rgb]{0,0,0}0}%
}}}}
\put(1501,-1336){\makebox(0,0)[rb]{\smash{{\SetFigFont{11}{13.2}{\familydefault}{\mddefault}{\updefault}{\color[rgb]{0,0,0}1}%
}}}}
\put(1501,-886){\makebox(0,0)[rb]{\smash{{\SetFigFont{11}{13.2}{\familydefault}{\mddefault}{\updefault}{\color[rgb]{0,0,0}1.5}%
}}}}
\put(1501,-2686){\makebox(0,0)[rb]{\smash{{\SetFigFont{11}{13.2}{\familydefault}{\mddefault}{\updefault}{\color[rgb]{0,0,0}-0.5}%
}}}}
\put(2551,-2411){\makebox(0,0)[b]{\smash{{\SetFigFont{11}{13.2}{\familydefault}{\mddefault}{\updefault}{\color[rgb]{0,0,0}1}%
}}}}
\put(3451,-2411){\makebox(0,0)[b]{\smash{{\SetFigFont{11}{13.2}{\familydefault}{\mddefault}{\updefault}{\color[rgb]{0,0,0}2}%
}}}}
\put(3501,-1636){\makebox(0,0)[lb]{\smash{{\SetFigFont{11}{13.2}{\familydefault}{\mddefault}{\updefault}{\color[rgb]{0,0,0}$1+2/n$}%
}}}}
\put(3801,-2611){\makebox(0,0)[b]{\smash{{\SetFigFont{11}{13.2}{\familydefault}{\mddefault}{\updefault}{\color[rgb]{0,0,0}$n\textrm{ odd}$}%
}}}}
\put(5701,-3141){\makebox(0,0)[b]{\smash{{\SetFigFont{12}{14.4}{\familydefault}{\mddefault}{\updefault}{\color[rgb]{0,0,0}(b)}%
}}}}
\put(2626,-3141){\makebox(0,0)[b]{\smash{{\SetFigFont{12}{14.4}{\familydefault}{\mddefault}{\updefault}{\color[rgb]{0,0,0}(a)}%
}}}}
\put(5251,-2411){\makebox(0,0)[b]{\smash{{\SetFigFont{11}{13.2}{\familydefault}{\mddefault}{\updefault}{\color[rgb]{0,0,0}1}%
}}}}
\put(5851,-2411){\makebox(0,0)[b]{\smash{{\SetFigFont{11}{13.2}{\familydefault}{\mddefault}{\updefault}{\color[rgb]{0,0,0}2}%
}}}}
\put(6451,-2411){\makebox(0,0)[b]{\smash{{\SetFigFont{11}{13.2}{\familydefault}{\mddefault}{\updefault}{\color[rgb]{0,0,0}3}%
}}}}
\put(6601,-2761){\makebox(0,0)[b]{\smash{{\SetFigFont{11}{13.2}{\familydefault}{\mddefault}{\updefault}{\color[rgb]{0,0,0}$n\textrm{ odd}$}%
}}}}
\put(5701,-6636){\makebox(0,0)[b]{\smash{{\SetFigFont{12}{14.4}{\familydefault}{\mddefault}{\updefault}{\color[rgb]{0,0,0}(d)}%
}}}}
\put(2626,-6636){\makebox(0,0)[b]{\smash{{\SetFigFont{12}{14.4}{\familydefault}{\mddefault}{\updefault}{\color[rgb]{0,0,0}(c)}%
}}}}
\put(2776,-736){\makebox(0,0)[b]{\smash{{\SetFigFont{11}{13.2}{\familydefault}{\mddefault}{\updefault}{\color[rgb]{0,0,0}$1+1/n$}%
}}}}
\put(3801,-1036){\makebox(0,0)[b]{\smash{{\SetFigFont{11}{13.2}{\familydefault}{\mddefault}{\updefault}{\color[rgb]{0,0,0}$n\textrm{ even}$}%
}}}}
\put(5726,-711){\makebox(0,0)[b]{\smash{{\SetFigFont{11}{13.2}{\familydefault}{\mddefault}{\updefault}{\color[rgb]{0,0,0}$2-1/n$}%
}}}}
\put(6801,-2061){\makebox(0,0)[b]{\smash{{\SetFigFont{11}{13.2}{\familydefault}{\mddefault}{\updefault}{\color[rgb]{0,0,0}$3+1/n$}%
}}}}
\put(5001,-711){\makebox(0,0)[b]{\smash{{\SetFigFont{11}{13.2}{\familydefault}{\mddefault}{\updefault}{\color[rgb]{0,0,0}$1/n$}%
}}}}
\put(6601,-886){\makebox(0,0)[b]{\smash{{\SetFigFont{11}{13.2}{\familydefault}{\mddefault}{\updefault}{\color[rgb]{0,0,0}$n\textrm{ even}$}%
}}}}
\put(2701,-3861){\makebox(0,0)[b]{\smash{{\SetFigFont{11}{13.2}{\familydefault}{\mddefault}{\updefault}{\color[rgb]{0,0,0}12 iterates of $-0.05+x-x^2$}%
}}}}
\put(3126,-6386){\makebox(0,0)[b]{\smash{{\SetFigFont{11}{13.2}{\familydefault}{\mddefault}{\updefault}{\color[rgb]{0,0,0}1}%
}}}}
\put(2226,-6386){\makebox(0,0)[b]{\smash{{\SetFigFont{11}{13.2}{\familydefault}{\mddefault}{\updefault}{\color[rgb]{0,0,0}0}%
}}}}
\put(1251,-6386){\makebox(0,0)[b]{\smash{{\SetFigFont{11}{13.2}{\familydefault}{\mddefault}{\updefault}{\color[rgb]{0,0,0}-1}%
}}}}
\put(4076,-6386){\makebox(0,0)[b]{\smash{{\SetFigFont{11}{13.2}{\familydefault}{\mddefault}{\updefault}{\color[rgb]{0,0,0}2}%
}}}}
\put(1251,-4936){\makebox(0,0)[rb]{\smash{{\SetFigFont{11}{13.2}{\familydefault}{\mddefault}{\updefault}{\color[rgb]{0,0,0}-1}%
}}}}
\put(1251,-6186){\makebox(0,0)[rb]{\smash{{\SetFigFont{11}{13.2}{\familydefault}{\mddefault}{\updefault}{\color[rgb]{0,0,0}-3}%
}}}}
\put(1251,-5611){\makebox(0,0)[rb]{\smash{{\SetFigFont{11}{13.2}{\familydefault}{\mddefault}{\updefault}{\color[rgb]{0,0,0}-2}%
}}}}
\put(2141,-4186){\makebox(0,0)[b]{\smash{{\SetFigFont{11}{13.2}{\familydefault}{\mddefault}{\updefault}{\color[rgb]{0,0,0}0}%
}}}}
\put(1251,-3636){\makebox(0,0)[rb]{\smash{{\SetFigFont{11}{13.2}{\familydefault}{\mddefault}{\updefault}{\color[rgb]{0,0,0}1}%
}}}}
\put(3676,-4761){\makebox(0,0)[b]{\smash{{\SetFigFont{11}{13.2}{\familydefault}{\mddefault}{\updefault}{\color[rgb]{0,0,0}1}%
}}}}
\put(2651,-5286){\makebox(0,0)[b]{\smash{{\SetFigFont{11}{13.2}{\familydefault}{\mddefault}{\updefault}{\color[rgb]{0,0,0}12}%
}}}}
\put(5776,-3861){\makebox(0,0)[b]{\smash{{\SetFigFont{11}{13.2}{\familydefault}{\mddefault}{\updefault}{\color[rgb]{0,0,0}12 iterates of $0.7+x+x^2$}%
}}}}
\put(5526,-5061){\makebox(0,0)[b]{\smash{{\SetFigFont{11}{13.2}{\familydefault}{\mddefault}{\updefault}{\color[rgb]{0,0,0}0}%
}}}}
\put(4876,-6361){\makebox(0,0)[b]{\smash{{\SetFigFont{11}{13.2}{\familydefault}{\mddefault}{\updefault}{\color[rgb]{0,0,0}$a$}%
}}}}
\put(6651,-6361){\makebox(0,0)[b]{\smash{{\SetFigFont{11}{13.2}{\familydefault}{\mddefault}{\updefault}{\color[rgb]{0,0,0}$c$}%
}}}}
\put(4401,-5811){\makebox(0,0)[rb]{\smash{{\SetFigFont{11}{13.2}{\familydefault}{\mddefault}{\updefault}{\color[rgb]{0,0,0}-1}%
}}}}
\put(4401,-4036){\makebox(0,0)[rb]{\smash{{\SetFigFont{11}{13.2}{\familydefault}{\mddefault}{\updefault}{\color[rgb]{0,0,0}1}%
}}}}
\put(4401,-4211){\makebox(0,0)[rb]{\smash{{\SetFigFont{11}{13.2}{\familydefault}{\mddefault}{\updefault}{\color[rgb]{0,0,0}$\alpha$}%
}}}}
\end{picture}%

%% file: biconv.pstex_t
\begin{picture}(0,0)%
\includegraphics{biconv.pstex}%
\end{picture}%
\setlength{\unitlength}{3947sp}%
\begingroup\makeatletter\ifx\SetFigFont\undefined%
\gdef\SetFigFont#1#2#3#4#5{%
  \reset@font\fontsize{#1}{#2pt}%
  \fontfamily{#3}\fontseries{#4}\fontshape{#5}%
  \selectfont}%
\fi\endgroup%
\begin{picture}(6494,3298)(804,-3176)
\put(2261,-3136){\makebox(0,0)[b]{\smash{{\SetFigFont{11}{13.2}{\familydefault}{\mddefault}{\updefault}{\color[rgb]{0,0,0}(a)}%
}}}}
\put(5701,-3136){\makebox(0,0)[b]{\smash{{\SetFigFont{11}{13.2}{\familydefault}{\mddefault}{\updefault}{\color[rgb]{0,0,0}(b)}%
}}}}
\put(3601,-2836){\makebox(0,0)[b]{\smash{{\SetFigFont{11}{13.2}{\familydefault}{\mddefault}{\updefault}{\color[rgb]{0,0,0}1}%
}}}}
\put(1576,-2811){\makebox(0,0)[b]{\smash{{\SetFigFont{11}{13.2}{\familydefault}{\mddefault}{\updefault}{\color[rgb]{0,0,0}$x_{1}$}%
}}}}
\put(2101,-2811){\makebox(0,0)[b]{\smash{{\SetFigFont{11}{13.2}{\familydefault}{\mddefault}{\updefault}{\color[rgb]{0,0,0}$x_{2}$}%
}}}}
\put(2551,-2811){\makebox(0,0)[b]{\smash{{\SetFigFont{11}{13.2}{\familydefault}{\mddefault}{\updefault}{\color[rgb]{0,0,0}$x_{3}$}%
}}}}
\put(3151,-2811){\makebox(0,0)[b]{\smash{{\SetFigFont{11}{13.2}{\familydefault}{\mddefault}{\updefault}{\color[rgb]{0,0,0}$x_{4}$}%
}}}}
\put(7051, 14){\makebox(0,0)[b]{\smash{{\SetFigFont{11}{13.2}{\familydefault}{\mddefault}{\updefault}{\color[rgb]{0,0,0}$n=1$}%
}}}}
\put(5776, 14){\makebox(0,0)[b]{\smash{{\SetFigFont{11}{13.2}{\familydefault}{\mddefault}{\updefault}{\color[rgb]{0,0,0}$n=2$}%
}}}}
\put(2326, 14){\makebox(0,0)[b]{\smash{{\SetFigFont{11}{13.2}{\familydefault}{\mddefault}{\updefault}{\color[rgb]{0,0,0}$n=2$}%
}}}}
\put(1176, 14){\makebox(0,0)[lb]{\smash{{\SetFigFont{11}{13.2}{\familydefault}{\mddefault}{\updefault}{\color[rgb]{0,0,0}$n=10$}%
}}}}
\put(4626, 14){\makebox(0,0)[lb]{\smash{{\SetFigFont{11}{13.2}{\familydefault}{\mddefault}{\updefault}{\color[rgb]{0,0,0}$n=10$}%
}}}}
\put(4351,-136){\makebox(0,0)[b]{\smash{{\SetFigFont{11}{13.2}{\familydefault}{\mddefault}{\updefault}{\color[rgb]{0,0,0}1}%
}}}}
\put(901,-2811){\makebox(0,0)[b]{\smash{{\SetFigFont{11}{13.2}{\familydefault}{\mddefault}{\updefault}{\color[rgb]{0,0,0}0}%
}}}}
\put(4351,-2811){\makebox(0,0)[b]{\smash{{\SetFigFont{11}{13.2}{\familydefault}{\mddefault}{\updefault}{\color[rgb]{0,0,0}0}%
}}}}
\put(901,-136){\makebox(0,0)[b]{\smash{{\SetFigFont{11}{13.2}{\familydefault}{\mddefault}{\updefault}{\color[rgb]{0,0,0}1}%
}}}}
\put(4751,-2811){\makebox(0,0)[b]{\smash{{\SetFigFont{12}{14.4}{\familydefault}{\mddefault}{\updefault}{\color[rgb]{0,0,0}$x_{1}$}%
}}}}
\put(5251,-2811){\makebox(0,0)[b]{\smash{{\SetFigFont{11}{13.2}{\familydefault}{\mddefault}{\updefault}{\color[rgb]{0,0,0}$x_{3}$}%
}}}}
\put(7051,-2836){\makebox(0,0)[b]{\smash{{\SetFigFont{11}{13.2}{\familydefault}{\mddefault}{\updefault}{\color[rgb]{0,0,0}1}%
}}}}
\put(4426,-586){\makebox(0,0)[rb]{\smash{{\SetFigFont{11}{13.2}{\familydefault}{\mddefault}{\updefault}{\color[rgb]{0,0,0}$f(x_{4})$}%
}}}}
\put(4426,-1186){\makebox(0,0)[rb]{\smash{{\SetFigFont{11}{13.2}{\familydefault}{\mddefault}{\updefault}{\color[rgb]{0,0,0}$f(x_{3})$}%
}}}}
\put(4426,-1636){\makebox(0,0)[rb]{\smash{{\SetFigFont{11}{13.2}{\familydefault}{\mddefault}{\updefault}{\color[rgb]{0,0,0}$f(x_{2})$}%
}}}}
\put(4426,-2161){\makebox(0,0)[rb]{\smash{{\SetFigFont{11}{13.2}{\familydefault}{\mddefault}{\updefault}{\color[rgb]{0,0,0}$f(x_{1})$}%
}}}}
\put(3526, 14){\makebox(0,0)[b]{\smash{{\SetFigFont{11}{13.2}{\familydefault}{\mddefault}{\updefault}{\color[rgb]{0,0,0}$n=1$}%
}}}}
\put(6816,-866){\makebox(0,0)[lb]{\smash{{\SetFigFont{11}{13.2}{\familydefault}{\mddefault}{\updefault}{\color[rgb]{0,0,0}$g$}%
}}}}
\put(1691,-496){\makebox(0,0)[rb]{\smash{{\SetFigFont{11}{13.2}{\familydefault}{\mddefault}{\updefault}{\color[rgb]{0,0,0}$f$}%
}}}}
\put(2891,-2066){\makebox(0,0)[lb]{\smash{{\SetFigFont{11}{13.2}{\familydefault}{\mddefault}{\updefault}{\color[rgb]{0,0,0}$g$}%
}}}}
\put(5551,-2811){\makebox(0,0)[b]{\smash{{\SetFigFont{11}{13.2}{\familydefault}{\mddefault}{\updefault}{\color[rgb]{0,0,0}$x_{4}$}%
}}}}
\put(5026,-2811){\makebox(0,0)[b]{\smash{{\SetFigFont{11}{13.2}{\familydefault}{\mddefault}{\updefault}{\color[rgb]{0,0,0}$x_{2}$}%
}}}}
\end{picture}%

%% file: tent4.pstex_t
\begin{picture}(0,0)%
\includegraphics{tent4.pstex}%
\end{picture}%
\setlength{\unitlength}{3828sp}%
\begingroup\makeatletter\ifx\SetFigFont\undefined%
\gdef\SetFigFont#1#2#3#4#5{%
  \reset@font\fontsize{#1}{#2pt}%
  \fontfamily{#3}\fontseries{#4}\fontshape{#5}%
  \selectfont}%
\fi\endgroup%
\begin{picture}(7065,3759)(549,-3565)
\put(2326,-3511){\makebox(0,0)[b]{\smash{{\SetFigFont{12}{14.4}{\familydefault}{\mddefault}{\updefault}{\color[rgb]{0,0,0}(a)}%
}}}}
\put(5926,-3511){\makebox(0,0)[b]{\smash{{\SetFigFont{12}{14.4}{\familydefault}{\mddefault}{\updefault}{\color[rgb]{0,0,0}(b)}%
}}}}
\put(626, 89){\makebox(0,0)[rb]{\smash{{\SetFigFont{11}{13.2}{\familydefault}{\mddefault}{\updefault}{\color[rgb]{0,0,0}1}%
}}}}
\put(4226, 89){\makebox(0,0)[rb]{\smash{{\SetFigFont{11}{13.2}{\familydefault}{\mddefault}{\updefault}{\color[rgb]{0,0,0}1}%
}}}}
\put(7576,-3211){\makebox(0,0)[b]{\smash{{\SetFigFont{11}{13.2}{\familydefault}{\mddefault}{\updefault}{\color[rgb]{0,0,0}1}%
}}}}
\put(2326,-3211){\makebox(0,0)[b]{\smash{{\SetFigFont{11}{13.2}{\familydefault}{\mddefault}{\updefault}{\color[rgb]{0,0,0}First 4 iterates of tent map}%
}}}}
\put(4251,-3211){\makebox(0,0)[b]{\smash{{\SetFigFont{11}{13.2}{\familydefault}{\mddefault}{\updefault}{\color[rgb]{0,0,0}0}%
}}}}
\put(5926,-3211){\makebox(0,0)[b]{\smash{{\SetFigFont{11}{13.2}{\familydefault}{\mddefault}{\updefault}{\color[rgb]{0,0,0}Graph of first 4 |sine| maps}%
}}}}
\put(3976,-3211){\makebox(0,0)[b]{\smash{{\SetFigFont{11}{13.2}{\familydefault}{\mddefault}{\updefault}{\color[rgb]{0,0,0}1}%
}}}}
\put(626,-3211){\makebox(0,0)[b]{\smash{{\SetFigFont{11}{13.2}{\familydefault}{\mddefault}{\updefault}{\color[rgb]{0,0,0}0}%
}}}}
\end{picture}%

%% file: Zorn.pstex_t
\begin{picture}(0,0)%
\includegraphics{Zorn.pstex}%
\end{picture}%
\setlength{\unitlength}{3552sp}%
\begingroup\makeatletter\ifx\SetFigFont\undefined%
\gdef\SetFigFont#1#2#3#4#5{%
  \reset@font\fontsize{#1}{#2pt}%
  \fontfamily{#3}\fontseries{#4}\fontshape{#5}%
  \selectfont}%
\fi\endgroup%
\begin{picture}(7063,5119)(1285,-4708)
\put(5851,-3686){\makebox(0,0)[b]{\smash{{\SetFigFont{12}{14.4}{\familydefault}{\mddefault}{\updefault}{\color[rgb]{0,0,0}Zorn Lemma}%
}}}}
\put(5851,-861){\makebox(0,0)[b]{\smash{{\SetFigFont{12}{14.4}{\familydefault}{\mddefault}{\updefault}{\color[rgb]{0,0,0}Tower Theorem}%
}}}}
\put(2576,-111){\makebox(0,0)[b]{\smash{{\SetFigFont{12}{14.4}{\familydefault}{\mddefault}{\updefault}{\color[rgb]{0,0,0}\( (X,\preceq ) \)}%
}}}}
\put(5851,-4461){\makebox(0,0)[b]{\smash{{\SetFigFont{12}{14.4}{\familydefault}{\mddefault}{\updefault}{\color[rgb]{0,0,0}\( (u\in X\succeq c)\textrm{ }(\forall c\in (C_{\leftarrow },\preceq )) \)}%
}}}}
\put(5851,-1661){\makebox(0,0)[b]{\smash{{\SetFigFont{12}{14.4}{\familydefault}{\mddefault}{\updefault}{\color[rgb]{0,0,0}\( \mathcal{C}_{\textrm{T}}=\textstyle \bigcap \{\mathcal{T}\subseteq (\mathcal{X},\subseteq )\! :\textrm{ }\mathcal{T}\textrm{ is a }C_{0}-\textrm{tower}\} \)}%
}}}}
\put(5851,-2886){\makebox(0,0)[b]{\smash{{\SetFigFont{12}{14.4}{\familydefault}{\mddefault}{\updefault}{\color[rgb]{0,0,0}\( \sup _{\mathcal{C}_{\textrm{T}}}(\mathcal{C}_{\textrm{T}})=C_{\leftarrow }=g(C_{\leftarrow })\in \mathcal{C}_{\textrm{T}}=\,_{\rightarrow}\mathcal{T} \subseteq (\mathcal{X},\subseteq ) \)}%
}}}}
\put(5851,-111){\makebox(0,0)[b]{\smash{{\SetFigFont{12}{14.4}{\familydefault}{\mddefault}{\updefault}{\color[rgb]{0,0,0}\( \mathcal{X}=\{C\subseteq X\! :C\textrm{ is a chain in }(X,\preceq )\} \)
}%
}}}}
\put(2901,-3161){\makebox(0,0)[rb]{\smash{{\SetFigFont{12}{14.4}{\familydefault}{\mddefault}{\updefault}{\color[rgb]{0,0,0}Hausdorff Maximal}%
}}}}
\put(2901,-3511){\makebox(0,0)[rb]{\smash{{\SetFigFont{12}{14.4}{\familydefault}{\mddefault}{\updefault}{\color[rgb]{0,0,0}Chain Theorem}%
}}}}
\end{picture}%

%% file: order.pstex_t
\begin{picture}(0,0)%
\includegraphics{order.pstex}%
\end{picture}%
\setlength{\unitlength}{3828sp}%
\begingroup\makeatletter\ifx\SetFigFont\undefined%
\gdef\SetFigFont#1#2#3#4#5{%
  \reset@font\fontsize{#1}{#2pt}%
  \fontfamily{#3}\fontseries{#4}\fontshape{#5}%
  \selectfont}%
\fi\endgroup%
\begin{picture}(6501,3535)(687,-3451)
\put(5888,-3411){\makebox(0,0)[b]{\smash{{\SetFigFont{11}{13.2}{\familydefault}{\mddefault}{\updefault}{\color[rgb]{0,0,0}(b)}%
}}}}
\put(2326,-3411){\makebox(0,0)[b]{\smash{{\SetFigFont{11}{13.2}{\familydefault}{\mddefault}{\updefault}{\color[rgb]{0,0,0}(a)}%
}}}}
\put(5888,-111){\makebox(0,0)[b]{\smash{{\SetFigFont{11}{13.2}{\familydefault}{\mddefault}{\updefault}{\color[rgb]{0,0,0}$\{a,b,c\}$}%
}}}}
\put(5888,-3021){\makebox(0,0)[b]{\smash{{\SetFigFont{11}{13.2}{\familydefault}{\mddefault}{\updefault}{\color[rgb]{0,0,0}$\emptyset$}%
}}}}
\put(5888,-1786){\makebox(0,0)[b]{\smash{{\SetFigFont{11}{13.2}{\familydefault}{\mddefault}{\updefault}{\color[rgb]{0,0,0}$\{b\}$}%
}}}}
\put(3226,-21){\makebox(0,0)[b]{\smash{{\SetFigFont{11}{13.2}{\familydefault}{\mddefault}{\updefault}{\color[rgb]{0,0,0}16}%
}}}}
\put(1426,-21){\makebox(0,0)[b]{\smash{{\SetFigFont{11}{13.2}{\familydefault}{\mddefault}{\updefault}{\color[rgb]{0,0,0}15}%
}}}}
\put(4603,-1936){\makebox(0,0)[rb]{\smash{{\SetFigFont{11}{13.2}{\familydefault}{\mddefault}{\updefault}{\color[rgb]{0,0,0}$\{a\}$}%
}}}}
\put(7188,-1111){\makebox(0,0)[lb]{\smash{{\SetFigFont{11}{13.2}{\familydefault}{\mddefault}{\updefault}{\color[rgb]{0,0,0}$\{b,c\}$}%
}}}}
\put(7188,-1936){\makebox(0,0)[lb]{\smash{{\SetFigFont{11}{13.2}{\familydefault}{\mddefault}{\updefault}{\color[rgb]{0,0,0}$\{c\}$}%
}}}}
\put(4603,-1111){\makebox(0,0)[rb]{\smash{{\SetFigFont{11}{13.2}{\familydefault}{\mddefault}{\updefault}{\color[rgb]{0,0,0}$\{a,b\}$}%
}}}}
\put(5888,-1336){\makebox(0,0)[b]{\smash{{\SetFigFont{11}{13.2}{\familydefault}{\mddefault}{\updefault}{\color[rgb]{0,0,0}$\{a,c\}$}%
}}}}
\put(1331,-2321){\makebox(0,0)[rb]{\smash{{\SetFigFont{11}{13.2}{\familydefault}{\mddefault}{\updefault}{\color[rgb]{0,0,0}1}%
}}}}
\put(951,-1621){\makebox(0,0)[rb]{\smash{{\SetFigFont{11}{13.2}{\familydefault}{\mddefault}{\updefault}{\color[rgb]{0,0,0}3}%
}}}}
\put(3706,-1621){\makebox(0,0)[lb]{\smash{{\SetFigFont{11}{13.2}{\familydefault}{\mddefault}{\updefault}{\color[rgb]{0,0,0}6}%
}}}}
\put(3341,-2321){\makebox(0,0)[lb]{\smash{{\SetFigFont{11}{13.2}{\familydefault}{\mddefault}{\updefault}{\color[rgb]{0,0,0}2}%
}}}}
\put(2326,-3071){\makebox(0,0)[b]{\smash{{\SetFigFont{11}{13.2}{\familydefault}{\mddefault}{\updefault}{\color[rgb]{0,0,0}0}%
}}}}
\put(2141,-836){\makebox(0,0)[b]{\smash{{\SetFigFont{11}{13.2}{\familydefault}{\mddefault}{\updefault}{\color[rgb]{0,0,0}10}%
}}}}
\put(2526,-836){\makebox(0,0)[b]{\smash{{\SetFigFont{11}{13.2}{\familydefault}{\mddefault}{\updefault}{\color[rgb]{0,0,0}11}%
}}}}
\put(3926,-836){\makebox(0,0)[b]{\smash{{\SetFigFont{11}{13.2}{\familydefault}{\mddefault}{\updefault}{\color[rgb]{0,0,0}14}%
}}}}
\put(726,-836){\makebox(0,0)[b]{\smash{{\SetFigFont{11}{13.2}{\familydefault}{\mddefault}{\updefault}{\color[rgb]{0,0,0}7}%
}}}}
\put(1226,-836){\makebox(0,0)[b]{\smash{{\SetFigFont{11}{13.2}{\familydefault}{\mddefault}{\updefault}{\color[rgb]{0,0,0}8}%
}}}}
\put(3466,-836){\makebox(0,0)[b]{\smash{{\SetFigFont{11}{13.2}{\familydefault}{\mddefault}{\updefault}{\color[rgb]{0,0,0}13}%
}}}}
\end{picture}%

%% file: logcob357_a.pstex_t
\begin{picture}(0,0)%
\includegraphics{logcob357_a.pstex}%
\end{picture}%
\setlength{\unitlength}{3710sp}%
\begingroup\makeatletter\ifx\SetFigFont\undefined%
\gdef\SetFigFont#1#2#3#4#5{%
  \reset@font\fontsize{#1}{#2pt}%
  \fontfamily{#3}\fontseries{#4}\fontshape{#5}%
  \selectfont}%
\fi\endgroup%
\begin{picture}(7395,3664)(202,-3415)
\put(2251,-3361){\makebox(0,0)[b]{\smash{{\SetFigFont{11}{13.2}{\familydefault}{\mddefault}{\updefault}{\color[rgb]{0,0,0}(a)}%
}}}}
\put(6001,-3361){\makebox(0,0)[b]{\smash{{\SetFigFont{11}{13.2}{\familydefault}{\mddefault}{\updefault}{\color[rgb]{0,0,0}(b)}%
}}}}
\put(526,-3136){\makebox(0,0)[b]{\smash{{\SetFigFont{11}{13.2}{\familydefault}{\mddefault}{\updefault}{\color[rgb]{0,0,0}.473}%
}}}}
\put(526,-2911){\makebox(0,0)[rb]{\smash{{\SetFigFont{11}{13.2}{\familydefault}{\mddefault}{\updefault}{\color[rgb]{0,0,0}.5}%
}}}}
\put(4166,114){\makebox(0,0)[rb]{\smash{{\SetFigFont{11}{13.2}{\familydefault}{\mddefault}{\updefault}{\color[rgb]{0,0,0}1}%
}}}}
\put(7551,-3136){\makebox(0,0)[b]{\smash{{\SetFigFont{11}{13.2}{\familydefault}{\mddefault}{\updefault}{\color[rgb]{0,0,0}1}%
}}}}
\put(2251,-3086){\makebox(0,0)[b]{\smash{{\SetFigFont{11}{13.2}{\familydefault}{\mddefault}{\updefault}{\color[rgb]{0,0,0}Order at $\lambda=3.569$}%
}}}}
\put(2251,-2836){\makebox(0,0)[b]{\smash{{\SetFigFont{11}{13.2}{\familydefault}{\mddefault}{\updefault}{\color[rgb]{0,0,0}Iterate 9000 of logistic map}%
}}}}
\put(6001,-3086){\makebox(0,0)[b]{\smash{{\SetFigFont{11}{13.2}{\familydefault}{\mddefault}{\updefault}{\color[rgb]{0,0,0}at $\lambda=3.569$}%
}}}}
\put(6001,-2836){\makebox(0,0)[b]{\smash{{\SetFigFont{11}{13.2}{\familydefault}{\mddefault}{\updefault}{\color[rgb]{0,0,0}9000 iterations on logistic map}%
}}}}
\put(4186,-3136){\makebox(0,0)[b]{\smash{{\SetFigFont{11}{13.2}{\familydefault}{\mddefault}{\updefault}{\color[rgb]{0,0,0}0}%
}}}}
\put(3901,-3136){\makebox(0,0)[rb]{\smash{{\SetFigFont{11}{13.2}{\familydefault}{\mddefault}{\updefault}{\color[rgb]{0,0,0}.488}%
}}}}
\put(4566,-3136){\makebox(0,0)[b]{\smash{{\SetFigFont{11}{13.2}{\familydefault}{\mddefault}{\updefault}{\color[rgb]{0,0,0}0.1}%
}}}}
\put(526, 89){\makebox(0,0)[rb]{\smash{{\SetFigFont{11}{13.2}{\familydefault}{\mddefault}{\updefault}{\color[rgb]{0,0,0}.507}%
}}}}
\end{picture}%

%% file: logcob357_b.pstex_t
\begin{picture}(0,0)%
\includegraphics{logcob357_b.pstex}%
\end{picture}%
\setlength{\unitlength}{3710sp}%
\begingroup\makeatletter\ifx\SetFigFont\undefined%
\gdef\SetFigFont#1#2#3#4#5{%
  \reset@font\fontsize{#1}{#2pt}%
  \fontfamily{#3}\fontseries{#4}\fontshape{#5}%
  \selectfont}%
\fi\endgroup%
\begin{picture}(7345,3639)(202,-6865)
\put(2326,-6811){\makebox(0,0)[b]{\smash{{\SetFigFont{11}{13.2}{\familydefault}{\mddefault}{\updefault}{\color[rgb]{0,0,0}(c)}%
}}}}
\put(5851,-6811){\makebox(0,0)[b]{\smash{{\SetFigFont{11}{13.2}{\familydefault}{\mddefault}{\updefault}{\color[rgb]{0,0,0}(d)}%
}}}}
\put(526,-3361){\makebox(0,0)[rb]{\smash{{\SetFigFont{11}{13.2}{\familydefault}{\mddefault}{\updefault}{\color[rgb]{0,0,0}.511}%
}}}}
\put(4126,-3361){\makebox(0,0)[rb]{\smash{{\SetFigFont{11}{13.2}{\familydefault}{\mddefault}{\updefault}{\color[rgb]{0,0,0}1}%
}}}}
\put(2326,-6286){\makebox(0,0)[b]{\smash{{\SetFigFont{11}{13.2}{\familydefault}{\mddefault}{\updefault}{\color[rgb]{0,0,0}Iterate 9000 of logistic map}%
}}}}
\put(5851,-6286){\makebox(0,0)[b]{\smash{{\SetFigFont{11}{13.2}{\familydefault}{\mddefault}{\updefault}{\color[rgb]{0,0,0}9000 iterations on logistic map}%
}}}}
\put(5851,-6536){\makebox(0,0)[b]{\smash{{\SetFigFont{11}{13.2}{\familydefault}{\mddefault}{\updefault}{\color[rgb]{0,0,0}at $\lambda_*=3.5699456$}%
}}}}
\put(4551,-6561){\makebox(0,0)[b]{\smash{{\SetFigFont{11}{13.2}{\familydefault}{\mddefault}{\updefault}{\color[rgb]{0,0,0}0.1}%
}}}}
\put(4201,-6561){\makebox(0,0)[b]{\smash{{\SetFigFont{11}{13.2}{\familydefault}{\mddefault}{\updefault}{\color[rgb]{0,0,0}0}%
}}}}
\put(7501,-6561){\makebox(0,0)[b]{\smash{{\SetFigFont{11}{13.2}{\familydefault}{\mddefault}{\updefault}{\color[rgb]{0,0,0}1}%
}}}}
\put(601,-6561){\makebox(0,0)[b]{\smash{{\SetFigFont{11}{13.2}{\familydefault}{\mddefault}{\updefault}{\color[rgb]{0,0,0}.472}%
}}}}
\put(526,-6361){\makebox(0,0)[rb]{\smash{{\SetFigFont{11}{13.2}{\familydefault}{\mddefault}{\updefault}{\color[rgb]{0,0,0}.493}%
}}}}
\put(3976,-6561){\makebox(0,0)[rb]{\smash{{\SetFigFont{11}{13.2}{\familydefault}{\mddefault}{\updefault}{\color[rgb]{0,0,0}.487}%
}}}}
\put(2326,-6536){\makebox(0,0)[b]{\smash{{\SetFigFont{11}{13.2}{\familydefault}{\mddefault}{\updefault}{\color[rgb]{0,0,0}"Edge of chaos" at $\lambda=\lambda_*$}%
}}}}
\end{picture}%

%% file: logcob357_c.pstex_t
\begin{picture}(0,0)%
\includegraphics{logcob357_c.pstex}%
\end{picture}%
\setlength{\unitlength}{3710sp}%
\begingroup\makeatletter\ifx\SetFigFont\undefined%
\gdef\SetFigFont#1#2#3#4#5{%
  \reset@font\fontsize{#1}{#2pt}%
  \fontfamily{#3}\fontseries{#4}\fontshape{#5}%
  \selectfont}%
\fi\endgroup%
\begin{picture}(7395,3664)(202,-10315)
\put(5851,-10261){\makebox(0,0)[b]{\smash{{\SetFigFont{11}{13.2}{\familydefault}{\mddefault}{\updefault}{\color[rgb]{0,0,0}(f)}%
}}}}
\put(2251,-10261){\makebox(0,0)[b]{\smash{{\SetFigFont{11}{13.2}{\familydefault}{\mddefault}{\updefault}{\color[rgb]{0,0,0}(e)}%
}}}}
\put(601,-10036){\makebox(0,0)[b]{\smash{{\SetFigFont{11}{13.2}{\familydefault}{\mddefault}{\updefault}{\color[rgb]{0,0,0}.472}%
}}}}
\put(526,-6811){\makebox(0,0)[rb]{\smash{{\SetFigFont{11}{13.2}{\familydefault}{\mddefault}{\updefault}{\color[rgb]{0,0,0}.511}%
}}}}
\put(526,-9811){\makebox(0,0)[rb]{\smash{{\SetFigFont{11}{13.2}{\familydefault}{\mddefault}{\updefault}{\color[rgb]{0,0,0}.493}%
}}}}
\put(2251,-9986){\makebox(0,0)[b]{\smash{{\SetFigFont{11}{13.2}{\familydefault}{\mddefault}{\updefault}{\color[rgb]{0,0,0}Chaos at $\lambda=3.57$}%
}}}}
\put(2251,-9736){\makebox(0,0)[b]{\smash{{\SetFigFont{11}{13.2}{\familydefault}{\mddefault}{\updefault}{\color[rgb]{0,0,0}Iterate 9000 of logistic map}%
}}}}
\put(3851,-10036){\makebox(0,0)[b]{\smash{{\SetFigFont{11}{13.2}{\familydefault}{\mddefault}{\updefault}{\color[rgb]{0,0,0}.487}%
}}}}
\put(4201,-10036){\makebox(0,0)[b]{\smash{{\SetFigFont{11}{13.2}{\familydefault}{\mddefault}{\updefault}{\color[rgb]{0,0,0}0}%
}}}}
\put(4151,-6786){\makebox(0,0)[rb]{\smash{{\SetFigFont{11}{13.2}{\familydefault}{\mddefault}{\updefault}{\color[rgb]{0,0,0}1}%
}}}}
\put(5851,-9736){\makebox(0,0)[b]{\smash{{\SetFigFont{11}{13.2}{\familydefault}{\mddefault}{\updefault}{\color[rgb]{0,0,0}9000 iterations on logistic map}%
}}}}
\put(5851,-9986){\makebox(0,0)[b]{\smash{{\SetFigFont{11}{13.2}{\familydefault}{\mddefault}{\updefault}{\color[rgb]{0,0,0}at $\lambda=3.57$}%
}}}}
\put(4551,-10036){\makebox(0,0)[b]{\smash{{\SetFigFont{11}{13.2}{\familydefault}{\mddefault}{\updefault}{\color[rgb]{0,0,0}0.1}%
}}}}
\put(7551,-10036){\makebox(0,0)[b]{\smash{{\SetFigFont{11}{13.2}{\familydefault}{\mddefault}{\updefault}{\color[rgb]{0,0,0}1}%
}}}}
\end{picture}%

%% file: log357_a.pstex_t
\begin{picture}(0,0)%
\includegraphics{log357_a.pstex}%
\end{picture}%
\setlength{\unitlength}{3828sp}%
\begingroup\makeatletter\ifx\SetFigFont\undefined%
\gdef\SetFigFont#1#2#3#4#5{%
  \reset@font\fontsize{#1}{#2pt}%
  \fontfamily{#3}\fontseries{#4}\fontshape{#5}%
  \selectfont}%
\fi\endgroup%
\begin{picture}(7095,3634)(549,-3415)
\put(2251,-3361){\makebox(0,0)[b]{\smash{{\SetFigFont{12}{14.4}{\familydefault}{\mddefault}{\updefault}{\color[rgb]{0,0,0}(a)}%
}}}}
\put(5926,-3361){\makebox(0,0)[b]{\smash{{\SetFigFont{12}{14.4}{\familydefault}{\mddefault}{\updefault}{\color[rgb]{0,0,0}(b)}%
}}}}
\put(626,-636){\makebox(0,0)[rb]{\smash{{\SetFigFont{11}{13.2}{\rmdefault}{\mddefault}{\itdefault}{\color[rgb]{0,0,0}2}%
}}}}
\put(626,-811){\makebox(0,0)[rb]{\smash{{\SetFigFont{11}{13.2}{\rmdefault}{\mddefault}{\itdefault}{\color[rgb]{0,0,0}6}%
}}}}
\put(626,-1211){\makebox(0,0)[rb]{\smash{{\SetFigFont{11}{13.2}{\rmdefault}{\mddefault}{\itdefault}{\color[rgb]{0,0,0}3}%
}}}}
\put(626,-1036){\makebox(0,0)[rb]{\smash{{\SetFigFont{11}{13.2}{\rmdefault}{\mddefault}{\itdefault}{\color[rgb]{0,0,0}7}%
}}}}
\put(4251,-3136){\makebox(0,0)[b]{\smash{{\SetFigFont{11}{13.2}{\familydefault}{\mddefault}{\updefault}{\color[rgb]{0,0,0}0}%
}}}}
\put(4226, 89){\makebox(0,0)[b]{\smash{{\SetFigFont{11}{13.2}{\familydefault}{\mddefault}{\updefault}{\color[rgb]{0,0,0} 1}%
}}}}
\put(651,114){\makebox(0,0)[rb]{\smash{{\SetFigFont{11}{13.2}{\familydefault}{\mddefault}{\updefault}{\color[rgb]{0,0,0}1}%
}}}}
\put(626,-1511){\makebox(0,0)[rb]{\smash{{\SetFigFont{11}{13.2}{\rmdefault}{\mddefault}{\itdefault}{\color[rgb]{0,0,0}1}%
}}}}
\put(1351,-3136){\makebox(0,0)[b]{\smash{{\SetFigFont{11}{13.2}{\familydefault}{\mddefault}{\updefault}{\color[rgb]{0,0,0}0.2}%
}}}}
\put(676,-3136){\makebox(0,0)[rb]{\smash{{\SetFigFont{11}{13.2}{\familydefault}{\mddefault}{\updefault}{\color[rgb]{0,0,0}0}%
}}}}
\put(3976,-3136){\makebox(0,0)[b]{\smash{{\SetFigFont{11}{13.2}{\familydefault}{\mddefault}{\updefault}{\color[rgb]{0,0,0}1}%
}}}}
\put(4251,-511){\makebox(0,0)[rb]{\smash{{\SetFigFont{11}{13.2}{\rmdefault}{\mddefault}{\itdefault}{\color[rgb]{0,0,0}9}%
}}}}
\put(4901,-686){\makebox(0,0)[rb]{\smash{{\SetFigFont{11}{13.2}{\rmdefault}{\mddefault}{\itdefault}{\color[rgb]{0,0,0}5}%
}}}}
\put(4901,-836){\makebox(0,0)[rb]{\smash{{\SetFigFont{11}{13.2}{\rmdefault}{\mddefault}{\itdefault}{\color[rgb]{0,0,0}2}%
}}}}
\put(7591,-3136){\makebox(0,0)[b]{\smash{{\SetFigFont{11}{13.2}{\familydefault}{\mddefault}{\updefault}{\color[rgb]{0,0,0} 1}%
}}}}
\put(2251,-2761){\makebox(0,0)[b]{\smash{{\SetFigFont{11}{13.2}{\familydefault}{\mddefault}{\updefault}{\color[rgb]{0,0,0}Graphical limit at 9001}%
}}}}
\put(2251,-2536){\makebox(0,0)[b]{\smash{{\SetFigFont{11}{13.2}{\familydefault}{\mddefault}{\updefault}{\color[rgb]{0,0,0}Stable 1-cycle, $\lambda=2.95$}%
}}}}
\put(4251,-1296){\makebox(0,0)[rb]{\smash{{\SetFigFont{11}{13.2}{\rmdefault}{\mddefault}{\itdefault}{\color[rgb]{0,0,0}10}%
}}}}
\put(5926,-2761){\makebox(0,0)[b]{\smash{{\SetFigFont{11}{13.2}{\familydefault}{\mddefault}{\updefault}{\color[rgb]{0,0,0}Graphical limit at 9001-9002}%
}}}}
\put(5926,-2536){\makebox(0,0)[b]{\smash{{\SetFigFont{11}{13.2}{\familydefault}{\mddefault}{\updefault}{\color[rgb]{0,0,0}Stable 2-cycle, $\lambda=3.4$}%
}}}}
\end{picture}%

%% file: log357_b.pstex_t
\begin{picture}(0,0)%
\includegraphics{log357_b.pstex}%
\end{picture}%
\setlength{\unitlength}{3828sp}%
\begingroup\makeatletter\ifx\SetFigFont\undefined%
\gdef\SetFigFont#1#2#3#4#5{%
  \reset@font\fontsize{#1}{#2pt}%
  \fontfamily{#3}\fontseries{#4}\fontshape{#5}%
  \selectfont}%
\fi\endgroup%
\begin{picture}(7092,3559)(497,-6790)
\put(6001,-6736){\makebox(0,0)[b]{\smash{{\SetFigFont{12}{14.4}{\familydefault}{\mddefault}{\updefault}{\color[rgb]{0,0,0}(d)}%
}}}}
\put(2326,-6736){\makebox(0,0)[b]{\smash{{\SetFigFont{12}{14.4}{\familydefault}{\mddefault}{\updefault}{\color[rgb]{0,0,0}(c)}%
}}}}
\put(651,-4711){\makebox(0,0)[rb]{\smash{{\SetFigFont{11}{13.2}{\rmdefault}{\mddefault}{\itdefault}{\color[rgb]{0,0,0}6}%
}}}}
\put(651,-5161){\makebox(0,0)[rb]{\smash{{\SetFigFont{11}{13.2}{\rmdefault}{\mddefault}{\itdefault}{\color[rgb]{0,0,0}8}%
}}}}
\put(651,-4936){\makebox(0,0)[rb]{\smash{{\SetFigFont{11}{13.2}{\rmdefault}{\mddefault}{\itdefault}{\color[rgb]{0,0,0}10}%
}}}}
\put(651,-3961){\makebox(0,0)[rb]{\smash{{\SetFigFont{11}{13.2}{\rmdefault}{\mddefault}{\itdefault}{\color[rgb]{0,0,0}5}%
}}}}
\put(651,-4131){\makebox(0,0)[rb]{\smash{{\SetFigFont{11}{13.2}{\rmdefault}{\mddefault}{\itdefault}{\color[rgb]{0,0,0}1}%
}}}}
\put(651,-4336){\makebox(0,0)[rb]{\smash{{\SetFigFont{11}{13.2}{\rmdefault}{\mddefault}{\itdefault}{\color[rgb]{0,0,0}2}%
}}}}
\put(2326,-5986){\makebox(0,0)[b]{\smash{{\SetFigFont{11}{13.2}{\familydefault}{\mddefault}{\updefault}{\color[rgb]{0,0,0}Stable 4-cycle, $\lambda=3.5$}%
}}}}
\put(2326,-6211){\makebox(0,0)[b]{\smash{{\SetFigFont{11}{13.2}{\familydefault}{\mddefault}{\updefault}{\color[rgb]{0,0,0}Graphical limit at 9001-9004}%
}}}}
\put(4251,-3361){\makebox(0,0)[b]{\smash{{\SetFigFont{11}{13.2}{\familydefault}{\mddefault}{\updefault}{\color[rgb]{0,0,0}1}%
}}}}
\put(4276,-6586){\makebox(0,0)[b]{\smash{{\SetFigFont{11}{13.2}{\familydefault}{\mddefault}{\updefault}{\color[rgb]{0,0,0}0}%
}}}}
\put(6001,-6211){\makebox(0,0)[b]{\smash{{\SetFigFont{11}{13.2}{\familydefault}{\mddefault}{\updefault}{\color[rgb]{0,0,0}Graphical limit at 9001-9008}%
}}}}
\put(6001,-5986){\makebox(0,0)[b]{\smash{{\SetFigFont{11}{13.2}{\familydefault}{\mddefault}{\updefault}{\color[rgb]{0,0,0}Stable 8-cycle, $\lambda=3.55$}%
}}}}
\put(3001,-6586){\makebox(0,0)[b]{\smash{{\SetFigFont{11}{13.2}{\familydefault}{\mddefault}{\updefault}{\color[rgb]{0,0,0}0.7}%
}}}}
\put(3976,-6586){\makebox(0,0)[b]{\smash{{\SetFigFont{11}{13.2}{\familydefault}{\mddefault}{\updefault}{\color[rgb]{0,0,0}1}%
}}}}
\put(6601,-6586){\makebox(0,0)[b]{\smash{{\SetFigFont{11}{13.2}{\familydefault}{\mddefault}{\updefault}{\color[rgb]{0,0,0}0.7}%
}}}}
\put(701,-3336){\makebox(0,0)[rb]{\smash{{\SetFigFont{11}{13.2}{\familydefault}{\mddefault}{\updefault}{\color[rgb]{0,0,0}1}%
}}}}
\put(651,-3811){\makebox(0,0)[rb]{\smash{{\SetFigFont{11}{13.2}{\rmdefault}{\mddefault}{\itdefault}{\color[rgb]{0,0,0}9}%
}}}}
\put(676,-6586){\makebox(0,0)[b]{\smash{{\SetFigFont{11}{13.2}{\familydefault}{\mddefault}{\updefault}{\color[rgb]{0,0,0}0}%
}}}}
\put(7551,-6586){\makebox(0,0)[b]{\smash{{\SetFigFont{11}{13.2}{\familydefault}{\mddefault}{\updefault}{\color[rgb]{0,0,0}1}%
}}}}
\end{picture}%

%% file: tent17.pstex_t
\begin{picture}(0,0)%
\includegraphics{tent17.pstex}%
\end{picture}%
\setlength{\unitlength}{3828sp}%
\begingroup\makeatletter\ifx\SetFigFont\undefined%
\gdef\SetFigFont#1#2#3#4#5{%
  \reset@font\fontsize{#1}{#2pt}%
  \fontfamily{#3}\fontseries{#4}\fontshape{#5}%
  \selectfont}%
\fi\endgroup%
\begin{picture}(7078,3634)(774,-3640)
\put(6101,-3586){\makebox(0,0)[b]{\smash{{\SetFigFont{12}{14.4}{\familydefault}{\mddefault}{\updefault}{\color[rgb]{0,0,0}(b)}%
}}}}
\put(2501,-3586){\makebox(0,0)[b]{\smash{{\SetFigFont{12}{14.4}{\familydefault}{\mddefault}{\updefault}{\color[rgb]{0,0,0}(a)}%
}}}}
\put(4451,-3286){\makebox(0,0)[b]{\smash{{\SetFigFont{11}{13.2}{\familydefault}{\mddefault}{\updefault}{\color[rgb]{0,0,0}0}%
}}}}
\put(6101,-3286){\makebox(0,0)[b]{\smash{{\SetFigFont{11}{13.2}{\familydefault}{\mddefault}{\updefault}{\color[rgb]{0,0,0}Graph of |$\sin(2^{16}\pi x)$|}%
}}}}
\put(2501,-3286){\makebox(0,0)[b]{\smash{{\SetFigFont{11}{13.2}{\familydefault}{\mddefault}{\updefault}{\color[rgb]{0,0,0}17th iterate of tent map}%
}}}}
\put(851,-3286){\makebox(0,0)[b]{\smash{{\SetFigFont{11}{13.2}{\familydefault}{\mddefault}{\updefault}{\color[rgb]{0,0,0}0}%
}}}}
\put(851,-111){\makebox(0,0)[rb]{\smash{{\SetFigFont{11}{13.2}{\familydefault}{\mddefault}{\updefault}{\color[rgb]{0,0,0}1}%
}}}}
\put(4426,-111){\makebox(0,0)[rb]{\smash{{\SetFigFont{11}{13.2}{\familydefault}{\mddefault}{\updefault}{\color[rgb]{0,0,0}1}%
}}}}
\end{picture}%

%% file: omega.pstex_t
\begin{picture}(0,0)%
\includegraphics{omega.pstex}%
\end{picture}%
\setlength{\unitlength}{3947sp}%
\begingroup\makeatletter\ifx\SetFigFont\undefined%
\gdef\SetFigFont#1#2#3#4#5{%
  \reset@font\fontsize{#1}{#2pt}%
  \fontfamily{#3}\fontseries{#4}\fontshape{#5}%
  \selectfont}%
\fi\endgroup%
\begin{picture}(5809,4529)(276,-3710)
\put(2776, 14){\makebox(0,0)[rb]{\smash{{\SetFigFont{11}{13.2}{\familydefault}{\mddefault}{\updefault}{\color[rgb]{0,0,0}0.8}%
}}}}
\put(2776,-1111){\makebox(0,0)[rb]{\smash{{\SetFigFont{11}{13.2}{\familydefault}{\mddefault}{\updefault}{\color[rgb]{0,0,0}0.4}%
}}}}
\put(2776,-1686){\makebox(0,0)[rb]{\smash{{\SetFigFont{11}{13.2}{\familydefault}{\mddefault}{\updefault}{\color[rgb]{0,0,0}0.2}%
}}}}
\put(2776,-561){\makebox(0,0)[rb]{\smash{{\SetFigFont{11}{13.2}{\familydefault}{\mddefault}{\updefault}{\color[rgb]{0,0,0}0.6}%
}}}}
\put(1751,-2361){\makebox(0,0)[b]{\smash{{\SetFigFont{11}{13.2}{\familydefault}{\mddefault}{\updefault}{\color[rgb]{0,0,0}$-0.4$}%
}}}}
\put(2276,-2361){\makebox(0,0)[b]{\smash{{\SetFigFont{11}{13.2}{\familydefault}{\mddefault}{\updefault}{\color[rgb]{0,0,0}$-0.2$}%
}}}}
\put(3366,-2361){\makebox(0,0)[b]{\smash{{\SetFigFont{11}{13.2}{\familydefault}{\mddefault}{\updefault}{\color[rgb]{0,0,0}0.2}%
}}}}
\put(4366,-2361){\makebox(0,0)[b]{\smash{{\SetFigFont{11}{13.2}{\familydefault}{\mddefault}{\updefault}{\color[rgb]{0,0,0}0.6}%
}}}}
\put(3866,-2361){\makebox(0,0)[b]{\smash{{\SetFigFont{11}{13.2}{\familydefault}{\mddefault}{\updefault}{\color[rgb]{0,0,0}0.4}%
}}}}
\put(1241,-2361){\makebox(0,0)[b]{\smash{{\SetFigFont{11}{13.2}{\familydefault}{\mddefault}{\updefault}{\color[rgb]{0,0,0}$-0.6$}%
}}}}
\put(2926,-2361){\makebox(0,0)[b]{\smash{{\SetFigFont{11}{13.2}{\familydefault}{\mddefault}{\updefault}{\color[rgb]{0,0,0}0}%
}}}}
\put(4901,-2361){\makebox(0,0)[b]{\smash{{\SetFigFont{11}{13.2}{\familydefault}{\mddefault}{\updefault}{\color[rgb]{0,0,0}0.8}%
}}}}
\put(731,-2361){\makebox(0,0)[b]{\smash{{\SetFigFont{11}{13.2}{\familydefault}{\mddefault}{\updefault}{\color[rgb]{0,0,0}$-0.8$}%
}}}}
\put(4801,-3661){\makebox(0,0)[b]{\smash{{\SetFigFont{11}{13.2}{\familydefault}{\mddefault}{\updefault}{\color[rgb]{0,0,0}$x_{-1}$}%
}}}}
\put(5311,-3661){\makebox(0,0)[b]{\smash{{\SetFigFont{11}{13.2}{\familydefault}{\mddefault}{\updefault}{\color[rgb]{0,0,0}$x_{-3}$}%
}}}}
\put(4516,-1186){\makebox(0,0)[rb]{\smash{{\SetFigFont{11}{13.2}{\familydefault}{\mddefault}{\updefault}{\color[rgb]{0,0,0}$f_{\textrm{B}}^{2}$}%
}}}}
\put(4691,-1761){\makebox(0,0)[rb]{\smash{{\SetFigFont{11}{13.2}{\familydefault}{\mddefault}{\updefault}{\color[rgb]{0,0,0}$f_{\textrm{B}}^{3}$}%
}}}}
\put(4626,-3476){\makebox(0,0)[b]{\smash{{\SetFigFont{11}{13.2}{\familydefault}{\mddefault}{\updefault}{\color[rgb]{0,0,0}$x_1$}%
}}}}
\put(5101,-3476){\makebox(0,0)[b]{\smash{{\SetFigFont{11}{13.2}{\familydefault}{\mddefault}{\updefault}{\color[rgb]{0,0,0}$x_2$}%
}}}}
\put(5501,-911){\makebox(0,0)[lb]{\smash{{\SetFigFont{11}{13.2}{\familydefault}{\mddefault}{\updefault}{\color[rgb]{0,0,0}$x$}%
}}}}
\put(5476,-2236){\makebox(0,0)[lb]{\smash{{\SetFigFont{11}{13.2}{\familydefault}{\mddefault}{\updefault}{\color[rgb]{0,0,0}1}%
}}}}
\put(5476,-2936){\makebox(0,0)[lb]{\smash{{\SetFigFont{11}{13.2}{\familydefault}{\mddefault}{\updefault}{\color[rgb]{0,0,0}$-0.25$}%
}}}}
\put(5501,-36){\makebox(0,0)[lb]{\smash{{\SetFigFont{11}{13.2}{\familydefault}{\mddefault}{\updefault}{\color[rgb]{0,0,0}$x_{-1}$}%
}}}}
\put(4426,-386){\makebox(0,0)[rb]{\smash{{\SetFigFont{11}{13.2}{\familydefault}{\mddefault}{\updefault}{\color[rgb]{0,0,0}$f_{\textrm{B}}$}%
}}}}
\put(1566,139){\makebox(0,0)[b]{\smash{{\SetFigFont{11}{13.2}{\familydefault}{\mddefault}{\updefault}{\color[rgb]{0,0,0}$f={\displaystyle \frac{4x^{2}-1}{3}}$}%
}}}}
\put(2776,-3511){\makebox(0,0)[b]{\smash{{\SetFigFont{11}{13.2}{\familydefault}{\mddefault}{\updefault}{\color[rgb]{0,0,0}$-0.4$}%
}}}}
\put(2841,714){\makebox(0,0)[b]{\smash{{\SetFigFont{11}{13.2}{\familydefault}{\mddefault}{\updefault}{\color[rgb]{0,0,0}1}%
}}}}
\put(2701,-2771){\makebox(0,0)[b]{\smash{{\SetFigFont{11}{13.2}{\rmdefault}{\mddefault}{\itdefault}{\color[rgb]{0,0,0}2}%
}}}}
\put(826,-2771){\makebox(0,0)[b]{\smash{{\SetFigFont{11}{13.2}{\rmdefault}{\mddefault}{\itdefault}{\color[rgb]{0,0,0}4}%
}}}}
\put(501,-2771){\makebox(0,0)[b]{\smash{{\SetFigFont{11}{13.2}{\rmdefault}{\mddefault}{\itdefault}{\color[rgb]{0,0,0}5}%
}}}}
\put(2701,-3196){\makebox(0,0)[b]{\smash{{\SetFigFont{11}{13.2}{\rmdefault}{\mddefault}{\itdefault}{\color[rgb]{0,0,0}1}%
}}}}
\put(1576,-3196){\makebox(0,0)[b]{\smash{{\SetFigFont{11}{13.2}{\rmdefault}{\mddefault}{\itdefault}{\color[rgb]{0,0,0}2}%
}}}}
\put(501,-3196){\makebox(0,0)[b]{\smash{{\SetFigFont{11}{13.2}{\rmdefault}{\mddefault}{\itdefault}{\color[rgb]{0,0,0}4}%
}}}}
\end{picture}%

%% file: attractor_a.pstex_t
\begin{picture}(0,0)%
\includegraphics{attractor_a.pstex}%
\end{picture}%
\setlength{\unitlength}{3828sp}%
\begingroup\makeatletter\ifx\SetFigFont\undefined%
\gdef\SetFigFont#1#2#3#4#5{%
  \reset@font\fontsize{#1}{#2pt}%
  \fontfamily{#3}\fontseries{#4}\fontshape{#5}%
  \selectfont}%
\fi\endgroup%
\begin{picture}(6937,3609)(527,-3415)
\put(2251,-3361){\makebox(0,0)[b]{\smash{{\SetFigFont{12}{14.4}{\familydefault}{\mddefault}{\updefault}{\color[rgb]{0,0,0}(a)}%
}}}}
\put(5776,-3286){\makebox(0,0)[b]{\smash{{\SetFigFont{12}{14.4}{\familydefault}{\mddefault}{\updefault}{\color[rgb]{0,0,0}(b)}%
}}}}
\put(7426,-3136){\makebox(0,0)[b]{\smash{{\SetFigFont{11}{13.2}{\familydefault}{\mddefault}{\updefault}{\color[rgb]{0,0,0}1}%
}}}}
\put(4126, 89){\makebox(0,0)[b]{\smash{{\SetFigFont{11}{13.2}{\familydefault}{\mddefault}{\updefault}{\color[rgb]{0,0,0}1}%
}}}}
\put(601,-3136){\makebox(0,0)[b]{\smash{{\SetFigFont{11}{13.2}{\familydefault}{\mddefault}{\updefault}{\color[rgb]{0,0,0}0}%
}}}}
\put(566, 89){\makebox(0,0)[b]{\smash{{\SetFigFont{11}{13.2}{\familydefault}{\mddefault}{\updefault}{\color[rgb]{0,0,0}1}%
}}}}
\put(3876,-3136){\makebox(0,0)[b]{\smash{{\SetFigFont{11}{13.2}{\familydefault}{\mddefault}{\updefault}{\color[rgb]{0,0,0}1}%
}}}}
\put(4186,-3136){\makebox(0,0)[b]{\smash{{\SetFigFont{11}{13.2}{\familydefault}{\mddefault}{\updefault}{\color[rgb]{0,0,0}0}%
}}}}
\put(2251,-2611){\makebox(0,0)[b]{\smash{{\SetFigFont{11}{13.2}{\familydefault}{\mddefault}{\updefault}{\color[rgb]{0,0,0}$\lambda=3.5699456$}%
}}}}
\put(5776,-2836){\makebox(0,0)[b]{\smash{{\SetFigFont{11}{13.2}{\familydefault}{\mddefault}{\updefault}{\color[rgb]{0,0,0}Iterates $=2001-2004$}%
}}}}
\put(5776,-2611){\makebox(0,0)[b]{\smash{{\SetFigFont{11}{13.2}{\familydefault}{\mddefault}{\updefault}{\color[rgb]{0,0,0}$\lambda=3.575$}%
}}}}
\put(2251,-2836){\makebox(0,0)[b]{\smash{{\SetFigFont{11}{13.2}{\familydefault}{\mddefault}{\updefault}{\color[rgb]{0,0,0}Iterates $=2001-2004$}%
}}}}
\end{picture}%

%% file: attractor_b.pstex_t
\begin{picture}(0,0)%
\includegraphics{attractor_b.pstex}%
\end{picture}%
\setlength{\unitlength}{3828sp}%
\begingroup\makeatletter\ifx\SetFigFont\undefined%
\gdef\SetFigFont#1#2#3#4#5{%
  \reset@font\fontsize{#1}{#2pt}%
  \fontfamily{#3}\fontseries{#4}\fontshape{#5}%
  \selectfont}%
\fi\endgroup%
\begin{picture}(6937,3634)(527,-3415)
\put(2251,-3361){\makebox(0,0)[b]{\smash{{\SetFigFont{12}{14.4}{\familydefault}{\mddefault}{\updefault}{\color[rgb]{0,0,0}(c)}%
}}}}
\put(5776,-3286){\makebox(0,0)[b]{\smash{{\SetFigFont{12}{14.4}{\familydefault}{\mddefault}{\updefault}{\color[rgb]{0,0,0}(d)}%
}}}}
\put(2251,-2611){\makebox(0,0)[b]{\smash{{\SetFigFont{11}{13.2}{\familydefault}{\mddefault}{\updefault}{\color[rgb]{0,0,0}$\lambda=3.66$}%
}}}}
\put(2251,-2836){\makebox(0,0)[b]{\smash{{\SetFigFont{11}{13.2}{\familydefault}{\mddefault}{\updefault}{\color[rgb]{0,0,0}Iterates $=2001-2004$}%
}}}}
\put(5776,-2611){\makebox(0,0)[b]{\smash{{\SetFigFont{11}{13.2}{\familydefault}{\mddefault}{\updefault}{\color[rgb]{0,0,0}$\lambda=3.8$}%
}}}}
\put(5776,-2836){\makebox(0,0)[b]{\smash{{\SetFigFont{11}{13.2}{\familydefault}{\mddefault}{\updefault}{\color[rgb]{0,0,0}Iterates $=2001-2002$}%
}}}}
\put(7426,-3136){\makebox(0,0)[b]{\smash{{\SetFigFont{11}{13.2}{\familydefault}{\mddefault}{\updefault}{\color[rgb]{0,0,0}1}%
}}}}
\put(4126, 89){\makebox(0,0)[b]{\smash{{\SetFigFont{11}{13.2}{\familydefault}{\mddefault}{\updefault}{\color[rgb]{0,0,0}1}%
}}}}
\put(4201,-3136){\makebox(0,0)[b]{\smash{{\SetFigFont{11}{13.2}{\familydefault}{\mddefault}{\updefault}{\color[rgb]{0,0,0}0}%
}}}}
\put(601,-3136){\makebox(0,0)[b]{\smash{{\SetFigFont{11}{13.2}{\familydefault}{\mddefault}{\updefault}{\color[rgb]{0,0,0}0}%
}}}}
\put(3876,-3136){\makebox(0,0)[b]{\smash{{\SetFigFont{11}{13.2}{\familydefault}{\mddefault}{\updefault}{\color[rgb]{0,0,0}1}%
}}}}
\put(566,114){\makebox(0,0)[b]{\smash{{\SetFigFont{11}{13.2}{\familydefault}{\mddefault}{\updefault}{\color[rgb]{0,0,0}1}%
}}}}
\end{picture}%

%% file: attractor_c.pstex_t
\begin{picture}(0,0)%
\includegraphics{attractor_c.pstex}%
\end{picture}%
\setlength{\unitlength}{3828sp}%
\begingroup\makeatletter\ifx\SetFigFont\undefined%
\gdef\SetFigFont#1#2#3#4#5{%
  \reset@font\fontsize{#1}{#2pt}%
  \fontfamily{#3}\fontseries{#4}\fontshape{#5}%
  \selectfont}%
\fi\endgroup%
\begin{picture}(6912,3754)(552,-3540)
\put(2326,-3486){\makebox(0,0)[b]{\smash{{\SetFigFont{12}{14.4}{\familydefault}{\mddefault}{\updefault}{\color[rgb]{0,0,0}(e)}%
}}}}
\put(5776,-3486){\makebox(0,0)[b]{\smash{{\SetFigFont{12}{14.4}{\familydefault}{\mddefault}{\updefault}{\color[rgb]{0,0,0}(f)}%
}}}}
\put(5776,-3136){\makebox(0,0)[b]{\smash{{\SetFigFont{11}{13.2}{\familydefault}{\mddefault}{\updefault}{\color[rgb]{0,0,0}First 12 iterates}%
}}}}
\put(601,-3136){\makebox(0,0)[b]{\smash{{\SetFigFont{11}{13.2}{\familydefault}{\mddefault}{\updefault}{\color[rgb]{0,0,0}0}%
}}}}
\put(4201,-3136){\makebox(0,0)[b]{\smash{{\SetFigFont{11}{13.2}{\familydefault}{\mddefault}{\updefault}{\color[rgb]{0,0,0}0}%
}}}}
\put(4126, 89){\makebox(0,0)[b]{\smash{{\SetFigFont{11}{13.2}{\familydefault}{\mddefault}{\updefault}{\color[rgb]{0,0,0}1}%
}}}}
\put(7426,-3136){\makebox(0,0)[b]{\smash{{\SetFigFont{11}{13.2}{\familydefault}{\mddefault}{\updefault}{\color[rgb]{0,0,0}1}%
}}}}
\put(1351,-3136){\makebox(0,0)[b]{\smash{{\SetFigFont{11}{13.2}{\familydefault}{\mddefault}{\updefault}{\color[rgb]{0,0,0}$\lambda_4$}%
}}}}
\put(1926,-3136){\makebox(0,0)[b]{\smash{{\SetFigFont{11}{13.2}{\familydefault}{\mddefault}{\updefault}{\color[rgb]{0,0,0}$\lambda_*$}%
}}}}
\put(3901,-3136){\makebox(0,0)[b]{\smash{{\SetFigFont{11}{13.2}{\familydefault}{\mddefault}{\updefault}{\color[rgb]{0,0,0}4}%
}}}}
\put(1351,-2836){\makebox(0,0)[b]{\smash{{\SetFigFont{11}{13.2}{\familydefault}{\mddefault}{\updefault}{\color[rgb]{0,0,0}$\lambda_4=3.449$}%
}}}}
\put(591,109){\makebox(0,0)[b]{\smash{{\SetFigFont{11}{13.2}{\familydefault}{\mddefault}{\updefault}{\color[rgb]{0,0,0}1}%
}}}}
\end{picture}%

%% file: Poison.pstex_t
\begin{picture}(0,0)%
\includegraphics{Poison.pstex}%
\end{picture}%
\setlength{\unitlength}{3947sp}%
\begingroup\makeatletter\ifx\SetFigFont\undefined%
\gdef\SetFigFont#1#2#3#4#5{%
  \reset@font\fontsize{#1}{#2pt}%
  \fontfamily{#3}\fontseries{#4}\fontshape{#5}%
  \selectfont}%
\fi\endgroup%
\begin{picture}(6922,3348)(693,-3576)
\put(5851,-3536){\makebox(0,0)[b]{\smash{{\SetFigFont{11}{13.2}{\familydefault}{\mddefault}{\updefault}{\color[rgb]{0,0,0}(b)}%
}}}}
\put(2451,-3536){\makebox(0,0)[b]{\smash{{\SetFigFont{11}{13.2}{\familydefault}{\mddefault}{\updefault}{\color[rgb]{0,0,0}(a)}%
}}}}
\put(5851,-3286){\makebox(0,0)[b]{\smash{{\SetFigFont{11}{13.2}{\familydefault}{\mddefault}{\updefault}{\color[rgb]{0,0,0}$-32$}%
}}}}
\put(4426,-1786){\makebox(0,0)[rb]{\smash{{\SetFigFont{11}{13.2}{\familydefault}{\mddefault}{\updefault}{\color[rgb]{0,0,0}$-0.5$}%
}}}}
\put(5926,-1936){\makebox(0,0)[lb]{\smash{{\SetFigFont{11}{13.2}{\familydefault}{\mddefault}{\updefault}{\color[rgb]{0,0,0}0}%
}}}}
\put(901,-3261){\makebox(0,0)[b]{\smash{{\SetFigFont{11}{13.2}{\familydefault}{\mddefault}{\updefault}{\color[rgb]{0,0,0}$-0.5$}%
}}}}
\put(7416,-1786){\makebox(0,0)[lb]{\smash{{\SetFigFont{11}{13.2}{\familydefault}{\mddefault}{\updefault}{\color[rgb]{0,0,0}0.5}%
}}}}
\put(4001,-3261){\makebox(0,0)[b]{\smash{{\SetFigFont{11}{13.2}{\familydefault}{\mddefault}{\updefault}{\color[rgb]{0,0,0}0.5}%
}}}}
\put(2451,-336){\makebox(0,0)[b]{\smash{{\SetFigFont{11}{13.2}{\familydefault}{\mddefault}{\updefault}{\color[rgb]{0,0,0}20}%
}}}}
\put(2451,-3286){\makebox(0,0)[b]{\smash{{\SetFigFont{11}{13.2}{\familydefault}{\mddefault}{\updefault}{\color[rgb]{0,0,0}0}%
}}}}
\put(5901,-336){\makebox(0,0)[b]{\smash{{\SetFigFont{11}{13.2}{\familydefault}{\mddefault}{\updefault}{\color[rgb]{0,0,0}$32$}%
}}}}
\end{picture}%

%% file: Case.pstex_t
\begin{picture}(0,0)%
\includegraphics{Case.pstex}%
\end{picture}%
\setlength{\unitlength}{3947sp}%
\begingroup\makeatletter\ifx\SetFigFont\undefined%
\gdef\SetFigFont#1#2#3#4#5{%
  \reset@font\fontsize{#1}{#2pt}%
  \fontfamily{#3}\fontseries{#4}\fontshape{#5}%
  \selectfont}%
\fi\endgroup%
\begin{picture}(6050,6115)(1176,-6551)
\put(5776,-6511){\makebox(0,0)[b]{\smash{{\SetFigFont{11}{13.2}{\familydefault}{\mddefault}{\updefault}{\color[rgb]{0,0,0}(b)}%
}}}}
\put(2626,-6511){\makebox(0,0)[b]{\smash{{\SetFigFont{11}{13.2}{\familydefault}{\mddefault}{\updefault}{\color[rgb]{0,0,0}(a)}%
}}}}
\put(3901,-811){\makebox(0,0)[rb]{\smash{{\SetFigFont{11}{13.2}{\familydefault}{\mddefault}{\updefault}{\color[rgb]{0,0,0}$c=0.3$}%
}}}}
\put(7126,-811){\makebox(0,0)[rb]{\smash{{\SetFigFont{11}{13.2}{\familydefault}{\mddefault}{\updefault}{\color[rgb]{0,0,0}$c=0.9$}%
}}}}
\put(3901,-3886){\makebox(0,0)[rb]{\smash{{\SetFigFont{11}{13.2}{\familydefault}{\mddefault}{\updefault}{\color[rgb]{0,0,0}$c=0.3$}%
}}}}
\put(7126,-3886){\makebox(0,0)[rb]{\smash{{\SetFigFont{11}{13.2}{\familydefault}{\mddefault}{\updefault}{\color[rgb]{0,0,0}$c=0.9$}%
}}}}
\put(3901,-4111){\makebox(0,0)[rb]{\smash{{\SetFigFont{11}{13.2}{\familydefault}{\mddefault}{\updefault}{\color[rgb]{0,0,0}$N=1000$}%
}}}}
\put(7126,-4111){\makebox(0,0)[rb]{\smash{{\SetFigFont{11}{13.2}{\familydefault}{\mddefault}{\updefault}{\color[rgb]{0,0,0}$N=1000$}%
}}}}
\end{picture}%

%% file: Appel.pstex_t
\begin{picture}(0,0)%
\includegraphics{Appel.pstex}%
\end{picture}%
\setlength{\unitlength}{2526sp}%
\begingroup\makeatletter\ifx\SetFigFont\undefined%
\gdef\SetFigFont#1#2#3#4#5{%
  \reset@font\fontsize{#1}{#2pt}%
  \fontfamily{#3}\fontseries{#4}\fontshape{#5}%
  \selectfont}%
\fi\endgroup%
\begin{picture}(8397,12254)(-187,-11215)
\put(1901,904){\makebox(0,0)[b]{\smash{{\SetFigFont{8}{9.6}{\familydefault}{\mddefault}{\updefault}{\color[rgb]{0,0,0}5}%
}}}}
\put(3651,-886){\makebox(0,0)[lb]{\smash{{\SetFigFont{8}{9.6}{\familydefault}{\mddefault}{\updefault}{\color[rgb]{0,0,0}3}%
}}}}
\put(3326,-361){\makebox(0,0)[b]{\smash{{\SetFigFont{8}{9.6}{\familydefault}{\mddefault}{\updefault}{\color[rgb]{0,0,0}$-1$}%
}}}}
\put(151,-886){\makebox(0,0)[rb]{\smash{{\SetFigFont{8}{9.6}{\familydefault}{\mddefault}{\updefault}{\color[rgb]{0,0,0}$-3$}%
}}}}
\put(2471,389){\makebox(0,0)[b]{\smash{{\SetFigFont{8}{9.6}{\familydefault}{\mddefault}{\updefault}{\color[rgb]{0,0,0}0}%
}}}}
\put(776,114){\makebox(0,0)[b]{\smash{{\SetFigFont{8}{9.6}{\familydefault}{\mddefault}{\updefault}{\color[rgb]{0,0,0}1}%
}}}}
\put(501,-296){\makebox(0,0)[b]{\smash{{\SetFigFont{8}{9.6}{\familydefault}{\mddefault}{\updefault}{\color[rgb]{0,0,0}2}%
}}}}
\put(451,-1346){\makebox(0,0)[b]{\smash{{\SetFigFont{8}{9.6}{\familydefault}{\mddefault}{\updefault}{\color[rgb]{0,0,0}$-1$}%
}}}}
\put(1336,-1711){\makebox(0,0)[b]{\smash{{\SetFigFont{8}{9.6}{\familydefault}{\mddefault}{\updefault}{\color[rgb]{0,0,0}0}%
}}}}
\put(3101,-1346){\makebox(0,0)[b]{\smash{{\SetFigFont{8}{9.6}{\familydefault}{\mddefault}{\updefault}{\color[rgb]{0,0,0}2}%
}}}}
\put(2801,-1711){\makebox(0,0)[b]{\smash{{\SetFigFont{8}{9.6}{\familydefault}{\mddefault}{\updefault}{\color[rgb]{0,0,0}1}%
}}}}
\put(6226,-2636){\makebox(0,0)[b]{\smash{{\SetFigFont{8}{9.6}{\familydefault}{\mddefault}{\updefault}{\color[rgb]{0,0,0}$-1$}%
}}}}
\put(6276,904){\makebox(0,0)[b]{\smash{{\SetFigFont{8}{9.6}{\familydefault}{\mddefault}{\updefault}{\color[rgb]{0,0,0}4}%
}}}}
\put(1831,-2636){\makebox(0,0)[b]{\smash{{\SetFigFont{8}{9.6}{\familydefault}{\mddefault}{\updefault}{\color[rgb]{0,0,0}$-5$}%
}}}}
\put(5301,339){\makebox(0,0)[b]{\smash{{\SetFigFont{8}{9.6}{\familydefault}{\mddefault}{\updefault}{\color[rgb]{0,0,0}1}%
}}}}
\put(6701,339){\makebox(0,0)[b]{\smash{{\SetFigFont{8}{9.6}{\familydefault}{\mddefault}{\updefault}{\color[rgb]{0,0,0}0}%
}}}}
\put(8026,-1861){\makebox(0,0)[lb]{\smash{{\SetFigFont{8}{9.6}{\familydefault}{\mddefault}{\updefault}{\color[rgb]{0,0,0}4}%
}}}}
\put(4526,-1861){\makebox(0,0)[rb]{\smash{{\SetFigFont{8}{9.6}{\familydefault}{\mddefault}{\updefault}{\color[rgb]{0,0,0}$-4$}%
}}}}
\put(151,-9086){\makebox(0,0)[rb]{\smash{{\SetFigFont{8}{9.6}{\familydefault}{\mddefault}{\updefault}{\color[rgb]{0,0,0}$-2$}%
}}}}
\put(1826,-10886){\makebox(0,0)[b]{\smash{{\SetFigFont{8}{9.6}{\familydefault}{\mddefault}{\updefault}{\color[rgb]{0,0,0}$-10$}%
}}}}
\put(401,-8861){\makebox(0,0)[b]{\smash{{\SetFigFont{8}{9.6}{\familydefault}{\mddefault}{\updefault}{\color[rgb]{0,0,0}2}%
}}}}
\put(701,-8161){\makebox(0,0)[b]{\smash{{\SetFigFont{8}{9.6}{\familydefault}{\mddefault}{\updefault}{\color[rgb]{0,0,0}0.5}%
}}}}
\put(401,-8561){\makebox(0,0)[b]{\smash{{\SetFigFont{8}{9.6}{\familydefault}{\mddefault}{\updefault}{\color[rgb]{0,0,0}1}%
}}}}
\put(3251,-9536){\makebox(0,0)[b]{\smash{{\SetFigFont{8}{9.6}{\familydefault}{\mddefault}{\updefault}{\color[rgb]{0,0,0}1}%
}}}}
\put(3251,-9286){\makebox(0,0)[b]{\smash{{\SetFigFont{8}{9.6}{\familydefault}{\mddefault}{\updefault}{\color[rgb]{0,0,0}2}%
}}}}
\put(8026,-9086){\makebox(0,0)[lb]{\smash{{\SetFigFont{8}{9.6}{\familydefault}{\mddefault}{\updefault}{\color[rgb]{0,0,0}10}%
}}}}
\put(4526,-9086){\makebox(0,0)[rb]{\smash{{\SetFigFont{8}{9.6}{\familydefault}{\mddefault}{\updefault}{\color[rgb]{0,0,0}$-10$}%
}}}}
\put(5751,-7836){\makebox(0,0)[b]{\smash{{\SetFigFont{8}{9.6}{\familydefault}{\mddefault}{\updefault}{\color[rgb]{0,0,0}1}%
}}}}
\put(6226,-10886){\makebox(0,0)[b]{\smash{{\SetFigFont{8}{9.6}{\familydefault}{\mddefault}{\updefault}{\color[rgb]{0,0,0}$-10$}%
}}}}
\put(1901,-7311){\makebox(0,0)[b]{\smash{{\SetFigFont{8}{9.6}{\familydefault}{\mddefault}{\updefault}{\color[rgb]{0,0,0}10}%
}}}}
\put(1901,-6936){\makebox(0,0)[b]{\smash{{\SetFigFont{8}{9.6}{\familydefault}{\mddefault}{\updefault}{\color[rgb]{0,0,0}(c)}%
}}}}
\put(6401,-7836){\makebox(0,0)[b]{\smash{{\SetFigFont{8}{9.6}{\familydefault}{\mddefault}{\updefault}{\color[rgb]{0,0,0}0.5}%
}}}}
\put(7601,-7836){\makebox(0,0)[b]{\smash{{\SetFigFont{8}{9.6}{\familydefault}{\mddefault}{\updefault}{\color[rgb]{0,0,0}$-0.5$}%
}}}}
\put(3251,-8861){\makebox(0,0)[b]{\smash{{\SetFigFont{8}{9.6}{\familydefault}{\mddefault}{\updefault}{\color[rgb]{0,0,0}$-0.5$}%
}}}}
\put(401,-9311){\makebox(0,0)[b]{\smash{{\SetFigFont{8}{9.6}{\familydefault}{\mddefault}{\updefault}{\color[rgb]{0,0,0}$-0.5$}%
}}}}
\put(6276,-2861){\makebox(0,0)[b]{\smash{{\SetFigFont{8}{9.6}{\familydefault}{\mddefault}{\updefault}{\color[rgb]{0,0,0}(b)}%
}}}}
\put(7701,-3986){\makebox(0,0)[b]{\smash{{\SetFigFont{8}{9.6}{\familydefault}{\mddefault}{\updefault}{\color[rgb]{0,0,0}$-1$}%
}}}}
\put(6051,-3636){\makebox(0,0)[b]{\smash{{\SetFigFont{8}{9.6}{\familydefault}{\mddefault}{\updefault}{\color[rgb]{0,0,0}1}%
}}}}
\put(7051,-3636){\makebox(0,0)[b]{\smash{{\SetFigFont{8}{9.6}{\familydefault}{\mddefault}{\updefault}{\color[rgb]{0,0,0}0}%
}}}}
\put(5126,-3636){\makebox(0,0)[b]{\smash{{\SetFigFont{8}{9.6}{\familydefault}{\mddefault}{\updefault}{\color[rgb]{0,0,0}2}%
}}}}
\put(4701,-4286){\makebox(0,0)[b]{\smash{{\SetFigFont{8}{9.6}{\familydefault}{\mddefault}{\updefault}{\color[rgb]{0,0,0}3}%
}}}}
\put(6051,-6686){\makebox(0,0)[b]{\smash{{\SetFigFont{8}{9.6}{\familydefault}{\mddefault}{\updefault}{\color[rgb]{0,0,0}$-1$}%
}}}}
\put(7051,-7836){\makebox(0,0)[b]{\smash{{\SetFigFont{8}{9.6}{\familydefault}{\mddefault}{\updefault}{\color[rgb]{0,0,0}0}%
}}}}
\put(6276,-11161){\makebox(0,0)[b]{\smash{{\SetFigFont{8}{9.6}{\familydefault}{\mddefault}{\updefault}{\color[rgb]{0,0,0}(f)}%
}}}}
\put(4526,-5711){\makebox(0,0)[rb]{\smash{{\SetFigFont{8}{9.6}{\familydefault}{\mddefault}{\updefault}{\color[rgb]{0,0,0}$-4$}%
}}}}
\put(8026,-5711){\makebox(0,0)[lb]{\smash{{\SetFigFont{8}{9.6}{\familydefault}{\mddefault}{\updefault}{\color[rgb]{0,0,0}5}%
}}}}
\put(1851,-6686){\makebox(0,0)[b]{\smash{{\SetFigFont{8}{9.6}{\familydefault}{\mddefault}{\updefault}{\color[rgb]{0,0,0}$-1$}%
}}}}
\put(1901,-3136){\makebox(0,0)[b]{\smash{{\SetFigFont{8}{9.6}{\familydefault}{\mddefault}{\updefault}{\color[rgb]{0,0,0}10}%
}}}}
\put(3301,-4111){\makebox(0,0)[b]{\smash{{\SetFigFont{8}{9.6}{\familydefault}{\mddefault}{\updefault}{\color[rgb]{0,0,0}0}%
}}}}
\put(1301,-4111){\makebox(0,0)[b]{\smash{{\SetFigFont{8}{9.6}{\familydefault}{\mddefault}{\updefault}{\color[rgb]{0,0,0}0.5}%
}}}}
\put(626,-5061){\makebox(0,0)[b]{\smash{{\SetFigFont{8}{9.6}{\familydefault}{\mddefault}{\updefault}{\color[rgb]{0,0,0}1}%
}}}}
\put(401,-5461){\makebox(0,0)[b]{\smash{{\SetFigFont{8}{9.6}{\familydefault}{\mddefault}{\updefault}{\color[rgb]{0,0,0}1.5}%
}}}}
\put(401,-5746){\makebox(0,0)[b]{\smash{{\SetFigFont{8}{9.6}{\familydefault}{\mddefault}{\updefault}{\color[rgb]{0,0,0}2}%
}}}}
\put(3301,-5486){\makebox(0,0)[b]{\smash{{\SetFigFont{8}{9.6}{\familydefault}{\mddefault}{\updefault}{\color[rgb]{0,0,0}$-0.5$}%
}}}}
\put(5076,-7836){\makebox(0,0)[b]{\smash{{\SetFigFont{8}{9.6}{\familydefault}{\mddefault}{\updefault}{\color[rgb]{0,0,0}1.5}%
}}}}
\put(7601,-8311){\makebox(0,0)[b]{\smash{{\SetFigFont{8}{9.6}{\familydefault}{\mddefault}{\updefault}{\color[rgb]{0,0,0}$-1$}%
}}}}
\put(5026,-8186){\makebox(0,0)[b]{\smash{{\SetFigFont{8}{9.6}{\familydefault}{\mddefault}{\updefault}{\color[rgb]{0,0,0}2}%
}}}}
\put(1901,-2861){\makebox(0,0)[b]{\smash{{\SetFigFont{8}{9.6}{\familydefault}{\mddefault}{\updefault}{\color[rgb]{0,0,0}(a)}%
}}}}
\put(6276,-6936){\makebox(0,0)[b]{\smash{{\SetFigFont{8}{9.6}{\familydefault}{\mddefault}{\updefault}{\color[rgb]{0,0,0}(d)}%
}}}}
\put(3301,-5786){\makebox(0,0)[b]{\smash{{\SetFigFont{8}{9.6}{\familydefault}{\mddefault}{\updefault}{\color[rgb]{0,0,0}$-1$}%
}}}}
\put(3251,-10061){\makebox(0,0)[b]{\smash{{\SetFigFont{8}{9.6}{\familydefault}{\mddefault}{\updefault}{\color[rgb]{0,0,0}0.5}%
}}}}
\put(701,-10061){\makebox(0,0)[b]{\smash{{\SetFigFont{8}{9.6}{\familydefault}{\mddefault}{\updefault}{\color[rgb]{0,0,0}0}%
}}}}
\put(1901,-11161){\makebox(0,0)[b]{\smash{{\SetFigFont{8}{9.6}{\familydefault}{\mddefault}{\updefault}{\color[rgb]{0,0,0}(e)}%
}}}}
\put(3116,-7811){\makebox(0,0)[b]{\smash{{\SetFigFont{8}{9.6}{\familydefault}{\mddefault}{\updefault}{\color[rgb]{0,0,0}0}%
}}}}
\put(3651,-9086){\makebox(0,0)[lb]{\smash{{\SetFigFont{8}{9.6}{\familydefault}{\mddefault}{\updefault}{\color[rgb]{0,0,0}2}%
}}}}
\put(6276,-7311){\makebox(0,0)[b]{\smash{{\SetFigFont{8}{9.6}{\familydefault}{\mddefault}{\updefault}{\color[rgb]{0,0,0}10}%
}}}}
\put(4851,-9886){\makebox(0,0)[b]{\smash{{\SetFigFont{8}{9.6}{\familydefault}{\mddefault}{\updefault}{\color[rgb]{0,0,0}$-1$}%
}}}}
\put(4876,-436){\makebox(0,0)[b]{\smash{{\SetFigFont{8}{9.6}{\familydefault}{\mddefault}{\updefault}{\color[rgb]{0,0,0}2}%
}}}}
\put(7576,-436){\makebox(0,0)[b]{\smash{{\SetFigFont{8}{9.6}{\familydefault}{\mddefault}{\updefault}{\color[rgb]{0,0,0}$-1$}%
}}}}
\put(4876,-10286){\makebox(0,0)[b]{\smash{{\SetFigFont{8}{9.6}{\familydefault}{\mddefault}{\updefault}{\color[rgb]{0,0,0}$-0.5$}%
}}}}
\put(6151,-10286){\makebox(0,0)[b]{\smash{{\SetFigFont{8}{9.6}{\familydefault}{\mddefault}{\updefault}{\color[rgb]{0,0,0}$0.5$}%
}}}}
\put(6766,-10286){\makebox(0,0)[b]{\smash{{\SetFigFont{8}{9.6}{\familydefault}{\mddefault}{\updefault}{\color[rgb]{0,0,0}1}%
}}}}
\put(7401,-10286){\makebox(0,0)[b]{\smash{{\SetFigFont{8}{9.6}{\familydefault}{\mddefault}{\updefault}{\color[rgb]{0,0,0}1.5}%
}}}}
\put(7401,-9911){\makebox(0,0)[b]{\smash{{\SetFigFont{8}{9.6}{\familydefault}{\mddefault}{\updefault}{\color[rgb]{0,0,0}2}%
}}}}
\put(3651,-6211){\makebox(0,0)[lb]{\smash{{\SetFigFont{8}{9.6}{\familydefault}{\mddefault}{\updefault}{\color[rgb]{0,0,0}3}%
}}}}
\put(151,-6211){\makebox(0,0)[rb]{\smash{{\SetFigFont{8}{9.6}{\familydefault}{\mddefault}{\updefault}{\color[rgb]{0,0,0}$-3$}%
}}}}
\put(5526,-10286){\makebox(0,0)[b]{\smash{{\SetFigFont{8}{9.6}{\familydefault}{\mddefault}{\updefault}{\color[rgb]{0,0,0}0}%
}}}}
\put(6086,-3136){\makebox(0,0)[b]{\smash{{\SetFigFont{8}{9.6}{\familydefault}{\mddefault}{\updefault}{\color[rgb]{0,0,0}3}%
}}}}
\end{picture}%

%% file: spectrum.pstex_t
\begin{picture}(0,0)%
\includegraphics{spectrum.pstex}%
\end{picture}%
\setlength{\unitlength}{2960sp}%
\begingroup\makeatletter\ifx\SetFigFont\undefined%
\gdef\SetFigFont#1#2#3#4#5{%
  \reset@font\fontsize{#1}{#2pt}%
  \fontfamily{#3}\fontseries{#4}\fontshape{#5}%
  \selectfont}%
\fi\endgroup%
\begin{picture}(9192,6305)(1,-6794)
\put(7801,-6736){\makebox(0,0)[b]{\smash{{\SetFigFont{10}{12.0}{\familydefault}{\mddefault}{\updefault}{\color[rgb]{0,0,0}(f)}%
}}}}
\put(4801,-6736){\makebox(0,0)[b]{\smash{{\SetFigFont{10}{12.0}{\familydefault}{\mddefault}{\updefault}{\color[rgb]{0,0,0}(e)}%
}}}}
\put(1801,-6736){\makebox(0,0)[b]{\smash{{\SetFigFont{10}{12.0}{\familydefault}{\mddefault}{\updefault}{\color[rgb]{0,0,0}(d)}%
}}}}
\put(7801,-3511){\makebox(0,0)[b]{\smash{{\SetFigFont{10}{12.0}{\familydefault}{\mddefault}{\updefault}{\color[rgb]{0,0,0}(c)}%
}}}}
\put(1801,-3511){\makebox(0,0)[b]{\smash{{\SetFigFont{10}{12.0}{\familydefault}{\mddefault}{\updefault}{\color[rgb]{0,0,0}(a)}%
}}}}
\put(4801,-3511){\makebox(0,0)[b]{\smash{{\SetFigFont{10}{12.0}{\familydefault}{\mddefault}{\updefault}{\color[rgb]{0,0,0}(b)}%
}}}}
\put(6876,-736){\makebox(0,0)[b]{\smash{{\SetFigFont{10}{12.0}{\familydefault}{\mddefault}{\updefault}{\color[rgb]{0,0,0}Resolvent}%
}}}}
\put(6876,-1011){\makebox(0,0)[b]{\smash{{\SetFigFont{10}{12.0}{\familydefault}{\mddefault}{\updefault}{\color[rgb]{0,0,0}set}%
}}}}
\put(6976,-2961){\makebox(0,0)[b]{\smash{{\SetFigFont{10}{12.0}{\familydefault}{\mddefault}{\updefault}{\color[rgb]{0,0,0}Point}%
}}}}
\put(926,-4636){\makebox(0,0)[b]{\smash{{\SetFigFont{10}{12.0}{\familydefault}{\mddefault}{\updefault}{\color[rgb]{0,0,0}set}%
}}}}
\put(5651,-4361){\makebox(0,0)[b]{\smash{{\SetFigFont{10}{12.0}{\familydefault}{\mddefault}{\updefault}{\color[rgb]{0,0,0}Residual}%
}}}}
\put(5651,-5836){\makebox(0,0)[b]{\smash{{\SetFigFont{10}{12.0}{\familydefault}{\mddefault}{\updefault}{\color[rgb]{0,0,0}spectrum}%
}}}}
\put(5651,-4636){\makebox(0,0)[b]{\smash{{\SetFigFont{10}{12.0}{\familydefault}{\mddefault}{\updefault}{\color[rgb]{0,0,0}spectrum}%
}}}}
\put(6876,-4636){\makebox(0,0)[b]{\smash{{\SetFigFont{10}{12.0}{\familydefault}{\mddefault}{\updefault}{\color[rgb]{0,0,0}spectrum}%
}}}}
\put(6876,-4361){\makebox(0,0)[b]{\smash{{\SetFigFont{10}{12.0}{\familydefault}{\mddefault}{\updefault}{\color[rgb]{0,0,0}Point}%
}}}}
\put(8701,-5836){\makebox(0,0)[b]{\smash{{\SetFigFont{10}{12.0}{\familydefault}{\mddefault}{\updefault}{\color[rgb]{0,0,0}spectrum}%
}}}}
\put(8651,-3236){\makebox(0,0)[b]{\smash{{\SetFigFont{10}{12.0}{\familydefault}{\mddefault}{\updefault}{\color[rgb]{0,0,0}spectrum}%
}}}}
\put(5651,-3236){\makebox(0,0)[b]{\smash{{\SetFigFont{10}{12.0}{\familydefault}{\mddefault}{\updefault}{\color[rgb]{0,0,0}spectrum}%
}}}}
\put(3901,-4636){\makebox(0,0)[b]{\smash{{\SetFigFont{10}{12.0}{\familydefault}{\mddefault}{\updefault}{\color[rgb]{0,0,0}set}%
}}}}
\put(2651,-5836){\makebox(0,0)[b]{\smash{{\SetFigFont{10}{12.0}{\familydefault}{\mddefault}{\updefault}{\color[rgb]{0,0,0}spectrum}%
}}}}
\put(2651,-5561){\makebox(0,0)[b]{\smash{{\SetFigFont{10}{12.0}{\familydefault}{\mddefault}{\updefault}{\color[rgb]{0,0,0}Continuous}%
}}}}
\put(8701,-4636){\makebox(0,0)[b]{\smash{{\SetFigFont{10}{12.0}{\familydefault}{\mddefault}{\updefault}{\color[rgb]{0,0,0}set}%
}}}}
\put(926,-1011){\makebox(0,0)[b]{\smash{{\SetFigFont{10}{12.0}{\familydefault}{\mddefault}{\updefault}{\color[rgb]{0,0,0}set}%
}}}}
\put(3901,-736){\makebox(0,0)[b]{\smash{{\SetFigFont{10}{12.0}{\familydefault}{\mddefault}{\updefault}{\color[rgb]{0,0,0}Resolvent}%
}}}}
\put(3901,-1011){\makebox(0,0)[b]{\smash{{\SetFigFont{10}{12.0}{\familydefault}{\mddefault}{\updefault}{\color[rgb]{0,0,0}set}%
}}}}
\put(2651,-3236){\makebox(0,0)[b]{\smash{{\SetFigFont{10}{12.0}{\familydefault}{\mddefault}{\updefault}{\color[rgb]{0,0,0}spectrum}%
}}}}
\put(2651,-2961){\makebox(0,0)[b]{\smash{{\SetFigFont{10}{12.0}{\familydefault}{\mddefault}{\updefault}{\color[rgb]{0,0,0}Continuous}%
}}}}
\put(3901,-2961){\makebox(0,0)[b]{\smash{{\SetFigFont{10}{12.0}{\familydefault}{\mddefault}{\updefault}{\color[rgb]{0,0,0}Residual}%
}}}}
\put(3901,-3236){\makebox(0,0)[b]{\smash{{\SetFigFont{10}{12.0}{\familydefault}{\mddefault}{\updefault}{\color[rgb]{0,0,0}spectrum}%
}}}}
\put(5651,-2961){\makebox(0,0)[b]{\smash{{\SetFigFont{10}{12.0}{\familydefault}{\mddefault}{\updefault}{\color[rgb]{0,0,0}Continuous}%
}}}}
\put(926,-736){\makebox(0,0)[b]{\smash{{\SetFigFont{10}{12.0}{\familydefault}{\mddefault}{\updefault}{\color[rgb]{0,0,0}Resolvent}%
}}}}
\put(6976,-3236){\makebox(0,0)[b]{\smash{{\SetFigFont{10}{12.0}{\familydefault}{\mddefault}{\updefault}{\color[rgb]{0,0,0}spectrum}%
}}}}
\put(8651,-2961){\makebox(0,0)[b]{\smash{{\SetFigFont{10}{12.0}{\familydefault}{\mddefault}{\updefault}{\color[rgb]{0,0,0}Continuous}%
}}}}
\put(926,-4361){\makebox(0,0)[b]{\smash{{\SetFigFont{10}{12.0}{\familydefault}{\mddefault}{\updefault}{\color[rgb]{0,0,0}Resolvent}%
}}}}
\put(3901,-4361){\makebox(0,0)[b]{\smash{{\SetFigFont{10}{12.0}{\familydefault}{\mddefault}{\updefault}{\color[rgb]{0,0,0}Resolvent}%
}}}}
\put(5651,-5561){\makebox(0,0)[b]{\smash{{\SetFigFont{10}{12.0}{\familydefault}{\mddefault}{\updefault}{\color[rgb]{0,0,0}Continuous}%
}}}}
\put(8701,-4361){\makebox(0,0)[b]{\smash{{\SetFigFont{10}{12.0}{\familydefault}{\mddefault}{\updefault}{\color[rgb]{0,0,0}Resolvent}%
}}}}
\put(8701,-5561){\makebox(0,0)[b]{\smash{{\SetFigFont{10}{12.0}{\familydefault}{\mddefault}{\updefault}{\color[rgb]{0,0,0}Continuous}%
}}}}
\put(  1,-3736){\makebox(0,0)[lb]{\smash{{\SetFigFont{10}{12.0}{\familydefault}{\mddefault}{\updefault}{\color[rgb]{0,0,0}$\lambda\textrm{-plane}$}%
}}}}
\end{picture}%

%% file: Initial-Final.pstex_t
\begin{picture}(0,0)%
\includegraphics{Initial-Final.pstex}%
\end{picture}%
\setlength{\unitlength}{3947sp}%
\begingroup\makeatletter\ifx\SetFigFont\undefined%
\gdef\SetFigFont#1#2#3#4#5{%
  \reset@font\fontsize{#1}{#2pt}%
  \fontfamily{#3}\fontseries{#4}\fontshape{#5}%
  \selectfont}%
\fi\endgroup%
\begin{picture}(4354,2944)(479,-2983)
\put(1226,-361){\makebox(0,0)[b]{\smash{{\SetFigFont{12}{14.4}{\familydefault}{\mddefault}{\updefault}{\color[rgb]{0,0,0}$(X,\mathcal{U})$}%
}}}}
\put(4076,-2436){\makebox(0,0)[b]{\smash{{\SetFigFont{12}{14.4}{\familydefault}{\mddefault}{\updefault}{\color[rgb]{0,0,0}$(Y_1,\mathcal{V}_1)$}%
}}}}
\put(1226,-2436){\makebox(0,0)[b]{\smash{{\SetFigFont{12}{14.4}{\familydefault}{\mddefault}{\updefault}{\color[rgb]{0,0,0}$(X_1,\textrm{FT}\{\mathcal{U};q\})$}%
}}}}
\put(4076,-361){\makebox(0,0)[b]{\smash{{\SetFigFont{12}{14.4}{\familydefault}{\mddefault}{\updefault}{\color[rgb]{0,0,0}$(Y,\mathcal{V})$}%
}}}}
\put(4076,-2811){\makebox(0,0)[b]{\smash{{\SetFigFont{12}{14.4}{\familydefault}{\mddefault}{\updefault}{\color[rgb]{0,0,0}$(Y_1,\textrm{IT}\{e;\mathcal{V}\})$}%
}}}}
\put(2851,-1361){\makebox(0,0)[b]{\smash{{\SetFigFont{12}{14.4}{\familydefault}{\mddefault}{\updefault}{\color[rgb]{0,0,0}$g$}%
}}}}
\put(2376,-1361){\makebox(0,0)[b]{\smash{{\SetFigFont{12}{14.4}{\familydefault}{\mddefault}{\updefault}{\color[rgb]{0,0,0}$h$}%
}}}}
\put(1226,-2811){\makebox(0,0)[b]{\smash{{\SetFigFont{12}{14.4}{\familydefault}{\mddefault}{\updefault}{\color[rgb]{0,0,0}$(X_1,\mathcal{U}_1)$}%
}}}}
\put(2626,-2486){\makebox(0,0)[b]{\smash{{\SetFigFont{12}{14.4}{\familydefault}{\mddefault}{\updefault}{\color[rgb]{0,0,0}$f$}%
}}}}
\put(2626,-2786){\makebox(0,0)[b]{\smash{{\SetFigFont{12}{14.4}{\familydefault}{\mddefault}{\updefault}{\color[rgb]{0,0,0}$f$}%
}}}}
\put(4066,-1426){\makebox(0,0)[b]{\smash{{\SetFigFont{12}{14.4}{\familydefault}{\mddefault}{\updefault}{\color[rgb]{0,0,0}$e$}%
}}}}
\put(1216,-1411){\makebox(0,0)[b]{\smash{{\SetFigFont{12}{14.4}{\familydefault}{\mddefault}{\updefault}{\color[rgb]{0,0,0}$q$}%
}}}}
\end{picture}%

%% file: DerSets.pstex_t
\begin{picture}(0,0)%
\includegraphics{DerSets.pstex}%
\end{picture}%
\setlength{\unitlength}{3947sp}%
\begingroup\makeatletter\ifx\SetFigFont\undefined%
\gdef\SetFigFont#1#2#3#4#5{%
  \reset@font\fontsize{#1}{#2pt}%
  \fontfamily{#3}\fontseries{#4}\fontshape{#5}%
  \selectfont}%
\fi\endgroup%
\begin{picture}(6194,6944)(1704,-6308)
\put(5561, 39){\makebox(0,0)[b]{\smash{{\SetFigFont{11}{13.2}{\familydefault}{\mddefault}{\updefault}{\color[rgb]{0,0,0}2. Selfish (Closed)}%
}}}}
\put(4551,-3961){\makebox(0,0)[rb]{\smash{{\SetFigFont{11}{13.2}{\familydefault}{\mddefault}{\updefault}{\color[rgb]{0,0,0}$X$}%
}}}}
\put(7701,-3961){\makebox(0,0)[rb]{\smash{{\SetFigFont{11}{13.2}{\familydefault}{\mddefault}{\updefault}{\color[rgb]{0,0,0}$X$}%
}}}}
\put(2701,-1036){\makebox(0,0)[b]{\smash{{\SetFigFont{11}{13.2}{\familydefault}{\mddefault}{\updefault}{\color[rgb]{0,0,0}1. Donor}%
}}}}
\put(2701,-2686){\makebox(0,0)[b]{\smash{{\SetFigFont{11}{13.2}{\familydefault}{\mddefault}{\updefault}{\color[rgb]{0,0,0}2. Selfish}%
}}}}
\put(2701,-2911){\makebox(0,0)[b]{\smash{{\SetFigFont{11}{13.2}{\familydefault}{\mddefault}{\updefault}{\color[rgb]{0,0,0}(Closed)}%
}}}}
\put(2701,-4611){\makebox(0,0)[b]{\smash{{\SetFigFont{11}{13.2}{\familydefault}{\mddefault}{\updefault}{\color[rgb]{0,0,0}3. Neutral}%
}}}}
\put(7126, 39){\makebox(0,0)[b]{\smash{{\SetFigFont{11}{13.2}{\familydefault}{\mddefault}{\updefault}{\color[rgb]{0,0,0}3. Neutral}%
}}}}
\put(3976, 39){\makebox(0,0)[b]{\smash{{\SetFigFont{11}{13.2}{\familydefault}{\mddefault}{\updefault}{\color[rgb]{0,0,0}1. Donor}%
}}}}
\put(5551,389){\makebox(0,0)[b]{\smash{{\SetFigFont{11}{13.2}{\familydefault}{\mddefault}{\updefault}{\color[rgb]{0,0,0}$X-A$}%
}}}}
\put(1951,-2836){\makebox(0,0)[b]{\smash{{\SetFigFont{11}{13.2}{\familydefault}{\mddefault}{\updefault}{\color[rgb]{0,0,0}$A$}%
}}}}
\put(6131,-2136){\makebox(0,0)[rb]{\smash{{\SetFigFont{11}{13.2}{\familydefault}{\mddefault}{\updefault}{\color[rgb]{0,0,0}$X-A$ open}%
}}}}
\put(7701,-2136){\makebox(0,0)[rb]{\smash{{\SetFigFont{11}{13.2}{\familydefault}{\mddefault}{\updefault}{\color[rgb]{0,0,0}$X-A$ open}%
}}}}
\put(4551,-361){\makebox(0,0)[rb]{\smash{{\SetFigFont{11}{13.2}{\familydefault}{\mddefault}{\updefault}{\color[rgb]{0,0,0}$X$}%
}}}}
\put(6131,-361){\makebox(0,0)[rb]{\smash{{\SetFigFont{11}{13.2}{\familydefault}{\mddefault}{\updefault}{\color[rgb]{0,0,0}$X$}%
}}}}
\put(6131,-3961){\makebox(0,0)[rb]{\smash{{\SetFigFont{11}{13.2}{\familydefault}{\mddefault}{\updefault}{\color[rgb]{0,0,0}$X$}%
}}}}
\put(7701,-361){\makebox(0,0)[rb]{\smash{{\SetFigFont{11}{13.2}{\familydefault}{\mddefault}{\updefault}{\color[rgb]{0,0,0}$X$}%
}}}}
\put(5026,-5961){\makebox(0,0)[lb]{\smash{{\SetFigFont{11}{13.2}{\familydefault}{\mddefault}{\updefault}{\color[rgb]{0,0,0}Der$(X-A)$}%
}}}}
\put(3476,-5961){\makebox(0,0)[lb]{\smash{{\SetFigFont{11}{13.2}{\familydefault}{\mddefault}{\updefault}{\color[rgb]{0,0,0}Bdy$_{X-A}(A)$}%
}}}}
\put(2326,-5961){\makebox(0,0)[lb]{\smash{{\SetFigFont{11}{13.2}{\familydefault}{\mddefault}{\updefault}{\color[rgb]{0,0,0}Der$(A)$}%
}}}}
\put(6576,-5961){\makebox(0,0)[lb]{\smash{{\SetFigFont{11}{13.2}{\familydefault}{\mddefault}{\updefault}{\color[rgb]{0,0,0}Bdy$_{A}(X-A)$}%
}}}}
\put(6686,-1571){\makebox(0,0)[lb]{\smash{{\SetFigFont{11}{13.2}{\familydefault}{\mddefault}{\updefault}{\color[rgb]{0,0,0}$A$}%
}}}}
\put(6686,-3371){\makebox(0,0)[lb]{\smash{{\SetFigFont{11}{13.2}{\familydefault}{\mddefault}{\updefault}{\color[rgb]{0,0,0}$A$}%
}}}}
\put(5111,-3371){\makebox(0,0)[lb]{\smash{{\SetFigFont{11}{13.2}{\familydefault}{\mddefault}{\updefault}{\color[rgb]{0,0,0}$A$ open}%
}}}}
\put(3536,-3371){\makebox(0,0)[lb]{\smash{{\SetFigFont{11}{13.2}{\familydefault}{\mddefault}{\updefault}{\color[rgb]{0,0,0}$A$}%
}}}}
\put(3536,-5171){\makebox(0,0)[lb]{\smash{{\SetFigFont{11}{13.2}{\familydefault}{\mddefault}{\updefault}{\color[rgb]{0,0,0}$A$}%
}}}}
\put(5111,-5171){\makebox(0,0)[lb]{\smash{{\SetFigFont{11}{13.2}{\familydefault}{\mddefault}{\updefault}{\color[rgb]{0,0,0}$A$ open}%
}}}}
\put(6686,-5171){\makebox(0,0)[lb]{\smash{{\SetFigFont{11}{13.2}{\familydefault}{\mddefault}{\updefault}{\color[rgb]{0,0,0}$A$}%
}}}}
\put(3536,-1571){\makebox(0,0)[lb]{\smash{{\SetFigFont{11}{13.2}{\familydefault}{\mddefault}{\updefault}{\color[rgb]{0,0,0}$A$}%
}}}}
\put(5111,-1571){\makebox(0,0)[lb]{\smash{{\SetFigFont{11}{13.2}{\familydefault}{\mddefault}{\updefault}{\color[rgb]{0,0,0}$A$ open}%
}}}}
\put(4551,-2136){\makebox(0,0)[rb]{\smash{{\SetFigFont{11}{13.2}{\familydefault}{\mddefault}{\updefault}{\color[rgb]{0,0,0}$X-A$ open}%
}}}}
\end{picture}%

%% file: cmpct_clsd.pstex_t
\begin{picture}(0,0)%
\includegraphics{cmpct_clsd.pstex}%
\end{picture}%
\setlength{\unitlength}{1776sp}%
\begingroup\makeatletter\ifx\SetFigFont\undefined%
\gdef\SetFigFont#1#2#3#4#5{%
  \reset@font\fontsize{#1}{#2pt}%
  \fontfamily{#3}\fontseries{#4}\fontshape{#5}%
  \selectfont}%
\fi\endgroup%
\begin{picture}(7164,5445)(176,-5533)
\put(1951,-3961){\makebox(0,0)[lb]{\smash{\SetFigFont{10}{12.0}{\familydefault}{\mddefault}{\updefault}{\color[rgb]{0,0,0}$x$}%
}}}
\put(6001,-5011){\makebox(0,0)[rb]{\smash{\SetFigFont{10}{12.0}{\familydefault}{\mddefault}{\updefault}{\color[rgb]{0,0,0}Hausdorff $X$}%
}}}
\put(1726,-2111){\makebox(0,0)[rb]{\smash{\SetFigFont{10}{12.0}{\familydefault}{\mddefault}{\updefault}{\color[rgb]{0,0,0}$V_{y}$}%
}}}
\put(6751,-2461){\makebox(0,0)[lb]{\smash{\SetFigFont{10}{12.0}{\familydefault}{\mddefault}{\updefault}{\color[rgb]{0,0,0}$U_{y}$}%
}}}
\put(5026,-1811){\makebox(0,0)[rb]{\smash{\SetFigFont{10}{12.0}{\familydefault}{\mddefault}{\updefault}{\color[rgb]{0,0,0}$y$}%
}}}
\put(176,-4886){\makebox(0,0)[rb]{\smash{\SetFigFont{10}{12.0}{\familydefault}{\mddefault}{\updefault}{\color[rgb]{0,0,0}$V$}%
}}}
\put(5401,-2686){\makebox(0,0)[rb]{\smash{\SetFigFont{10}{12.0}{\familydefault}{\mddefault}{\updefault}{\color[rgb]{0,0,0}compact $K$}%
}}}
\end{picture}

%% file: DerSets1.pstex_t
\begin{picture}(0,0)%
\includegraphics{DerSets1.pstex}%
\end{picture}%
\setlength{\unitlength}{3947sp}%
\begingroup\makeatletter\ifx\SetFigFont\undefined%
\gdef\SetFigFont#1#2#3#4#5{%
  \reset@font\fontsize{#1}{#2pt}%
  \fontfamily{#3}\fontseries{#4}\fontshape{#5}%
  \selectfont}%
\fi\endgroup%
\begin{picture}(7019,2546)(279,-8060)
\put(6601,-7486){\makebox(0,0)[b]{\smash{{\SetFigFont{11}{13.2}{\familydefault}{\mddefault}{\updefault}{\color[rgb]{0,0,0}(d) $A$ is Cantor}%
}}}}
\put(4376,-6961){\makebox(0,0)[b]{\smash{{\SetFigFont{11}{13.2}{\familydefault}{\mddefault}{\updefault}{\color[rgb]{0,0,0}$A$}%
}}}}
\put(5251,-5761){\makebox(0,0)[b]{\smash{{\SetFigFont{11}{13.2}{\familydefault}{\mddefault}{\updefault}{\color[rgb]{0,0,0}$X$}%
}}}}
\put(7126,-5761){\makebox(0,0)[b]{\smash{{\SetFigFont{11}{13.2}{\familydefault}{\mddefault}{\updefault}{\color[rgb]{0,0,0}$X$}%
}}}}
\put(6251,-6961){\makebox(0,0)[b]{\smash{{\SetFigFont{11}{13.2}{\familydefault}{\mddefault}{\updefault}{\color[rgb]{0,0,0}$A$}%
}}}}
\put(1501,-5761){\makebox(0,0)[b]{\smash{{\SetFigFont{11}{13.2}{\familydefault}{\mddefault}{\updefault}{\color[rgb]{0,0,0}$X$}%
}}}}
\put(3376,-5761){\makebox(0,0)[b]{\smash{{\SetFigFont{11}{13.2}{\familydefault}{\mddefault}{\updefault}{\color[rgb]{0,0,0}$X$}%
}}}}
\put(2826,-7486){\makebox(0,0)[b]{\smash{{\SetFigFont{11}{13.2}{\familydefault}{\mddefault}{\updefault}{\color[rgb]{0,0,0}(b) $A$ is nwd}%
}}}}
\put(626,-6961){\makebox(0,0)[b]{\smash{{\SetFigFont{11}{13.2}{\familydefault}{\mddefault}{\updefault}{\color[rgb]{0,0,0}$A$}%
}}}}
\put(2501,-6961){\makebox(0,0)[b]{\smash{{\SetFigFont{11}{13.2}{\familydefault}{\mddefault}{\updefault}{\color[rgb]{0,0,0}$A$}%
}}}}
\put(4701,-7486){\makebox(0,0)[b]{\smash{{\SetFigFont{11}{13.2}{\familydefault}{\mddefault}{\updefault}{\color[rgb]{0,0,0}(c) Bdy($A$) is nwd}%
}}}}
\put(4616,-8011){\makebox(0,0)[lb]{\smash{{\SetFigFont{11}{13.2}{\familydefault}{\mddefault}{\updefault}{\color[rgb]{0,0,0}= Der$(\mathscr{C}(A))$}%
}}}}
\put(4701,-7761){\makebox(0,0)[b]{\smash{{\SetFigFont{11}{13.2}{\familydefault}{\mddefault}{\updefault}{\color[rgb]{0,0,0}Der($A$) = Bdy($A$)}%
}}}}
\put(951,-7761){\makebox(0,0)[b]{\smash{{\SetFigFont{11}{13.2}{\familydefault}{\mddefault}{\updefault}{\color[rgb]{0,0,0}Der$(A)=\emptyset$}%
}}}}
\put(951,-7486){\makebox(0,0)[b]{\smash{{\SetFigFont{11}{13.2}{\familydefault}{\mddefault}{\updefault}{\color[rgb]{0,0,0}(a) $A$ is isolated}%
}}}}
\put(2826,-7761){\makebox(0,0)[b]{\smash{{\SetFigFont{11}{13.2}{\familydefault}{\mddefault}{\updefault}{\color[rgb]{0,0,0}Bdy$(\mathscr{C}(\mathrm{Cl}(A)))=\mathrm{Cl}(A)$}%
}}}}
\put(951,-8011){\makebox(0,0)[b]{\smash{{\SetFigFont{11}{13.2}{\familydefault}{\mddefault}{\updefault}{\color[rgb]{0,0,0}Der$(\mathscr{C}(A))\subseteq \mathscr{C}(A)$}%
}}}}
\put(6421,-8011){\makebox(0,0)[lb]{\smash{{\SetFigFont{11}{13.2}{\familydefault}{\mddefault}{\updefault}{\color[rgb]{0,0,0}= Bdy$(\mathscr{C}(A))$}%
}}}}
\put(6751,-7761){\makebox(0,0)[rb]{\smash{{\SetFigFont{11}{13.2}{\familydefault}{\mddefault}{\updefault}{\color[rgb]{0,0,0}Der$(A)=A$}%
}}}}
\put(2641,-8011){\makebox(0,0)[lb]{\smash{{\SetFigFont{11}{13.2}{\familydefault}{\mddefault}{\updefault}{\color[rgb]{0,0,0}= Bdy$(\mathrm{Cl}(A))$}%
}}}}
\end{picture}%

%% file: Cantor.pstex_t
\begin{picture}(0,0)%
\includegraphics{Cantor.pstex}%
\end{picture}%
\setlength{\unitlength}{3552sp}%
\begingroup\makeatletter\ifx\SetFigFont\undefined%
\gdef\SetFigFont#1#2#3#4#5{%
  \reset@font\fontsize{#1}{#2pt}%
  \fontfamily{#3}\fontseries{#4}\fontshape{#5}%
  \selectfont}%
\fi\endgroup%
\begin{picture}(5985,2799)(435,-2269)
\put(5851,189){\makebox(0,0)[lb]{\smash{{\SetFigFont{12}{14.4}{\familydefault}{\mddefault}{\updefault}{\color[rgb]{0,0,0}$C_0$}%
}}}}
\put(4076,-76){\makebox(0,0)[b]{\smash{{\SetFigFont{12}{14.4}{\familydefault}{\mddefault}{\updefault}{\color[rgb]{0,0,0}$x_{3}^{(1)}$}%
}}}}
\put(5851,-1161){\makebox(0,0)[lb]{\smash{{\SetFigFont{12}{14.4}{\familydefault}{\mddefault}{\updefault}{\color[rgb]{0,0,0}$C_3$}%
}}}}
\put(5851,-711){\makebox(0,0)[lb]{\smash{{\SetFigFont{12}{14.4}{\familydefault}{\mddefault}{\updefault}{\color[rgb]{0,0,0}$C_2$}%
}}}}
\put(5851,-1611){\makebox(0,0)[lb]{\smash{{\SetFigFont{12}{14.4}{\familydefault}{\mddefault}{\updefault}{\color[rgb]{0,0,0}$C_4$}%
}}}}
\put(5851,-261){\makebox(0,0)[lb]{\smash{{\SetFigFont{12}{14.4}{\familydefault}{\mddefault}{\updefault}{\color[rgb]{0,0,0}$C_1$}%
}}}}
\put(5576,374){\makebox(0,0)[b]{\smash{{\SetFigFont{12}{14.4}{\familydefault}{\mddefault}{\updefault}{\color[rgb]{0,0,0}$x_{2}^{(0)}$}%
}}}}
\put(1601,-531){\makebox(0,0)[b]{\smash{{\SetFigFont{12}{14.4}{\familydefault}{\mddefault}{\updefault}{\color[rgb]{0,0,0}$x_{2}^{(2)}$}%
}}}}
\put(5076,-531){\makebox(0,0)[b]{\smash{{\SetFigFont{12}{14.4}{\familydefault}{\mddefault}{\updefault}{\color[rgb]{0,0,0}$x_{7}^{(2)}$}%
}}}}
\put(4576,-531){\makebox(0,0)[b]{\smash{{\SetFigFont{12}{14.4}{\familydefault}{\mddefault}{\updefault}{\color[rgb]{0,0,0}$x_{6}^{(2)}$}%
}}}}
\put(1051,374){\makebox(0,0)[b]{\smash{{\SetFigFont{12}{14.4}{\familydefault}{\mddefault}{\updefault}{\color[rgb]{0,0,0}$x_{1}^{(0)}$}%
}}}}
\put(5851,-2211){\makebox(0,0)[lb]{\smash{{\SetFigFont{12}{14.4}{\familydefault}{\mddefault}{\updefault}{\color[rgb]{0,0,0}$\mathcal{C}$}%
}}}}
\put(2626,-76){\makebox(0,0)[b]{\smash{{\SetFigFont{12}{14.4}{\familydefault}{\mddefault}{\updefault}{\color[rgb]{0,0,0}$x_{2}^{(1)}$}%
}}}}
\put(1051,-61){\makebox(0,0)[b]{\smash{{\SetFigFont{12}{14.4}{\familydefault}{\mddefault}{\updefault}{\color[rgb]{0,0,0}$x_{1}^{(1)}$}%
}}}}
\put(2626,-511){\makebox(0,0)[b]{\smash{{\SetFigFont{12}{14.4}{\familydefault}{\mddefault}{\updefault}{\color[rgb]{0,0,0}$x_{4}^{(2)}$}%
}}}}
\put(4076,-511){\makebox(0,0)[b]{\smash{{\SetFigFont{12}{14.4}{\familydefault}{\mddefault}{\updefault}{\color[rgb]{0,0,0}$x_{5}^{(2)}$}%
}}}}
\put(5576,-511){\makebox(0,0)[b]{\smash{{\SetFigFont{12}{14.4}{\familydefault}{\mddefault}{\updefault}{\color[rgb]{0,0,0}$x_{8}^{(2)}$}%
}}}}
\put(5576,-976){\makebox(0,0)[b]{\smash{{\SetFigFont{12}{14.4}{\familydefault}{\mddefault}{\updefault}{\color[rgb]{0,0,0}$x_{16}^{(3)}$}%
}}}}
\put(2076,-531){\makebox(0,0)[b]{\smash{{\SetFigFont{12}{14.4}{\familydefault}{\mddefault}{\updefault}{\color[rgb]{0,0,0}$x_{3}^{(2)}$}%
}}}}
\put(1051,-511){\makebox(0,0)[b]{\smash{{\SetFigFont{12}{14.4}{\familydefault}{\mddefault}{\updefault}{\color[rgb]{0,0,0}$x_{1}^{(2)}$}%
}}}}
\put(1616,-976){\makebox(0,0)[b]{\smash{{\SetFigFont{12}{14.4}{\familydefault}{\mddefault}{\updefault}{\color[rgb]{0,0,0}$x_{4}^{(3)}$}%
}}}}
\put(2051,-976){\makebox(0,0)[b]{\smash{{\SetFigFont{12}{14.4}{\familydefault}{\mddefault}{\updefault}{\color[rgb]{0,0,0}$x_{5}^{(3)}$}%
}}}}
\put(4601,-976){\makebox(0,0)[b]{\smash{{\SetFigFont{12}{14.4}{\familydefault}{\mddefault}{\updefault}{\color[rgb]{0,0,0}$x_{12}^{(3)}$}%
}}}}
\put(1051,-976){\makebox(0,0)[b]{\smash{{\SetFigFont{12}{14.4}{\familydefault}{\mddefault}{\updefault}{\color[rgb]{0,0,0}$x_{1}^{(3)}$}%
}}}}
\put(5101,-976){\makebox(0,0)[b]{\smash{{\SetFigFont{12}{14.4}{\familydefault}{\mddefault}{\updefault}{\color[rgb]{0,0,0}$x_{13}^{(3)}$}%
}}}}
\put(5576,-61){\makebox(0,0)[b]{\smash{{\SetFigFont{12}{14.4}{\familydefault}{\mddefault}{\updefault}{\color[rgb]{0,0,0}$x_{4}^{(1)}$}%
}}}}
\put(2626,-976){\makebox(0,0)[b]{\smash{{\SetFigFont{12}{14.4}{\familydefault}{\mddefault}{\updefault}{\color[rgb]{0,0,0}$x_{8}^{(3)}$}%
}}}}
\put(4076,-976){\makebox(0,0)[b]{\smash{{\SetFigFont{12}{14.4}{\familydefault}{\mddefault}{\updefault}{\color[rgb]{0,0,0}$x_{9}^{(3)}$}%
}}}}
\end{picture}%